\documentclass{JHEP3}
\usepackage{amsmath,amssymb}
\usepackage[dvips]{graphicx}
\usepackage{epsfig}
\usepackage{psfrag}
\usepackage{array}
\title{Boundary state of multiple D0-branes and closed string coupling}
\author{Yoshinao Sato\\
Institute of Physics, University of Tokyo\\
Komaba, Meguro-ku, Tokyo 153-8902, Japan
\\E-mail: \email{ysato@hep1.c.u-tokyo.ac.jp}}

\preprint{UT-Komaba/07-3}
 
\abstract{In this paper we study a boundary state of multiple D0-branes with spacetime dependent scalar fields in the $\alpha'$ expansion. We calculate a formulas for non-Abelian boundary state defined by using a Wilson loop factor and path ordering. The boundary state contains divergences which vanish when the scalar fields on D0-branes satisfy the equation of motion. Hence the boundary state is well-defined if the configuration of D0-branes in on-shell. We can show the constructed boundary state is BRST invariant. From the constructed boundary state we extract couplings of the scalar fields to closed string fields. Our results reproduce the correct formulas for supergravity current distribution obtained from the disk amplitudes, Matrix theory potential, non-Abelian DBI action. Our calculations are performed up to order ${\alpha'}^2$ both in the bosonic string and the IIA superstring theory. Furthermore we confirm that our boundary state is identical to the correct one in the case of a single boosted D0-brane and a noncommutative D2-brane. These results support the correctness of the formulas for non-Abelian boundary states.}

\begin{document}
\section{Introduction}\label{paper:section:1}
In this paper we consider a boundary state of multiple D0-branes of an arbitrary spacetime dependent configuration in the $\alpha'$ expansion. We calculate a formula for non-Abelian boundary states defined by using a Wilson loop factor and path ordering \cite{NPB288525, NPB29383, NPB308221}. The boundary state is BRST invariant and includes no singularity when the scalar fields on D0-branes satisfy the equation of motion as in the case of Abelian boundary state \cite{9909027,9909095}. Hence the boundary state is well-defined if the configuration of D0-branes is on-shell. From the constructed boundary state we can extract the correct formulas for supergravity current distribution derived from the disk amplitude \cite{0103124}, Matrix theory potentail \cite{9711078,9712185,9904095}, and non-Abelian DBI action \cite{9910053}.  Furthermore we confirm that our boundary state coincides with the correct one in the cases of a single boosted D0-brane \cite{9701190} and a noncommutative D2-branes \cite{9906214}. These results support the correctness of the formula for non-Abelian boundary states with the Wilson loop factor and the path ordering prescription. 

%%%%%%%

In the case that there exist multiple D-branes, we cannot define definite positions of D-branes in general. This can be seen from that gauge and scalar fields on multiple D-branes are non-Abelian and hence not necessarily commutative. Such a noncommutivity causes an interesting phenomenon, formulation of higher dimensional D-branes. However, we face the ordering problem dealing with multiple D-branes. It is known that symmetrized trace for a non-Abelian DBI action is correct only up to order ${\alpha'}^4$ \cite{9701125,9703217,0002180}. 
%For a non-Abelian DBI action, symmetrized trace has been proposed \cite{9701125}. At order ${\alpha'}^6$, however, the non-Abelian DBI action with symmetrized trace is not able to reproduce the correct spectrum of open strings obtained by the worldsheet analysis \cite{9703217,0002180}. 
Less is known about a non-Abelian boundary state. It is discussed in \cite{NPB308221} that a contribution of non-Abelian open string fields is incorporated into a boundary state by including the Wilson loop factor and taking trace with path ordering. 

A boundary state is a closed string description of D-branes, and includes information about closed strings emitted from D-branes \cite{NPB288525,NPB29383,NPB308221,NPB431131} (see also \cite{9912161,9912275} and references therein). From a boundary state we can extract couplings of opens strings on D-branes to closed strings, a potential between D-branes, and long distance behavior of closed string fields \cite{9707068}. We can take account of an effect of open strings on closed string emission from a D-brane. In other words, we can incorporate the open string background fields into a boundary states of a D-brane. In the case that there exist multiple D-branes, we cannot determine definite positions of D-branes because non-Abelian fields on them are noncommutative in general. Such fuzziness of D-brane worldvolume makes unclear what a boundary condition is imposed on open strings, or the boundary state in terms of closed strings. Furthermore shape of worldsheets representing closed string emission becomes complicated due to the existence of off-diagonal components in opens string fields, or open strings stretched between D-branes. Therefore it is not easy to investigate a non-Abelian boundary state.
%a interesting phenomenon, formulation of a higher dimensional D-brane, and the ordering problem. 
%For a non-Abelian DBI action, symmetrized trace has been proposed \cite{9701125}. At order ${\alpha'}^6$, however, the non-Abelian DBI action with symmetrized trace is not able to reproduce the correct spectrum of open strings obtained by the worldsheet analysis \cite{9703217,0002180}. Less is known about non-Abelian boundary states.
It is discussed in \cite{NPB308221} that a closed string state, which is given by including a Wilson loop factor into the boundary state of a single D-brane and taking trace with path ordering, reproduces the correct disk amplitudes with an arbitrary closed string vertex and the D-brane boundary.  This formula for non-Abelian boundary states represents that effects of non-Abelian fields on closed string emission from D-branes are accounted for by including the Wilson loop factor and using path ordering prescription as explained in appendix \ref{paper:part:review}.
%. However, we do not have rigorous proof of the correctness of this formulas.

In this paper we investigate a boundary state of multiple D0-branes in the $\alpha'$ expansion by using the formula presented in \cite{NPB308221} to confirm its correctness. We focus on  multiple D0-branes with an arbitrary time-dependent configuration of scalar fields. Our boundary state is BRST invariant for an arbitrary configuration of D0-branes formally. However, it contains divergences which vanish when the configuration of D0-branes is on-shell. Hence the boundary state is well-defined when the scalar field on D0-branes satisfied the equation of motion. In other words we can derive the correct equation of motion from the boundary state by requiring its finiteness. We derive couplings to closed strings from the constructed boundary state. Our results in the DKPS limit \cite{9608024,9903165} reproduce the correct non-Abelian DBI action and the correct formulas for supergravity charge density \cite{9711078,9712185,9904095,0103124,9910053}. The constructed boundary state is identical to the already known one in the cases of a single boosted D0-brane and a noncommutative D2-brane. These results show the correctness of path ordering in non-Abelian boundary states. We perform calculations up to order ${\alpha'}^2$ both in the bosonic string and the type IIA superstring theory. In the case of IIA superstring theory we need to introduce a worldsheet fermion. Further calculations enable us to derive higher $\alpha'$ corrections to non-Abelian DBI action. 
We focus on multiple D0-branes in this paper, since characteristic behavior caused by noncommutivity of open string fields is incorporated. Extension to higher dimensional D-branes can be done straightforwardly by introducing a non-Abelian gauge field. The study in this paper is non-Abelian extension of \cite{9909027,9909095}. We note that backreaction from emitted closed strings to D-branes, or $g_s$ corrections are ignored in this paper. In other words, we regard D-branes as infinitely massive objects. 

We investigate a non-Abelian boundary state in the operator formalism, or equivalently in terms of creation and annihilation operators of closed strings in this paper. This enable to us to calculate the general boundary state in $\alpha'$ expansion. In the case that multiple D-branes formulate a single higher dimensional flat D-brane with a constant magnetic flux, we can use the path integral formalism \cite{9804163}. However, in presence of nontrivial excitation of open strings, it is difficult to calculate the boundary state in such a formalism. Furthermore it is difficult to take use of such an approach in the general D-brane system. A method to construct D0-brane matrices from higher dimensional D-brane in a constant background B-field is studies in \cite{0501086}. 

The outline of this paper is as follows. 
%In section \ref{paper:part:3} we investigate a boundary state of multiple D-branes. 
%In chapter \ref{paper:part:review} we review a boundary state of a single D-brane and consider non-Abelian extension of boundary states. In chapter \ref{paper:part:5} we summarize our study on non-Abelian boundary states in this paper and discuss future directions. 
In section \ref{paper:section:2}, we briefly review boundary states and couplings to closed strings. In section \ref{paper:section:3} we calculate the boundary state of multiple D0-branes and show our results of couplings to closed strings.  The results of closed string couplings extracted the boundary state is shown in \eqref{boundary:eq:main-result-coupling-bosonic}, \eqref{boundary:eq:main-result-coupling-bosonic2} for the bosonic string theory, and in \eqref{boundary:eq:main-result-coupling-super} for the type IIA superstring theory. In section \ref{paper:section:4}, we consider two nontrivial cases: a single boosted D0-brane, and a noncommutative D2-brane. In section \ref{paper:section:5} we summarize our study on non-Abelian boundary states in this paper and discuss future directions. In appendix \ref{boundary:appendix:a}, the results of calculations of creation operators operating on a D0-brane boundary state which are used to calculate the boundary state with scalar fields are shown. In appendix \ref{boundary:appendix:b}, detailed calculations in the proof of BRST invariance of the boundary state are shown. In appendix \ref{paper:section:emtensor}, we show the couplings of multiple D0-branes to massless closed strings derived from disk amplitudes \cite{0103124}, Matrix theory potential \cite{9711078,9712185,9904095}, and non-Abelian DBI action \cite{9910053}. In appendix \ref{paper:part:review} the reasons why we think a non-Abelian boundary state is given by using a Wilson-loop factor and path ordering.

%%%%%%%%%%%%%%%%%%%%%%%%%%%%%%%%%%%%%%%%%%%

\section{Review of boundary state and closed string coupling}\label{paper:section:2}
In this section we review basics of a boundary state of a single D-brane and couplings to closed strings in subsection \ref{boundary:subsection:2-1}, and explain how to extract couplings to closed strings from a boundary state in \ref{boundary:subsection:2-2}. 

\subsection{Boundary state of a single D$p$-brane}\label{boundary:subsection:2-1}
Firstly a bosonic boundary state is reviewed \cite{NPB288525, NPB29383, NPB308221, 9912161, 9912275}. We expand a string embedding function in terms of creation and annihilation operators as
%\begin{equation*}
%  \hat{X}(\tau,\sigma)=\hat{x}^\mu + \alpha' \tau  g^{\mu\nu} \hat{p}_\nu + i\sqrt{\frac{\alpha'}{2}}\sum_{n\ne0}\left(\frac{\alpha^\mu_{n}}{n}e^{-i(\tau-\sigma)}+\frac{\tilde{\alpha}^\mu_{n}}{n}e^{-i(\tau+\sigma)}\right).
%\end{equation*}
\begin{equation*}
  \hat{X}(z,\bar{z})=\hat{x}^\mu - i \frac{\alpha'}{2} \hat{p}^\mu \log |z|^2 + i\sqrt{\frac{\alpha'}{2}}\sum_{n\ne0}\frac{1}{n}\left(\frac{\alpha^\mu_{n}}{z^n}+\frac{\tilde{\alpha}^\mu_{n}}{\bar{z}^n}\right).
\end{equation*}
We also expand ghost fields in mode operators as
%\begin{alignat*}{4}
%    b(\tau,\sigma)&=i^{-2}\sum_{n} b_n e^{-i n(\tau-\sigma)},& \quad &  \tilde{b}(\tau,\sigma)&= i ^2 \sum_{n} \tilde{b}_n e^{-i n(\tau+\sigma)}\\
%    c(\tau,\sigma)&=i \sum_{n} c_n e^{-i n(\tau-\sigma)},& \quad &  \tilde{c}(\tau,\sigma)&=i^{-1} \sum_{n} \tilde{c}_n e^{-i n(\tau+\sigma)}.
%\end{alignat*}
\begin{alignat*}{4}
    b(z)&=\sum_{n} \frac{b_n}{z^{n+2}},& \quad &  \tilde{b}(\bar{z})&= \sum_{n} \frac{\tilde{b}_n}{\bar{z}^{n+2}} \\
    c(z)&=\sum_{n} \frac{c_n}{z^{n-1}},& \quad &  \tilde{c}(\bar{z})&= \sum_{n} \frac{\tilde{c}_n}{\bar{z}^{n-1}}.
\end{alignat*}
The mode operators satisfy (anti)commutation relations
\begin{align*}
   [\alpha_m^\mu,\alpha_n^\nu]&=[\tilde{\alpha}_m^\mu,\tilde{\alpha}_n^\nu]=mg^{\mu\nu}\delta_{m+n,0}\\
   \{b_m,c_n\}&=\{\tilde{b}_m,\tilde{c}_n\}=\delta_{m+n,0}.
\end{align*} 
It is convenient to introduce $b_0^\pm, c_0^\pm$
\begin{alignat*}{3}
   b_0^+&=b_0+\tilde{b}_0 & \quad & b_0^- =\frac{1}{2}(b_0-\tilde{b}_0)\\
   c_0^+&=\frac{1}{2}(c_0+\tilde{c}_0)& \quad & c_0^-=c_0-\tilde{c}_0
\end{alignat*}
which satisfy
\begin{equation*}
   \{b_0^\pm,c_0^\pm\}=1.
\end{equation*}
$\mathrm{SL}(2,\mathbb{C})$-vacuum $|0\rangle$ is defined by
\begin{align*}
   \hat{p}|0\rangle& =0 \\
   \alpha_n^\mu|0\rangle &= \tilde{\alpha}_n^\mu|0\rangle =0 \quad n \ge 1\\
   b_n|0\rangle &= \tilde{b}_n|0\rangle =0 \quad n\ge -1\\
   c_n|0\rangle &= \tilde{c}_n|0\rangle =0 \quad n\ge 2.
\end{align*}
We also denote $|0\rangle=|p=0\rangle$ in some cases. Eigenstates of $\hat{x},\hat{p}$ are related by the Fourier transformation each other, explicitly
\begin{equation*}
   |x\rangle=\int\!\!\frac{d^Dp}{(2\pi)^D}\ e^{ipx}|p\rangle\ , \quad
   |p\rangle=\int\!\!d^Dx\ e^{-ipx}|x\rangle 
\end{equation*}
where $D$ is the critical dimension. $|x\rangle, |p\rangle$ can be represented by
\begin{alignat*}{2}
   |x\rangle&=e^{i\hat{p}x\phantom{-}}|x=0\rangle=&\delta(\hat{x}-x)|p=0\rangle\\
   |p\rangle&=e^{-ip\hat{x}}|p=0\rangle=&(2\pi)^D\delta(\hat{p}-p)|x=0\rangle.
\end{alignat*}
We use the normalization such that
\begin{align*}
   \langle x'|x \rangle &= \delta^D(x'-x)\\
   \langle p'|p \rangle &= (2\pi)^D\delta^D(p'-p)
\end{align*}
and
\begin{equation*}
   \langle p' | \tilde{c}_{-1}c_{-1}c_0^-c_0^+c_1\tilde{c}_1|p\rangle = (2\pi)^D\delta^D(p'-p).
\end{equation*}
A boundary state of a single D$p$-brane is given by
\begin{align*}
 |Dp\rangle & = |Dp\rangle_{\alpha} |B\rangle_\mathrm{gh} \\
   |Dp\rangle_\alpha & = \frac{T_p}{2}\delta^D(\hat{x}^i-\xi^i)\exp\left\{-\sum_{n>0}\frac{1}{n}\alpha^\mu_{-n} S_{\mu\nu} \tilde{\alpha}_{-n}^\nu \right\}|0\rangle \\
   |B\rangle_\mathrm{gh} &= \exp\left\{-\sum_{n>0}(\tilde{b}_{-n}c_{-n}+b_{-n}\tilde{c}_{-n})\right\}c_0^{+}c_1\tilde{c_1}|0\rangle
\end{align*}
where $ S_{\mu\nu}  = g (\eta_{ab}, -\delta_{ij})$, and $T_p$ is tension of a D$p$-brane. $a=0,1,\cdots,p$ represents Neumann directions and $i=p+1,\cdots ,D-1$ indicates Dirichlet directions. As is easily seen from the equation above, a boundary state of a D-brane takes the form of a coherent state of closed strings. The boundary state satisfies boundary conditions below.
\begin{align*}
   \hat{p}^a|Dp\rangle&=0, \quad \hat{x}^i|Dp\rangle = \xi^i\\
   (\alpha_n^a+\tilde{\alpha}_{-n}^a)|Dp\rangle &=0, \quad (\alpha_n^i-\tilde{\alpha}_n^i)|Dp\rangle=0 \quad n \ne 0
\end{align*}
\begin{align*}
   (b_n-\tilde{b}_{-n})|Dp\rangle&=0\\
   (c_n+\tilde{c}_{-n})|Dp\rangle&=0.
\end{align*}

Secondly, we explain boundary states in superstring theories. We focus on the NS sector, because we are interested in couplings to NS-NS fields in this paper. A worldsheet fermion can be expanded as
\begin{align*}
  \psi^\mu(z)&=\sum_{r\in \mathbb{Z}+\frac{1}{2}}\frac{\psi_r^\mu}{z^{r+1/2}}\\
   \tilde{\psi}^\mu(\bar{z})&=\sum_{r\in \mathbb{Z}+\frac{1}{2}}\frac{\tilde{\psi}_r^\mu}{\bar{z}^{r+1/2}}.
\end{align*} 
%\begin{align*}
%   \psi^\mu(\tau,\sigma)&=i^{-1/2}\sum_{r\in \mathbb{Z}+\frac{1}{2}}\psi_r^\mu e^{-ir(\tau-\sigma)}\\
%   \tilde{\psi}^\mu(\tau,\sigma)&=i^{1/2\phantom{-}}\sum_{r\in \mathbb{Z}+\frac{1}{2}}\tilde{\psi}_r^\mu e^{-ir(\tau+\sigma)}
%\end{align*} 
In addition we write down superghost fields in mode expansion as
%\begin{alignat*}{4}
%    \beta(\tau,\sigma)&=i^{-3/2} \sum_{r\in\mathbb{Z}+\frac{1}{2}} \beta_r e^{-ir(\tau-\sigma)},& \quad &  \tilde{\beta}(\tau,\sigma)&=i^{3/2} \sum_{r\in\mathbb{Z}+\frac{1}{2}} \tilde{\beta}_r e^{-i r(\tau+\sigma)}\\
%    \gamma(\tau,\sigma)&=i^{1/2} \sum_{r\in\mathbb{Z}+\frac{1}{2}} \gamma_r e^{-i r(\tau-\sigma)},& \quad &  \tilde{\gamma}(\tau,\sigma)&=i^{-1/2} \sum_{r\in\mathbb{Z}+\frac{1}{2}} \tilde{\gamma}_r e^{-i r(\tau+\sigma)}.
%\end{alignat*}
\begin{alignat*}{4}
    \beta(z)&= \sum_{r\in\mathbb{Z}+\frac{1}{2}} \frac{\beta_r}{z^{r+3/2}},& \quad &  \tilde{\beta}(\bar{z})&=\sum_{r\in\mathbb{Z}+\frac{1}{2}} \frac{\tilde{\beta}_r}{\bar{z}^{r+3/2}}\\
    \gamma(\tau,\sigma)&=\sum_{r\in\mathbb{Z}+\frac{1}{2}} \frac{\gamma_r}{z^{r-1/2}},& \quad &  \tilde{\gamma}(\bar{z})&=\sum_{r\in\mathbb{Z}+\frac{1}{2}} \frac{\tilde{\gamma}_r}{\bar{z}^{r-1/2}}.
\end{alignat*}
Mode operators satisfy (anti)commutation relations
\begin{align*}
  \{\psi^\mu_r,\psi^\nu_s\}&=  \{\tilde{\psi}^\mu_r,\tilde{\psi}^\nu_s\}=  g^{\mu\nu}\delta_{r+s,0}\\
  [\beta_r,\gamma_s]&=[\tilde{\beta}_r,\tilde{\gamma}_s]=\delta_{r+s,0}.
\end{align*}
The boundary state of a D$p$-brane takes the form of
\begin{equation*}
    |Dp\rangle = |Dp\rangle_\alpha|B\rangle_\mathrm{gh}|B\rangle_\mathrm{NS}
\end{equation*}
where $|Dp\rangle_\alpha, |B\rangle_\mathrm{gh}$ is same as that of the bosonic boundary state. $|Dp\rangle_\psi, |B\rangle_\mathrm{sgh}$ can be expressed as
\begin{align*}
    |B\rangle_\mathrm{NS}&=\frac{1}{2}\left(|Dp,+\rangle_\psi|B,+\rangle_\mathrm{sgh}-|Dp,-\rangle_\psi|B,-\rangle_\mathrm{sgh} \right)\\
    |Dp,\eta\rangle_\psi &=-i\exp\left\{\sum_{r=1/2}^\infty i\eta  \psi^\mu_{-r} S_{\mu\nu} \tilde{\psi}^\nu_{-r}\right\}|0\rangle \\
    |B,\eta\rangle_\mathrm{sgh}&=\exp\left\{\sum_{r=1/2}^\infty i\eta (\gamma_{-r}\tilde{\beta}_{-r}-\beta_{-r}\tilde{\gamma}_{-r})\right\}e^{-\phi(0)}e^{-\tilde{\phi}(0)}|0\rangle
\end{align*}
where $\eta=\pm 1$ and $\phi(z)$ is related to the superghosts by bosonization
\begin{alignat*}{4}
   \beta(z)&=e^{-\phi(z)}\partial \xi(z),& \quad & \tilde{\beta}(\bar{z})&=e^{-\tilde{\phi}(\bar{z})}\bar{\partial} \tilde{\xi}(\bar{z}) \\
   \gamma(z)&=e^{\phi(z)}\eta(z),& \quad & \tilde{\gamma}(\bar{z})&=e^{\tilde{\phi}(\bar{z})}\tilde{\eta}(\bar{z}).
\end{alignat*}
The boundary conditions can be written in terms of mode operators as
\begin{align*}
    (\psi_r^\mu-i\eta S^\mu_\nu \tilde{\psi}_r^\nu)|Dp,\eta\rangle_\psi & = 0\\
    (\beta_r+i\eta\tilde{\beta}_{-r})|B,\eta\rangle_\mathrm{sgh} &= 0 \\
    (\gamma_r+i\eta\tilde{\gamma}_{-r})|B,\eta\rangle_\mathrm{sgh} &= 0. 
\end{align*}
$\mathrm{SL}(2,\mathbb{C})$-vacuum is defined by
\begin{align*}
    \psi^\mu_{r}|0\rangle& =\tilde{\psi}^\mu_{r}|0\rangle =0 \quad r \ge 1/2 \\
    \beta_{r}|0\rangle & = \tilde{\beta}_{r}|0\rangle =0 \quad r \ge -1/2 \\
    \gamma_{r}|0\rangle & = \tilde{\gamma}_{r}|0\rangle =0 \quad r \ge 3/2. 
\end{align*}

\subsection{Coupling to closed string}\label{boundary:subsection:2-2}
A boundary state $|B\rangle$ couples to the closed string field
\begin{equation*}
\begin{split}
   |\Phi\rangle = \int\!\!\frac{d^Dk}{(2\pi)^D} & \Bigl\{T(k)+\frac{1}{2}(h_{\mu\nu}(k)+b_{\mu\nu}(k))\alpha^\mu_{-1}\tilde{\alpha}^\nu_{-1}\\
&+\left(-\phi(k)+\frac{1}{4}g^{\mu\nu}h_{\mu\nu}(k)\right)(\tilde{c}_{-1}b_{-1}+c_{-1}\tilde{b}_{-1})+\cdots\Bigr\}c_1\tilde{c}_1|k\rangle
\end{split}
\end{equation*}
through a source term
\begin{equation*}
   \langle \Phi | c_0^- | B\rangle.
\end{equation*}
In this way a boundary state acts as a source for the closed strings \cite{9704125}. $T(k),h_{\mu\nu}(k)=h_{\nu\mu}(k), b_{\mu\nu}=-b_{\nu\mu}(k)$ and $\phi(k)$ are the Fourier transform of closed string tachyon, graviton, antisymmetric tensor, and dilaton fields. Note that we use the convention for the Fourier transformation such that
\begin{equation*}
   f(x)=\int\!\!\frac{d^Dk}{(2\pi)^D}\ f(k) e^{ikx}, \quad f(k)=\int\!\!\ d^Dx\ f(x) e^{-ikx}. 
\end{equation*}  
Then we can write
\begin{equation*}
\begin{split} 
   f(\hat{x})|k=0\rangle = & \int\!\! d^Dx f(x)|x\rangle \\
 = & \int\!\!\frac{d^Dk}{(2\pi)^D}\ f(k)|k\rangle  = f(\hat{k})|x=0\rangle.
\end{split}
\end{equation*}
Suppose that a boundary state takes the form of
\begin{equation}
\begin{split}
   |B\rangle=\int\!\!d^{p+1}x\int\!\!\frac{d^{D-p-1}k}{(2\pi)^{D-p-1}}& \Bigl\{F(x^a,k^i)+\left(A_{\mu\nu}(x^a,k^i)+C_{\mu\nu}(x^a,k^i)\right)\alpha_{-1}^\mu\tilde{\alpha}_{-1}^\nu \\
   & +B(x^a,k^i)(b_{-1}\tilde{c}_{-1}+\tilde{b}_{-1}c_{-1})+\cdots\Bigr\}c_0^{+}c_1\tilde{c_1}|x^a,k^i\rangle. \label{boundary:eq:form}
\end{split}
\end{equation}
where $A_{\mu\nu(k)}=A_{\nu\mu}(k), B_{\mu\nu}(k)=-B_{\nu\mu}(k)$. In the presence of a D$p$-brane, we need to add a source term  
\begin{equation}
\begin{split}
S_\mathrm{source}=\langle \Phi |c_0^- | B\rangle=&\\
\int\!\!d^{p+1}x\int\!\!\frac{d^{D-p-1}k}{(2\pi)^{D-p-1}}\Bigl\{ & T(x^a,-k^i)F(x^a,k^i) + \frac{1}{2}h_{\mu\nu}(x^a,-k^i)\left(A^{\mu\nu}(x^a,k^i)+B(x^a,k^i)g^{\mu\nu}\right)\\
&+\frac{1}{2}b_{\mu\nu}(x^a,-k^i)C^{\mu\nu}(x^a,k^i)-2\phi(x^a,-k^i)B(x^a,k^i)+\cdots\Bigr\}.  \label{boundary:eq:coupling}\end{split}
\end{equation}
We can extract from this an energy momentum tensor as follows.
\begin{equation*}
   T^{\mu\nu}(x^a,k^i)=\left(A^{\mu\nu}(x^a,k^i)+B(x^a,k^i)g^{\mu\nu}\right).
\end{equation*}

\psfrag{mathbf{A}^a,mathbf{X}^i}{$\mathrm{A}^a,\mathrm{X}^i$}
\psfrag{phi,h^{munu},b^{munu}}{$\phi,h^{\mu\nu},b^{\mu\nu}$}
\EPSFIGURE[ht]{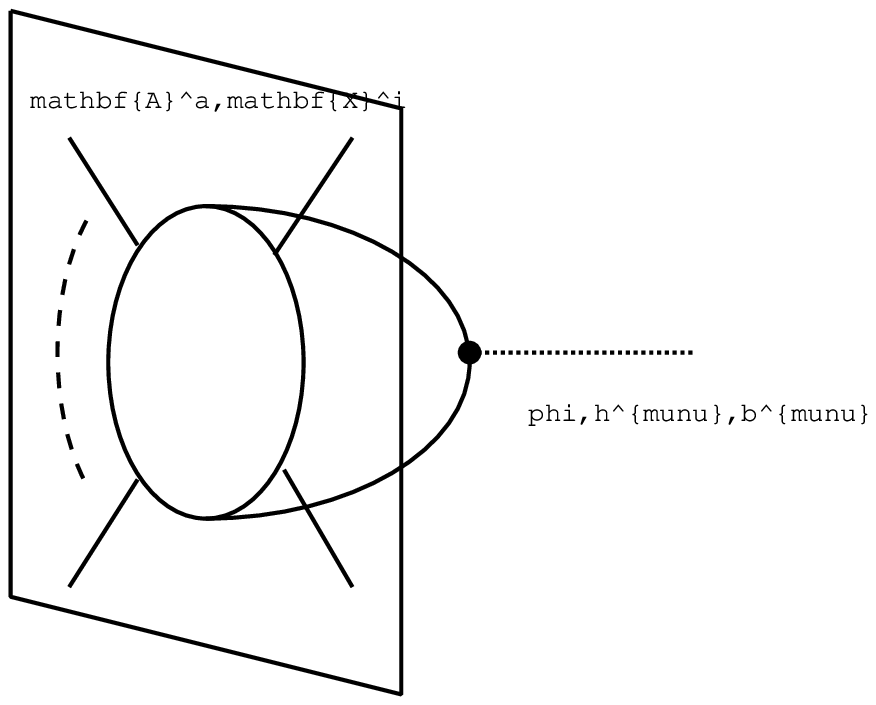,width=0.6\textwidth}{Emission of closed string from D-brane}

\section{Boundary state of multiple D0-branes}\label{paper:section:3}
A boundary state of a single D-brane with an arbitrary time-dependent configuration of gauge and scalar fields on it has been investigated \cite{9909027,9909095}. In contrast, boundary states of multiple D-branes are not well understood\footnote{An equation of motion of Abelian gauge field is derived in \cite{0312260} using an another definition of boundary state.}. In this section we construct a boundary state of multiple D0-branes of an arbitrary time-dependent configuration by using path ordering in the $\alpha'$ expansion. We derive linear couplings to closed string fields up to $\mathcal{O}({\alpha'}^2)$. Our results reproduce the linear part in closed string fields of the non-Abelian DBI action which are derived from scattering amplitudes of string theories. In addition the extracted linear coupling is same as the one derived from disk amplitude \cite{0103124}, long distance Matrix theory potential \cite{9711078,9712185,9904095}, and non-Abelian DBI action \cite{9910053}. The constructed boundary state is BRST invariant for arbitrary configurations formally, however, includes divergences which remain after the zeta function regularization. These singularities vanish when the scalar fields on D0-branes satisfy the equation of motion. Hence the boundary states is well-defined and BRST invariant when the configuration of D0-branes is on-shell. We consider the bosonic string theory first, and then the type IIA superstring theory by. In this chapter we focus on multiple D0-branes, since the essential behavior of noncommutivity is incorporated. By introducing a non-Abelian gauge filed, extension to higher dimensional D-branes is done straightforwardly.

\subsection{Construction of boundary state}\label{boundary:subsection:3-1}
A boundary state of D$p$-brane with gauge and scalar fields $\mathbf{A}^a(\hat{X}^a),\mathbf{X}^i(\hat{X}^a)$ is of the form
\begin{equation}
 |Dp[\mathbf{A}^a,\mathbf{X}^i]\rangle=\mathrm{trP}\exp\left[ i \int_0^{2\pi} d\sigma \left(\partial_\sigma \hat{X}^a(\sigma) \mathbf{A}_a(\hat{X}^a(\sigma)) + \mathbf{X}^i ({\hat{X}^a(\sigma)})\hat{\Pi}_i(\sigma) \right)\right]|Dp\rangle . \label{paper:eq:pdim-nonabelian-boundarystate}
\end{equation}
according to and \cite{NPB308221} and \eqref{review:eq:nonabelian-boundarystate} in appendix \ref{paper:part:review}. Here $(a,i)$ is Neumann and Dirichlet directions respectively. $\mathrm{P}$ means path ordering of open string fields $\mathbf{A}^a(\hat{X}^a(\sigma)),\mathbf{X}^i(\hat{X}^a(\sigma))$ with respect to $\sigma$. $\hat{\Pi}^i(\sigma)$ is the conjugate momentum to the embedding function $\hat{X}^i(\tau,\sigma)$ at the boundary $\tau=0$ along the Dirichlet directions, and $\hat{X}^a(\sigma)$ is the embedding function at the boundary $\tau=0$ along the Neumann directions. In general, open string fields on D-branes are functions of string embedding at the boundary $\tau=0$ along Neumann directions. We do not need to take care of ordering of closed string operators because $\hat{X}^a(\sigma), \partial_\sigma\hat{X}^a(\sigma), \hat{\Pi}_i(\sigma)$ are commutative each other. Even if open string fields are independent of $\hat{X}^a(\sigma)$, we keep to assigning $\sigma$ to $\mathbf{A}^a, \mathbf{X}^i$ implicitly to determine the ordering. In this section we denote operators on closed string state by hatted symbols like $\hat{X}^\mu, \hat{\Pi}^\mu$, and matrix-valued scalar fields on D0-branes by boldface symbols like $\mathbf{A}^a, \mathbf{X}^i$. More strictly, a matrix configuration of D0-brane $\mathbf{X}^i$ is related to the a transverse scalar field $\Phi^i$ through a relation
\begin{equation*}
     \mathbf{X}^i = 2\pi\alpha' \mathbf{\Phi}^i .
\end{equation*}
Take care that matrices $\mathbf{X}^i$ and operators $\hat{X}^mu(\tau,\sigma)$ are different quantities. In remaining all we will not use $X^\mu(\tau,\sigma)$ without explicit attention to avoid confusing. 

It is worthwhile to see a boundary state as superposition of closed string states. A boundary state of a D-brane without open string excitation satisfied 
\begin{equation*}
      \hat{\Pi}^a(\sigma)|Dp\rangle = 0, \quad \hat{X}^i(\sigma)|Dp\rangle=0. 
\end{equation*}
Thus $|Dp\rangle$ can be rewritten as
\begin{equation*}
     |Dp\rangle = |\Pi^a(\sigma)=0,X^i(\sigma)=0\rangle = \int\!\!\mathcal{D}X^a(\sigma)|X^a(\sigma),X^i(\sigma)=0\rangle.
\end{equation*}
This represents that a boundary state of a D-brane is superposition of closed strings of various shapes on the D-brane. Here $|X^a(\sigma)\rangle$ and $|\Pi^i(\sigma)\rangle $is the eigenstate of the embedding function $\hat{X}^a(\sigma)$ and the conjugate momentum $\hat{\Pi}^i(\sigma)$ respectively:
\begin{equation*}
    \hat{X}^a(\sigma)|X^a(\sigma)\rangle = X^a(\sigma)|X^a(\sigma)\rangle, \quad
    \hat{\Pi}^i(\sigma)|\Pi^i(\sigma)\rangle = \Pi^i(\sigma)|\Pi^i(\sigma)\rangle.
\end{equation*}
In presence of nontrivial excitation of opens strings on a D-brane, the boundary state is given by \eqref{boundary:eq:boundary}. Therefore we can say that the Wilson loop factor
\begin{equation*}
  W[\hat{X}^\mu(\sigma)]=\mathrm{trP}\exp\left[ i \int_0^{2\pi} d\sigma \left(\partial_\sigma \hat{X}^a(\sigma) \mathbf{A}_a(\hat{X}^a(\sigma)) + \mathbf{X}^i ({\hat{X}^a(\sigma)})\hat{\Pi}_i(\sigma) \right)\right]
\end{equation*}  
gives a weight function for closed string states to incorporate the effects of $\mathbf{A}^a,\mathbf{X}^i$ into the boundary state. In other words, an open string background modifies closed string emission from D-branes.

In this section we consider the boundary state
\begin{equation}
   |B\rangle=\mathrm{trP}\exp\left[ i \int_0^{2\pi} d\sigma \mathbf{X}^i ({\hat{X}^0(\sigma)})\hat{\Pi}_i(\sigma) \right]|D0\rangle \quad (i=1,\cdots, D-1). \label{boundary:eq:boundary}
\end{equation}
%Here $\mathrm{P}$ means path ordering with respect to $\sigma$, $|D0\rangle$ is a boundary state of D0-brane, and $\hat{\Pi}^i(\sigma)$ is a conjugate momentum of string embedding $\hat{X}^i(\tau,\sigma)$ at the boundary $\tau=0$. $\hat{X}^0(\sigma)$ is string embedding along the time direction. In general, open string fields on D-brane are functions of string embedding at the boundary $\tau=0$ along Neumann directions. 
$\hat{X}^\mu(\tau,\sigma)$ and $\hat{\Pi}^\mu(\tau,\sigma)$ can be written by using creation and annihilation operators as
\begin{align*}
   \hat{X}^\mu(\sigma,\tau)&=\hat{x}^\mu+g^{\mu\nu}\alpha'\hat{p}_\nu\tau+i\sqrt{\frac{\alpha'}{2}}\sum_{n\ne 0}\left(\frac{\alpha_n^\mu}{n}e^{-in(\tau-\sigma)}+\frac{\tilde{\alpha}_n^\mu}{n}e^{-in(\tau+\sigma)}\right)\\
   \hat{\Pi}_\mu(\sigma,\tau)&=\frac{1}{2\pi}\hat{p}_\mu+\frac{g_{\mu\nu}}{2\pi\sqrt{2\alpha'}}\sum_{n\ne 0}\left(\alpha_n^\nu e^{-in(\tau-\sigma)}+\tilde{\alpha}_n^\nu e^{-in(\tau+\sigma)}\right)
\end{align*}
where $g_{\mu\nu}=g\eta_{\mu\nu}$ and $g^{\mu\nu}=\frac{1}{g}\eta^{\mu\nu}$. Especially at the boundary $\tau=0$ we have
\begin{equation}
\begin{split}
   \hat{X}^\mu(\sigma)&=\hat{x}^\mu+i\sqrt{\frac{\alpha'}{2}}\sum_{n\ne 0}\frac{1}{n}\left(\alpha_n^\mu-\tilde{\alpha}^\mu_{-n}\right)e^{in\sigma}\\
   \hat{\Pi}_\mu(\sigma) &=\frac{1}{2\pi}\hat{p}_\mu+\frac{g_{\mu\nu}}{2\pi\sqrt{2\alpha'}}\sum_{n\ne 0}(\alpha_n^\nu+\tilde{\alpha}_{-n}^\nu)e^{in\sigma}.
\end{split}\label{paper:section3:expansion-1}
\end{equation}
$|\mathrm{D0}\rangle$ can be written as
\begin{equation*}
   |\mathrm{D0}\rangle=\frac{T_0}{2}\delta^{d-1}(\hat{x})\exp\left[\sum_{n>0}\frac{g}{n}(\alpha^0_{-n}\tilde\alpha^0_{-n}+\alpha^i_{-n}\tilde{\alpha}^i_{-n})\right]|0\rangle .
\end{equation*}
where $g$ is defined by $g_{\mu\nu}=g\eta_{\mu\nu}$. For diagonal matrices $\mathbf{X}^i=\mathrm{diag}(\xi_1^i,\cdots,\xi_N^i)$, the boundary state \eqref{boundary:eq:boundary} gives a summation of boundary states of D0-branes located at $x^i=\xi_a^i\ (a=1,\cdots,N)$, where the matrix size $N$ represents the number of D0-branes. In presence of off-diagonal components in $\mathbf{X}^i$, however, these matrices becomes noncommutative in general and cannot be interpreted as positions of D0-branes. It is important to clarify a correct ordering prescription of $\mathbf{X}^i$.

\paragraph*{Our strategy}~\\
In this section we investigate boundary states in the operator formalism. This enable to us to expand the general boundary state in powers of $\alpha'$. Our strategy to calculate the boundary state is as follows. First we represent $\hat{X}^0(\sigma),\hat{\Pi}^i(\sigma)$ in terms of creation and annihilation operators as shown in \eqref{paper:section3:expansion-1}. Then we expand the operator $e^{i\int\!\!d\sigma \mathbf{X}\hat{\Pi}}$ in $\alpha'$ expansion taking the DKPS limit \eqref{boundary:eq:DKPS}. Move all annihilation operators to right using the commutation relation of creation and annihilation operators. By utilizing the boundary conditions \eqref{boundary:eq:d0-boundarycondition} on $|D0\rangle$, we replace all annihilation operators by creation operators. We perform the integral which causes in path ordering prescription if possible. After all the resulting boundary state consists of creation operators and $\mathbf{X}^i(t), \dot{\mathbf{X}}^i,\cdots$. The results are presented in subsection \ref{boundary:subsection:3-result}. Readers who are interested only in the results can skip to the subsection \ref{boundary:subsection:3-result}.
\\ \\

We begin to calculate \eqref{boundary:eq:boundary} in the operator formalism. The scalar field $\mathbf{X}^i$ depends on the worldsheet field $\hat{X}^0(\sigma)$ at the boundary. In order to calculate the boundary state in terms of oscillators, we divide $\hat{X}^0(\sigma)$ into the zero mode part and the oscillator part:
\begin{equation*}
\begin{split}
    \hat{X}^0(\sigma)&=t+\tilde{X}^0(\sigma)\\
  \tilde{X}^0(\sigma)&=i\sqrt{\frac{\alpha'}{2}}\sum_{n\ne0}\frac{1}{n}(\alpha_n^0-\tilde{\alpha}_{-n}^0)e^{in\sigma}.  
\end{split}
\end{equation*}
where $t=\hat{x}^0$. The scalar field $\mathbf{X}^i(\hat{X}^0(\sigma))$ can be expanded in the oscillator part $\tilde{X}^0(\sigma)$ as
\begin{equation}
\begin{split}
     \mathbf{X}^i(\hat{X}^0(\sigma))&=\sum_{k=0}^\infty \frac{d \mathbf{X}^i(t)}{dt^n}\left(\tilde{X}^0(\sigma)\right)^k\\
&=\mathbf{X}^i(t)+\dot{\mathbf{X}}^i(t)i\sqrt{\frac{\alpha'}{2}}\sum_{n\ne0}\frac{1}{n}(\alpha_n^0-\tilde{\alpha}_{-n}^0)e^{in\sigma}+ \cdots.
\end{split}\label{paper:section3:expansion-2}
\end{equation}
By substituting \eqref{paper:section3:expansion-1} and \eqref{paper:section3:expansion-1}, $U=i\int_0^{2\pi}d\sigma \mathbf{X}^i(\hat{X}^0(\sigma))\hat{\Pi}_i(\sigma)$ can be written in terms of creation and annihilation operators as
\begin{equation*}
\begin{split} 
U=&i\int_0^1 d\sigma \left(\hat{p}_i+\frac{g\delta_{ij}}{\sqrt{2\alpha'}}\sum_{n \ne 0} (\alpha^j_n+\tilde{\alpha}^j_{-n})e^{2\pi in\sigma}\right)\\
&\left(\mathbf{X}^i(t)+\dot{\mathbf{X}}^i(t)\ i\sqrt{\frac{\alpha'}{2}}\sum_{m\ne 0}\frac{1}{m}(\alpha^0_{m_1}-\tilde{\alpha}^0_{-m_1})e^{2\pi i m \sigma}\right.\\
&\phantom{(\mathbf{X}^i(t)\ \ }+\frac{1}{2!}\ddot{\mathbf{X}}^i(t)\left(i\sqrt{\frac{\alpha'}{2}}\right)^2\sum_{m_1, m_2 \ne 0}\frac{1}{m_1m_2}(\alpha^0_{m_1}-\tilde{\alpha}^0_{-m_1})(\alpha^0_{m_2}-\tilde{\alpha}^0_{-m_2})e^{2\pi i (m_1 \sigma + m_2 \sigma)}\\
&\phantom{(\mathbf{X}^i(t)\ \ }+\cdots\Biggr).
\end{split}
\end{equation*}
Here we have changed variables as $\sigma \to 2\pi \sigma$. This can be rewritten as
\begin{equation}
\begin{split}
%&i\int_0^{2\pi}d\sigma \mathbf{X}^i(\hat{X}^0(\sigma))\hat{\Pi}_i(\sigma) =\\
U&=\int_0^1 d\sigma \Biggl(i\hat{p}\mathbf{X}(t)\\
&+\frac{ig}{\sqrt{2\alpha'}}\mathbf{X}^i(t)\sum_{n \ne 0} (\alpha^i_n+\tilde{\alpha}^i_{-n})e^{2\pi in\sigma}\\
&+\ i\sqrt{\frac{\alpha'}{2}}i\hat{p}\dot{\mathbf{X}}(t)\sum_{m\ne 0}\frac{1}{m}(\alpha^0_{m_1}-\tilde{\alpha}^0_{-m_1})e^{2\pi i m \sigma}\\
&+\frac{ig}{\sqrt{2\alpha'}}\ i\sqrt{\frac{\alpha'}{2}}\dot{\mathbf{X}}^i(t)\sum_{n \ne 0} (\alpha^i_n+\tilde{\alpha}^i_{-n})e^{2\pi in\sigma}\sum_{m\ne 0}\frac{1}{m}(\alpha^0_{m_1}-\tilde{\alpha}^0_{-m_1})e^{2\pi i m \sigma}\\
&+\left(i\sqrt{\frac{\alpha'}{2}}\right)^2\frac{1}{2!}i\hat{p}\ddot{\mathbf{X}}(t)\sum_{m_1, m_2 \ne 0}\frac{1}{m_1m_2}(\alpha^0_{m_1}-\tilde{\alpha}^0_{-m_1})(\alpha^0_{m_2}-\tilde{\alpha}^0_{-m_2})e^{2\pi i (m_1 \sigma + m_2 \sigma)}\\
&+\cdots\Biggr).
\end{split}\label{paper:section3:expansion-3}
\end{equation}
In preparation to expand the boundary state in $\alpha'$, we consider $\mathrm{P}e^{\int_0^1 d\sigma (A_\sigma+B(\sigma))}$. Here $\mathrm{P}$ means path ordering with respect to $\sigma$. Suppose that $[A_\sigma, B(\sigma)] \ne0$, and $A_\sigma $ does not depend on $\sigma$ as function. Nevertheless the subscript $\sigma$ in $A_\sigma$ is needed in order to determined the ordering. Then $\mathrm{P}e^{\int_0^1 d\sigma (A_\sigma+B(\sigma))}$ can be expanded in powers of $B(\sigma)$ as
\begin{equation} 
\begin{split}
\mathrm{P}e^{\int_0^1 d\sigma (A+B(\sigma))}
=e^{A}
&+\int_0^1\!\! d\sigma_1\ e^{(1-\sigma_1)A}B(\sigma_1)e^{\sigma_1 A}\\
&+ \int_0^1\!\! d\sigma_1 \int_0^{\sigma_1}\!\!\!\! d\sigma_2\ e^{(1-\sigma_1)A}B(\sigma_1)e^{(\sigma_1-\sigma_2)A}B(\sigma_2)e^{\sigma_2 A}+\cdots.
\end{split}\label{paper:section3:expansion-4}
\end{equation}
In our case we can identify $A=i\hat{p}_i\mathbf{X}^i(t), B=\mbox{(the other terms)}$.

Using \eqref{paper:section3:expansion-3} and \eqref{paper:section3:expansion-4}, the boundary state \eqref{boundary:eq:boundary} can be expanded as 
\begin{equation*}
\begin{split}
 |B\rangle&=\left\{\ \mathrm{tr}\left[e^{i\hat{p}_i\mathbf{X}^i}\right] \right. \\
&+\frac{ig}{\sqrt{2\alpha'}} \sum_{n_1 \ne 0} \int_0^1\!\! d\sigma_1\ \mathrm{tr}\left[\mathbf{X}^{i_1} e^{i\hat{p}\mathbf{X}}\right](\alpha^{i_1}_{n_1}+\tilde{\alpha}^{i_1}_{-n_1}) e^{2\pi i n_1\sigma_1}\\
&+i\sqrt{\frac{\alpha'}{2}} \sum_{n_1 \ne 0} \int_0^1\!\! d\sigma_1\ \mathrm{tr}\left[i\hat{p}\dot{\mathbf{X}} e^{i\hat{p}\mathbf{X}}\right](\alpha^0_{n_1}+\tilde{\alpha}^0_{-n_1}) e^{2\pi i n_1\sigma_1}\\
&+\left(\frac{ig}{\sqrt{2\alpha'}}\right)^2 \sum_{n_1, n_2 \ne 0} \int_0^1\!\! d\sigma_1 \!\! \int_0^{\sigma_1}\!\!\!\! d\sigma_2\ \mathrm{tr}\left[e^{i(1-\sigma_{12})\hat{p}\mathbf{X}}\mathbf{X}^{i_1}e^{i\sigma_{12}\hat{p}\mathbf{X}}\mathbf{X}^{i_2}\right]\\
& \hspace{0.4\textwidth}(\alpha^i_{n_1}+\tilde{\alpha}^{i_1}_{-n_1})(\alpha^i_{n_2}+\tilde{\alpha}^{i_2}_{-n_2})e^{2\pi i (n_1\sigma_1+n_2\sigma_2)}\\
&+ \cdots \left. \right\}|D0\rangle
\end{split}
\end{equation*}
where we define $\sigma_{pq}=\sigma_{p}-\sigma_{q}$, and take use cyclicity of trace. 

We take the DKPS limit \cite{9608024,9903165}:
\begin{equation}
      \hat{p}_i \sim \mathcal{O}(1/\alpha'), \quad \mathbf{X}^i \sim \mathcal{O}(\alpha'). \label{boundary:eq:DKPS}
\end{equation}
%The latter is equivalent to $\mathbf{\Phi} \sim \mathcal{O}(1)$. %Later we take the Seiberg-Witten limit when we examine a noncommutative D2-brane. 
In this limit  we have
\begin{alignat*}{3}
i\hat{p}\mathbf{X} &\sim \mathcal{O}(1)
&\quad\quad 
& \\
i\sqrt{\frac{\alpha'}{2}}i\hat{p}\dot{\mathbf{X}} &\sim \mathcal{O}({\alpha'}^{1/2}) 
&\quad\quad 
\frac{ig}{\sqrt{2\alpha'}}\mathbf{X}^i &\sim \mathcal{O}({\alpha'}^{1/2}) \\
\frac{1}{2!}\left(i\sqrt{\frac{\alpha'}{2}}\right)^2 i\hat{p}\ddot{\mathbf{X}} &\sim \mathcal{O}(\alpha')
&\quad\quad 
\frac{ig}{\sqrt{2\alpha'}}i\sqrt{\frac{\alpha'}{2}}\dot{\mathbf{X}}^i &\sim \mathcal{O}(\alpha') \\
\frac{1}{3!}\left(i\sqrt{\frac{\alpha'}{2}}\right)^3 i\hat{p}\mathbf{X}^{(3)} &\sim \mathcal{O}({\alpha'}^{3/2})
&\quad\quad 
\frac{1}{2!}\frac{ig}{\sqrt{2\alpha'}}\left(i\sqrt{\frac{\alpha'}{2}}\right)^2\sqrt{\alpha'}\ddot{\mathbf{X}}^i &\sim \mathcal{O}({\alpha'}^{3/2}) \\
\frac{1}{4!}\left(i\sqrt{\frac{\alpha'}{2}}\right)^4 i\hat{p}\mathbf{X}^{(4)} &\sim \mathcal{O}({\alpha'}^2)
&\quad\quad 
\frac{1}{3!}\frac{ig}{\sqrt{2\alpha'}}\left(i\sqrt{\frac{\alpha'}{2}}\right)^3\sqrt{\alpha'}\mathbf{X}^{i(3)} &\sim \mathcal{O}({\alpha'}^2) \\
&\vdots
&\quad\quad
&\vdots.
\end{alignat*}
In this way we can expand the boundary state in $\alpha'$. Table \ref{boundary:table:1} represents all terms which appear at each order in $\alpha'$, where we ignore a factor $e^{i\hat{p}\mathbf{X}}$ and ordering. If plural $i$'s appear in a single term, they are actually different indexes $(i_1,i_2,\cdots)$. We denote them all by the same symbol $i$ for simplicity. For example $\ddot{\mathbf{X}}^i\mathbf{X}^i$ in the table represents that a term $\mathrm{trP'}[\ddot{\mathbf{X}}^{i_1}\mathbf{X}^{i_2}]$ appears in the boundary state. $\mathrm{P'}[\cdots]$ is an appropriately ordered product which is defined in \eqref{boundary:eq:P'} on page \pageref{boundary:page:P'}.
\TABLE[hb]{
\begin{tabular}{|l|m{0.8\textwidth}|}
\hline
$\mathcal{O}({\alpha'}^0)$ &\vspace{0.5ex}  1 \vspace{0.5ex}\\
\hline
$\mathcal{O}({\alpha'}^{1/2})$ &\vspace{0.5ex} $i\hat{p}\dot{\mathbf{X}}\ ,\  \mathbf{X}^i$\vspace{0.5ex}\\
\hline
$\mathcal{O}({\alpha'}^1)$ &\vspace{0.5ex} $(i\hat{p}\dot{\mathbf{X}})^2\ ,\  i\hat{p}\dot{\mathbf{X}}\mathbf{X}^i\ ,\  (\mathbf{X}^i)^2\ ,\  i\hat{p}\ddot{\mathbf{X}}\ ,\  \dot{\mathbf{X}}^i$\vspace{0.5ex}\\
\hline
$\mathcal{O}({\alpha'}^{3/2})$ &\vspace{0.5ex} $(i\hat{p}\dot{\mathbf{X}})^3\ ,\  (i\hat{p}\dot{\mathbf{X}})^2\mathbf{X}^i\ ,\  i\hat{p}\dot{\mathbf{X}}(\mathbf{X}^i)^2\ ,\ (\mathbf{X}^i)^3\ ,\  i\hat{p}\ddot{\mathbf{X}}i\hat{p}\dot{\mathbf{X}}\ ,\ i\hat{p}\ddot{\mathbf{X}}\mathbf{X}^i\ ,\ \dot{\mathbf{X}}^i i\hat{p}\dot{\mathbf{X}}\ ,\ $\vspace{0.5ex}\newline
$\dot{\mathbf{X}}^i\mathbf{X}^i\ ,\ i\hat{p}\mathbf{X}^{i(3)}\ ,\ \ddot{\mathbf{X}}^i$\vspace{0.5ex}\\
\hline
$\mathcal{O}({\alpha'}^2)$ &\vspace{0.5ex} $(i\hat{p}\dot{\mathbf{X}})^4\ ,\  (i\hat{p}\dot{\mathbf{X}})^3(\mathbf{X}^i)^1\ ,\  (i\hat{p}\dot{\mathbf{X}})^2(\mathbf{X}^i)^2\ ,\  i\hat{p}\dot{\mathbf{X}}(\mathbf{X}^i)^3\ ,\  (\mathbf{X}^i)^4\ ,\ $\vspace{0.5ex}\newline
$i\hat{p}\ddot{\mathbf{X}}(i\hat{p}\dot{\mathbf{X}})^2\ ,\ i\hat{p}\ddot{\mathbf{X}}i\hat{p}\dot{\mathbf{X}}\mathbf{X}^i\ ,\ i\hat{p}\ddot{\mathbf{X}}(\mathbf{X}^i)^2\ ,\ \dot{\mathbf{X}}^i (i\hat{p}\dot{\mathbf{X}})^2\ ,\ \dot{\mathbf{X}}^i i\hat{p}\dot{\mathbf{X}}\mathbf{X}^i\ ,\ \dot{\mathbf{X}}^i(\mathbf{X}^i)\ ,\ $\vspace{0.5ex}\newline
$i\hat{p}\mathbf{X}^{(3)}i\hat{p}\dot{\mathbf{X}}\ ,\ i\hat{p}\mathbf{X}^{(3)}\mathbf{X}^i\ ,\ \ddot{\mathbf{X}^i}i\hat{p}\dot{\mathbf{X}}\ ,\ \ddot{\mathbf{X}}^i\mathbf{X}^i\ ,\ (i\hat{p}\ddot{\mathbf{X}})^2\ ,\ i\hat{p}\ddot{\mathbf{X}}\dot{\mathbf{X}}^i\ ,\ (\dot{\mathbf{X}}^i)^2\ ,\ $\vspace{0.5ex}\newline
$i\hat{p}\mathbf{X}^{(4)}\ ,\ \mathbf{X}^{i(3)}$\vspace{0.5ex}\\
\hline
\end{tabular}
\caption{All terms which appear in the boundary state up to order ${\alpha'}^2$}
\label{boundary:table:1}
}

Before calculating the boundary state in detail, we define and calculate some in preparation. A boundary state of D0-brane $|D0\rangle$ satisfies boundary conditions
\begin{alignat}{3}
   \hat{x}^i |D0\rangle &=0 &\quad,\quad (\alpha^i_n-\tilde{\alpha}^i_{-n})|D0\rangle&=0 \nonumber\\
   \hat{p}^0 |D0\rangle &=0 &\quad,\quad (\alpha^0_n+\tilde{\alpha}^0_{-n})|D0\rangle&=0 \quad n \ne 0. \label{boundary:eq:d0-boundarycondition}
\end{alignat}
By utilizing these conditions, we can convert annihilation operators which operates directly on $|D0\rangle$ into creation operators:
\begin{alignat*}{3} 
  \alpha^i_n &\to \tilde{\alpha}^i_{-n} &\quad,\quad \tilde{\alpha}^i_n &\to \alpha^i_{-n} \\
  \alpha^0_n &\to -\tilde{\alpha}^0_{-n} &\quad,\quad \tilde{\alpha}^0_n &\to -\alpha^0_{-n} \quad n>0.
\end{alignat*}
For a given closed string state which is constructed by operating $\alpha^\mu_{n},\tilde{\alpha}^\mu_{n} (n\in\mathbb{Z})$ on $|D0\rangle$, move all annihilation operators to the right side by using commutation relation $[\alpha^\mu_n,\alpha^\nu_m]=g^{\mu\nu}\delta_{n+m,0},[\tilde{\alpha}^\mu_n,\tilde{\alpha}^\nu_m]=g^{\mu\nu}\delta_{n+m,0}$, convert them into creation operators, and then we get a state of the form $|D0\rangle$ only with creation operators on it. In such a manner the boundary state can be reduced to the form easy to interpret. Following this procedure, we need to consider
\begin{align*}
   A_p^{\mu_1\cdots \mu_p}(\sigma_1, \cdots, \sigma_p) &= \prod_{q=1}^p a^{\mu_q}(\sigma_p) \\
   a^{\mu}(\sigma) &= 
\begin{cases}
i\sqrt{\frac{\alpha'}{2}}\sum_{n\ne 0}\frac{1}{n}(\alpha^a_n - \tilde{\alpha}^a_{-n})e^{2\pi i n \sigma} \quad \mathrm{Neumann} \\
\frac{ig}{\sqrt{2\alpha'}}\sum_{n\ne 0}(\alpha^i_n + \tilde{\alpha}^i_{-n})e^{2\pi i n \sigma} \quad \mathrm{Dirichlet}
\end{cases}.
\end{align*}
In the remaining of this section, we always think that $A_p^{\mu_1\cdots \mu_p}(\sigma_1, \cdots, \sigma_p)$ operates on $|Dp\rangle$, and omit $|Dp\rangle$ in calculations for simplicity. We need to pay attention to $A_{p+q}\ne A_{p}A_{q}$ in general. Because creation and annihilation operators along different directions are commutative, $A$ can be divided into a product of Neumann and Dirichlet parts as 
\begin{equation*}
\begin{split}
   A_{p+q}^{a_1\cdots a_p i_1 \cdots i_q }(\sigma_1, \cdots, \sigma_p, \sigma'_1, \cdots, \sigma'_q)
&= A_{p}^{a_1 \cdots a_p}(\sigma_1, \cdots, \sigma_p)A_{q}^{i_1 \cdots i_q}(\sigma'_1, \cdots, \sigma_q').
\end{split}
\end{equation*}
$a$ and $i$ mean Neumann and Dirichlet directions respectively. In our case $a=0, i=1,\cdots ,d-1$. $A^{\mu_1\cdots\mu_p}_p(\sigma_1,\cdots,\sigma_p)$ has a property 
\begin{equation} 
 A_p^{\mu_1\cdots\mu_p}(\sigma_1,\cdots,\sigma_p) = A_p^{\mu_{\tau(1)}\cdots\mu_{\tau(p)}}(\sigma_{\tau(1)},\cdots,\sigma_{\tau(p)}) \quad \tau\mbox{: permutation}. \label{boundary:eq:property}
\end{equation}
It is convenient for simplicity of equations to abbreviate so that 
\begin{equation}
\begin{split}
\left(i\sqrt{\frac{\alpha'}{2}}\right)^{p} A^{a_1\cdots a_p}_p &\to  A_p^{a_1\cdots a_p}\\
\left(\frac{ig}{\sqrt{2\alpha'}}\right)^{p} A^{i_1\cdots i_p}_p &\to A_p^{i_1\cdots i_p}\\
\frac{1}{n!}\frac{d^n}{dt^n}\mathbf{X} &\to \frac{d^n}{dt^n}\mathbf{X}.
\end{split}\label{boundary:eq:rescale}
\end{equation}
In the final results, we should recover this abbreviation. For example $A^{00}_2(\sigma_1,\sigma_2)$ and $ A^{i_1i_2}_2(\sigma_1,\sigma_2)$ are
\begin{align*}
 A^{00}_2(\sigma_1,\sigma_2) &= \sum_{n_1,n_2\ne 0}\frac{1}{n_1n_2}(\alpha^0_{n_1} - \tilde{\alpha}^0_{-n_1})(\alpha^0_{n_2} - \tilde{\alpha}^0_{-n_2})e^{2\pi i (n_1 \sigma_1 + n_2 \sigma_2)}\\
 A^{i_1i_2}_2(\sigma_1,\sigma_2) &= \sum_{n_1,n_2\ne 0}(\alpha^{i_1}_{n_1} + \tilde{\alpha}^{i_1}_{-n_1})(\alpha^{i_2}_{n_2} + \tilde{\alpha}^{i_2}_{-n_2})e^{2\pi i (n_1 \sigma_1 + n_2 \sigma_2)}.
\end{align*}
The number of creation operators $\alpha^\mu_{-n}, \tilde{\alpha}^\mu_{-n} (n>0)$ in each term of $A_p^{\mu_1\cdots\mu_p}$ after eliminating all annihilation operators $\alpha^\mu_{n}, \tilde{\alpha}^\mu_{n} (n>0)$ is at most $p$. By construction it contains $p$ creation and annihilation operators at first. Each time when we move a annihilation operator to the right side over a creation operator, the number of creation and annihilation operators does not change or decreases by 2. Conversion of an annihilation operator to a creation operator does not change the total number of creation and operation operators in each term. After all $A_p^{\mu_1\cdots\mu_p}$ can be divided into a sum of the form
\begin{equation*} 
   A_p^{\mu_1\cdots\mu_p} = A_{p,p}^{\mu_1\cdots\mu_p} + A_{p,p-2}^{\mu_1\cdots\mu_p} + \cdots + A_{p,1\mbox{ or }0}^{\mu_1\cdots\mu_p} 
\end{equation*}
where $A_{p,q}^{\mu_1\cdots\mu_p}$ represents terms which includes $q$ creation operators and no annihilation operator arises from $A_p^{\mu_1\cdots\mu_p}$. For example, operating $A^{00}_2(\sigma_1,\sigma_2), A^{i_1i_2}_2(\sigma_1,\sigma_2)$ on $|D0\rangle$ and utilizing the commutation relation and the boundary conditions, we have
\begin{align*}
A^{00}_2(\sigma_1,\sigma_2)=&\sum_{n_1,n_2>0}\frac{4}{n_1n_2}\left(\tilde{\alpha}_{-n_1}^0\tilde{\alpha}_{-n_2}^0e^{2\pi i (n_1\sigma_1+n_2\sigma_2)}+\alpha_{-n_1}^0\tilde{\alpha}_{-n_2}^0e^{2\pi i (-n_1\sigma_1+n_2\sigma_2)}\right.\\
&\phantom{=\sum_{n_1,n2>0}\frac{4}{n_1n_2}(}
\left. +\ \tilde{\alpha}_{-n_1}^0\alpha_{-n_2}^0e^{2\pi i (n_1\sigma_1-n_2\sigma_2)}+\alpha_{-n_1}^0\alpha_{-n_2}^0e^{2\pi i (-n_1\sigma_1-n_2\sigma_2)}\right)\\
&+\sum_{n>0}\frac{4}{ng}\cos 2\pi n \sigma_{12},
\end{align*}
and
\begin{align*}
A^{i_1i_2}_2(\sigma_1,\sigma_2)=&\sum_{n_1,n_2>0}4\left(\tilde{\alpha}_{-n_1}^{i_1}\tilde{\alpha}_{-n_2}^{i_2}e^{2\pi i (n_1\sigma_1+n_2\sigma_2)}+\alpha_{-n_1}^{i_1}\tilde{\alpha}_{-n_2}^{i_2}e^{2\pi i (-n_1\sigma_1+n_2\sigma_2)}\right.\\
&\phantom{=\sum_{n_1,n2>0}4(}
\left. +\ \tilde{\alpha}_{-n_1}^{i_1}\alpha_{-n_2}^{i_2}e^{2\pi i (n_1\sigma_1-n_2\sigma_2)}+\alpha_{-n_1}^{i_1}\alpha_{-n_2}^{i_2}e^{2\pi i (-n_1\sigma_1-n_2\sigma_2)}\right)\\
&+\sum_{n>0}\frac{4n}{g}\delta^{i_1i_2}\cos 2\pi n \sigma_{12}.
\end{align*}
Results of $A_p^{\mu_1\cdots\mu_p}$ which is used in this section are shown in appendix \ref{boundary:appendix:a}. 

In this way, we can list up all terms in the boundary state  up to order ${\alpha'}^2$ as shown in table \ref{boundary:table:2} Here $\mathbf{X}^{i(n)}=\frac{d^n}{dt^n}\mathbf{X}^i$.
Every index of $\mathbf{X}^i$ and its derivatives should be contracted with that of $\alpha^i$. We abbreviate creation operators $(\alpha^i_{-n}, \tilde{\alpha}^i_{-n})$ as $\alpha^i$, and  $(\alpha^0_{-n}, \tilde{\alpha}^0_{-n})$ as $\alpha^0$.  In a similar way every time derivative should make a pair with $\alpha^0_{-n},\tilde{\alpha}^0_{-n}$. Again we neglect a factor $e^{i\hat{p}\mathbf{X}}$ and ordering.

\TABLE[hbt]{
\begin{tabular}{|l|m{12cm}|}
\hline
$\mathcal{O}({\alpha'}^0)$ &\vspace{0.5ex} 1 \vspace{0.5ex}\\
\hline
$\mathcal{O}({\alpha'}^{1/2})$ &\vspace{0.5ex} $i\hat{p}\dot{\mathbf{X}}(\alpha^0)\ ,\  \mathbf{X}^i(\alpha^i)$\vspace{0.5ex}\\
\hline
$\mathcal{O}({\alpha'}^1)$ &\vspace{0.5ex}$(i\hat{p}\dot{\mathbf{X}})^2(\alpha^0\alpha^0+1)\ ,\  i\hat{p}\dot{\mathbf{X}}\mathbf{X}^i(\alpha^0\alpha^i)\ ,\  (\mathbf{X}^i)^2(\alpha^i\alpha^i+1)\ ,\  i\hat{p}\ddot{\mathbf{X}}(\alpha^0\alpha^0+1)\ ,\ $\vspace{0.5ex}\newline
$\dot{\mathbf{X}}^i(\alpha^0\alpha^i)$\\
\hline
$\mathcal{O}({\alpha'}^{3/2})$&\vspace{0.5ex}  $(i\hat{p}\dot{\mathbf{X}})^3(\alpha^0\alpha^0\alpha^0+\alpha^0)\ ,\  (i\hat{p}\dot{\mathbf{X}})^2\mathbf{X}^i(\alpha^0\alpha^0\alpha^i+\alpha^i)\ ,\ i\hat{p}\dot{\mathbf{X}}(\mathbf{X}^i)^2(\alpha^0\alpha^i\alpha^i+\alpha^0)\ ,\ $\vspace{0.5ex}\newline
$(\mathbf{X}^i)^3(\alpha^i\alpha^i\alpha^i+\alpha^i)\ ,\ i\hat{p}\ddot{\mathbf{X}}i\hat{p}\dot{\mathbf{X}}(\alpha^0\alpha^0\alpha^0+\alpha^0)\ ,\  i\hat{p}\ddot{\mathbf{X}}\mathbf{X}^i(\alpha^0\alpha^0\alpha^i+\alpha^i)\ ,\ $\vspace{0.5ex}\newline
$\dot{\mathbf{X}}^i i\hat{p}\dot{\mathbf{X}}(\alpha^0\alpha^0\alpha^i+\alpha^i)\ ,\ \dot{\mathbf{X}}^i\mathbf{X}^i(\alpha^0\alpha^i\alpha^i)\ ,\ i\hat{p}\mathbf{X}^{i(3)}(\alpha^0\alpha^0\alpha^0+\alpha^0)\ ,\ $\vspace{0.5ex}\newline
$\ddot{\mathbf{X}}^i(\alpha^0\alpha^0\alpha^i+\alpha^i)$ \vspace{0.5ex}\\
\hline
$\mathcal{O}({\alpha'}^2)$ &\vspace{0.5ex} $(i\hat{p}\dot{\mathbf{X}})^4(\alpha^0\alpha^0\alpha^0\alpha^0+\alpha^0\alpha^0+1)\ ,\  (i\hat{p}\dot{\mathbf{X}})^3(\mathbf{X}^i)^1(\alpha^0\alpha^0\alpha^0\alpha^i+\alpha^0\alpha^i)\ ,\ $\vspace{0.5ex}\newline
$(i\hat{p}\dot{\mathbf{X}})^2(\mathbf{X}^i)^2(\alpha^0\alpha^0\alpha^i\alpha^i+\alpha^0\alpha^0+\alpha^i\alpha^i+1)\ ,\  i\hat{p}\dot{\mathbf{X}}(\mathbf{X}^i)^3(\alpha^0\alpha^i\alpha^i\alpha^i+\alpha^0\alpha^i)\ ,\  $\vspace{0.5ex}\newline
$(\mathbf{X}^i)^4(\alpha^i\alpha^i\alpha^i\alpha^i+\alpha^i\alpha^i+1)\ ,\ i\hat{p}\ddot{\mathbf{X}}(i\hat{p}\dot{\mathbf{X}})^2(\alpha^0\alpha^0\alpha^0\alpha^0+\alpha^0\alpha^0+1)\ ,\ $\vspace{0.5ex}\newline
$i\hat{p}\ddot{\mathbf{X}}i\hat{p}\dot{\mathbf{X}}\mathbf{X}^i(\alpha^0\alpha^0\alpha^0\alpha^i+\alpha^0\alpha^i)\ ,\ i\hat{p}\ddot{\mathbf{X}}(\mathbf{X}^i)^2(\alpha^0\alpha^0\alpha^i\alpha^i+\alpha^0\alpha^0+\alpha^i\alpha^i+1)\ ,\ $\vspace{0.5ex}\newline
$\dot{\mathbf{X}}^i (i\hat{p}\dot{\mathbf{X}})^2(\alpha^0\alpha^0\alpha^0\alpha^i+\alpha^0\alpha^i)\ ,\ $\vspace{0.5em}\newline
$\dot{\mathbf{X}}^i i\hat{p}\dot{\mathbf{X}}\mathbf{X}^i(\alpha^0\alpha^0\alpha^i\alpha^i+\alpha^0\alpha^0+\alpha^i\alpha^i+1)\ ,\ $\vspace{0.5em}\newline
$\dot{\mathbf{X}}^i(\mathbf{X}^i)^2(\alpha^0\alpha^i\alpha^i\alpha^i+\alpha^0\alpha^i)\ ,\ i\hat{p}\mathbf{X}^{(3)}i\hat{p}\dot{\mathbf{X}}(\alpha^0\alpha^0\alpha^0\alpha^0+\alpha^0\alpha^0+1)\ ,\ $\vspace{0.5ex}\newline
$i\hat{p}\mathbf{X}^{(3)}\mathbf{X}^i(\alpha^0\alpha^0\alpha^0\alpha^1+\alpha^0\alpha^1)\ ,\ \ddot{\mathbf{X}}^i i\hat{p}\dot{\mathbf{X}}(\alpha^0\alpha^0\alpha^0\alpha^i+\alpha^0\alpha^i)\ ,\ $\vspace{0.5ex}\newline
$\ddot{\mathbf{X}}^i\mathbf{X}^i(\alpha^0\alpha^0\alpha^i\alpha^i+\alpha^0\alpha^0+\alpha^i\alpha^i+1)\ ,\ (i\hat{p}\ddot{\mathbf{X}})^2(\alpha^0\alpha^0\alpha^0\alpha^0+\alpha^0\alpha^0+1)\ ,\ $\vspace{0.5ex}\newline
$i\hat{p}\ddot{\mathbf{X}}\dot{\mathbf{X}}^i(\alpha^0\alpha^0\alpha^0\alpha^i+\alpha^0\alpha^i)\ ,\ (\dot{\mathbf{X}}^i)^2(\alpha^0\alpha^0\alpha^i\alpha^i+\alpha^0\alpha^0+\alpha^i\alpha^i+1)\ ,\ $\vspace{0.5ex}\newline
$i\hat{p}\mathbf{X}^{(4)}(\alpha^0\alpha^0\alpha^0\alpha^0+\alpha^0\alpha^0+1)\ ,\ \mathbf{X}^{i(3)}(\alpha^0\alpha^0\alpha^0\alpha^i+\alpha^0\alpha^i)$.\\
\hline
\end{tabular}
\caption{All terms which appear in the boundary state up to order ${\alpha'}^2$}
\label{boundary:table:2}
}

In what follows, we explain how to derive strict form of a given term in the table \ref{boundary:table:1}, or equivalently, the table \ref{boundary:table:2}. Each term in the table is a product of 
\begin{equation*}
\{ i\hat{p}\dot{\mathbf{X}}, \mathbf{X}^i, i\hat{p}\ddot{\mathbf{X}}, \dot{\mathbf{X}}^i, \ddot{\mathbf{X}}^i, \cdots\}.
\end{equation*}
We assign indexes $\{a_1, i_1, a_1a_1, a_1i_1, i_1i_2, \cdots\}$ to them. Note that a time derivatives has an index $a=0$. %This is reasonable since in case of higher dimensional D-brane we have derivatives in several Neumann directions. 
Then we can denote any given term in the table as a product of
\begin{equation*}
   Y_p^{\mu^{p}_1\cdots\mu^{p}_{m(Y_p)}} \in \{ i\hat{p}\dot{\mathbf{X}}, \mathbf{X}^i, i\hat{p}\ddot{\mathbf{X}}, \dot{\mathbf{X}}^i, \ddot{\mathbf{X}}^i, \cdots\} \quad (\mu = 0, 1, \cdots ,d-1)
\end{equation*} 
where $m(Y_p)$ is the number of indexes of $Y_p$. Suppose that $p\ne q \Rightarrow Y_p \ne Y_q$. What we need to consider is 
\begin{equation*}
  \mathcal{O}= \prod_{q=1}^N Z_q^{\mu^{q}_1\cdots\mu^{q}_{m(Z_q)}} = \prod_{p=1}^M \left(Y_p^{\mu^{p}_1\cdots\mu^{p}_{m(Y_p)}}\right)^{n(Y_p)}.
\end{equation*}
Here $n(Y_p)$ is multiplicity of $Y_p$, $M$ is the number of different kinds of $Y_p$, and $N=\sum_{p=1}^M n(Y_p)$ is the total number of $\{ i\hat{p}\dot{\mathbf{X}}, \mathbf{X}^i, i\hat{p}\ddot{\mathbf{X}}, \dot{\mathbf{X}}^i, \ddot{\mathbf{X}}^i, \cdots\}$. $\mathcal{O}$ is represented as
\begin{equation*}
\begin{split}
   \mathcal{O}&=Z_1\cdots Z_N\\
&=\underbrace{Y_1 \cdots Y_1}_{n(Y_1)}\underbrace{Y_2 \cdots Y_2}_{n(Y_2)}\underbrace{Y_M \cdots Y_M}_{n(Y_M)}.
\end{split}
\end{equation*}
We define $\mathrm{P'}$ by \label{boundary:page:P'}
\begin{multline}
\mathrm{trP'}[Z_1\cdots Z_N]
=\sum_{\tau\in\Sigma}\Biggl\{
\int_{1>\sigma_N>\cdots>\sigma_1>0}\!\!\!\!\!\! d\sigma\ \mathrm{tr}\Biggl[
e^{i\hat{p}\mathbf{X}}\prod_{q=1}^N\left(e^{-i\sigma_q\hat{p\mathbf{X}}}Z_{\tau(q)}^{\mu^{\tau(q)}_1\cdots\mu^{\tau(q)}_{n(Z_q)}}e^{i\sigma_q\hat{p}\mathbf{X}}\right)\\
A_L^{\mu^{\tau(1)}_1\cdots\mu^{\tau(1)}_{n(Z_{\tau(1)})}\cdots\mu^{\tau(N)}_1\cdots\mu^{\tau(N)}_{n(Z_{\tau(N)})}}(\underbrace{\sigma_1,\cdots,\sigma_1}_{m(Z_{\tau(1)})},\cdots,\underbrace{\sigma_N,\cdots,\sigma_N}_{m(Z_{\tau(N)})})\Biggr]\Biggr\}. \label{boundary:eq:P'}
\end{multline}
Here $\Sigma$ is a set of all possible permutation of $\{Y_p\}$. Thus $\tau$ runs $\frac{N!}{n(Y_1)!\cdots n(Y_M)!}$ different permutations. Any given term $\mathcal{O}$ in \ref{boundary:table:1} represents $P'[\mathcal{O}]$. In table \ref{boundary:table:2}, the creation operator part of each term are explicitly shown. 

After the variable change $\sigma_q \to \sigma_{\tau(q)}$ \eqref{boundary:eq:P'} can be rewritten as
\begin{multline*}
%\mathrm{trP'}[Z_1\cdots Z_N]=
%\mathrm{trP''}[Z_{1,\sigma_1}\cdots Z_{N,\sigma_N}e^{i\int_0^1d\sigma \hat{p}\mathbf{X}}A_L(\sigma_1,\cdots,\sigma_N)]=\\
\sum_{\tau\in\Sigma}\Biggl\{
\int_{1>\sigma_{\tau(N)}>\cdots>\sigma_{\tau(1)}>0}\!\!\!\!\!\! d\sigma\ \mathrm{tr}\Biggl[
e^{i\hat{p}\mathbf{X}}\prod_{q=1}^N\left(e^{-i\sigma_q\hat{p\mathbf{X}}}Z_{\tau(q)}^{\mu^{\tau(q)}_1\cdots\mu^{\tau(q)}_{n(Z_q)}}e^{i\sigma_q\hat{p}\mathbf{X}}\right)\\
A_L^{\mu^{1}_1\cdots\mu^{1}_{n(Z_{1})}\cdots\mu^{N}_1\cdots\mu^{N}_{n(Z_{N})}}(\underbrace{\sigma_1,\cdots,\sigma_1}_{n(Z_1)},\cdots,\underbrace{\sigma_N,\cdots,\sigma_N}_{n(Z_N)})\Biggr]\Biggr\}.
\end{multline*}
Using this equation we define $\mathrm{P}''$ by
\begin{multline}
\mathrm{trP''}[Z_{1,\sigma_1}\cdots Z_{N,\sigma_N}e^{i\int_0^1d\sigma \hat{p}\mathbf{X}}A_L(\sigma_1,\cdots,\sigma_N)]\\
= \sum_{\tau\in\Sigma}\Biggl\{
\int_{1>\sigma_{\tau(N)}>\cdots>\sigma_{\tau(1)}>0}\!\!\!\!\!\! d\sigma\ \mathrm{tr}\Biggl[
e^{i\hat{p}\mathbf{X}}\prod_{q=1}^N\left(e^{-i\sigma_q\hat{p\mathbf{X}}}Z_{\tau(q)}^{\mu^{\tau(q)}_1\cdots\mu^{\tau(q)}_{n(Z_q)}}e^{i\sigma_q\hat{p}\mathbf{X}}\right)\\
A_L^{\mu^{1}_1\cdots\mu^{1}_{n(Z_{1})}\cdots\mu^{N}_1\cdots\mu^{N}_{n(Z_{N})}}(\underbrace{\sigma_1,\cdots,\sigma_1}_{n(Z_1)},\cdots,\underbrace{\sigma_N,\cdots,\sigma_N}_{n(Z_N)})\Biggr]\Biggr\}.\label{boundary:eq:P''}
\end{multline}
Here $\sigma_q$ is assigned to $Z_q$ in order to determine the ordering, although they are not depend on $\sigma$'s.  
For example we examine $\dot{\mathbf{X}}^i\mathbf{X}^i$. This term in the table means there exist following terms at $\alpha^{'}$ in the boundary state:   
\begin{equation*}
\begin{split}
\mbox{(what }\dot{\mathbf{X}}^i\mathbf{X}^i \mbox{ means)}&\\
=\mathrm{trP'}[\dot{\mathbf{X}}^{i_1}\mathbf{X}^{i_2} &]\\
&=\int_0^1\!\! d\sigma_1 \int_0^{\sigma_1}\!\!\!\! d\sigma_2\ \mathrm{tr}[e^{i(1-\sigma_{12})\hat{p\mathbf{X}}}\dot{\mathbf{X}}^{i_1}e^{i\sigma_{12}\hat{p}\mathbf{X}}\mathbf{X}^{i_2}]A_3^{0i_1i_2}(\sigma_1,\sigma_1,\sigma_2)\\
&+\int_0^1\!\! d\sigma_1 \int_0^{\sigma_1}\!\!\!\! d\sigma_2\ \mathrm{tr}[e^{i(1-\sigma_{12})\hat{p\mathbf{X}}}\mathbf{X}^{i_2}e^{i\sigma_{12}\hat{p}\mathbf{X}}\dot{\mathbf{X}}^{i_1}]A_3^{i_20i_1}(\sigma_1,\sigma_2,\sigma_2)\\
=\mathrm{trP''}[\dot{\mathbf{X}}^{i_1}_{\sigma_1}\mathbf{X}^{i_2}_{\sigma_2} & e^{i \int_0^1 d\sigma \hat{p}\mathbf{X}}A_3^{i_10i_2}(\sigma_1,\sigma_1,\sigma_2)]\\
&=\int_0^1\!\! d\sigma_1 \int_0^{\sigma_1}\!\!\!\! d\sigma_2\ \mathrm{tr}[e^{i(1-\sigma_{12})\hat{p\mathbf{X}}}\dot{\mathbf{X}}^{i_1}e^{i\sigma_{12}\hat{p}\mathbf{X}}\mathbf{X}^{i_2}]A_3^{0i_1i_2}(\sigma_1,\sigma_1,\sigma_2)\\
&+\int_0^1\!\! d\sigma_2 \int_0^{\sigma_2}\!\!\!\! d\sigma_1\ \mathrm{tr}[e^{i(1-\sigma_{21})\hat{p\mathbf{X}}}\mathbf{X}^{i_2}e^{i\sigma_{21}\hat{p}\mathbf{X}}\dot{\mathbf{X}}^{i_1}]A_3^{i_20i_1}(\sigma_2,\sigma_1,\sigma_1).
\end{split}
\end{equation*}

We note that because the boundary state satisfies the level matching condition, we can eliminate those terms which do not satisfy this condition without explicit calculations. Such disappearance happens due to an integration like $\int_0^1 d\sigma e^{2\pi i n \sigma} = 0$. In the methods shown above in this section, we can explicitly write down the boundary state at any given order in $\alpha'$.

In what follows, we calculate the boundary state at each order in $\alpha'$ in detail. 
\subsubsection*{\underline{order ${\alpha'}^0$}}
At zeroth order the boundary state gives simply 
\begin{equation}
   |B\rangle_{{\alpha'}^0} = \mathrm{tr}e^{i\hat{p}\mathbf{X}}|D0\rangle. \label{boundary:eq:alpha0}
\end{equation}

\subsubsection*{\underline{order ${\alpha'}^{1/2}$}}
All terms at this order vanish. 
\begin{align*}
\frac{ig}{\sqrt{2\alpha'}} \sum_{n_1 \ne 0} \int_0^1\!\! d\sigma_1\ \mathrm{tr}\left[\mathbf{X}^{i_1} e^{i\hat{p}\mathbf{X}}\right](\alpha^{i_1}_{n_1}+\tilde{\alpha}^{i_1}_{-n_1}) e^{2\pi i n_1\sigma_1}&=0\\
i\sqrt{\frac{\alpha'}{2}}\sum_{n_1 \ne 0} \int_0^1\!\! d\sigma_1\ \mathrm{tr}\left[i\hat{p}\mathbf{X} e^{i\hat{p}\mathbf{X}}\right]\frac{1}{n_1}(\alpha^0_{n_1}-\tilde{\alpha}^0_{-n_1}) e^{2\pi i n_1\sigma_1}&=0.
\end{align*}
These can be seen easily by that $\int_0^1 d\sigma e^{2\pi i n \sigma}=0 (n\ne0)$. Hence we have
\begin{equation*}
   |B\rangle_{{\alpha'}^{1/2}} = 0.
\end{equation*}

\subsubsection*{\underline{order ${\alpha'}^1$}}
The boundary state at this order includes information about the energy-momentum tensor. There is singularities which remain after the zeta function regularization. These divergences vanish if an equation of motion is satisfied as shown in subsection \ref{boundary:subsection:3-3}. At order ${\alpha'}^1$ there exist terms listed below:
\begin{subequations}
\begin{align}
&\bullet\ \int_0^1\!\!d\sigma\int_0^{\sigma_1}\!\!\!\! d\sigma_2\ \mathrm{tr}[e^{i(1-\sigma_{12})\hat{p}\mathbf{X}}(i\hat{p}\dot{\mathbf{X}})e^{i\sigma_{12}i\hat{p}\mathbf{X}}(i\hat{p}\dot{\mathbf{X}})]A_2^{00}(\sigma_1,\sigma_2)\label{boundary:eq:alpha1-1}\\
\begin{split}
&\bullet\ \int_0^1\!\!d\sigma\int_0^{\sigma_1}\!\!\!\! d\sigma_2\ \mathrm{tr}[e^{i(1-\sigma_{12})\hat{p}\mathbf{X}}(i\hat{p}\dot{\mathbf{X}})e^{i\sigma_{12}\hat{p}\mathbf{X}}\mathbf{X}^i]A_2^{0i}(\sigma_1,\sigma_2)\\
& \hspace{1.5em}+\int_0^1\!\!d\sigma\int_0^{\sigma_1}\!\!\!\! d\sigma_2\ \mathrm{tr}[e^{i(1-\sigma_{12})\hat{p}\mathbf{X}}\mathbf{X}^i e^{i\sigma_{12}\hat{p}\mathbf{X}}(i\hat{p}\dot{\mathbf{X}})]A_2^{i0}(\sigma_1,\sigma_2)
\end{split}\label{boundary:eq:alpha1-2}\\
&\bullet\ \int_0^1\!\!d\sigma\int_0^{\sigma_1}\!\!\!\! d\sigma_2\ \mathrm{tr}[e^{i(1-\sigma_{12})\hat{p}\mathbf{X}}\mathbf{X}^{i_1}e^{i\sigma_{12}\hat{p}\mathbf{X}}\mathbf{X}^{i_2}]A_2^{i_1i_2}(\sigma_1,\sigma_2)\label{boundary:eq:alpha1-3}\\
&\bullet\ \int_0^1\!\!d\sigma\ \mathrm{tr}[(i\hat{p}\ddot{\mathbf{X}})e^{i\hat{p}\mathbf{X}}]A_2^{00}(\sigma,\sigma)\label{boundary:eq:alpha1-4}
\\&\bullet\ \int_0^1\!\!d\sigma\ \mathrm{tr}[\dot{\mathbf{X}}^i e^{i\hat{p}\mathbf{X}}]A_2^{0i}(\sigma,\sigma)\label{boundary:eq:alpha1-5}
\end{align}
\end{subequations}
$A^{\mu_1\mu_2}(\sigma_1,\sigma_2)$ is (see appendix \ref{boundary:appendix:a})
\begin{align*}
A^{00}_2(\sigma_1,\sigma_2)=&\sum_{n_1,n_2>0}\frac{4}{n_1n_2}\left(\tilde{\alpha}_{-n_1}^0\tilde{\alpha}_{-n_2}^0e^{2\pi i (n_1\sigma_1+n_2\sigma_2)}+\alpha_{-n_1}^0\tilde{\alpha}_{-n_2}^0e^{2\pi i (-n_1\sigma_1+n_2\sigma_2)}\right.\\
&\phantom{=\sum_{n_1,n2>0}\frac{4}{n_1n_2}(}
\left. +\ \tilde{\alpha}_{-n_1}^0\alpha_{-n_2}^0e^{2\pi i (n_1\sigma_1-n_2\sigma_2)}+\alpha_{-n_1}^0\alpha_{-n_2}^0e^{2\pi i (-n_1\sigma_1-n_2\sigma_2)}\right)\\
&+\sum_{n>0}\frac{4}{ng}\cos 2\pi n \sigma_{12},
\end{align*}
\begin{align*}
A^{0i}_2(\sigma_1,\sigma_2)=&\sum_{n_1,n_2>0}\frac{-4}{n_1}\left(\tilde{\alpha}_{-n_1}^{0}\tilde{\alpha}_{-n_2}^{i}e^{2\pi i (n_1\sigma_1+n_2\sigma_2)}+\alpha_{-n_1}^{0}\tilde{\alpha}_{-n_2}^{i}e^{2\pi i (-n_1\sigma_1+n_2\sigma_2)}\right.\\
&\phantom{=\sum_{n_1,n2>0}4(}
\left. +\ \tilde{\alpha}_{-n_1}^{0}\alpha_{-n_2}^{i}e^{2\pi i (n_1\sigma_1-n_2\sigma_2)}+\alpha_{-n_1}^{0}\alpha_{-n_2}^{i}e^{2\pi i (-n_1\sigma_1-n_2\sigma_2)}\right),
\end{align*}
\begin{align*}
A^{i0}_2(\sigma_1,\sigma_2)=&\sum_{n_1,n_2>0}\frac{-4}{n_2}\left(\tilde{\alpha}_{-n_1}^{i}\tilde{\alpha}_{-n_2}^{0}e^{2\pi i (n_1\sigma_1+n_2\sigma_2)}+\alpha_{-n_1}^{i}\tilde{\alpha}_{-n_2}^{0}e^{2\pi i (-n_1\sigma_1+n_2\sigma_2)}\right.\\
&\phantom{=\sum_{n_1,n2>0}4(}
\left. +\ \tilde{\alpha}_{-n_1}^{i}\alpha_{-n_2}^{0}e^{2\pi i (n_1\sigma_1-n_2\sigma_2)}+\alpha_{-n_1}^{i}\alpha_{-n_2}^{0}e^{2\pi i (-n_1\sigma_1-n_2\sigma_2)}\right),
\end{align*}
\begin{align*}
A^{i_1i_2}_2(\sigma_1,\sigma_2)=&\sum_{n_1,n_2>0}4\left(\tilde{\alpha}_{-n_1}^{i_1}\tilde{\alpha}_{-n_2}^{i_2}e^{2\pi i (n_1\sigma_1+n_2\sigma_2)}+\alpha_{-n_1}^{i_1}\tilde{\alpha}_{-n_2}^{i_2}e^{2\pi i (-n_1\sigma_1+n_2\sigma_2)}\right.\\
&\phantom{=\sum_{n_1,n2>0}4(}
\left. +\ \tilde{\alpha}_{-n_1}^{i_1}\alpha_{-n_2}^{i_2}e^{2\pi i (n_1\sigma_1-n_2\sigma_2)}+\alpha_{-n_1}^{i_1}\alpha_{-n_2}^{i_2}e^{2\pi i (-n_1\sigma_1-n_2\sigma_2)}\right)\\
&+\sum_{n>0}\frac{4n}{g}\delta^{i_1i_2}\cos 2\pi n \sigma_{12}.
\end{align*}
First we consider \eqref{boundary:eq:alpha1-1}, \eqref{boundary:eq:alpha1-2},and \eqref{boundary:eq:alpha1-3}. Changing variables as (i) $\sigma_1 = \sigma_2'+1, \sigma_2 =  \sigma_1', (\sigma_{12} = 1-\sigma_{12}')$. Under this transformation
\begin{multline*}
\int_0^1\!\!d\sigma_1\int_0^{\sigma_1}\!\!\!\! d\sigma_2\ \mathrm{tr}[e^{i(1-\sigma_{12})\hat{p}\mathbf{X}}(i\hat{p}\dot{\mathbf{X}})e^{i\sigma_{12} \hat{p}\mathbf{X}}(i\hat{p}\dot{\mathbf{X}})]A_2^{00}(\sigma_1,\sigma_2)\\
\to \int_0^1\!\!d\sigma_1\int_{-1+\sigma_1}^0 \!\!\!\! d\sigma_2\ \mathrm{tr}[e^{i(1-\sigma_{12})\hat{p}\mathbf{X}}(i\hat{p}\dot{\mathbf{X}})e^{i\sigma_{12} \hat{p}\mathbf{X}}(i\hat{p}\dot{\mathbf{X}})]A_2^{00}(\sigma_1,\sigma_2),
\end{multline*}
\begin{multline*}
\int_0^1\!\!d\sigma_1\int_0^{\sigma_1}\!\!\!\! d\sigma_2\ \mathrm{tr}[e^{i(1-\sigma_{12})\hat{p}\mathbf{X}}(i\hat{p}\dot{\mathbf{X}})e^{i\sigma_{12}\hat{p}\mathbf{X}}\mathbf{X}^i]A_2^{0i}(\sigma_1,\sigma_2)\\
\to \int_0^1\!\!d\sigma\int_{-1+\sigma_1}^0\!\!\!\! d\sigma_2\ \mathrm{tr}[e^{i(1-\sigma_{12})\hat{p}\mathbf{X}}\mathbf{X}^i e^{i\sigma_{12}\hat{p}\mathbf{X}}(i\hat{p}\dot{\mathbf{X}})]A_2^{i0}(\sigma_1,\sigma_2),
\end{multline*}
\begin{subequations}
\begin{multline}
\int_0^1\!\!d\sigma_1\int_0^{\sigma_1}\!\!\!\! d\sigma_2\ \mathrm{tr}[e^{i(1-\sigma_{12})\hat{p}\mathbf{X}}\mathbf{X}^i e^{i\sigma_{12}\hat{p}\mathbf{X}}(i\hat{p}\dot{\mathbf{X}})]A_2^{i0}(\sigma_1,\sigma_2)\\
\to \int_0^1\!\!d\sigma_1\int_{-1+\sigma_1}^0\!\!\!\! d\sigma_2\ \mathrm{tr}[e^{i(1-\sigma_{12})\hat{p}\mathbf{X}}(i\hat{p}\dot{\mathbf{X}})e^{i\sigma_{12}\hat{p}\mathbf{X}}\mathbf{X}^i]A_2^{0i}(\sigma_1,\sigma_2), \label{boundary:eq:specialcase1}
\end{multline}
\begin{multline}
\int_0^1\!\!d\sigma_1\int_0^{\sigma_1}\!\!\!\! d\sigma_2\ \mathrm{tr}[e^{i(1-\sigma_{12})\hat{p}\mathbf{X}}\mathbf{X}^{i_1}e^{i\sigma_{12}\hat{p}\mathbf{X}}\mathbf{X}^{i_2}]A_2^{i_1i_2}(\sigma_1,\sigma_2)\\
\to \int_0^1\!\!d\sigma_1\int_{-1+\sigma_1}^0\!\!\!\! d\sigma_2\ \mathrm{tr}[e^{i(1-\sigma_{12})\hat{p}\mathbf{X}}\mathbf{X}^{i_2}e^{i\sigma_{12}\hat{p}\mathbf{X}}\mathbf{X}^{i_1}]A_2^{i_1i_2}(\sigma_1,\sigma_2) \label{boundary:eq:specialcase2}
\end{multline}
\end{subequations}
by using cyclicity of trace and the property \eqref{boundary:eq:property}. We combine two domains of integration as    
\begin{equation*}
   \frac{1}{2}\int_0^1\!\!d\sigma_1 \left\{\int_0^{\sigma_1}\!\!\!\! + \int_{-1+\sigma_1}^0\!\!\ \right\} d\sigma_2 f(\sigma_1,\sigma_2).
= \frac{1}{2}\int_0^1 d\sigma_{12} \int_0^1 d\sigma_1 f(\sigma_1, \sigma_2)
\end{equation*}
We note that change of variables (ii) $\sigma_1' = 1+\sigma_2, \sigma_2' = \sigma_1$ leads the same results. In this case, we have
\begin{equation*}
   \frac{1}{2}\int_0^1\!\!d\sigma_2 \left\{\int_{\sigma_2}^1\!\!\!\! + \int_1^{1+\sigma_2}\!\!\ \right\} d\sigma_1 f(\sigma_1,\sigma_2)
= \frac{1}{2}\int_0^1 d\sigma_{12} \int_0^1 d\sigma_2 f(\sigma_1, \sigma_2).
\end{equation*}
The integration domains are shown in figure \ref{boundary:fig:2sigma}.

\psfrag{sigma_1}{$\sigma_1$}\psfrag{sigma_2}{$\sigma_2$}\psfrag{sigma_3}{$\sigma_3$}\psfrag{1}{$1$}\psfrag{-1}{$-1$}\psfrag{(i)}{(i)}\psfrag{(ii)}{(ii)}\psfrag{(iii)}{(iii)}
\EPSFIGURE[hb]{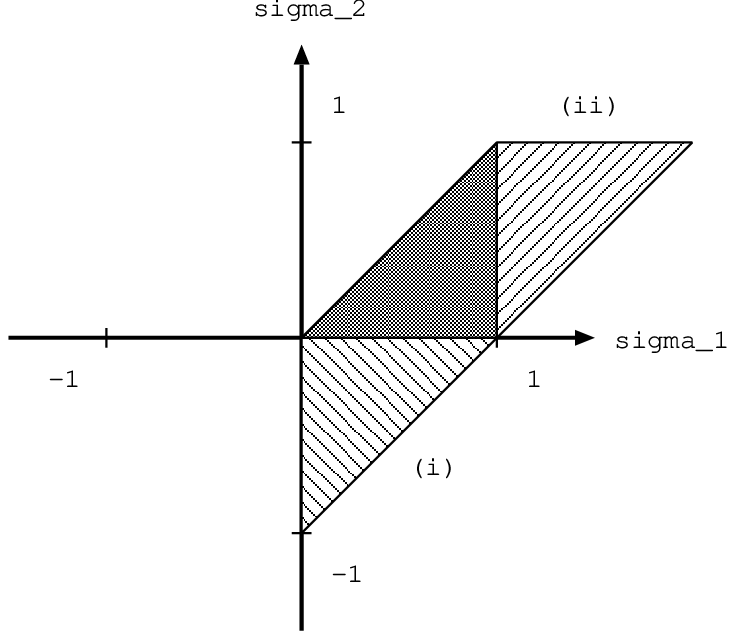,width=0.6\textwidth}{Change of variables $\sigma_1,\sigma_2$\label{boundary:fig:2sigma}}
Then it is possible to integrate over $\sigma_1$ to get
\begin{align*} 
&\int_0^1\!\! d\sigma_1\  A_2^{00}(\sigma_1,\sigma_2)\Bigr|_{\sigma_{12}\ \mathrm{fixed}}\\
=&\sum_{n>0}\frac{4}{n^2}\left(\alpha_{-n}^0\tilde{\alpha}_{-n}^0e^{- 2\pi i n\sigma_{12}}+ \tilde{\alpha}_{-n}^0\alpha_{-n}^0e^{2\pi i n \sigma_{12}}\right)+\sum_{n>0}\frac{4}{ng}\cos 2\pi n \sigma_{12},
\end{align*}
\begin{align*} 
&\int_0^1\!\! d\sigma_1\  A_2^{0i}(\sigma_1,\sigma_2)\Bigr|_{\sigma_{12}\ \mathrm{fixed}}\\
=&\sum_{n>0}\frac{-4}{n}\left(\alpha_{-n}^0\tilde{\alpha}_{-n}^ie^{- 2\pi i n\sigma_{12}}+ \tilde{\alpha}_{-n}^0\alpha_{-n}^ie^{2\pi i n \sigma_{12}}\right)\phantom{+\sum_{n>0}\frac{4}{ng}\cos 2\pi n \sigma_{12}},
\end{align*}
\begin{align*} 
&\int_0^1\!\! d\sigma_1\  A_2^{i0}(\sigma_1,\sigma_2)\Bigr|_{\sigma_{12}\ \mathrm{fixed}}\\
=&\sum_{n>0}\frac{-4}{n}\left(\alpha_{-n}^i\tilde{\alpha}_{-n}^0e^{- 2\pi i n\sigma_{12}}+ \tilde{\alpha}_{-n}^i\alpha_{-n}^0e^{2\pi i n \sigma_{12}}\right)\phantom{+\sum_{n>0}\frac{4}{ng}\cos 2\pi n \sigma_{12}},
\end{align*}
\begin{align*}
&\int_0^1\!\! d\sigma_1\  A_2^{i_1i_2}(\sigma_1,\sigma_2)\Bigr|_{\sigma_{12}\ \mathrm{fixed}}\\
=&\sum_{n>0}4\left(\alpha_{-n}^{i_1}\tilde{\alpha}_{-n}^{i_2}e^{- 2\pi i n\sigma_{12}}+ \tilde{\alpha}_{-n}^{i_1}\alpha_{-n}^{i_2}e^{2\pi i n \sigma_{12}}\right)+\sum_{n>0}\frac{4n}{g}\delta^{i_1i_2}\cos 2\pi n \sigma_{12}.
\end{align*}
Here we have used
\begin{align*}
 \int_0^1\!\! d\sigma_{1}\ e^{2\pi i (n_1\sigma_1+n_2\sigma_2)}\bigr|_{\sigma_{12}\ \mathrm{fixed}} & = \int_0^1\!\! d\sigma\ e^{2\pi i(-n_2\sigma_{12} + (n_1+n_2)\sigma)} = 0 \\
 \int_0^1\!\! d\sigma_{1}\ e^{2\pi i (- n_1\sigma_1 + n_2\sigma_2)}\bigr|_{\sigma_{12}\ \mathrm{fixed}} & = \int_0^1\!\! d\sigma\ e^{-2\pi i(n_2\sigma_{12} -(n_1-n_2)\sigma)} = e^{-2\pi i n_2 \sigma_{12}} \delta_{n_1,n_2} \\
 \int_0^1\!\! d\sigma_{1}\ e^{2\pi i (n_1\sigma_1-n_2\sigma_2)}\bigr|_{\sigma_{12}\ \mathrm{fixed}} & = \int_0^1\!\! d\sigma\ e^{2\pi i(n_2\sigma_{12} + (n_1-n_2)\sigma)} = e^{2\pi i n_2 \sigma_{12}} \delta_{n_1,n_2} \\
 \int_0^1\!\! d\sigma_{1}\ e^{- 2\pi i (n_1\sigma_1 + n_2\sigma_2)}\bigr|_{\sigma_{12}\ \mathrm{fixed}} & = \int_0^1\!\! d\sigma\ e^{2\pi i(n_2\sigma_{12} - (n_1+n_2)\sigma)} = 0  \\
&(n_1, n_2 >0).
\end{align*}
After all \eqref{boundary:eq:alpha1-1}, \eqref{boundary:eq:alpha1-2}, \eqref{boundary:eq:alpha1-3} become
\begin{subequations}
\begin{equation}
\begin{split} 
\left(i\sqrt{\frac{\alpha'}{2}}\right)^2\frac{1}{2}\int_0^1\!\!d\sigma\ \mathrm{tr}&\left[e^{i(1-\sigma)\hat{p}\mathbf{X}}(i\hat{p}\dot{\mathbf{X}})e^{i\sigma \hat{p}\mathbf{X}}(i\hat{p}\dot{\mathbf{X}})\right]\\
&\left\{\sum_{n>0}\frac{4}{n^2}\left(\alpha_{-n}^0\tilde{\alpha}_{-n}^0e^{- 2\pi i n\sigma }+ \tilde{\alpha}_{-n}^0\alpha_{-n}^0e^{2\pi i n \sigma }\right)+\sum_{n>0}\frac{4}{ng}\cos 2\pi n \sigma \right\},
\end{split}\label{boundary:eq:alpha1-1'}
\end{equation}
\begin{equation}
\frac{ig}{\sqrt{2\alpha'}}i\sqrt{\frac{\alpha'}{2}}\int_0^1\!\!d\sigma\ \mathrm{tr}\left[e^{i(1-\sigma)\hat{p}\mathbf{X}}(i\hat{p}\dot{\mathbf{X}})e^{i\sigma\hat{p}\mathbf{X}}\mathbf{X}^i\right]\sum_{n>0}\frac{-4}{n}\left(\alpha_{-n}^0\tilde{\alpha}_{-n}^ie^{- 2\pi i n\sigma }+ \tilde{\alpha}_{-n}^0\alpha_{-n}^ie^{2\pi i n \sigma }\right),\label{boundary:eq:alpha1-2'}
\end{equation}
\begin{equation}
\begin{split}
\left(\frac{ig}{\sqrt{2\alpha'}}\right)^2\frac{1}{2}\int_0^1\!\!d\sigma\ \mathrm{tr}&\left[e^{i(1-\sigma)\hat{p}\mathbf{X}}\mathbf{X}^{i_1}e^{i\sigma\hat{p}\mathbf{X}}\mathbf{X}^{i_2}\right]\\
&\left\{\sum_{n>0}4\left(\alpha_{-n}^{i_1}\tilde{\alpha}_{-n}^{i_2}e^{- 2\pi i n\sigma }+ \tilde{\alpha}_{-n}^{i_1}\alpha_{-n}^{i_2}e^{2\pi i n \sigma }\right)+\sum_{n>0}\frac{4n}{g}\delta^{i_1i_2}\cos 2\pi n \sigma\right\}
\end{split}\label{boundary:eq:alpha1-3'}
\end{equation}
where we restore the abbreviation \eqref{boundary:eq:rescale}. 

Similarly \eqref{boundary:eq:alpha1-4}, \eqref{boundary:eq:alpha1-5} become
\begin{align}
&\left(i\sqrt{\frac{\alpha'}{2}}\right)^2\frac{1}{2!}\mathrm{tr}\left[(i\hat{p}\ddot{\mathbf{X}})e^{i\hat{p}\mathbf{X}}\right]\left\{\sum_{n>0}\frac{8}{n^2}(\alpha^0_{-n}\tilde{\alpha}^0_{-n})+\sum_{n>0}\frac{4}{ng} \right\},\label{boundary:eq:alpha1-4'}\\
&\frac{ig}{\sqrt{2\alpha'}}i\sqrt{\frac{\alpha'}{2}}\mathrm{tr}\left[\dot{\mathbf{X}}^ie^{i\hat{p}\mathbf{X}}\right]\left\{\sum_{n>0}\frac{-4}{n}(\alpha^0_{-n}\tilde{\alpha}^i_{-n}+\tilde{\alpha}^0_{-n}\alpha^i_{-n}) \right\}.\label{boundary:eq:alpha1-5'}
\end{align} 
\end{subequations}
The numerical factor $1/2!$ in \eqref{boundary:eq:alpha1-4'} arises because we restore the abbreviation \eqref{boundary:eq:rescale}. We note that the last term in \eqref{boundary:eq:alpha1-1'} at $\sigma=0$ and the last term in \eqref{boundary:eq:alpha1-4'} diverse. These singularities vanish when the equation of motions for $\mathbf{X}^i$ is satisfied as shown in subsection \ref{boundary:subsection:3-3}. 

%Note that these all terms satisfy $\frac{d^n}{d\hat{p}^n}[\cdots]\Big|_{\hat{p}=0} = 0 $ for $n \ge 3$.

\subsubsection*{\underline{${\alpha'}^{3/2}$ order}}
In this subsection we focus on $(\alpha)^0, (\alpha)^2$ parts, and ignore $(\alpha)^p\ (p\ge4)$ parts of the boundary state. The $(\alpha)^1$ part does not satisfy the level matching condition, and thus should vanish. Under this restriction, there is no term which contributes to the boundary state at ${\alpha'}^{3/2}$ order. We consider $(\mathbf{X}^i)^3$ as an example. $(\mathbf{X}^i)^3$ in the table represents
\begin{equation*}
\begin{split}
   \int_0^1\!\!d\sigma_1\int_0^1\!\!d\sigma_2\int_0^1\!\!d\sigma_3\ \mathrm{trP} & \left[e^{i(1-\sigma_{13})\hat{p}\mathbf{X}}\mathbf{X}^{i_1}e^{i\sigma_{12}\hat{p}\mathbf{X}}\mathbf{X}^{i_2}e^{i\sigma_{23}\hat{p}\mathbf{X}}\mathbf{X}^{i_3}\right]  \sum_{n,m>0}\frac{8m}{g}\\
&\left\{ \delta^{i_2i_3}\cos 2\pi m \sigma_{23}(\tilde{\alpha}_{-n}^{i_1}e^{2\pi i n \sigma_1}+\alpha_{-n}^{i_1}e^{-2\pi i n \sigma_1}) \right.\\
&+ \delta^{i_1i_3}\cos 2\pi m \sigma_{13}(\tilde{\alpha}_{-n}^{i_2}e^{2\pi i n \sigma_2}+\alpha_{-n}^{i_2}e^{-2\pi i n \sigma_2})\\
&+\left. \delta^{i_1i_2}\cos 2\pi m \sigma_{12}(\tilde{\alpha}_{-n}^{i_3}e^{2\pi i n \sigma_3}+\alpha_{-n}^{i_3}e^{-2\pi i n \sigma_3}) \right\}
\end{split}
\end{equation*}
where we ignore $(\alpha)^3$ terms. Change variables in the second line change as (ii) $\sigma_1=\sigma_2'+1. \sigma_2=\sigma_3'+1, \sigma_3=\sigma_1'$, and in the third line as (iii) $\sigma_1=\sigma_3'+1. \sigma_2=\sigma_1', \sigma_3=\sigma_2'$, then we have 
\begin{equation*}
\begin{split}
    \int_0^1\!\!d\sigma_1\int_0^1\!\!d\sigma_{12}\int_0^{1-\sigma_{12}}\!\!d\sigma_{23}\ 
&\mathrm{trP}\left[e^{i(1-\sigma_{13})\hat{p}\mathbf{X}}\mathbf{X}^{i_1}e^{i\sigma_{12}\hat{p}\mathbf{X}}\mathbf{X}^{i_2}e^{i\sigma_{23}\hat{p}\mathbf{X}}\mathbf{X}^{i_3}\right] \sum_{n,m>0}\frac{8m}{g}\\
&\left\{ \delta^{i_2i_3}\cos 2\pi m \sigma_{23}(\tilde{\alpha}_{-n}^{i_1}e^{2\pi i n \sigma_1}+\alpha_{-n}^{i_1}e^{-2\pi i n \sigma_1}) \right\} =0. 
\end{split}
\end{equation*}
The integration domains are shown in figure \ref{boundary:fig:3sigma}, where (i) represents the original domain of integration, (ii) and (iii) displays domains of integration after the variable changes (ii) and (iii) respectively.

\EPSFIGURE[ht]{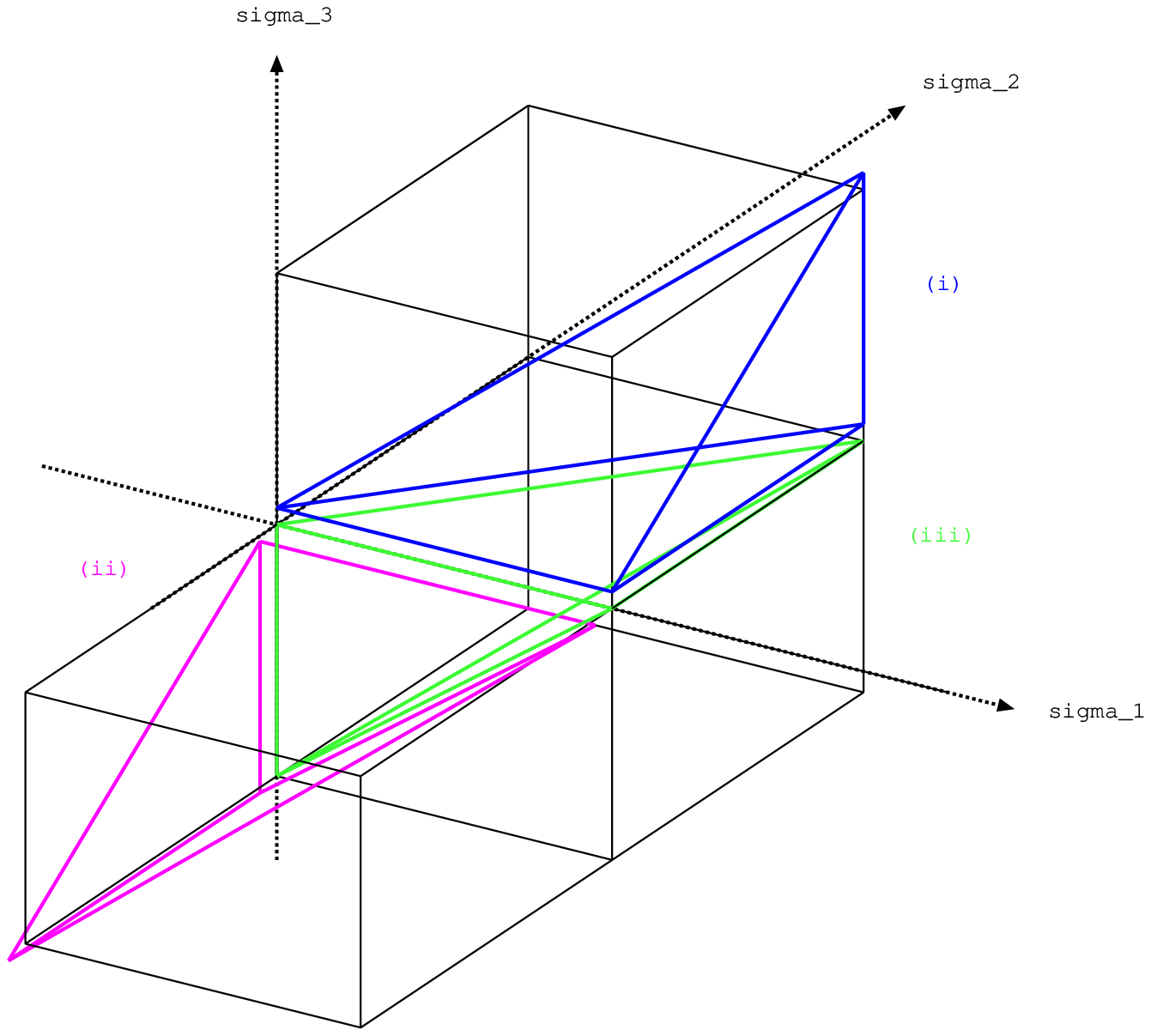,width=0.6\textwidth}{Change of integral variables $\sigma_1,\sigma_2, \sigma_3$\label{boundary:fig:3sigma}}

We can see the other terms vanish by similar calculations.

\subsubsection*{\underline{order ${\alpha'}^2$}}
At order ${\alpha'}^2$ we can see from table \ref{boundary:table:2} that the $(\alpha\tilde{\alpha})^0$ part includes 11 terms, and the $\alpha_{-n}\tilde{\alpha}_{-n}$ part contains 25 terms.  In a way similar to the case at order ${\alpha'}^{1/2}$, those terms which break the level matching condition become zero after integration. We show all possible terms of the boundary state at order ${\alpha'}^2$ below.

\paragraph*{$(\alpha)^2$ term}
\begin{itemize}
\item $\mathrm{trP''}[(i\hat{p}\dot{\mathbf{X}})_{\sigma_1}(i\hat{p}\dot{\mathbf{X}})_{\sigma_2}(i\hat{p}\dot{\mathbf{X}})_{\sigma_3}(i\hat{p}\dot{\mathbf{X}})_{\sigma_4}e^{i\int_0^1d\sigma \hat{p}\mathbf{X}}A^{0000}_{4,2}(\sigma_1,\sigma_2,\sigma_3,\sigma_4)]$
\item $\mathrm{trP''}[(i\hat{p}\dot{\mathbf{X}})_{\sigma_1}(i\hat{p}\dot{\mathbf{X}})_{\sigma_2}(i\hat{p}\dot{\mathbf{X}})_{\sigma_3}\mathbf{X}^{i_1}_{\sigma_4}e^{i\int_0^1d\sigma \hat{p}\mathbf{X}}A^{000}_{3,1}(\sigma_1,\sigma_2,\sigma_3)A_{1,1}^{i_1}(\sigma_4)]$
\item $\mathrm{trP''}[(i\hat{p}\dot{\mathbf{X}})_{\sigma_1}(i\hat{p}\dot{\mathbf{X}})_{\sigma_2}\mathbf{X}^{i_1}_{\sigma_3}\mathbf{X}^{i_2}_{\sigma_4}e^{i\int_0^1d\sigma \hat{p}\mathbf{X}}A^{00}_{2,0}(\sigma_1,\sigma_2)A_{2,2}^{i_1i_2}(\sigma_3,\sigma_4)]$
\item $\mathrm{trP''}[(i\hat{p}\dot{\mathbf{X}})_{\sigma_1}(i\hat{p}\dot{\mathbf{X}})_{\sigma_2}\mathbf{X}^{i_1}_{\sigma_3}\mathbf{X}^{i_2}_{\sigma_4}e^{i\int_0^1d\sigma \hat{p}\mathbf{X}}A^{00}_{2,2}(\sigma_1,\sigma_2)A_{2,0}^{i_1i_2}(\sigma_3,\sigma_4)]$
\item $\mathrm{trP''}[(i\hat{p}\dot{\mathbf{X}})_{\sigma_1}\mathbf{X}^{i_1}_{\sigma_2}\mathbf{X}^{i_2}_{\sigma_3}\mathbf{X}^{i_3}_{\sigma_4}e^{i\int_0^1d\sigma \hat{p}\mathbf{X}}A_{1,1}^{0}(\sigma_1)A_{3,1}^{i_1i_2i_3}(\sigma_2,\sigma_3,\sigma_4)]$  
\item $\mathrm{trP''}[\mathbf{X}^{i_1}_{\sigma_1}\mathbf{X}^{i_2}_{\sigma_2}\mathbf{X}^{i_3}_{\sigma_3}\mathbf{X}^{i_4}_{\sigma_4}e^{i\int_0^1d\sigma \hat{p}\mathbf{X}}A_{4,2}^{i_1i_2i_3i_4}(\sigma_1, \sigma_2,\sigma_3,\sigma_4)]$
\item $\mathrm{trP''}[(i\hat{p}\ddot{\mathbf{X}})_{\sigma_1}(i\hat{p}\dot{\mathbf{X}})_{\sigma_2}(i\hat{p}\dot{\mathbf{X}})_{\sigma_3}e^{i\int_0^1d\sigma \hat{p}\mathbf{X}}A_{4,2}^{0000}(\sigma_1, \sigma_1,\sigma_2,\sigma_3)]$
\item $\mathrm{trP''}[(i\hat{p}\ddot{\mathbf{X}})_{\sigma_1}(i\hat{p}\dot{\mathbf{X}})_{\sigma_2}\mathbf{X}^{i_1}_{\sigma_3}e^{i\int_0^1d\sigma \hat{p}\mathbf{X}}A_{3,1}^{000}(\sigma_1, \sigma_1,\sigma_2)A_{1,1}^{i_1}(\sigma_3)]$
\item $\mathrm{trP''}[(i\hat{p}\ddot{\mathbf{X}})_{\sigma_1}\mathbf{X}^{i_1}_{\sigma_2}\mathbf{X}^{i_2}_{\sigma_3}e^{i\int_0^1d\sigma \hat{p}\mathbf{X}}A_{2,2}^{00}(\sigma_1, \sigma_1)A_{2,0}^{i_1i_2}(\sigma_2,\sigma_3)]$
\item $\mathrm{trP''}[(i\hat{p}\ddot{\mathbf{X}})_{\sigma_1}\mathbf{X}^{i_1}_{\sigma_2}\mathbf{X}^{i_2}_{\sigma_3}e^{i\int_0^1d\sigma \hat{p}\mathbf{X}}A_{2,0}^{00}(\sigma_1, \sigma_1)A_{2,2}^{i_1i_2}(\sigma_2,\sigma_3)]$
\item $\mathrm{trP''}[\dot{\mathbf{X}}^{i_1}_{\sigma_1}(i\hat{p}\dot{\mathbf{X}})_{\sigma_2}(i\hat{p}\dot{\mathbf{X}})_{\sigma_3}e^{i\int_0^1d\sigma \hat{p}\mathbf{X}}A_{3,1}^{00}(\sigma_1, \sigma_2,\sigma_3)A_{1,1}^{i_1}(\sigma_1)]$
\item $\mathrm{trP''}[\dot{\mathbf{X}}^{i_1}_{\sigma_1}(i\hat{p}\dot{\mathbf{X}})_{\sigma_2}\mathbf{X}^{i_2}_{\sigma_3}e^{i\int_0^1d\sigma \hat{p}\mathbf{X}}A_{2,2}^{00}(\sigma_1, \sigma_2)A_{2,0}^{i_1i_2}(\sigma_1,\sigma_3)]$
\item $\mathrm{trP''}[\dot{\mathbf{X}}^{i_1}_{\sigma_1}(i\hat{p}\dot{\mathbf{X}})_{\sigma_2}\mathbf{X}^{i_2}_{\sigma_3}e^{i\int_0^1d\sigma \hat{p}\mathbf{X}}A_{2,0}^{00}(\sigma_1, \sigma_2)A_{2,2}^{i_1i_2}(\sigma_1,\sigma_3)]$
\item $\mathrm{trP''}[\dot{\mathbf{X}}^{i_1}_{\sigma_1}\mathbf{X}^{i_2}_{\sigma_2}\mathbf{X}^{i_3}_{\sigma_3}e^{i\int_0^1d\sigma \hat{p}\mathbf{X}}A_{1,1}^{0}(\sigma_1)A_{3,1}^{i_1i_2i_3}(\sigma_1,\sigma_2,\sigma_3)]$
\item $\mathrm{trP''}[(i\hat{p}\mathbf{X}^{(3)})_{\sigma_1}(i\hat{p}\dot{\mathbf{X}})_{\sigma_2}e^{i\int_0^1d\sigma \hat{p}\mathbf{X}}A_{4,2}^{0000}(\sigma_1,\sigma_1,\sigma_1,\sigma_2)]$
\item $\mathrm{trP''}[(i\hat{p}\mathbf{X}^{(3)})_{\sigma_1}\mathbf{X}^{i_1}_{\sigma_2}e^{i\int_0^1d\sigma \hat{p}\mathbf{X}}A_{3,1}^{000}(\sigma_1,\sigma_1,\sigma_1)A_1^{i_1}(\sigma_2)]$
\item $\mathrm{trP''}[\ddot{\mathbf{X}}^{i_1}_{\sigma_1}(i\hat{p}\dot{\mathbf{X}})_{\sigma_2}e^{i\int_0^1d\sigma \hat{p}\mathbf{X}}A_{3,1}^{000}(\sigma_1,\sigma_1,\sigma_2)A_1^{i_1}(\sigma_1)]$
\item $\mathrm{trP''}[\ddot{\mathbf{X}}^{i_1}_{\sigma_1}\mathbf{X}^{i_2}_{\sigma_2}e^{i\int_0^1d\sigma \hat{p}\mathbf{X}}A_{2,2}^{00}(\sigma_1,\sigma_1)A_{2,0}^{i_1i_2}(\sigma_1,\sigma_2)]$
\item $\mathrm{trP''}[\ddot{\mathbf{X}}^{i_1}_{\sigma_1}\mathbf{X}^{i_2}_{\sigma_2}e^{i\int_0^1d\sigma \hat{p}\mathbf{X}}A_{2,0}^{00}(\sigma_1,\sigma_1)A_{2,2}^{i_1i_2}(\sigma_1,\sigma_2)]$
\item $\mathrm{trP''}[(i\hat{p}\ddot{\mathbf{X}})_{\sigma_1}(i\hat{p}\ddot{\mathbf{X}})_{\sigma_2}e^{i\int_0^1d\sigma \hat{p}\mathbf{X}}A_{4,2}^{0000}(\sigma_1,\sigma_1,\sigma_2,\sigma_2)]$
\item $\mathrm{trP''}[(i\hat{p}\ddot{\mathbf{X}})_{\sigma_1}\dot{\mathbf{X}}^{i_1}_{\sigma_2}e^{i\int_0^1d\sigma \hat{p}\mathbf{X}}A_{3,1}^{000}(\sigma_1,\sigma_1,\sigma_2)A_1^{i_1}(\sigma_2)]$
\item $\mathrm{trP''}[\dot{\mathbf{X}}^{i_1}_{\sigma_1}\dot{\mathbf{X}}^{i_2}_{\sigma_2}e^{i\int_0^1d\sigma \hat{p}\mathbf{X}}A_{2,2}^{00}(\sigma_1,\sigma_2)A_{2,0}^{i_1i_2}(\sigma_1,\sigma_2)]$
\item $\mathrm{trP''}[\dot{\mathbf{X}}^{i_1}_{\sigma_1}\dot{\mathbf{X}}^{i_2}_{\sigma_2}e^{i\int_0^1d\sigma \hat{p}\mathbf{X}}A_{2,0}^{00}(\sigma_1,\sigma_2)A_{2,2}^{i_1i_2}(\sigma_1,\sigma_2)]$
\item $\mathrm{trP''}[(i\hat{p}\mathbf{X}^{(4)})_{\sigma_1}e^{i\int_0^1d\sigma \hat{p}\mathbf{X}}A_{4,2}^{0000}(\sigma_1,\sigma_1,\sigma_1,\sigma_1)]$
\item $\mathrm{trP''}[\mathbf{X}^{i(3)}_{\sigma_1}e^{i\int_0^1d\sigma \hat{p}\mathbf{X}}A_{3,1}^{000}(\sigma_1,\sigma_1,\sigma_1)A_1^{i_1}(\sigma_1)]$
\end{itemize}
\paragraph*{$(\alpha)^0$ term}
\begin{itemize}
\item $\mathrm{trP''}[(i\hat{p}\dot{\mathbf{X}})_{\sigma_1}(i\hat{p}\dot{\mathbf{X}})_{\sigma_2}(i\hat{p}\dot{\mathbf{X}})_{\sigma_3}(i\hat{p}\dot{\mathbf{X}})_{\sigma_4}e^{i\int_0^1d\sigma \hat{p}\mathbf{X}}A^{0000}_{4,0}(\sigma_1,\sigma_2,\sigma_3,\sigma_4)]$
\item $\mathrm{trP''}[(i\hat{p}\dot{\mathbf{X}})_{\sigma_1}(i\hat{p}\dot{\mathbf{X}})_{\sigma_2}\mathbf{X}^{i_1}_{\sigma_3}\mathbf{X}^{i_2}_{\sigma_4}e^{i\int_0^1d\sigma \hat{p}\mathbf{X}}A^{00}_{2,0}(\sigma_1,\sigma_2)A_{2,0}^{i_1i_2}(\sigma_3,\sigma_4)]$
\item $\mathrm{trP''}[\mathbf{X}^{i_1}_{\sigma_1}\mathbf{X}^{i_2}_{\sigma_2}\mathbf{X}^{i_3}_{\sigma_3}\mathbf{X}^{i_4}_{\sigma_4}e^{i\int_0^1d\sigma \hat{p}\mathbf{X}}A_{4,0}^{i_1i_2i_3i_4}(\sigma_1, \sigma_2,\sigma_3,\sigma_4)]$
\item $\mathrm{trP''}[(i\hat{p}\ddot{\mathbf{X}})_{\sigma_1}(i\hat{p}\dot{\mathbf{X}})_{\sigma_2}(i\hat{p}\dot{\mathbf{X}})_{\sigma_3}e^{i\int_0^1d\sigma \hat{p}\mathbf{X}}A_{4,0}^{0000}(\sigma_1, \sigma_1,\sigma_2,\sigma_3)]$
\item $\mathrm{trP''}[(i\hat{p}\ddot{\mathbf{X}})_{\sigma_1}\mathbf{X}^{i_1}_{\sigma_2}\mathbf{X}^{i_2}_{\sigma_3}e^{i\int_0^1d\sigma \hat{p}\mathbf{X}}A_{2,0}^{00}(\sigma_1, \sigma_1)A_{2,0}^{i_1i_2}(\sigma_2,\sigma_3)]$
\item $\mathrm{trP''}[\dot{\mathbf{X}}^{i_1}_{\sigma_1}(i\hat{p}\dot{\mathbf{X}})_{\sigma_2}\mathbf{X}^{i_2}_{\sigma_3}e^{i\int_0^1d\sigma \hat{p}\mathbf{X}}A_{2,0}^{00}(\sigma_1, \sigma_2)A_{2,0}^{i_1i_2}(\sigma_1,\sigma_3)]$
\item $\mathrm{trP''}[(i\hat{p}\mathbf{X}^{(3)})_{\sigma_1}(i\hat{p}\dot{\mathbf{X}})_{\sigma_2}e^{i\int_0^1d\sigma \hat{p}\mathbf{X}}A_{4,0}^{0000}(\sigma_1,\sigma_1,\sigma_1,\sigma_2)]$
\item $\mathrm{trP''}[\ddot{\mathbf{X}}^{i_1}_{\sigma_1}\dot{\mathbf{X}}^{i_2}_{\sigma_2}e^{i\int_0^1d\sigma \hat{p}\mathbf{X}}A_{2,0}^{00}(\sigma_1,\sigma_1)A_{2,0}^{i_1i_2}(\sigma_1,\sigma_2)]$
\item $\mathrm{trP''}[(i\hat{p}\ddot{\mathbf{X}})_{\sigma_1}(i\hat{p}\ddot{\mathbf{X}})_{\sigma_2}e^{i\int_0^1d\sigma \hat{p}\mathbf{X}}A_{4,0}^{0000}(\sigma_1,\sigma_1,\sigma_2,\sigma_2)]$
\item $\mathrm{trP''}[\dot{\mathbf{X}}_{\sigma_1}\dot{\mathbf{X}}^{i_2}_{\sigma_2}e^{i\int_0^1d\sigma \hat{p}\mathbf{X}}A_{2,0}^{00}(\sigma_1,\sigma_2)A_{2,0}^{i_1i_2}(\sigma_1,\sigma_2)]$
\item $\mathrm{trP''}[(i\hat{p}\mathbf{X})^{(4)}_{\sigma_1}e^{i\int_0^1d\sigma \hat{p}\mathbf{X}}A_{4,0}^{0000}(\sigma_1,\sigma_1,\sigma_1,\sigma_1)]$
\end{itemize}
%We note that all these terms satisfy $\frac{d^n}{d\hat{p}^n}[\cdots]_{\hat{p}=0} = 0 $ for $n \ge 5$. 

In what follows we calculate the boundary state estimated at $\hat{p}=0$, namely $\langle p=0| B\rangle$. This gives coupling to closed string fields which are constant in Dirichlet directions, and total charges integrated over the Dirichlet directions as seen from \eqref{boundary:eq:derivative-coupling}. and \eqref{boundary:eq:total-charge} respectively. At $\hat{p}=0$ we have the following terms.\\
%The boundary state estimated at $\hat{p}=0$ gives linear couplings to closed string fields independent of the coordinates along Dirichlet directions. These couplings contain information about an action of multiple D0-brane in a flat background, and the total energy-momentum tensor. 
\paragraph*{$(\alpha)^2$ term at $\hat{p}=0$}
\begin{equation*}
 \mathrm{tr}[\mathbf{X}^{i_1}\mathbf{X}^{i_2}\mathbf{X}^{i_3}\mathbf{X}^{i_4}]\int_0^1\!\!d\sigma_1\int_0^{\sigma_1}\!\!d\sigma_2\int_0^{\sigma_2}d\sigma_3\int_0^{\sigma_3}\!\!d\sigma_4\ A_{4,2}^{i_1i_2i_3i_4}(\sigma_1, \sigma_2,\sigma_3,\sigma_4)\ ,
\end{equation*}
\begin{equation*}
\begin{split}
&\mathrm{tr}[\dot{\mathbf{X}}^{i_1}\mathbf{X}^{i_2}\mathbf{X}^{i_3}]\int_0^1\!\!d\sigma_1\int_0^{\sigma_1}\!\!d\sigma_2\int_0^{\sigma_2}\!\!d\sigma_3\ A_{1,1}^{0}(\sigma_1)A_{3,1}^{i_1i_2i_3}(\sigma_1,\sigma_2,\sigma_3)\\
&+\mathrm{tr}[\mathbf{X}^{i_3}\dot{\mathbf{X}}^{i_1}\mathbf{X}^{i_2}]\int_0^1\!\!d\sigma_1\int_0^{\sigma_1}\!\!d\sigma_2\int_0^{\sigma_2}\!\!d\sigma_3\ A_{1,1}^{0}(\sigma_2)A_{3,1}^{i_1i_2i_3}(\sigma_2,\sigma_3,\sigma_1)\\
&+\mathrm{tr}[\mathbf{X}^{i_2}\mathbf{X}^{i_3}\dot{\mathbf{X}}^{i_1}]\int_0^1\!\!d\sigma_1\int_0^{\sigma_1}\!\!d\sigma_2\int_0^{\sigma_2}\!\!d\sigma_3\ A_{1,1}^{0}(\sigma_3)A_{3,1}^{i_1i_2i_3}(\sigma_3,\sigma_1,\sigma_2)\ ,
\end{split}
\end{equation*}
\begin{equation*}
\begin{split}
&\mathrm{tr}[\ddot{\mathbf{X}}^{i_1}\mathbf{X}^{i_2}]\int_0^1\!\!d\sigma_1\int_0^{\sigma_1}\!\!d\sigma_2\ \left(A_{2,2}^{00}(\sigma_1,\sigma_1)A_{2,0}^{i_1i_2}(\sigma_1,\sigma_2) + A_{2,0}^{00}(\sigma_1,\sigma_1)A_{2,2}^{i_1i_2}(\sigma_1,\sigma_2)\right)\\
&+\mathrm{tr}[\mathbf{X}^{i_2}\ddot{\mathbf{X}}^{i_1}]\int_0^1\!\!d\sigma_1\int_0^{\sigma_1}\!\!d\sigma_2\ \left(A_{2,2}^{00}(\sigma_2,\sigma_2)A_{2,0}^{i_1i_2}(\sigma_2,\sigma_1)+A_{2,0}^{00}(\sigma_2,\sigma_2)A_{2,2}^{i_1i_2}(\sigma_2,\sigma_1)\right)\ ,
\end{split}
\end{equation*}
\begin{equation*}
\mathrm{tr}[\dot{\mathbf{X}}^{i_1}\dot{\mathbf{X}}^{i_2}]\int_0^1\!\!d\sigma_1\int_0^{\sigma_1}\!\!d\sigma_2\ \left(A_{2,2}^{00}(\sigma_1,\sigma_2)A_{2,0}^{i_1i_2}(\sigma_1,\sigma_2)+A_{2,0}^{00}(\sigma_1,\sigma_2)A_{2,2}^{i_1i_2}(\sigma_1,\sigma_2)\right)\ ,
\end{equation*}
\begin{equation*}
\mathrm{tr}[\mathbf{X}^{i(3)}]\int_0^1\!\!d\sigma_1\ A_{3,1}^{000}(\sigma_1,\sigma_1,\sigma_1)A_1^{i_1}(\sigma_1).
\end{equation*}
\paragraph*{$(\alpha)^0$ term  at $\hat{p}=0$}
\begin{equation*}
 \mathrm{tr}[\mathbf{X}^{i_1}\mathbf{X}^{i_2}\mathbf{X}^{i_3}\mathbf{X}^{i_4}]\int_0^1\!\!d\sigma_1\int_0^{\sigma_1}\!\!d\sigma_2\int_0^{\sigma_2}d\sigma_3\int_0^{\sigma_3}\!\!d\sigma_4\ A_{4,0}^{i_1i_2i_3i_4}(\sigma_1, \sigma_2,\sigma_3,\sigma_4)\ ,
\end{equation*}
\begin{equation*}
\begin{split}
&\mathrm{tr}[\ddot{\mathbf{X}}^{i_1}\mathbf{X}^{i_2}]\int_0^1\!\!d\sigma_1\int_0^{\sigma_1}\!\!d\sigma_2\ A_{2,0}^{00}(\sigma_1,\sigma_1)A_{2,0}^{i_1i_2}(\sigma_1,\sigma_2)\\
&+\mathrm{tr}[\mathbf{X}^{i_2}\ddot{\mathbf{X}}^{i_1}]\int_0^1\!\!d\sigma_1\int_0^{\sigma_1}\!\!d\sigma_2\ A_{2,0}^{00}(\sigma_2,\sigma_2)A_{2,0}^{i_1i_2}(\sigma_2,\sigma_1)\ ,
\end{split}
\end{equation*}
\begin{equation*}
\mathrm{tr}[\dot{\mathbf{X}}^{i_1}\dot{\mathbf{X}}^{i_2}]\int_0^1\!\!d\sigma_1\int_0^{\sigma_1}\!\!d\sigma_2\ A_{2,0}^{00}(\sigma_1,\sigma_2)A_{2,0}^{i_1i_2}(\sigma_1,\sigma_2).
\end{equation*}
Because we need not to take care of $e^{i\hat{p}\mathbf{X}}$ in ordering at $\hat{p}=0$, all $\sigma$'s disappear from matrix parts. Therefore it is sufficient to integrate oscillation operator parts over $\sigma$'s. For $(\alpha)^0$ terms we need integrations: 
\begin{align*}
   \int_0^1\!\!d\sigma_1\int_0^{\sigma_1}\!\!d\sigma_2\int_0^{\sigma_2}\!\!d\sigma_3\int_0^{\sigma_3}\!\!d\sigma_4\ \cos2\pi n \sigma_{12}\cos2\pi m \sigma_{34} & = 0 \\
   \int_0^1\!\!d\sigma_1\int_0^{\sigma_1}\!\!d\sigma_2\int_0^{\sigma_2}\!\!d\sigma_3\int_0^{\sigma_3}\!\!d\sigma_4\ \cos2\pi n \sigma_{13}\cos2\pi m \sigma_{24} & = \frac{1}{16n^2\pi^2}\delta_{n,m}\\
   \int_0^1\!\!d\sigma_1\int_0^{\sigma_1}\!\!d\sigma_2\int_0^{\sigma_2}\!\!d\sigma_3\int_0^{\sigma_3}\!\!d\sigma_4\ \cos2\pi n \sigma_{14}\cos2\pi m \sigma_{23} & = - \frac{1}{16n^2\pi^2}\delta_{n,m}
\end{align*}
\begin{equation*}
   \int_0^1d\sigma \cos2\pi n\sigma \cos2\pi m\sigma =\frac{1}{2}\delta_{n,m}.
\end{equation*}
We omit to write down integrations which are needed in calculations of $(\alpha)^2$ terms. After calculations the results at order ${\alpha'}^2$ at $\hat{p}=0$ are
\begin{subequations}
\begin{equation}
\left(\frac{ig}{\sqrt{2\alpha'}}\right)^4\mathrm{tr}[\mathbf{X}^i,\mathbf{X}^k][\mathbf{X}^j,\mathbf{X}^k]\sum_{n>0}\frac{1}{gn\pi^2}(\tilde{\alpha}^i_{-n}\alpha^j_{-n}+\alpha^i_{-n}\tilde{\alpha}^j_{-n})\label{boundary:eq:p=0-1}
\end{equation}
\begin{equation}
   i\sqrt{\frac{\alpha'}{2}}\left(\frac{ig}{\sqrt{2\alpha'}}\right)^3\mathrm{tr}\left[[\mathbf{X}^i,\mathbf{X}^j]\dot{\mathbf{X}}^j\right]\sum_{n>0}\frac{4i}{gn\pi}(\tilde{\alpha}^0_{-n}\alpha^i_{-n}-\alpha^0_{-n}\tilde{\alpha}^i_{-n})\label{boundary:eq:p=0-2}
\end{equation}
\begin{equation}
   \left(i\sqrt{\frac{\alpha'}{2}}\right)^2\left(\frac{ig}{\sqrt{2\alpha'}}\right)^2\mathrm{tr}[\dot{\mathbf{X}}^{i_1}\dot{\mathbf{X}}^{i_2}]
\sum_{n>0}\frac{4}{gn}(2\delta^{i_1i_2}\tilde{\alpha}^0_{-n}\alpha^0_{-n}+\tilde{\alpha}^{i_1}_{-n}\alpha^{i_2}_{-n}+\alpha^{i_1}_{-n}\tilde{\alpha}^{i_2}_{-n})\label{boundary:eq:p=0-3}
\end{equation}
\begin{equation}
   \frac{1}{3!}\left(i\sqrt{\frac{\alpha'}{2}}\right)^3\frac{ig}{\sqrt{2\alpha'}}\mathrm{tr}[\mathbf{X}^{i(3)}]\zeta(1)\sum_{n>0}\frac{-48}{gn}(\tilde{\alpha}^0_{-n}\alpha^i_{-n}+\alpha^0_{-n}\tilde{\alpha}^i_{-n})\label{boundary:eq:p=0-4}
\end{equation}
\begin{equation}
   \left(\frac{ig}{\sqrt{2\alpha'}}\right)^4\mathrm{tr}\left[\mathbf{X}^i,\mathbf{X}^j\right]^2\zeta(0)\frac{1}{2g^2\pi^2}\label{boundary:eq:p=0-5}
\end{equation}
\begin{equation}
   \left(i\sqrt{\frac{\alpha'}{2}}\right)^2\left(\frac{ig}{\sqrt{2\alpha'}}\right)^2\mathrm{tr}[\dot{\mathbf{X}}^i\dot{\mathbf{X}}^i]\frac{4}{g^2}\zeta(0)\label{boundary:eq:p=0-6}
\end{equation}
\end{subequations}
where we restore the abbreviation \eqref{boundary:eq:rescale}. $\zeta(p)$ is the Riemann zeta function
\begin{equation*}
   \zeta(p)=\sum_{n=1}^\infty \frac{1}{n^p}.
\end{equation*} 
$\zeta(p)$ is finite except at $p=1$. For example $\zeta(0)=-\frac{1}{2}, \zeta(-1)=-\frac{1}{12}$. 

\subsection*{Remark}
We calculate the boundary state up to order ${\alpha'}^2$ in this paper. It is possible to derive higher order terms by further computations straightforwardly. Higher order terms in the boundary state include information about $\alpha'$ corrections to an action, and an energy-momentum tensor of multiple D0-branes.

\subsection{Result of calculation of boundary state}\label{boundary:subsection:3-result}
In subsection \ref{boundary:subsection:3-1} we have calculated the boundary state \eqref{boundary:eq:boundary} of multiple D0-branes with an arbitrary configuration of the scalar field defined by using the Wilson loop factor. In this subsection we summarize the results obtained in subsection \ref{boundary:subsection:3-1}. Readers who are interested only in our results of coupling to closed string can skip to section \ref{paper:section:4} where our main results of the closed string couplings are presented. 

The resulting boundary state has the form of
\begin{equation}
   |B\rangle = \Bigl\{f(t,\hat{p})+\left(a_{\mu\nu}(t,\hat{p})+c_{\mu\nu}(t,\hat{p})\right)\alpha_{-1}^\mu\tilde{\alpha}_{-1}^\nu+\cdots\Bigr\}|D0\rangle \label{boundary:eq:boundary-result}
\end{equation}
where $a_{\mu\nu}=a_{\mu\nu}, c_{\mu\nu}=-c_{\mu\nu}$. Explicit formulas for $f(t,k),a_{\mu\nu}(t,k)$ and $c_{\mu\nu}(t,k)$ are shown below. At order ${\alpha'}^0$ we have
\begin{equation}
\begin{split}
   f(t,k)_{{\alpha'}^0} & = \mathrm{tr}[1] \\
   a_{\mu\nu}(t,k)_{{\alpha'}^0}  &= 0\\
   c_{\mu\nu}(t,k)_{{\alpha'}^0}  &= 0. 
\end{split}\label{boundary:eq:result-alpha0}
\end{equation}
This can be seen from \eqref{boundary:eq:alpha0}. At order $\alpha'$ we find\\
\begin{equation}
\begin{split}
f(t,k)_{\alpha'}
= &  -\frac{\alpha'}{g}\int_0^1\!\!d\sigma\ \mathrm{tr}\left[e^{i(1-\sigma)\hat{p}\mathbf{X}}(i\hat{p}\dot{\mathbf{X}})e^{i\sigma\hat{p}\mathbf{X}}(i\hat{p}\dot{\mathbf{X}})\right]\sum_{n>0}\frac{\cos2\pi n\sigma}{n} \\
& -\frac{g}{(2\pi)^2\alpha'}\int_0^1\!\!d\sigma\ \mathrm{tr}\left[e^{i(1-\sigma)\hat{p}\mathbf{X}}[\mathbf{X}^i,i\hat{p}\mathbf{X}]e^{i\sigma\hat{p}\mathbf{X}}[\mathbf{X}^i,i\hat{p}\mathbf{X}]\right]\sum_{n>0}\frac{\cos2\pi n\sigma}{n}
\\
a_{00}(t,k)_{\alpha'}
= & - 2 \alpha'\int_0^1\!\!d\sigma\ \mathrm{tr}[e^{i(1-\sigma)\hat{p}\mathbf{X}}(i\hat{p}\dot{\mathbf{X}})e^{i\sigma\hat{p}\mathbf{X}}(i\hat{p}\dot{\mathbf{X}})]\cos2\pi\sigma  \\
& - 2 \alpha' \mathrm{tr}\left[(i\hat{p}\ddot{\mathbf{X}})e^{i\hat{p}\mathbf{X}}\right]
\\
a_{0i}(t,k)_{\alpha'}
= &\ g\int_0^1\!\!d\sigma\ \mathrm{tr}\left[e^{i(1-\sigma)\hat{p}\mathbf{X}}(i\hat{p}\dot{\mathbf{X}})e^{i\sigma\hat{p}\mathbf{X}}\mathbf{X}^i\right]\cos 2\pi \sigma \\
& + 2g\ \mathrm{tr}\left[\dot{\mathbf{X}}^ie^{i\hat{p}\mathbf{X}}\right]
\\
a_{ij}(t,k)_{\alpha'}
= & -\frac{2g^2}{\alpha'}\int_0^1\!\!d\sigma\ \mathrm{tr}\left[e^{i(1-\sigma)\hat{p}\mathbf{X}}\mathbf{X}^{i}e^{i\sigma\hat{p}\mathbf{X}}\mathbf{X}^{j}\right]\cos 2\pi\sigma 
\\
c_{0i}(t,k)_{\alpha'}
= & -ig\int_0^1\!\!d\sigma\ \mathrm{tr}\left[e^{i(1-\sigma)\hat{p}\mathbf{X}}(i\hat{p}\dot{\mathbf{X}})e^{i\sigma\hat{p}\mathbf{X}}\mathbf{X}^i\right]\sin 2\pi \sigma
\\
c_{ij}(t,k)_{\alpha'}
= & \frac{2ig^2}{\alpha'}\int_0^1\!\!d\sigma\ \mathrm{tr}\left[e^{i(1-\sigma)\hat{p}\mathbf{X}}\mathbf{X}^{i}e^{i\sigma\hat{p}\mathbf{X}}\mathbf{X}^{j}\right]\sin 2\pi\sigma .
\end{split}\label{boundary:eq:result-alpha1}
\end{equation}
This can be seen from \eqref{boundary:eq:alpha1-1'}, \eqref{boundary:eq:alpha1-2'}, \eqref{boundary:eq:alpha1-3'}, \eqref{boundary:eq:alpha1-4'} and \eqref{boundary:eq:alpha1-5'}. Here we have eliminated the singular terms which are considered in subsection \ref{boundary:subsection:3-3}. In particular we can see that the second term in $f(t,k)_{\alpha'}$ arises from \eqref{boundary:eq:decompose-divergence}.

At $\hat{p}=0$ we find
\begin{equation}
\begin{split}
f(t,k=0)
= &\ \mathrm{tr}[1] -\frac{1}{2}\mathrm{tr}\left[\dot{\mathbf{X}}^i\dot{\mathbf{X}}^i\right]-\frac{1}{4}\left(\frac{g}{2\pi\alpha'}\right)^2\mathrm{tr}\left[\mathbf{X}^i,\mathbf{X}^j\right]^2\\
a_{00}(t,k=0)
= &\ 2g\mathrm{tr}[\dot{\mathbf{X}}^{i}\dot{\mathbf{X}}^{i}] \\
a_{0i}(t,k=0)
= &\ 2g\mathrm{tr}\left[\dot{\mathbf{X}}^i\right]\\
a_{ij}(t,k=0)
= &\ 2g\left(\frac{g}{2\pi\alpha'}\right)^2\mathrm{tr}[\mathbf{X}^i,\mathbf{X}^k][\mathbf{X}^j,\mathbf{X}^k] + 2g\mathrm{tr}[\dot{\mathbf{X}}^{i}\dot{\mathbf{X}}^{j}] \\
c_{0i}(t,k=0)
= & \frac{ig^2}{\pi\alpha'}\mathrm{tr}[[\mathbf{X}^i,\mathbf{X}^j]\dot{\mathbf{X}}^j]\\
c_{ij}(t,k=0)
= & \frac{ig^2}{\pi\alpha'}\mathrm{tr}[\mathbf{X}^i,\mathbf{X}^j]
\end{split}\label{boundary:eq:p=0}
\end{equation}
up to order ${\alpha'}^2$. This can be seen from \eqref{boundary:eq:p=0-1}, \eqref{boundary:eq:p=0-2}, \eqref{boundary:eq:p=0-3}, \eqref{boundary:eq:p=0-4}, \eqref{boundary:eq:p=0-5} and \eqref{boundary:eq:p=0-6}. $c^{ij}(t,k)$ arises only in the case that the number of D0-branes is infinite. In such a case we cannot use cyclicity of trace, and thus \eqref{boundary:eq:specialcase1} and \eqref{boundary:eq:specialcase2} do not hold. After calculations without utilizing the cyclicity of trace, we find $c^{ij}\simeq \mathrm{tr}[\mathbf{X}^i\mathbf{X}^j]$. Except in the case the size of $\mathbf{X}^i$ is infinite, this term vanishes.

\subsection{BRST invariance}\label{boundary:subsection:3-2}
In this subsection, we confirm that the boundary state \eqref{boundary:eq:boundary} is BRST invariant. The boundary state is BRST invariant for an arbitrary configuration of the non-Abelian scalar field $\mathbf{X}^i$. However, in this proof the existence of singularities in the boundary state is not accounted for. As shown in subsection \ref{boundary:subsection:3-3}, the boundary state is well-defined when $\mathbf{X}^i$ is on-shell. From another viewpoint, the divergences can be absorbed by a field redefinition \eqref{boundary:eq:redef}. This redefinition makes the boundary state finite, while it breaks the BRST invariance when $\mathbf{X}^i$ is off-shell. 

We denote that
\begin{align*}
 \hat{b}(\sigma) & = b_0^+ +\sum_{n\ne0}(b_n+\tilde{b}_{-n})e^{in\sigma} \\
 \hat{c}(\sigma) & = c_0^- +\sum_{n\ne0}(c_n-\tilde{c}_{-n})e^{in\sigma}.
\end{align*}
The BRST charge can be written as
\begin{equation*}
\begin{split}
   Q_B & = -\int_0^{2\pi}\!\!d\sigma\ \hat{c}\hat{\Pi}_\mu\partial\hat{X}^\mu(\sigma)
-\frac{1}{2}\int_0^{2\pi}\!\!d\sigma\ \hat{\pi}_b(\sigma)\left(\hat{\Pi}^\mu\hat{\Pi}_\mu(\sigma)+\partial_\sigma\hat{X}^\mu\partial_\sigma\hat{X}_\mu(\sigma)\right)
+Q_B^\mathrm{ghost}
\end{split}
\end{equation*}
where $Q_B^\mathrm{ghost}$ consists only of ghost fields. Because the ghost part of $|B\rangle$ is same as that of $|D0\rangle$, we can see $\hat{\pi}_{b}(\sigma)|B\rangle=0$, and $Q_B^\mathrm{ghost}|B\rangle=0 $. Considering that the boundary state of D-brane without excitation is BRST invariant, what we have to prove is
\begin{equation*}
   \left[\int_0^{2\pi}\!\!d\sigma\ \hat{c}\hat{\Pi}_\mu\partial\hat{X}^\mu(\sigma) , \mathrm{P}e^{i \int d\sigma' \hat{\Pi}\mathbf{\Phi}(\hat{X}^0)(\sigma')}\right]=0.
\end{equation*}
This leads to the BRST invariance of the boundary state
\begin{equation*}
      Q_B|B\rangle=0.
\end{equation*} 
We note that we need not to take trace to prove the BRST invariance because the BRST operator does not include open string fields. 
To avoid confusing, we denote transverse scalars on D0-branes by $\mathbf{\Phi}^i$ in this subsection. Note that $\hat{X}^\mu(\sigma)$ is the string embedding function. First we prove the following theorem by induction on power of $\hat{\Pi}\mathbf{\Phi}$:

\newtheorem{theorem}{Theorem}[section]
\begin{theorem}\label{boundary:theorem}
The equation \eqref{boundary:eq:theorem} holds for $n \in \mathbb{N}$, where $f(\sigma_n)$ is an arbitrary function which may not commute with $\hat{c}\hat{\Pi}_i\partial\hat{X}^i(\sigma)$.  
\begin{equation}
\begin{split}
&   \left[\int_0^{2\pi}\!\!d\sigma\ \hat{c}\hat{\Pi}_\mu\partial\hat{X}^\mu(\sigma), 
\int_0^{2\pi}\!\!d\sigma_1\ i\hat{\Pi}\mathbf{\Phi}(\hat{X}^0)(\sigma_1)\cdots\int_0^{\sigma_{n-1}}\!\!d\sigma_{n}\ i\hat{\Pi}\mathbf{\Phi}(\hat{X}^0)(\sigma_n) f(\sigma_n)\right]\\
=& -\int_0^{2\pi}\!\!d\sigma_1\ i\hat{\Pi}\mathbf{\Phi}(\hat{X}^0)(\sigma_1)\cdots\int_0^{\sigma_{n-1}}\!\!d\sigma_n\ \hat{c}\hat{\Pi}\mathbf{\Phi}(\hat{X}^0)(\sigma_n)\partial_{\sigma_n}f(\sigma_n)\\
&+\int_0^{2\pi}\!\!d\sigma_1\ i\hat{\Pi}\mathbf{\Psi}(\hat{X}^0)(\sigma_1)\cdots\int_0^{\sigma_{n-1}}\!\!d\sigma_{n}\ i\hat{\Pi}\mathbf{\Psi}(\hat{X}^0)(\sigma_n)\left[\int_0^{2\pi}\!\!d\sigma\ \hat{c}\hat{\Pi}_\mu\partial\hat{X}^\mu(\sigma), f(\sigma_n)\right]. 
\end{split}\label{boundary:eq:theorem}
\end{equation}
\end{theorem}
Here we abbreviate $i\hat{\Pi}_i(\sigma)\mathbf{\Psi}^i(\hat{X}^0(\sigma))$ as $i\hat{\Pi}\mathbf{\Psi}(\sigma)$, and  $\hat{c}(\sigma)\hat{\Pi}_i(\sigma)\mathbf{\Psi}^i(\hat{X}^0(\sigma))$ by $\hat{c}\hat{\Pi}\mathbf{\Psi}(\sigma)$. In preparation to prove the theorem we see that
\begin{equation}
\begin{split}
&\left[\int_0^{2\pi}\!\!d\sigma\ \hat{c}\hat{\Pi}_\mu\partial\hat{X}^\mu(\sigma), \int_0^{\sigma_{n-1}}\!\!d\sigma_n\ i\hat{\Pi}_i\mathbf{\Phi}^i (\sigma_n)f(\sigma_n)\right]\\ 
&\hspace{9em}=\ \hat{c}\hat{\Pi}\mathbf{\Psi} (\sigma_{n-1})f(\sigma_{n-1})\\
&\hspace{10em}-\int_0^{\sigma_{n-1}}\!\!d\sigma_n\ \hat{c}\hat{\Pi}\mathbf{\Psi} (\sigma_n)\partial_{\sigma_{n}} f(\sigma_n)\\
&\hspace{10em}-\int_0^{\sigma_{n-1}}\!\!d\sigma_n\ \hat{c}(2\pi)\hat{\Pi}_i(2\pi)\delta(\sigma_n-2\pi)\mathbf{\Psi}(\hat{X}^0(\sigma_n))f(\sigma_n)\\
&\hspace{10em}+\int_0^{\sigma_{n-1}}\!\!d\sigma_n\ i\hat{\Pi}\mathbf{\Psi} (\sigma_n)\left[\int_0^{2\pi}\!\!d\sigma\ \hat{c}\hat{\Pi}_\mu\partial\hat{X}^\mu,f(\sigma_n)\right].
\end{split}\label{boundary:eq:remma}
\end{equation}

Detailed calculations are shown in appendix \ref{boundary:appendix:b}. Then we shall prove theorem \ref{boundary:theorem}. Suppose that \eqref{boundary:eq:theorem} holds for a particular value of $n(\ge 2)$. We choose
\begin{equation*}
   f(\sigma_n)=\int_0^{\sigma_{n}}\!\!d\sigma_{n+1}\ i\hat{\Pi}\mathbf{\Psi}(\sigma_{n+1})g(\sigma_{n+1})
\end{equation*}
where $g(\sigma_{n+1})$ is an arbitrary function. Then the supposed equation leads
\begin{equation*}
\begin{split}
&   \left[\int_0^{2\pi}\!\!d\sigma\ \hat{c}\hat{\Pi}_\mu\partial\hat{X}^\mu(\sigma), 
\int_0^{2\pi}\!\!d\sigma_1\ i\hat{\Pi}\mathbf{\Phi} (\sigma_1)\cdots\int_0^{\sigma_{n}}\!\!d\sigma_{n+1}\ i\hat{\Pi}\mathbf{\Phi} (\sigma_{n+1}) g(\sigma_{n+1})\right]\\
=& -\int_0^{2\pi}\!\!d\sigma_1\ i\hat{\Pi}\mathbf{\Phi}(\sigma_1)\cdots\int_0^{\sigma_{n-1}}\!\!d\sigma_{n}\ \hat{c}\hat{\Pi}\mathbf{\Phi}(\sigma_{n})\\
&\hspace{10em}\partial_{\sigma_{n}}\left(\int_0^{\sigma_{n}}\!\!d\sigma_{n+1}\ i\hat{\Pi}\mathbf{\Phi}(\sigma_{n+1})g(\sigma_{n+1})\right)\\
&+\int_0^{2\pi}\!\!d\sigma_1\ i\hat{\Pi}\mathbf{\Phi}(\sigma_1)\cdots\int_0^{\sigma_{n-1}}\!\!d\sigma_{n}\ i\hat{\Pi}\mathbf{\Phi}(\sigma_{n})\\
&\hspace{10em}\left[\int_0^{2\pi}\!\!d\sigma\ \hat{c}\hat{\Pi}_\mu\partial\hat{X}^\mu(\sigma), \int_0^{\sigma_n}d\sigma_{n+1}\ i\hat{\Pi}\mathbf{\Phi}(\sigma_{n+1}) (\sigma_{n+1})\right]\\
=& -\int_0^{2\pi}\!\!d\sigma_1\ i\hat{\Pi}\mathbf{\Phi}(\sigma_1)\cdots\int_0^{\sigma_{n-1}}\!\!d\sigma_{n}\ \hat{c}\hat{\Pi}\mathbf{\Phi}(\sigma_{n})\  i\hat{\Pi}\mathbf{\Phi}(\sigma_{n})\ g(\sigma_{n})\\
&+\int_0^{2\pi}\!\!d\sigma_1\ i\hat{\Pi}\mathbf{\Phi}(\sigma_1)\cdots\int_0^{\sigma_{n-1}}\!\!d\sigma_{n}\ i\hat{\Pi}\mathbf{\Phi}(\sigma_{n})\ \hat{c}\hat{\Pi}\mathbf{\Phi}(\sigma_{n})\ g(\sigma_{n})\\
&-\int_0^{2\pi}\!\!d\sigma_1\ i\hat{\Pi}\mathbf{\Phi}(\sigma_1)\cdots\int_0^{\sigma_n}\!\!d\sigma_{n+1}\ \hat{c}\hat{\Pi}\mathbf{\Phi}(\sigma_{n+1})\partial_{\sigma_{n+1}} g(\sigma_{n+1}) \\
&+\int_0^{2\pi}\!\!d\sigma_1\ i\hat{\Pi}\mathbf{\Phi}(\sigma_1)\cdots\int_0^{\sigma_{n}}\!\!d\sigma_{n+1}\ \hat{c}(2\pi)\hat{\Pi}_i(2\pi)\delta(\sigma_{n+1}-2\pi)\mathbf{\Phi}(\hat{X}^0(\sigma_{n+1}))\partial_{\sigma_{n+1}} g(\sigma_{n+1})\\
&+\int_0^{2\pi}\!\!d\sigma_1\ i\hat{\Pi}\mathbf{\Phi}(\sigma_1)\cdots\int_0^{\sigma_{n}}\!\!d\sigma_{n+1}\ i\hat{\Pi}\mathbf{\Phi}(\sigma_{n+1})\partial \left[\int_0^{2\pi}\!\!d\sigma\ \hat{c}\hat{\Pi}\partial\hat{X}(\sigma), g(\sigma_{n+1})\right] \\
=&-\int_0^{2\pi}\!\!d\sigma_1\ i\hat{\Pi}\mathbf{\Phi}(\sigma_1)\cdots\int_0^{\sigma_n}\!\!d\sigma_{n+1}\ \hat{c}\hat{\Pi}\mathbf{\Phi}(\sigma_{n+1})\partial_{\sigma_{n+1}} g(\sigma_{n+1}) \\
&+\int_0^{2\pi}\!\!d\sigma_1\ i\hat{\Pi}\mathbf{\Phi}(\sigma_1)\cdots\int_0^{\sigma_{n}}\!\!d\sigma_{n+1}\ i\hat{\Pi}\mathbf{\Phi}(\sigma_{n+1})\left[\int_0^{2\pi}\!\!d\sigma\ \hat{c}\hat{\Pi}\partial\hat{X}(\sigma), g(\sigma_{n+1})\right].
\end{split}
\end{equation*}
This is precisely \eqref{boundary:eq:theorem} for $n+1$. 
\begin{equation*}
\int_0^{\sigma_n}\!\!d\sigma_{n+1}\ \hat{c}(2\pi)\hat{\Pi}_i(2\pi)\delta(2\pi\sigma_{n+1})\mathbf{\Phi}^i (\sigma_{n+1})g(\sigma_{n+1})
\end{equation*}
vanishes because the integrand has non-zero value only at the point $\sigma_1=\cdots=\sigma_{n+1}=2\pi$. What remains in order to prove the theorem is to confirm that \eqref{boundary:eq:theorem} is satisfied for $n=1,2$. These confirmations are shown in \ref{boundary:appendix:b}. Put these all together, the theorem \ref{boundary:theorem} has been proven. Finally we set $f(\sigma)=1$ for each value of $n\in\mathbb{N}$, and then we get
\begin{equation*}
      \left[\int_0^{2\pi}\!\!d\sigma\ \hat{c}\hat{\Pi}_\mu\partial\hat{X}^\mu(\sigma), \mathrm{P}e^{i \int d\sigma \hat{\Pi}(\sigma)\mathbf{X}(\hat{X}^0(\sigma))}\right]=0.
\end{equation*}
Hence the boundary state \eqref{boundary:eq:boundary} is BRST invariant.

\subsection{Divergence and equation of motion}\label{boundary:subsection:3-3}
The boundary state includes singularities which remain after the zeta function regularization. Such divergences vanish if an equation of motion is satisfied. In other words, we can derive an equation of motion by requiring finiteness of the boundary state. 

We rewrite the second term in \eqref{boundary:eq:alpha1-3'} and \eqref{boundary:eq:alpha1-4'} as
\begin{align*}
&\left(\frac{ig}{\sqrt{2\alpha'}}\right)^2\frac{1}{2}\int_0^1\!\!d\sigma\ \mathrm{tr}\left[e^{i(1-\sigma)\hat{p}\mathbf{X}}\mathbf{X}^ie^{i\sigma \hat{p}\mathbf{X}}\mathbf{X}^i\right]  
\sum_{n>0}\frac{4n}{g}\cos 2\pi n \sigma\\
+& \left(i\sqrt{\frac{\alpha'}{2}}\right)^2\frac{1}{2!}\mathrm{tr}\left[(i\hat{p}\ddot{\mathbf{X}})e^{i\hat{p}\mathbf{X}}\right]\sum_{n>0}\frac{4}{ng}.
\end{align*}

Integrate by part the first term twice to have
\begin{equation} 
\begin{split}
-\int_0^1\!\!d\sigma\ \mathrm{tr} & \left[e^{i(1-\sigma)\hat{p}\mathbf{X}}\mathbf{X}^ie^{i\sigma \hat{p}\mathbf{X}}\mathbf{X}^i\right]\sum_{n>0}\frac{4}{(2\pi)^2 n g}\frac{d^2}{d\sigma^2}(\cos 2\pi n \sigma)\\
=&\int_0^1\!\!d\sigma\ \mathrm{tr}\left[e^{i(1-\sigma)\hat{p}\mathbf{X}}[\mathbf{X}^i,i\hat{p}\mathbf{X}]e^{i\sigma \hat{p}\mathbf{X}}\mathbf{X}^i\right]\sum_{n>0}\frac{4}{(2\pi)^2 ng}\frac{d}{d\sigma}(\cos 2\pi n \sigma)\\
=&\ \mathrm{tr}\left[e^{i\hat{p}\mathbf{X}}[[\mathbf{X}^i,i\hat{p}\mathbf{X}],\mathbf{X}^i]\right]\sum_{n>0}\frac{4}{(2\pi)^2 ng}\\
&\ +\int_0^1\!\!d\sigma\ \mathrm{tr}\left[e^{i(1-\sigma)\hat{p}\mathbf{X}}[\mathbf{X}^i,i\hat{p}\mathbf{X}]e^{i\sigma \hat{p}\mathbf{X}}[\mathbf{X}^i,i\hat{p}\mathbf{X}]\right]\sum_{n>0}\frac{4}{(2\pi)^2 n g}\cos 2\pi n \sigma\ .
\end{split}\label{boundary:eq:decompose-divergence}   
\end{equation}
Here we have used an identity
\begin{equation*}
   \frac{d}{d\sigma}\mathrm{tr}\left[e^{i(1-\sigma)\hat{p}\mathbf{X}}Ae^{i\sigma\hat{p}\mathbf{X}}B\right] = \mathrm{tr}\left[e^{i(1-\sigma)\hat{p}\mathbf{X}}[A,i\hat{p}\mathbf{X}]e^{i\sigma\hat{p}\mathbf{X}}B\right]\ .
\end{equation*}

The boundary state includes the following singularity after the zeta function regularization at $\alpha'$ order: 
\begin{equation*}
% \left(\frac{ig}{\sqrt{2\alpha'}}\right)^2\frac{1}{2}\mathrm{tr}\left[e^{i\hat{p}\mathbf{X}}[[\mathbf{X}^i,i\hat{p}\mathbf{X}],\mathbf{X}^i]\right]\frac{4}{(2\pi)^2 g}
%+ \left(i\sqrt{\frac{\alpha'}{2}}\right)^2\frac{1}{2!}\mathrm{tr}\left[(i\hat{p}\ddot{\mathbf{X}})e^{i\hat{p}\mathbf{X}}\right]\frac{4}{g}
 -i\hat{p}_j\mathrm{tr}\left[e^{i\hat{p}\mathbf{X}}\left(\frac{g}{(2\pi)^2\alpha'}[[\mathbf{X}^i,\mathbf{X}^j],\mathbf{X}^i]+\frac{\alpha'}{g}\ddot{\mathbf{X}}^j\right)\right]\zeta(1)|D0\rangle
\end{equation*}
where 
\begin{equation*}
     \zeta(1)=\sum_{n=1}^{\infty} \frac{1}{n}.
\end{equation*} 
This singularity is proportional to an equation of motion
\begin{equation*}
   \ddot{\mathbf{X}}^j=\frac{g^2}{(2\pi\alpha')^2}[[\mathbf{X}^i,\mathbf{X}^j,]\mathbf{X}^i]\ .
 \end{equation*}
Hence if the equation of motion is satisfied, the boundary state is constructed out of creation operators on the vacuum with finite coefficients. We can say that our boundary state of multiple D0-branes is well-defined when the scalar field on D0-branes is on-shell.

From another point of view, the divergences in the boundary state can be absorbed by a field redefinition
\begin{equation}
   \mathbf{X}^i \to \mathbf{X}^i + \zeta(1)\left(\frac{g}{(2\pi)\alpha'}[[\mathbf{X}^j,\mathbf{X}^i],\mathbf{X}^j]+\frac{\alpha'}{g}\ddot{\mathbf{X}}^i\right) + \mathcal{O}({\alpha'}^2).\label{boundary:eq:redef}
\end{equation} 
The boundary state contains no singularity after the field definition. However, this redefinition breaks the BRST invariance when $\mathbf{X}^i$ is off-shell. As easily seen, \eqref{boundary:eq:redef} is the identity when $\mathbf{X}^i$ satisfied the equation of motion. 

It is not sure that all divergences appearing at higher orders in $\alpha'$ can be absorbed by a field redefinition. We can see that a singularity which arises at order ${\alpha'}^2$ can be absorbed by the field redefinition at least in the case that $\hat{p}=0$. At order ${\alpha'}^2$ the only term which includes divergence is \eqref{boundary:eq:p=0-4}
\begin{equation*}
   -2\alpha'\mathrm{tr}[\mathbf{X}^{i(3)}]\zeta(1)\sum_{n>0}\frac{1}{n}(\tilde{\alpha}^0_{-n}\alpha^i_{-n}+\alpha^0_{-n}\tilde{\alpha}^i_{-n}).
\end{equation*}
This is canceled out by the term 
%\begin{equation*}
%  2\alpha'\left[\dot{\mathbf{X}}^{i(3)}e^{i\hat{p}\mathbf{X}}\right]\sum_{n>0}\frac{1}{n}(\tilde{\alpha}^0_{-n}\alpha^i_{-n}+\alpha^0_{-n}\tilde{\alpha}^i_{-n})
%\end{equation*}
which appears when we redefine $\mathbf{X}^i$ according to \eqref{boundary:eq:redef} from the term \eqref{boundary:eq:alpha1-5'}
\begin{equation*}
  2g\left[\dot{\mathbf{X}}^ie^{i\hat{p}\mathbf{X}}\right]\sum_{n>0}\frac{1}{n}(\tilde{\alpha}^0_{-n}\alpha^i_{-n}+\alpha^0_{-n}\tilde{\alpha}^i_{-n}).
\end{equation*}
Note that the term $\frac{d}{dt}\mathrm{tr}[\mathbf{X}^j,[\mathbf{X}^i,\mathbf{X}^j]]\zeta(1)$ becomes zero due to the cyclicity of trace. Hence any other divergence does not happen at order ${\alpha'}^2$ in the case that $\hat{p}=0$. After all the boundary state is regular up to order ${\alpha'}^2$ at $\hat{p}=0$. We omit to determine the formula for the field redefinition at higher orders than ${\alpha'}^2$. This can be obtained by requiring cancellation of divergence at higher orders in the case that $\hat{p}\ne 0$.

In this paper we follow \cite{9909027,9909095} concerning the definition of non-Abelian boundary state. \cite{0312260} takes an another approach: the boundary state is normalized by construction, however, is not BRST invariant for an arbitrary excitation of open string fields. By requiring BRST invariance an equation of motion for the gauge field at the leading order in $\alpha'$ is derived. The divergences in our boundary state vanish after a field redefinition. However, this redefinition breaks the BRST invariance when a configuration of D-branes is off-shell. It seems that our boundary state is essentially same to that in \cite{0312260} after the field redefinition.

To summarize this subsection, our boundary state of multiple D0-branes contains no singularity and BRST invariant when $\mathbf{X}^i$ is on-shell. 

\subsection{Coupling to closed string}\label{boundary:subsection:3-4}
We have constructed the boundary state in subsection \ref{boundary:subsection:3-1}. We extract couplings to closed strings from the constructed boundary state. Our results of coupling to massless closed string fields are identical to that presented in section \ref{paper:section:2}. Although open and closed string fields and closed string couplings are functions of $\hat{x}^0=t$, we suppress the argument $t$ of them in this subsection. 

As shown in \eqref{boundary:eq:boundary-result}, the boundary state can be written as
\begin{equation*}
   |B\rangle = \Bigl\{f(\hat{p})+\left(a_{\mu\nu}(\hat{p})+c_{\mu\nu}(\hat{p})\right)\alpha_{-1}^\mu\tilde{\alpha}_{-1}^\nu+\cdots\Bigr\}|D0\rangle
\end{equation*}
where $a_{\mu\nu}=a_{\mu\nu}, c_{\mu\nu}=-c_{\mu\nu}$. $|D0\rangle$ is
\begin{equation*}
   |D0\rangle= \frac{T_0}{2}\exp\left\{-\sum_{n>0}\frac{1}{n}S_{\mu\nu}\alpha^\mu_{-n}\tilde{\alpha}^\nu_{-n}-\sum_{n>0}(\tilde{b}_{-n}c_{-n}+b_{-n}\tilde{c}_{-n})\right\}c_0^+c_1\tilde{c}_1\delta^{25}(\hat{x}^i)|0\rangle. 
\end{equation*}
Then the boundary state can be rewritten as
\begin{equation*}
\begin{split}
%&\frac{T_0}{2} \int\!\!\frac{d^{25}k}{(2\pi)^{25}} \Bigl\{f(k)+\left(a_{\mu\nu}(k)+c_{\mu\nu}(k)\right)\alpha_{-1}^\mu\tilde{\alpha}_{-1}^\nu+\cdots\Bigr\}\\
%&\phantom{\frac{T_0}{2} \int\!\!\frac{d^{25}k}{(2\pi)^{25}} \Bigl\{f(k)+}
%\Bigl\{1-\alpha_{-1}^\mu\tilde{\alpha}_{-1}^\nu S_{\mu\nu}-(\tilde{b}_{-1}c_{-1}+b_{-1}\tilde{c}_{-1})+\cdots\Bigr\}c_0^+c_1\tilde{c}_1|k\rangle\\
&\frac{T_0}{2} \int\!\!\frac{d^{25}k}{(2\pi)^{25}} \Bigl\{f(k)+\left(a_{\mu\nu}(k)+c_{\mu\nu}(k)-S_{\mu\nu}f(k)\right)\alpha_{-1}^\mu\tilde{\alpha}_{-1}^\nu\\
&\phantom{=\frac{T_0}{2} \int\!\!\frac{d^{25}k}{(2\pi)^{25}} \Bigl\{f(k)+(a_{\mu\nu}(k)}
-f(k) (\tilde{b}_{-1}c_{-1}+b_{-1}\tilde{c}_{-1}) +\cdots\Bigr\}c_0^+c_1\tilde{c}_1|k\rangle.
\end{split}
\end{equation*}
We can identify $F, A, B$ and $C$ in \eqref{boundary:eq:form} as follows:
\begin{align*}
  F(k)&=\phantom{-}\frac{T_0}{2}f(k)\\
  A_{\mu\nu}(k)&=\phantom{-}\frac{T_0}{2}(a_{\mu\nu}(k)-f(k)S_{\mu\nu})\\
  B(k)&=-\frac{T_0}{2}f(k) \\
  C_{\mu\nu}(k)&=\phantom{-}\frac{T_0}{2}c_{\mu\nu}(k).
\end{align*}
Thus the source term \eqref{boundary:eq:coupling} becomes
\begin{equation}
\begin{split}
S_\mathrm{source}= \int\!\!dt\int\!\!\frac{d^{25}k}{(2\pi)^{25}}\frac{T_0}{2}\Bigl\{ & T(-k)f(k) + \frac{1}{2}h_{\mu\nu}(-k)\left(a^{\mu\nu}(k)-f(k)(S^{\mu\nu}+g^{\mu\nu})\right)\\
&+\frac{1}{2}b_{\mu\nu}(-k)c^{\mu\nu}(k)+2\phi(-k)f(k)+\cdots\Bigr\}.
\end{split}
\end{equation}
%An energy-momentum tensor can be extracted as
%\begin{equation*}
%   T_{\mu\nu}(k)=\frac{T_0}{2}\left(a_{\mu\nu}(k)+2gf(k)\delta_{\mu 0}\delta_{\nu0}\right)
%\end{equation*}
Considering $S_{\mu\nu}=g (\eta_{00},-\delta_{ij})$, we can see $S_{\mu\nu}+g_{\mu\nu}=- 2g\ \delta_{\mu0}\delta_{\nu0}$. 
Explicit formulas for $f(k),a^{\mu\nu}(k),c^{\mu\nu}(k)$ are shown in \eqref{boundary:eq:result-alpha0}, \eqref{boundary:eq:result-alpha1} and  \eqref{boundary:eq:p=0}.
$a_{\mu\nu}(k), c_{\mu\nu}(k)$ in \eqref{boundary:eq:result-alpha1} are identical (up to the the overall normalization) to the energy-momentum tensor of bosonic strings derived from the disk amplitudes shown in \eqref{paper:eq:emtensor-bosonic}:
\begin{equation}
\begin{split}
  T^{00}(k) & = - \frac{2\alpha'}{g}\int_0^1\!\!d\sigma\ \mathrm{tr}[e^{i(1-\sigma)\hat{p}\mathbf{X}}(i\hat{p}\dot{\mathbf{X}})e^{i\sigma\hat{p}\mathbf{X}}(i\hat{p}\dot{\mathbf{X}})]\cos 2\pi \sigma - \frac{2 \alpha'}{g} \mathrm{tr}\left[(i\hat{p}\ddot{\mathbf{X}})e^{i\hat{p}\mathbf{X}}\right] \\
T^{0i}(k) & = \int_0^1\!\!d\sigma\ \mathrm{tr}\left[e^{i(1-\sigma)\hat{p}\mathbf{X}}(i\hat{p}\dot{\mathbf{X}})e^{i\sigma\hat{p}\mathbf{X}}\mathbf{X}^i\right]e^{2\pi i \sigma}+ 2\ \mathrm{tr}\left[\dot{\mathbf{X}}^ie^{i\hat{p}\mathbf{X}}\right]\\
  T^{ij}(k) & = -\frac{2g}{\alpha'}\int_0^1\!\!d\sigma\ \mathrm{tr}\left[e^{i(1-\sigma)k\mathbf{X}}\mathbf{X}^ie^{i\sigma k \mathbf{X}}\mathbf{X}^j \right]e^{2\pi i \sigma}.
\end{split}\label{boundary:eq:main-result-coupling-bosonic}
\end{equation}

In what follows we consider the case $T(x)$, $h^{\mu\nu}(x)$, $b^{\mu\nu}(x)$, $\phi(x)$ are independent of the coordinates along Dirichlet directions $x^i$. We denote the closed string fields as 
\begin{align*}
T(k)&=(2\pi)^{25}\delta^{25}(k)T\\ h^{\mu\nu}(k)&=(2\pi)^{25}\delta^{25}(k) h^{\mu\nu}\\ b^{\mu\nu}(k)&=(2\pi)^{25}\delta^{25}(k) b^{\mu\nu}\\ \phi^{\mu\nu}(k)&=(2\pi)^{25}\delta^{25}(k) \phi.
\end{align*}
Take care that $T,h^{\mu\nu},b^{\mu\nu},\phi$ depends on the coordinates along Neumann directions $\hat{x}^a$. In case of D0-branes, the closed string fields depends only on $x^0=t$.%In this case we need only $f(k=0)$, $a^{\mu\nu}(k=0)$, $c^{\mu\nu}(k=0)$ in order to derive the couplings. 
In this case we find the couplings to closed strings \eqref{boundary:eq:coupling} become
\begin{equation*}
\begin{split}
 S_\mathrm{source}=   \int\!\!dt%\int\!\!\frac{d^{25}k}{(2\pi)^{25}}
\Biggl\{
%&T \frac{T_0}{2}\left(\mathrm{tr}[1] -\frac{1}{2}\mathrm{tr}\left[\dot{\mathbf{X}}^i\dot{\mathbf{X}}^i\right]-\frac{1}{4}\left(\frac{g}{2\pi\alpha'}\right)^2\mathrm{tr}\left[\mathbf{X}^i,\mathbf{X}^j\right]^2\right) \\
&\phi T_0\left(\mathrm{tr}[1] -\frac{1}{2}\mathrm{tr}\left[\dot{\mathbf{X}}^i\dot{\mathbf{X}}^i\right]-\frac{1}{4}\left(\frac{g}{2\pi\alpha'}\right)^2\mathrm{tr}\left[\mathbf{X}^i,\mathbf{X}^j\right]^2\right) \\
& +\frac{1}{2g}h_{00}T_0\left(\mathrm{tr}[1] + \frac{1}{2}\mathrm{tr}\left[\dot{\mathbf{X}}^i\dot{\mathbf{X}}^i\right]-\frac{1}{4}\left(\frac{g}{2\pi\alpha'}\right)^2\mathrm{tr}\left[\mathbf{X}^i,\mathbf{X}^j\right]^2\right) \\
& + \frac{1}{2g}(h_{0i}+h_{i0})T_0\left(\mathrm{tr}[\dot{\mathbf{X}}^i]\right)\\
& +\frac{1}{2g}(b_{0i}-b_{i0})T_0\left(\frac{ig}{2\pi\alpha'}\mathrm{tr}[[\mathbf{X}^i,\mathbf{X}^j]\dot{\mathbf{X}}^j]\right)\\
& +\frac{1}{2g} h_{ij}T_0\left(\mathrm{tr}\left[\dot{\mathbf{X}}^i\dot{\mathbf{X}}^j\right]+\left(\frac{g}{2\pi\alpha'}\right)^2\mathrm{tr}[\mathbf{X}^i,\mathbf{X}^k][\mathbf{X}^j,\mathbf{X}^k]\right)\\
& +\frac{1}{2g} b_{ij}T_0\left(\frac{ig}{2\pi\alpha'}\mathrm{tr}[\mathbf{X}^i,\mathbf{X}^j]\right)
\Biggr\}.
\end{split}
\end{equation*}
From this source terms, we can extract the couplings to massless closed string fields as follows.
%The graviton coupling, namely the energy-momentum tensor becomes
\begin{equation}
\begin{split}
   I_h^{00} & = T_0g\left(\mathrm{tr}[1] + \frac{1}{2}\mathrm{tr}\left[\dot{\mathbf{X}}^i\dot{\mathbf{X}}^i\right]-\frac{1}{4}\left(\frac{g}{2\pi\alpha'}\right)^2\mathrm{tr}\left[\mathbf{X}^i,\mathbf{X}^j\right]^2\right)\\
   I_h^{0i} &= T_0g\left(\mathrm{tr}[\dot{\mathbf{X}}^i]\right)\\
   I_h^{ij} & = T_0g\left(\mathrm{tr}\left[\dot{\mathbf{X}}^i\dot{\mathbf{X}}^j\right]+\left(\frac{g}{2\pi\alpha'}\right)^2\mathrm{tr}\left[\mathbf{X}^i,\mathbf{X}^k\right]\left[\mathbf{X}^j,\mathbf{X}^k\right]\right).
\\%\intertext{The dilaton coupling becomes}
   I_\phi&= T_0 \left(\mathrm{tr}[1] -\frac{1}{2}\mathrm{tr}\left[\dot{\mathbf{X}}^i\dot{\mathbf{X}}^i\right]-\frac{1}{4}\left(\frac{g}{2\pi\alpha'}\right)^2\mathrm{tr}\left[\mathbf{X}^i,\mathbf{X}^j\right]^2\right).
\\%\intertext{The NS-NS B-field coupling becomes}
   I_b^{0i} & = T_0g\left(\frac{ig}{2\pi\alpha'}\mathrm{tr}\left[[\mathbf{X}^i,\mathbf{X}^j]\dot{\mathbf{X}^j}\right]\right) \\
   I_b^{ij} & = T_0g\left(\frac{ig}{2\pi\alpha'}\mathrm{tr}\left[\mathbf{X}^i,\mathbf{X}^j\right]\right). 
\end{split}\label{boundary:eq:main-result-coupling-bosonic2}
\end{equation}
These results are identical to the closed string coupling derived from Matrix theory potential, and non-Abelian DBI action of multiple D0-branes \eqref{paper:eq:emtensor-super} estimated at $k=0$. We can say our boundary state reproduces the non-Abelian DBI action linear in massless closed string fields correctly.

We note that the $n$-th derivatives of closed string fields $h_{\mu\nu}(x),b_{\mu\nu}(x), \phi(x)$ with respect to $x^i$ couple to the $n$-th derivatives of the corresponding couplings $I_{\mu\nu}(k),I_b{\mu\nu}(k),\phi(k)$ with respect to $k^i$. The source term \eqref{boundary:eq:coupling} takes the form of 
\begin{equation}
\begin{split}
  S_\mathrm{source}\Big|_{h,b,\phi}= T_0\int\!\!dt\Bigl\{
& \phi(t,0) I_\phi(t) + \frac{1}{2}h_{\mu\nu}(t,0)I_h^{\mu\nu}(t)+\frac{1}{2}b_{\mu\nu}(t,0)I_b^{\mu\nu}(t,0)\\
&+ \sum_{n=1}^\infty\frac{1}{n!}(\partial_{i_1}\cdots\partial_{i_n}\phi)(t,0)I_\phi^{i_1\cdots i_n}(t)\\
&+ \sum_{n=1}^\infty\frac{1}{n!}(\partial_{i_1}\cdots\partial_{i_n}h_{\mu\nu})(t,0)I_h^{\mu\nu(i_1\cdots i_n)}(t)\\
&+ \sum_{n=1}^\infty\frac{1}{n!}(\partial_{i_1}\cdots\partial_{i_n}b_{\mu\nu})(t,0)I_n^{\mu\nu(i_1\cdots i_n)}(t)\Bigr\}
\end{split} \label{boundary:eq:derivative-coupling}
\end{equation}
%\begin{equation}
%    h^{ij}_{(i_1\cdots i_n)}  (t) a_{ij}^{(i_1\cdots i_n)}(t), 
%\end{equation}
where
\begin{equation*}
\begin{split}
    h^{ij}_{(i_1\cdots i_n)}(t) = & (\partial_{x^{i_1}}\cdots \partial_{x^{i_n}})h^{ij}(t,x)\Big|_{x^i=0}\\
    b^{ij}_{(i_1\cdots i_n)}(t) = & (\partial_{x^{i_1}}\cdots \partial_{x^{i_n}})b^{ij}(t,x)\Big|_{x^i=0}\\
    \phi_{(i_1\cdots i_n)}(t) = & (\partial_{x^{i_1}}\cdots \partial_{x^{i_n}})\phi(t,x)\Big|_{x^i=0}\\
    I_{ij}^{(i_1\cdots i_n)}(t) = & (\partial_{k_{i_1}}\cdots \partial_{k_{i_n}}) I_{ij}(t,k)\Big|_{\hat{k}_i=0}.
\end{split} 
\end{equation*}
This corresponds to the non-Abelian Taylor expansion of closed string fields in powers of scalar fields $\mathbf{\Phi}(x^a)$ in a non-Abelian DBI action \cite{9910053}: 
\begin{equation*}
\begin{split}
    h^{ij}(x^a,\mathbf{\Phi}^i(x^a)) &= \sum_{n=0}^{\infty}\frac{1}{n!}\mathbf{\Phi}^{i_1}(x^a)\cdots\mathbf{\Phi}^{i_n}(x^a)(\partial_{x^{i_1}}\cdots \partial_{x^{i_n}})h^{ij}(x^a,x^i)\Big|_{x^i=0}\\
    b^{ij}(x^a,\mathbf{\Phi}^i(x^a)) &= \sum_{n=0}^{\infty}\frac{1}{n!}\mathbf{\Phi}^{i_1}(x^a)\cdots\mathbf{\Phi}^{i_n}(x^a)(\partial_{x^{i_1}}\cdots \partial_{x^{i_n}})b^{ij}(x^a,x^i)\Big|_{x^i=0}\\
    \phi(x^a,\mathbf{\Phi}^i(x^a)) &= \sum_{n=0}^{\infty}\frac{1}{n!}\mathbf{\Phi}^{i_1}(x^a)\cdots\mathbf{\Phi}^{i_n}(x^a)(\partial_{x^{i_1}}\cdots \partial_{x^{i_n}})\phi(x^a,x^i)\Big|_{x^i=0}. 
\end{split}
\end{equation*}
From this point of view,  $I(k=0)$ in \eqref{boundary:eq:main-result-coupling-bosonic} represents the total charge integrated over the Dirichlet directions. This can be seen as follows.
\begin{equation}
   I(k^i=0)=\int\!\!dx^i\ e^{ik_ix^i}I(x^i)\big|_{k^i=0} = \int\!\!dx^i\ I(x^i). \label{boundary:eq:total-charge}
\end{equation}
In a similar way, we can see that derivatives of $I(k^i)$ with respect to $k^i$ contains information about the charge distribution. 
%
%As shown by the study in this section, the boundary state reproduces the correct couplings to closed strings. Hence our definition of 
%non-Abelian boundary states is correct at least up to ${\alpha'}^2$ order. 
\\ \\
To summarize this subsection, we have extracted the closed string couplings \eqref{boundary:eq:main-result-coupling-bosonic}, \eqref{boundary:eq:main-result-coupling-bosonic2} from the constructed boundary state up to $\mathcal{O}({\alpha'}^2)$. Our results are identical to those derived from a Matrix theory potential \cite{9711078,9712185, 9904095}, disk one-point amplitudes with a D-brane boundary \cite{0103124} which are presented in \eqref{paper:eq:emtensor-bosonic}, \eqref{paper:eq:emtensor-super}. In addition we have confirmed up to $\mathcal{O}({\alpha'}^2)$ that our results realize the linear part in closed string fields of the non-Abelian DBI action \cite{9910053} presented in \eqref{paper:eq:emtensor-bosonic}, \eqref{paper:eq:emtensor-super}. This represents that the formulas for boundary state \eqref{review:eq:nonabelian-boundarystate}, or equivalently \eqref{boundary:eq:boundary}, satisfies the requirement \eqref{review:eq:disk-nonabelian} that the boundary state should reproduce the correct disk amplitude at least for massless closed strings.

%We expect the boundary state gives the correct action of multiple D0-branes also at higher orders in $\alpha'$, in contrast to the DBI action with symmetrized trace, which is valid only up to $\mathcal{O}({\alpha'}^4)$.

\subsection{Extension to type IIA superstring}\label{boundary:subsection:3-5}
We have investigated boundary states in the bosonic string theory. Extension to the type IIA superstring theory is given in this subsection. In addition to a worldsheet boson, we need to take account of a worldsheet fermion. 

A worldsheet fermion can be expanded as     
\begin{align*}
   \psi^\mu(\tau,\sigma)&=i^{-1/2}\sum_{r\in \mathbb{Z}+\frac{1}{2}}\psi_r^\mu e^{-ir(\tau-\sigma)}\\
   \tilde{\psi}^\mu(\tau,\sigma)&=i^{1/2\phantom{-}}\sum_{r\in \mathbb{Z}+\frac{1}{2}}\tilde{\psi}_r^\mu e^{-ir(\tau+\sigma)}.
\end{align*} 
$\psi^\mu_s, \tilde{\psi}^\nu_t$ satisfy anticommutation relations
\begin{equation*}
   \{\psi^\mu_s,\psi^\nu_t\}=\{\tilde{\psi}^\mu_s,\tilde{\psi}^\nu_t\}=g^{\mu\nu}\delta_{r+s,0}.
\end{equation*}
It is convenient to define
\begin{align*}
   \hat{\Psi}^a(\sigma) & =\frac{1}{2}(\psi^a(\sigma)+\tilde{\psi}^a(\sigma)) =  \frac{1}{2i^{1/2}}\sum_r (\psi_r^a+i\tilde{\psi}_{-r}^a)e^{ir\sigma}\\
   \hat{\Psi}^i(\sigma) & =\frac{1}{2}(\psi^i(\sigma)-\tilde{\psi}^i(\sigma)) = \frac{1}{2i^{1/2}}\sum_r (\psi_r^i-i\tilde{\psi}_{-r}^i)e^{ir\sigma}.
\end{align*}
Then the GSO projected boundary state $|B\rangle$ can be written as
\begin{equation*}
   |B\rangle = \mathrm{GSO}|B,+\rangle 
\end{equation*}
where
\begin{equation*}
\begin{split}
|B,+\rangle & = \mathrm{trP}\exp \left[ \int_0^{2\pi} d\sigma \left(i\mathbf{X}^i ({\hat{X}^0})\hat{\Pi}_i(\sigma) \right.\right.\\
&\left.\left. \hspace{4em}-\frac{1}{\pi}\dot{\mathbf{X}}^i(\hat{X}^0)\hat{\Psi}^0\hat{\Psi}_i(\sigma) -\frac{i}{(2\pi)^2\alpha'}[\mathbf{X}^i(\hat{X}^0),\mathbf{X}^j(\hat{X}^0)]\hat{\Psi}_i\hat{\Psi}_j(\sigma) \right)\right]|D0,+\rangle\\
 |D0,+\rangle & = |D0\rangle_\alpha|B\rangle_\mathrm{gh}|D0,+\rangle_\psi|B,+\rangle_\mathrm{sgh}.
\end{split}
\end{equation*}
Note that we define $\psi,\tilde{\psi}$ to be dimensionless, while they have dimension of $\alpha'$ in \cite{0103124}. After the Wick rotation $\tau=it$ and exchange $\tau \leftrightarrow \sigma$, the boundary action under consideration is same as that given in \cite{0103124}.

In a manner similar to the case of $A_p^{\mu_1\cdots \mu_p}(\sigma_1, \cdots, \sigma_p)$ in the bosonic case, we define 
\begin{align*}
   B_p^{\mu_1\cdots \mu_p}(\sigma_1, \cdots, \sigma_p) &= \prod_{q=1}^p b^{\mu_q}(\sigma_p) \\
   b^{\mu}(\sigma) &= 
\begin{cases}
 \frac{\sqrt{2\pi}}{2i^{1/2}}\sum_{r}(\psi^a_r - \tilde{\psi}^a_{-r})e^{2\pi i r \sigma} \quad \mathrm{Neumann} \\
  \frac{\sqrt{2\pi}}{2i^{1/2}}\sum_{r}(\psi^i_n + \tilde{\psi}^i_{-r})e^{2\pi i r \sigma} \quad \mathrm{Dirichlet}
\end{cases}.
\end{align*}
In the remaining of this subsection, we always think that $B_p^{\mu_1\cdots \mu_p}(\sigma_1, \cdots, \sigma_p)$ operates on $|Dp\rangle,+$ and omit $|Dp\rangle,+$ as in the case of the bosonic string. 
$B^{\mu_1\cdots\mu_p}_p(\sigma_1,\cdots,\sigma_p)$ has a property 
\begin{equation} 
 B_p^{\mu_1\cdots\mu_p}(\sigma_1,\cdots,\sigma_p) = \mathrm{sgn}(\sigma) B_p^{\mu_{\tau(1)}\cdots\mu_{\tau(p)}}(\sigma_{\tau(1)},\cdots,\sigma_{\tau(p)}) \quad \tau\mbox{: permutation}. \label{boundary:eq:property-fermion}
\end{equation}
Here we have an additional numerical factor $\mathrm{sgn}(\sigma)$ because $\psi_{-r}^\mu,\tilde{\psi}_{-r}^\mu$ are anticommutative rather than commutative. It is convenient to abbreviate so that 
\begin{equation}
\begin{split}
\left(\frac{\sqrt{2\pi}}{2i^{1/2}}\right)^{p}B^{\mu_1\cdots \mu_p}_p &\to B_p^{\mu_1\cdots \mu_p}\\
\frac{1}{n!}\frac{d^n}{dt^n}\mathbf{X} &\to \frac{d^n}{dt^n}\mathbf{X}.
\end{split}\label{boundary:eq:rescale-fermion}
\end{equation}
What is different from the bosonic case is that $\hat{\Psi}$ always appears in the form of $\hat{\Psi}^0(\sigma)\hat{\Psi}^i(\sigma)$ or $\hat{\Psi}^i(\sigma)\hat{\Psi}^j(\sigma)$. Thus $B$ should be a product of an arbitrary power of $b^{0}(\sigma)b^{i}(\sigma)$ and $b^{i}(\sigma)b^{j}(\sigma)$ before operation on $|D0\rangle$. Results of calculating $B$ are shown in appendix \ref{boundary:appendix:a}. In calculations of the boundary state we find a factor
\begin{equation*}
   \sum_{n=0}^\infty\sin 2\pi \left(n+\frac{1}{2}\right)\sigma.
\end{equation*}
which diverges naively. We regularize this by restoration of the worldsheet time $t  =i \tau \ne 0$ in summation, and take the limit of $t \to 0$ finally. By this regularization at the worldsheet boundary we can calculate this factor as
\begin{equation*} 
\begin{split}
&   \lim_{t \to 0}\sum_{n=0}^\infty \frac{1}{2i}\left\{e^{2\pi i (n+1/2) (\sigma+it)}-e^{-2\pi i (n+1/2)(\sigma-it)}\right\}\\
&= \lim_{t \to 0} \frac{1}{2i}\left\{\frac{e^{\pi i (\sigma+it)}}{1-e^{2\pi i (\sigma+it)}}-\frac{e^{-\pi i (\sigma-it)}}{1-e^{-2\pi i (\sigma-it)}}\right\}\\
&=\frac{1}{2 \sin \pi \sigma}.
\end{split}
\end{equation*}
In computation of the boundary state we need to deal with singularities by the zeta function regularization: \cite{NPB308221}
\begin{equation*}
   \sum_{r=\textstyle\frac{1}{2},\frac{1}{3},\cdots}^\infty 1 = \lim_{s\to 0}\sum_{n=0}^\infty\left(n+\textstyle\frac{1}{2}\right)^{-s} = \zeta(0,\textstyle\frac{1}{2}) = 0
\end{equation*}
where $\zeta(s,a)=\sum_{n=0}^\infty (n+a)^{-s}$ is the Hurwitz zeta function.

In the type IIA superstring theory, calculations of boundary state and closed string coupling can be done in a way similar to the bosonic case. Although we omit details of such calculations, the result of closed string coupling to massless closed strings become
\begin{equation}
\begin{split}
    I_\phi(t,k) & = T_0\,\mathrm{Str}\left[\left(1-\frac{1}{2}\dot{\mathbf{X}}^i\dot{\mathbf{X}}^i-\frac{1}{4(2\pi\alpha')^2}[\mathbf{X}^i,\mathbf{X}^j][\mathbf{X}^i,\mathbf{X}^j]\right)e^{ik\mathbf{X}}\right] \\
    I_h^{00}(t,k) & = T_0\,\mathrm{Str}\left[\left(1+\frac{1}{2}\dot{\mathbf{X}}^i\dot{\mathbf{X}}^i-\frac{1}{4(2\pi\alpha')^2}[\mathbf{X}^i,\mathbf{X}^j][\mathbf{X}^i,\mathbf{X}^j]\right)e^{ik\mathbf{X}}\right]\\
    I_h^{0i}(t,k) & = T_0\,\mathrm{Str}\left[\left(\dot{\mathbf{X}}^i\right)e^{ik\mathbf{X}} \right]\\
    I_h^{ij}(t,k) & = T_0\,\mathrm{Str}\left[\left(\dot{\mathbf{X}}^i\dot{\mathbf{X}}^i-\frac{1}{4}[\mathbf{X}^i,\mathbf{X}^j][\mathbf{X}^i,\mathbf{X}^j]\right)e^{ik\mathbf{X}}\right]\\
    I_b^{0i}(t,k) &= T_0\,\mathrm{Str}\left[\left(\frac{i}{2\pi\alpha'}[\mathbf{X}^i,\mathbf{X}^j]\dot{\mathbf{X}}^j\right)e^{ik\mathbf{X}}\right]\\
    I_b^{ij}(t,k) &= T_0\,\mathrm{Str}\left[\left(\frac{-i}{2\pi\alpha'}[\mathbf{X}^i,\mathbf{X}^j]\right)e^{ik\mathbf{X}}\right].
\end{split}\label{boundary:eq:main-result-coupling-super}
\end{equation}

%As the result of such computations, a NS-NS massless part of the boundary state becomes
%\begin{equation*}
% |B\rangle = (T_{\mu\nu}(t,\hat{p})+I_{\mu\nu}(t,\hat{p}))\psi_{-1/2}^\mu\tilde{\psi}_{-1/2}^\nu \delta(x^i)|p^i=0\rangle|B\rangle_\mathrm{gh}|B\rangle_\mathrm{sgh}
%\end{equation*}
%where $T_{\mu\nu}(t,\hat{p})$ and $I_{\mu\nu}(t,\hat{p})$ are symmetric and antisymmetric tensors given up to $\mathcal{O}({\alpha'}^2)$ by
%\begin{align*}
%T_{00} &= T_0 g\ \mathrm{tr}\Bigl[e^{i\hat{p}\mathbf{X}}+\frac{1}{2}\int_0^1\!\! d\sigma\ e^{i(1-\sigma)\hat{p}\mathbf{X}}\dot{\mathbf{X}}^i e^{i\sigma \hat{p}\mathbf{X}}\dot{\mathbf{X}}^i \\ 
%&\phantom{=T_0 g\ \mathrm{tr}\Bigl[} -\frac{1}{4}\left(\frac{g}{2\pi\alpha'}\right)^2\int_0^1\!\! d\sigma\ e^{i(1-\sigma)\hat{p}\mathbf{X}}[\mathbf{X}^i,\mathbf{X}^j]e^{i\sigma \hat{p}\mathbf{X}}[\mathbf{X}^i,\mathbf{X}^j] \Bigr] \\
%T_{0i} &= T_0g\ \mathrm{tr}\Bigl[\dot{\mathbf{X}^i}e^{i\hat{p}\mathbf{X}}\Bigr] \\
%T_{ij} &= T_0g\ \mathrm{tr}\Bigl[\int_0^1\!\! d\sigma\ e^{i(1-\sigma)\hat{p}\mathbf{X}}\dot{\mathbf{X}}^i e^{i\sigma \hat{p}\mathbf{X}}\dot{\mathbf{X}}^j \\
%&\phantom{=T_0g\ \mathrm{tr}\Bigl[} -\left(\frac{g}{2\pi\alpha'}\right)^2\int_0^1\!\! d\sigma\ e^{i(1-\sigma)\hat{p}\mathbf{X}}[\mathbf{X}^i,\mathbf{X}^k]e^{i\sigma \hat{p}\mathbf{X}}[\mathbf{X}^j,\mathbf{X}^k] \Bigr]\\ 
%I_{0i} &= T_0g\ \Bigl[\frac{ig}{2\pi\alpha'}\int_0^1\!\! d\sigma\ e^{i(1-\sigma)\hat{p}\mathbf{X}}[\mathbf{X}^i,\mathbf{X}^j]e^{i\sigma \hat{p}\mathbf{X}}\dot{\mathbf{X}}^j\Bigr].
%\end{align*}

These are identical the couplings to massless closed strings derived from the disk amplitude, Matrix theory potentail, and non-Abelian DBI action shown in \eqref{paper:eq:emtensor-super}. It is worthwhile to note that these couplings take the form of symmetrized trace in the superstring theory \eqref{boundary:eq:main-result-coupling-super}, but not in the bosonic theory \eqref{boundary:eq:main-result-coupling-bosonic}. The same behavior can be seen in the result obtained from disk amplitudes as shown in \eqref{paper:eq:emtensor-super} and \eqref{paper:eq:emtensor-bosonic}. 

\section{Coincidence of boundary state in nontrivial case}\label{paper:section:4}
We confirm that our boundary state realizes the correct one in two nontrivial cases in this section. The first case is a single boosted D0-brane, and the second case is a noncommutative D2-brane. In both cases our boundary state is exactly identical to the previously known one. 

\subsection{Single boosted D0-brane}\label{boundary:subsection:4-1}
We consider a single boosted D0-brane with velocity $v$ along a direction $x^1$. In this subsection we focus on a matter part of directions $x^0,x^1$ in the boundary state. 

It is known that a boundary state of the D0-brane takes the form of \cite{9701190} 
\begin{equation}
\begin{split}
    |D0\rangle_\mathrm{boost} & =\frac{T_0}{2}\sqrt{1-v^2}\exp\left\{-\sum_{n>0}\frac{1}{n}\alpha^\mu_{-n}M_{\mu\nu}(v)\tilde{\alpha}^\nu_{-n}\right\}\delta(\hat{x}^i-vt)|0\rangle  \label{boundary:eq:boost1}\\
   M_{\mu\nu}(v)& = 
\begin{pmatrix}
   -\frac{1-v^2}{1+v^2} & -\frac{2v}{1-v^2} \\
   -\frac{2v}{1-v^2} & -\frac{1-v^2}{1+v^2} 
\end{pmatrix} \quad \mu,\nu=0,1 \quad i=1.
\end{split} 
\end{equation}
We expand $M_{\mu\nu}$ in powers of $v$ to have
\begin{equation*}
     M_{\mu\nu}(v) = S_{\mu\nu} - g
\begin{pmatrix}
   2v^2 & 2v \\
   2v & 2v^2 
\end{pmatrix} + \mathcal{O}(v^3).
\end{equation*}

In calculations of our boundary state \eqref{boundary:eq:boundary} for a single boosted D0-brane, we need not to take account of ordering because $\mathbf{X}^i= vt$ is not a matrix but just a number. Then all terms in the boundary state are reduced to the form of 
\begin{equation*}
\mathrm{tr}[(\dot{\mathbf{X}})^ne^{i\hat{p}\mathbf{X}}]|x=0\rangle = v^n \delta(\hat{x}^i-vt)|0\rangle,
\end{equation*}
All computations in order to determine numerical factors of these terms are same as that of the general boundary state at $\hat{p}=0$. Then we find from \eqref{boundary:eq:p=0}
\begin{align}
   |B\rangle & = \frac{T_0}{2}\Bigl\{f+\sum_{n>0}\frac{1}{n}a_{\mu\nu}\alpha_{-1}^\mu\tilde{\alpha}_{-1}^\nu+C+\mathcal{O}(v^3)\Bigr\}e^{i\hat{p}\mathbf{X}}|D0\rangle \quad \mu,\nu=0,1 \label{boundary:eq:boost2}\\
   |D0\rangle & =\frac{T_0}{2}\exp\left\{-\sum_{n>0}\frac{1}{n}\alpha^\mu_{-n}S_{\mu\nu}\tilde{\alpha}^\nu_{-n}\right\}|0\rangle \nonumber
\end{align}
where $f,a_{\mu\nu}$ and $c$ are
\begin{align*}
 f & =1-\frac{1}{2}v^2\\
a_{\mu\nu} &= g 
\begin{pmatrix}
   2v^2 & 2v \\
   2v & 2v^2
\end{pmatrix},\\
 C&=2g^2v^2 \sum_{n,m>0}\frac{1}{nm}(\alpha^0_{-n}\alpha_{-m}^0\tilde{\alpha}_{-n}^1\tilde{\alpha}_{-m}^1\\
  &\phantom{2v^2\sum_{n,m>0}(} + 2\alpha^0_{-n}\alpha_{-m}^1\tilde{\alpha}_{-m}^0\tilde{\alpha}_{-m}^1\\
  &\phantom{2v^2\sum_{n,m>0}(} +\alpha^1_{-n}\alpha_{-m}^1\tilde{\alpha}_{-n}^0\tilde{\alpha}_{-m}^0). 
\end{align*}
We can see \eqref{boundary:eq:boost1} and \eqref{boundary:eq:boost2} are identical. Hence our boundary state \eqref{boundary:eq:boundary} is identical to the already known one \eqref{boundary:eq:boost1} at least up to $\mathcal{O}(v^2)$.

\subsection{Noncommutative D2-brane}\label{boundary:subsection:4-2}
In this subsection we calculate the boundary state \eqref{boundary:eq:boundary} for a noncommutative D2-brane. we focus on a matter part of directions $x^1,x^2$ in the boundary state. We start from two matrices $\mathrm{X}^i\ (i=1,2)$ describing a two-dimensional noncommutative plane which satisfies   
\begin{equation}
  [\mathbf{X}^1,\mathbf{X}^2]=i\theta. \label{d2boundary:eq:1}
\end{equation}
Here $\theta$ is a noncommutivity identified as 
\begin{equation*}
 \theta=\frac{2\pi\alpha'}{gb},
\end{equation*}
where $b$ is strength of magnetic flux on the D2-brane. A boundary state to be considered in this subsection is represented by
\begin{equation}
   |B\rangle = \mathrm{tr}\mathrm{P}\exp\left[i \int_0^{2\pi}d\sigma\ \mathbf{X}^i \hat{\Pi}_i(\sigma) \right]|\mathrm{D0}\rangle. \label{d2boundary:eq:2}
\end{equation}
For diagonal matrices $\mathbf{X}^i=\mathrm{diag}(\xi_1^i,\cdots,\xi_N^i)$, the boundary state \eqref{d2boundary:eq:2} gives a summation of D0-brane boundary states located at $x^i=\xi_a^i\ (a=1,\cdots,N)$, where matrix size $N$ corresponds to the number of D0-branes. In presence of off-diagonal components in $\mathbf{X}^i$, they are noncommutative in general and cannot be interpreted as positions of D0-branes. Therefore what the boundary state \eqref{d2boundary:eq:2} represents is a nontrivial problem for the general $\mathbf{X}^i$. 
 
The boundary state \eqref{d2boundary:eq:2} can be computed by using a path integral, and the result is identical to the boundary state of a D2-brane with a constant field strength up to overall normalization \cite{9804163}. It is difficult, however, to adopt such a method to the general case.
%in presence of an arbitrary excitation of transverse scalar and gauge fields on the D2-brane. 
In this subsection we calculate the boundary state in the operator formalism in powers of $\frac{g\theta}{2\pi\alpha'}$. This expansion is essentially the $\alpha'$ expansion in the Seiberg-Witten limit \cite{9908142}:
%Open strings on the D2-brane gives a 2+1 dimensional noncommutative gauge theory in the Seiberg-Witten limit. In the noncommutative gauge theory, coordinates $x^i$ become noncommutative and satisfy an relation
%\begin{equation*}
%   [x^1,x^2]=i\theta.
%\end{equation*}
%This is identical to the relation \eqref{d2boundary:eq:1}. Then we can say that the matrices $\mathrm{X}^i$ are infinite dimensional representations of the noncommutative coordinates $x^i$. 
%The Seiberg-Witten limit is a zero slope limit so that
\begin{equation*} 
   \alpha'\to \epsilon^{1/2}, \quad g\to\epsilon, \quad \theta: \mathrm{fixed}.
\end{equation*}
%Therefore the expansion parameter $\frac{g\theta}{2\pi\alpha'}$ goes to zero, and so our expansion can be considered essentially as the $\alpha'\to 0$ limit. 

In the remaining of this subsection we calculate the boundary state \eqref{d2boundary:eq:2} of infinitely many D0-branes of the configuration \eqref{d2boundary:eq:1}. We consider order by order in powers of $\mathcal{X}^i$ other than a factor $e^{i\hat{p}\mathbf{X}}$.

\paragraph*{zeroth order}\ \\
First, we consider the zeroth order term. The following identity holds: 
\begin{equation*}
\mathrm{tr}e^{i\mathbf{X}^i\hat{p}_i}=\frac{2\pi}{\theta}\delta^2(\hat{p}).
\end{equation*}
The numerical factor can be confirmed by using a coherent state. Thus at zeroth order the boundary state becomes
\begin{equation*}
   \frac{2\pi}{\theta}\delta^2(\hat{p})\delta^2(\hat{x})|p_i=0\rangle=\frac{1}{\theta}\delta^2(\hat{p})|x^i=0\rangle=\frac{1}{2\pi\theta}|p_i=0\rangle.
\end{equation*}

\paragraph*{first order}\ \\
Second, we confirm the first order terms vanish. By using cyclicity of trace, a trace part of the first order terms becomes
\begin{equation*}
\mathrm{tr}\left[e^{i(1-\sigma_1)\mathbf{X}\cdot\hat{p}}\mathbf{X}^{i_1} e^{i\sigma_1\mathbf{X}\cdot\hat{p}}\right]      = \mathrm{tr}\left[e^{i\mathbf{X}\cdot\hat{p}}\mathbf{X}^{i_1}\right].
\end{equation*}
$\mathbf{X}^i\hat{p}_i\ (i=1,2)$ is abbreviated as $\mathbf{X}\cdot\hat{p}$. We can see the trace part is independent of $\sigma_1$. Integrate a oscillation operator part over $\sigma$, then the first oder terms vanish as follows.
\begin{equation*}
   \frac{ig}{\sqrt{2\alpha'}}\mathrm{tr}\left[e^{i\mathbf{X}\cdot\hat{p}}\mathbf{X}^{i_1}\right]\sum_{n_1\ne 0}(\alpha_{n_1}^{i_1}+\tilde{\alpha}_{-n_1}^{i_1})\int_0^1 d\sigma_1 e^{2\pi i n_1\sigma_1} =0.
\end{equation*}
This result is consistent to the level matching condition for closed string states. 

\paragraph*{second order}\ \\
At second order what we have to consider is
\begin{equation*}
\begin{split}
   \int_0^1 d\sigma_1\ \int_0^{\sigma_1}d\sigma_2\ \left(\frac{ig}{\sqrt{2\alpha'}}\right)^2 & \mathrm{tr}\left[e^{i(1-\sigma_{12})\mathbf{X}\cdot\hat{p}}\mathbf{X}^{i_1}e^{i\sigma_{12}\mathbf{X}\cdot\hat{p}}\mathbf{X}^{i_2}\right] \\
& \sum_{n_1,n_2\ne 0}(\alpha_{n_1}^{i_1}+\tilde{\alpha}_{n_1}^{i_1})(\alpha_{n_2}^{i_2}+\tilde{\alpha}_{n_2}^{i_2})e^{2\pi i(n_1\sigma_1+n_2\sigma_2)}.
\end{split}
\end{equation*}
Here we have used cyclicity of trace, and $\sigma_{12}=\sigma_1-\sigma_2$. We divide our calculations into three steps. First we calculate the trace part. Next we compute the string oscillation part. Finally we perform the integrals. 

In order to calculate the trace part, we consider the matrices $\mathbf{X}^i$ as representation matrices of operators $X^i$ acting on a Hilbert space. We introduce operators $X^i$ and a Hilbert space\footnote{We denote a state in the Hilbert space by double angle-brackets $|\cdot\rangle\rangle$ to distinguish from a closed string state denoted by a single angle-bracket $|\cdot\rangle$.} $\mathcal{H}=\{|n\rangle\rangle\ |\ n=0,1,2,\cdots\}$ such that
\begin{equation*}
   \langle\langle n| X^i | m \rangle \rangle = (\mathbf{X}^i)_{nm}. 
\end{equation*}
Such operators can be constructed by considering a harmonic oscillator. The operators $X^i$ should satisfy an identity
\begin{equation*}
   [X^1,X^2]=i\theta.
\end{equation*}
This is identically the commutator relation of position and momentum operators in the one-dimensional quantum mechanics. On the basis of this observation, we define creation and annihilation operators by
\begin{equation*}
   a^\dagger=\frac{1}{\sqrt{2\theta}}(X^1-iX^2) \qquad a=\frac{1}{\sqrt{2\theta}}(X^1+iX^2).
\end{equation*}
Then we identify $|n\rangle\rangle$ as a number state, or explicitly
\begin{equation*}
   a|0\rangle\rangle=0,\quad |n\rangle\rangle=\frac{(a^\dagger)^n}{\sqrt{n!}}|0\rangle\rangle.
\end{equation*}
In addition, it is convenient to introduce a coherent state $|\mu \rangle\rangle =e^{-\frac{|\mu|^2}{2}}e^{\mu a^\dagger}|0\rangle\rangle$ where $\mu$ is a complex number. A coherent state has properties
\begin{equation*}
  a |\mu \rangle\rangle = \mu |\mu \rangle\rangle 
\quad
 \langle \langle m | \mu \rangle \rangle = e^{-\frac{|\mu|^2}{2}}\frac{\mu^m}{\sqrt{m!}} 
\quad
  1  = \frac{1}{\pi}\int\!\! d^2\mu \ |\mu \rangle\rangle \langle\langle \mu| 
\end{equation*}
where $d^2\mu$ is an integral over real and imaginary parts of $\mu$. Note that coherent states are not orthonormal, but satisfy
\begin{equation*}
  \langle\langle \mu' | \mu \rangle\rangle = e^{-\frac{|\mu'|^2}{2}-\frac{|\mu|^2}{2}+\mu^{'*}\mu}.
\end{equation*}

In what follows, we calculate a trace part of the second order terms by using coherent states. Note that the trace over matrix indices is equivalent to the one over the Hilbert space. The trace part is written as
\begin{align*}
 F_{\mathrm{2nd}}^{i_1i_2}
&=\mathrm{tr}\left[e^{i(1-\sigma_{12})\mathbf{X}\cdot\hat{p}}\mathbf{X}^{i_1}e^{i\sigma_{12}\mathbf{X}\cdot\hat{p}}\mathbf{X}^{i_2}\right]\\
%&=F_{\mathrm{2nd}}(k_1^i=(1-\sigma_{12})\hat{p}^i,k_3^i=\sigma_{12}\hat{p}^i;k_2^i=\delta^{i_1 i},k_4^i=\delta^{i_2i}),
&=F_{\mathrm{2nd}}\left((1-\sigma_{12})\hat{p},\sigma_{12}\hat{p}\ ;\ \delta^{i_1 i},\delta^{i_2i}\right),
\end{align*}
where 
\begin{equation*}
F_{\mathrm{2nd}}(k_1,k_3\ ;\ k_2,k_4)=\mathrm{tr}\left[e^{ik_1X}k_2X e^{ik_3X}k_4X\right]. 
\end{equation*}
Note that we denote indexes of space $i=1,2$ by superscripts $k^i$, while different variables are distinguished by subscripts $k_m$ . In other words $k_m^i$ represents the $i$-th space component of a variable $k_m$. Then we calculate $F_{\mathrm{2nd}}(k_1,k_3;k_2,k_4)$. An relation
\begin{equation*}
e^{ik_aX}=e^{i\lambda_a^* a}e^{i\lambda_a a^\dagger}e^{\frac{|\lambda_a|^2}{2}}
\end{equation*}
holds from the Baker-Campbell-Hausdorff formula where $\lambda_a=\sqrt{\frac{\theta}{2}}(k_a^1+ik_a^2)$. By using this relation we have
\begin{equation*}
\begin{split}
&F_{\mathrm{2nd}}(k_1,k_3;k_2,k_4)\\
&=\frac{1}{\pi^3}e^{\frac{|\lambda_1|^2}{2}+\frac{|\lambda_3|^2}{2}}\sum_n \iiint d^2\mu_1 d^3 \mu_3 d^2 \mu_5 \langle\langle n| e^{i\lambda_1^*a} |\mu_1\rangle\rangle\langle\langle \mu_1|e^{i\lambda_1a^\dagger}(\lambda_2a^\dagger+\lambda_2^*a)e^{i\lambda_3^*a}|\mu_3\rangle\rangle \\
&\hspace{10em}\langle\langle \mu_3|e^{i\lambda_3a^\dagger}(\lambda_4a^\dagger+\lambda_4^*a)|\mu_5\rangle\rangle\langle\langle \mu_5| n\rangle\rangle \\
&=\frac{1}{\pi^3}e^{\frac{|\lambda_1|^2}{2}+\frac{|\lambda_3|^2}{2}}\iiint d^2\mu_1 d^2 \mu_3 d^2\mu_5\ (\lambda_2\mu_1^*+\lambda_2\mu_3)(\lambda_4\mu_3^*+\lambda_4^*\mu_5)\\
&\hspace{10em}e^{-|\mu_1|^2-|\mu_3|^2-|\mu_5|^2+\mu_5^*\mu_1+\mu_1^*\mu_3+\mu_3^*\mu_5+i\lambda_1^*\mu_1+i\lambda_1\mu_1^*+i\lambda_3^*\mu_3+i\lambda_3\mu_3^*}\\
&=\frac{1}{\pi}e^{\frac{|\lambda_1|^2}{2}-\frac{|\lambda_3|^2}{2}}\int d^2\mu_1 \Bigl((\lambda_2\lambda_4^*+\lambda_2^*\lambda_4)|\mu_1|^2+\lambda_2^*\lambda_4^*(\mu_1)^2+\lambda_2\lambda_4(\mu_1^*)^2\\
&\hspace{10em}+i(\lambda_2^*\lambda_3\lambda_4^*+\lambda_2^*\lambda_3^*\lambda_4)\mu_1+i(\lambda_2\lambda_3^*\lambda_4+\lambda_2^*\lambda_3\lambda_4)\mu_1^*\\
&\hspace{20em}\ \ +\lambda_2^*\lambda_4(1-|\lambda_3|^2) \Bigr)e^{i(\lambda_1^*+\lambda_3^*)\mu_1+i(\lambda_1+\lambda_3)\mu_1^*}.
\end{split}
\end{equation*}
Substituting $k_1=(1-\sigma_{12})\hat{p},k_3=\sigma_{12}\hat{p}$ into this, we have
\begin{align*}
\frac{(2\pi)^2}{\pi}e^{\frac{1}{4}(1-2\sigma_{12})\theta|p|^2}& \Bigl\{\\
&(\lambda_2\lambda_4^*+\lambda_2^*\lambda_4)\frac{1}{(i\sqrt{2\theta})^2}\left(\frac{\partial^2}{\partial p_1p_1}+\frac{\partial^2}{\partial p_2p_2}\right)\\
&+\lambda_2^*\lambda_4^*\frac{1}{(i\sqrt{2\theta})^2}\left(\frac{\partial^2}{\partial p_1p_1}+2i\frac{\partial^2}{\partial p_1p_2}-\frac{\partial^2}{\partial p_2p_2}\right)\\
&+\lambda_2\lambda_4\frac{1}{(i\sqrt{2\theta})^2}\left(\frac{\partial^2}{\partial p_1p_1}-2i\frac{\partial^2}{\partial p_1p_2}-\frac{\partial^2}{\partial p_2p_2}\right)\\
&+i\sqrt{\frac{\theta}{2}}(\lambda_2^*(p_1+ip_2)\lambda_4^*+\lambda_2^*(p_1-ip_2)\lambda_4)\frac{1}{i\sqrt{2\theta}}\left(\frac{\partial}{\partial p_1}+i\frac{\partial}{\partial p_2}\right)\\
&+i\sqrt{\frac{\theta}{2}}(\lambda_2(p_1-ip_2)\lambda_4+\lambda_2^*(p_1+ip_2)\lambda_4)\frac{1}{i\sqrt{2\theta}}\left(\frac{\partial}{\partial p_1}-i\frac{\partial}{\partial p_2}\right)\\
&+\lambda_2^*\lambda_4(1-|\lambda_3|^2) \Bigr\}\delta^2(\sqrt{2\theta}p).
\end{align*}
Integrating by part, this equation becomes
\begin{equation*}
  \frac{\pi}{\theta}(\lambda_2^*\lambda_4-\lambda_2\lambda_4^*)(1-2\sigma_{12})\delta^2(\hat{p}).
\end{equation*}
Into this we substitute $\lambda_2=\sqrt{\frac{\theta}{2}}(\delta^{i_1 1}+i \delta^{i_1 2}),\lambda_4=\sqrt{\frac{\theta}{2}}(\delta^{i_2 1}+i \delta^{i_2 2})$, then the trace part becomes
\begin{equation*}
   F_{\mathrm{2nd}}^{i_1i_2}=i\pi(\delta^{i_1 1}\delta^{i_2 2}-\delta^{i_1 2}\delta^{i_2 1})(1-2\sigma_{12})\delta^2(\hat{p})=i\pi(1-2\sigma_{12})\delta^2(\hat{p})\epsilon^{i_1i_2}.
\end{equation*}
Here we set $\epsilon^{12}=1$. 

The string oscillation part becomes $A_2^{i_1i_2}(\sigma_1,\sigma_2)$ given in appendix \ref{boundary:appendix:a}. Finally we integrate a product of the trace and oscillation parts over $\sigma_1$ and $\sigma_2$. After integration non-vanishing terms are
\begin{equation*}
\begin{split}
 &\int_0^1 d\sigma_1 \int_0^{\sigma_1} d\sigma_2\ A^{i_1i_2}_{2,2}(\sigma_1,\sigma_2) i\pi(1-2\sigma_1+2\sigma_2)\epsilon^{i_1i_2}
=\sum_{n>0}\frac{4}{n}\alpha_{-n_1}^{i_1}\tilde{\alpha}_{-n_2}^{i_2}\epsilon^{i_1i_2}.
\end{split}
\end{equation*}
%Pay attention that in integration the case of $n_1=n_2$ should be treated separately from general values of $n_1,n_2$.
After all the boundary state at second order becomes
\begin{equation*}
 \left(\frac{ig}{\sqrt{2\alpha'}}\right)^2\sum_{n>0}\frac{4}{n}\alpha_{-n_1}^{i_1}\tilde{\alpha}_{-n_2}^{i_2}\epsilon^{i_1i_2}\delta^2(\hat{p})=\frac{2\pi}{\theta}\delta^2(\hat{p})\left(-\frac{2g\theta}{2\pi\alpha'}\right)\sum_{n>0}\frac{1}{n}g\alpha_{-n_1}^{i_1}\tilde{\alpha}_{-n_2}^{i_2}\epsilon^{i_1i_2}
\end{equation*}
where we restore the abbreviation of $A_2^{i_1i_2}(\sigma_1,\sigma_2)$.

\paragraph*{third order}\ \\
We can see that the third order terms vanish in a similar way. We omit detailed calculations. Note that disappearance of the third order terms is consistent to the level matching condition for closed string states.

\paragraph*{fourth order}\ \\
At fourth order we should calculate
\begin{multline*}
   \int_0^1 d\sigma_1 \int_0^{\sigma_1}d\sigma_2 \int_0^{\sigma_2}d\sigma_3 \int_0^{\sigma_3}d\sigma_4\ \left(\frac{ig}{\sqrt{2\alpha'}}\right)^4\\
\mathrm{tr}\left[e^{i(1-\sigma_{14})\mathbf{X}\cdot\hat{p}}\mathbf{X}^{i_1}e^{i\sigma_{12}\mathbf{X}\cdot\hat{p}}\mathbf{X}^{i_2}e^{i\sigma_{23}\mathbf{X}\cdot\hat{p}}\mathbf{X}^{i_3}e^{i\sigma_{34}\mathbf{X}\cdot\hat{p}}\mathbf{X}^{i_4}\right]\\
\sum_{n_1,n_2,n_3,n_4\ne 0}(\alpha_{n_1}^{i_1}+\tilde{\alpha}_{n_1}^{i_1})(\alpha_{n_2}^{i_2}+\tilde{\alpha}_{n_2}^{i_2})(\alpha_{n_3}^{i_3}+\tilde{\alpha}_{n_3}^{i_3})(\alpha_{n_4}^{i_4}+\tilde{\alpha}_{n_4}^{i_4})e^{2\pi i(n_1\sigma_1+n_2\sigma_2+n_3\sigma_3+n_4\sigma_4)}.
\end{multline*}
In a way similar to the case at second order, we divide calculations into three steps: calculations of the trace part, the string oscillation part, and the integrations over $\sigma$'s.

The trace part at fourth order can be calculated in a manner similar to the case at second oder, and the result is
\begin{equation*}  
\begin{split}
&F_\mathrm{4th}(k_1,k_2,k_3,k_4)\\
=&\mathrm{tr}\left[e^{i(1-\sigma_{14})\mathbf{X}\cdot\hat{p}}k_1\mathbf{X}e^{i\sigma_{12}\mathbf{X}\cdot\hat{p}}k_2\mathbf{X}e^{i\sigma_{23}\mathbf{X}\cdot\hat{p}}k_3\mathbf{X}e^{i\sigma_{34}k_4\mathbf{X}\cdot\hat{p}}k_4\mathbf{X}\right]\\
=&\frac{\theta}{\pi}\Bigl[(\lambda_{2r}\lambda_{4r}\lambda_{6i}\lambda_{8i}+\lambda_{2i}\lambda_{4i}\lambda_{6r}\lambda_{8r})\\
&\hspace{4em}(-4+8(\sigma_{1}+\sigma_2-\sigma_{3}-\sigma_4)+8(\sigma_1+\sigma_2)(\sigma_3+\sigma_4)-16(\sigma_1\sigma_2+\sigma_3\sigma_4))\\
&+(\lambda_{2r}\lambda_{4i}\lambda_{6r}\lambda_{8i}+\lambda_{2i}\lambda_{4r}\lambda_{6i}\lambda_{8r})\\
&\hspace{4em}(-8(\sigma_{2}-\sigma_3)+8(\sigma_1+\sigma_3)(\sigma_2+\sigma_4)-16(\sigma_1\sigma_3+\sigma_2\sigma_4))\\
&+(\lambda_{2r}\lambda_{4i}\lambda_{6i}\lambda_{8r}+\lambda_{2i}\lambda_{4r}\lambda_{6r}\lambda_{8i})\\
&\hspace{4em}(4-8(\sigma_{1}-\sigma_4)+8(\sigma_1+\sigma_4)(\sigma_2+\sigma_3)-16(\sigma_1\sigma_4+\sigma_2\sigma_3))\Bigr],
\end{split}
\end{equation*}
where $\lambda_2=\sqrt{\frac{\theta}{2}}(k_1^1+ik_1^2),\lambda_4=\sqrt{\frac{\theta}{2}}(k_2^1+ik_2^2),\lambda_6=\sqrt{\frac{\theta}{2}}(k_3^1+ik_3^2),\lambda_8=\sqrt{\frac{\theta}{2}}(k_4^1+ik_4^2)$. The subscripts $r$ and $i$ represent real and imaginary parts. Substitute $k_1^i=\delta^{i_1 i},k_2^i=\delta^{i_2 i},k_3^i=\delta^{i_3 i},k_4^i=\delta^{i_4 i}$ into $F_{\mathrm{4th}}$, and then we get the trace part $F_{\mathrm{4th}}^{i_1i_2i_3i_4}$.

The string oscillator part is $A_4^{i_1i_2i_3i_4}(\sigma_1,\sigma_2,\sigma_3\,\sigma_4)$ given in appendix \ref{boundary:appendix:a}. Finally we integrate a product of the trace and string oscillation parts. Although we omit details of the integrations, the boundary state at fourth order becomes
\begin{equation*}
\begin{split}
   &\left(\frac{ig}{\sqrt{2\alpha'}}\right)^4\delta^2({\hat{p}})\left(-\frac{2\theta}{g^2\pi}\zeta(0)-\frac{4\theta}{g\pi}\sum_{n>0}\frac{1}{n}\alpha_{-n}^i\tilde{\alpha}_{-n}^i\right)\\
%&=\frac{2\pi}{\theta}\delta^2(\hat{p})\left(-\frac{\zeta(0)g^2\theta^2}{4\pi^2\alpha{'^2}}-\frac{2g^2\theta^2}{4\pi^2{\alpha'}^2}\sum_{n>0}\frac{1}{n}g\alpha_{-n}^{i_1}\tilde{\alpha}_{-n}^{i_2}\delta^{i_1i_2})\right)
&=\frac{2\pi}{\theta}\delta^2(\hat{p})\left(-\left(\frac{g\theta}{2\pi\alpha'}\right)^2\zeta(0)-2\left(\frac{g\theta}{2\pi\alpha'}\right)^2\sum_{n>0}\frac{1}{n}g\alpha_{-n}^{i_1}\tilde{\alpha}_{-n}^{i_2}\delta^{i_1i_2})\right)
\end{split}
\end{equation*}
where we restore the abbreviation of $A_4^{i_1i_2i_3i_4}(\sigma_1,\sigma_2,\sigma_3\,\sigma_4)$.

\paragraph*{Equivalence between boundary states of D2-brane and multiple D0-branes}\ \\
Note that we focus on a matter part of directions $x^1,x^2$ in the boundary state. We have calculated the boundary state \eqref{d2boundary:eq:2} up to second order of $\frac{1}{b}=\frac{g\theta}{2\pi\alpha'}$, and the results is
\begin{equation}
\begin{split}
 |B\rangle=& \frac{T_0}{2}\frac{2\pi}{\theta}\delta^2(\hat{p})\delta^2(\hat{x})\left(1-\frac{\zeta(0)}{b^2}-\left(\frac{2}{b}\epsilon^{i_1i_2}+\frac{2}{b^2}\delta^{i_1i_2}\right)\sum_{n>0}\frac{1}{n}g\alpha_{-n}^{i_1}\tilde{\alpha}_{-n}^{i_2}+\mathcal{O}(b^{-3})\right)\\
&\hspace{10em}\exp\left\{\sum_{n>0}\frac{1}{n}g\alpha^{i}_{-n}\tilde{\alpha}^{i}_{-n}\right\}|0\rangle\\
=&\frac{T_0}{4\pi\theta}\left(1-\frac{\zeta(0)}{b^2}+\sum_{n>0}\frac{1}{n}g\alpha_{-n}^{i_1}\tilde{\alpha}_{-n}^{i_2}
\begin{pmatrix}
 -\frac{2}{b^2} & -\frac{2}{b} \\ \frac{2}{b} & -\frac{2}{b^2}
\end{pmatrix}^{i_1i_2}
+\mathcal{O}(b^{-3})\right)e^{\sum_{n>0}\frac{1}{n}g\alpha^{i}_{-n}\tilde{\alpha}^{i}_{-n}}|0\rangle.
\end{split} \label{boundary:eq:noncommutative1}
\end{equation}
Therefore, it is sufficient to consider only $\hat{p}^1=\hat{p}^2=0$ terms because the boundary state is proportional to $\delta(\hat{p}^1)\delta(\hat{p}^2)$. By using this result, we can easily confirm that our result \eqref{boundary:eq:p=0} of the general boundary state at $\hat{p}=0$ realizes the boundary state derived in this subsection \eqref{boundary:eq:noncommutative1}. 
  
It is known that a boundary state of a D2-brane with a background B-field $B_{ij}=B\epsilon_{ij}$ is written as \cite{9906214} 
\begin{align*}
   |D2\rangle_B &=\frac{T_2}{2}(\det(g+B))^{-\zeta(0)}\exp \left[-\sum_{n>0}\frac{1}{n}\alpha^{i}_{-n}\tilde{\alpha}^{j}_{-n}M_{ij}\right]|0\rangle \\ 
                          M_{ij}&=(g-B)_{ik}(g+B)^{-1\ kl}g_{lj}=\frac{g}{g^2+B^2}\begin{pmatrix}g^2-B^2 & 2gB \\  -2gB &g^2-B^2 \end{pmatrix}_{ij}.
%\left(\frac{1+B}{1-B}\right)^{ij}=\frac{1}{1+B^2}\begin{pmatrix}1-B^2 & 2B \\ -2B & 1-B^2 \end{pmatrix}^{ij}. 
\end{align*}
We expand this boundary state in $\frac{1}{B}$ to have
\begin{equation}
\begin{split}
  |D2\rangle_B &=\frac{T_2}{2}B^{-2\zeta(0)}\left(1-\frac{\zeta(0)g^2}{B^2}+\sum_{n>0}\frac{1}{n}g\alpha_{-n}^{i_1}\tilde{\alpha}_{-n}^{i_2} 
\begin{pmatrix}
 -\frac{2g^2}{B^2} & -\frac{2g}{B} \\ \frac{2g}{B} & -\frac{2g^2}{B^2}
\end{pmatrix}^{i_1i_2}
+\mathcal{O}((B/g)^{-3}\right)\\
&\hspace{10em}\exp\left\{\sum_{n>0}\frac{1}{n}g\alpha^{j_1}_{-n}\tilde{\alpha}^{j2}_{-n}\delta^{j_1j_2}\right\}|0\rangle.
\end{split} \label{boundary:eq:noncommutative2}
\end{equation}
With the identification
\begin{equation*}
   b=\frac{B}{g},
\end{equation*}
\eqref{boundary:eq:noncommutative1} and \eqref{boundary:eq:noncommutative2} are identical. Thus we can say two boundary states $|B\rangle$ and $|D2\rangle_B$, the former is constructed from D0-brane matrices and the latter is the D2-brane boundary state, are equivalent at least up to $\mathcal{O}(b^{-3})$. We can see that overall normalization is identical:
\begin{equation*}
    \frac{T_2}{2}B^{-2\zeta(0)}=\frac{T_0}{4\pi\theta}
\end{equation*}
where $T_2=\frac{T_0}{4\pi^2\alpha'}$, and $\zeta(0)=-\frac{1}{2}$. Although we have omitted $(\alpha_{-n})^4$ terms, they become
\begin{equation*}
    \sum_{n,m>0}g^2\frac{2}{b^2}\left(-2\alpha_{-n}^1\alpha_{-m}^2\tilde{\alpha}_{-n}^1\tilde{\alpha}_{-m}^2+\alpha_{-n}^1\alpha_{-m}^1\tilde{\alpha}_{-n}^2\tilde{\alpha}_{-m}^2+\alpha_{-n}^2\alpha_{-m}^2\tilde{\alpha}_{-n}^1\tilde{\alpha}_{-m}^1)\right)
\end{equation*}
both from $|B\rangle$ and $|D2\rangle_B$.

To summarize this subsection, we have confirmed that the boundary state \eqref{d2boundary:eq:2} constructed from the matrices \eqref{d2boundary:eq:1} reproduces the boundary state of a D2-brane with a constant background B-field. 

%%%%%%%%%%%%%%%
% paper:part:5
%%%%%%%%%%%%%%%

\section{Conclusion and discussion}\label{paper:section:5}

We have studied the boundary state of multiple D0-branes with an arbitrary configuration of the scalar field in $\alpha'$ expansion both in bosonic string and superstring theories. The boundary state is BRST invariant for an arbitrary configuration formally. However, the boundary state includes singularities when the scalar field is off-shell. Hence our boundary state is well-defined only when the scalar filed satisfies the equation of motion. 
%These divergences are absorbed by a field redefinition which is singular if the equation of motion is satisfied. 
In other words the on-shell boundary state contains no singularity and BRST invariant. In this way we have extracted the correct equation of motions of multiple D0-branes from the boundary state by requiring finiteness of the boundary state. Furthermore we have investigated couplings of massless open string fields to NS-NS massless closed string fields. Our results \eqref{boundary:eq:main-result-coupling-bosonic}, \eqref{boundary:eq:main-result-coupling-bosonic2}, \eqref{boundary:eq:main-result-coupling-super} realize the correct formulas for supergravity current distribution and a linear part  of the non-Abelian DBI action in closed string fields at least up to order ${\alpha'}^2$ . In addition, we have confirmed our boundary state is identical to the previously known one in the cases of a single boosted D0-brane and a noncommutative D2-brane. To summarize, our results support the formula for non-Abelian boundary states defined by using the Wilson loop factor is the correct one. 

At the end of this paper, we discuss future directions and related topics. We have focused on couplings of massless open string fields to massless NS-NS closed fields. We can derive couplings to massless R-R closed string fields to extract D-charge density, couplings to massive closed string fields, and couplings of fermionic and massive open string fields. It is interesting problem to reveal an role of massive open strings in noncommutative and fuzzy D-brane systems.

We can calculate the boundary state and closed string coupling at higher orders in $\alpha'$. If our boundary state is the correct one, higher $\alpha'$ corrections to non-Abelian DBI action can be derived by developing our study in this paper. In a similar way we can calculate $\alpha'$ corrections to closed string couplings such as energy-momentum tensor, F-charge density and D-charge density. Note that the non-Abelian DBI action with symmetrized trace \cite{9910053} is valid only up to order ${\alpha'}^4$ \cite{9703217,0002180}. 
%We expect that our boundary state is correct at all orders in $\alpha'$. If this is the case, the boundary state gives $\alpha'$ corrections to the non-Abelian DBI action correctly in contrast that symmetrized trace is valid only up to ${\alpha'}^4$. With the method used in this chapter, we can derive $\alpha'$ corrections at higher orders to the non-Abelian DBI action, energy-momentum tensor, F-charge density, and D-charge density.

One of important problems is to find a way to know boundary conditions a non-Abelian boundary state satisfies, and how open string excitations influence to them. %We cannot determine definite positions where D-branes are located in multiple D-brane system because they are noncommutative in general. 
Such a method will give us a geometrical interpretation of the general non-Abelian D-brane system. 
A boundary state for a D-brane satisfying the Dirichlet boundary condition on a curved submanifold embedded in the flat space is studied in \cite{0505184}.

Another unsolved problem is to find a way to include $g_s$ correction. A boundary state represents only closed string state emitted from D-branes, and effects of them on the D-branes are ignored. We can also say that D-branes are regarded as infinitely massive objects. In other words we cannot deal with a worldsheet which has many separated boundaries on D-branes by a boundary state. %Less is known about backreaction. 
$g_s$ corrections in a closed string field theory with a dynamical D-brane are considered in \cite{0309074}. Scattering of quantized two D0-branes is studies in \cite{9705111}. It may be useful to introduce a concept like a solitonic operator to D-brane systems \cite{0603152}. 

It is also interesting to consider the idempotency relation \cite{0306189} of non-Abelian boundary states with an arbitrary open string excitation. 

%%%%%%%%%%%%%
\vfill

\acknowledgments{I would like to thank Mitsuhiro Kato and Koji Hashimoto for valuable advice and discussion.}

\appendix
%%%%%%%%%%%%%%%
% paper:part:review
%%%%%%%%%%%%%%%

\section{Creation operators on D0-brane boundary state}\label{boundary:appendix:a}
In this appendix results of $A_{p,q}^{\mu_1\cdots\mu_p}(\sigma_1,\cdots, \sigma_p), B_{p,q}^{\mu_1\cdots\mu_p}(\sigma_1,\cdots, \sigma_p)$ are shown. They are abbreviated according to \eqref{boundary:eq:rescale} in this appendix for simplicity. In the final results, restore these factors all.

\subsection*{Worldsheet boson}
\begin{align*}
A^{0}_1(\sigma_1)&=\sum_{n>0} \frac{-2}{n}(\tilde{\alpha}^0_{-n}e^{2\pi i n_1\sigma_1}+\alpha^0_{-n}e^{-2\pi i n_1\sigma_1})\\
A^{i_1}_1(\sigma_1)&=\sum_{n>0}2(\tilde{\alpha}^{i_1}_{-n}e^{2\pi i n_1\sigma_1}+\alpha^{i_1}_{-n}e^{-2\pi i n_1\sigma_1})
\end{align*}

\begin{align*}
A^{00}_2(\sigma_1,\sigma_2)=&\sum_{n_1,n_2>0}\frac{4}{n_1n_2}\left(\tilde{\alpha}_{-n_1}^0\tilde{\alpha}_{-n_2}^0e^{2\pi i (n_1\sigma_1+n_2\sigma_2)}+\alpha_{-n_1}^0\tilde{\alpha}_{-n_2}^0e^{2\pi i (-n_1\sigma_1+n_2\sigma_2)}\right.\\
&\phantom{=\sum_{n_1,n2>0}\frac{4}{n_1n_2}(}
\left. +\ \tilde{\alpha}_{-n_1}^0\alpha_{-n_2}^0e^{2\pi i (n_1\sigma_1-n_2\sigma_2)}+\alpha_{-n_1}^0\alpha_{-n_2}^0e^{2\pi i (-n_1\sigma_1-n_2\sigma_2)}\right)\\
&+\sum_{n>0}\frac{4}{ng}\cos 2\pi n \sigma_{12}
\end{align*}
\begin{align*}
A^{i_1i_2}_2(\sigma_1,\sigma_2)=&\sum_{n_1,n_2>0}4\left(\tilde{\alpha}_{-n_1}^{i_1}\tilde{\alpha}_{-n_2}^{i_2}e^{2\pi i (n_1\sigma_1+n_2\sigma_2)}+\alpha_{-n_1}^{i_1}\tilde{\alpha}_{-n_2}^{i_2}e^{2\pi i (-n_1\sigma_1+n_2\sigma_2)}\right.\\
&\phantom{=\sum_{n_1,n2>0}4(}
\left. +\ \tilde{\alpha}_{-n_1}^{i_1}\alpha_{-n_2}^{i_2}e^{2\pi i (n_1\sigma_1-n_2\sigma_2)}+\alpha_{-n_1}^{i_1}\alpha_{-n_2}^{i_2}e^{2\pi i (-n_1\sigma_1-n_2\sigma_2)}\right)\\
&+\sum_{n>0}\frac{4n}{g}\delta^{i_1i_2}\cos 2\pi n \sigma_{12}
\end{align*}

\begin{equation*}
\begin{split}
&A^{000}_{3,3}(\sigma_1,\sigma_2,\sigma_3)=\sum_{n_1,n_2,n_3>0}\frac{-8}{n_1n_2n_3}\\
&\left\{\tilde{\alpha}_{-n_1}^{0}\tilde{\alpha}_{-n_2}^{0}\tilde{\alpha}_{-n_3}^{0}e^{2\pi i (+n_1\sigma_1+n_2\sigma_2+n_3\sigma_3)}+\tilde{\alpha}_{-n_1}^{0}\tilde{\alpha}_{-n_2}^{0}\alpha_{-n_3}^{0}e^{2\pi i (+n_1\sigma_1+n_2\sigma_2-n_3\sigma_3)}\right.\\
&+\tilde{\alpha}_{-n_1}^{0}\alpha_{-n_2}^{0}\tilde{\alpha}_{-n_3}^{0}e^{2\pi i (+n_1\sigma_1-n_2\sigma_2+n_3\sigma_3)}+\alpha_{-n_1}^{0}\tilde{\alpha}_{-n_2}^{0}\alpha_{-n_3}^{0}e^{2\pi i (-n_1\sigma_1+n_2\sigma_2-n_3\sigma_3)}\\
&+\tilde{\alpha}_{-n_1}^{0}\alpha_{-n_2}^{0}\alpha_{-n_3}^{0}e^{2\pi i (+n_1\sigma_1-n_2\sigma_2+n_3\sigma_3)}+\alpha_{-n_1}^{0}\tilde{\alpha}_{-n_2}^{0}\alpha_{-n_3}^{0}e^{2\pi i (-n_1\sigma_1+n_2\sigma_2-n_3\sigma_3)}\\
&\left.+\alpha_{-n_1}^{0}\alpha_{-n_2}^{0}\tilde{\alpha}_{-n_3}^{0}e^{2\pi i (-n_1\sigma_1-n_2\sigma_2+n_3\sigma_3)}+\alpha_{-n_1}^{0}\alpha_{-n_2}^{0}\alpha_{-n_3}^{0}e^{2\pi i (-n_1\sigma_1-n_2\sigma_2-n_3\sigma_3)}\right\}\\
\end{split}
\end{equation*}
\begin{equation*}
\begin{split}
&A^{000}_{3,1}(\sigma_1,\sigma_2,\sigma_3)=\sum_{n,m>0}\frac{-8}{gnm}\left\{ \cos 2\pi m\sigma_{23}(\tilde{\alpha}_{-n}^0 e^{2\pi i n \sigma_1}+\alpha_{-n}^0 e^{-2\pi i n\sigma_1}) \right.\\
&\left.+\cos 2\pi m\sigma_{13}(\tilde{\alpha}_{-n}^0 e^{2\pi i n \sigma_2}+\alpha_{-n}^0e^{-2\pi i n\sigma_2})
+\cos 2\pi m\sigma_{12}(\tilde{\alpha}_{-n}^0e^{2\pi i n \sigma_3}+\alpha_{-n}^0e^{-2\pi i n\sigma_3}) \right\}
\end{split}
\end{equation*}
\begin{equation*}
\begin{split}
&A^{i_1i_2i_3}_{3,3}(\sigma_1,\sigma_2,\sigma_3)=\sum_{n_1,n_2,n_3>0}8\\
&\left\{\tilde{\alpha}_{-n_1}^{i_1}\tilde{\alpha}_{-n_2}^{i_2}\tilde{\alpha}_{-n_3}^{i_3}e^{2\pi i (+n_1\sigma_1+n_2\sigma_2+n_3\sigma_3)}+\tilde{\alpha}_{-n_1}^{i_1}\tilde{\alpha}_{-n_2}^{i_2}\alpha_{-n_3}^{i_3}e^{2\pi i (+n_1\sigma_1+n_2\sigma_2-n_3\sigma_3)}\right.\\
&+\tilde{\alpha}_{-n_1}^{i_1}\alpha_{-n_2}^{i_2}\tilde{\alpha}_{-n_3}^{i_3}e^{2\pi i (+n_1\sigma_1-n_2\sigma_2+n_3\sigma_3)}+\alpha_{-n_1}^{i_1}\tilde{\alpha}_{-n_2}^{i_2}\alpha_{-n_3}^{i_3}e^{2\pi i (-n_1\sigma_1+n_2\sigma_2-n_3\sigma_3)}\\
&+\tilde{\alpha}_{-n_1}^{i_1}\alpha_{-n_2}^{i_2}\alpha_{-n_3}^{i_3}e^{2\pi i (+n_1\sigma_1-n_2\sigma_2+n_3\sigma_3)}+\alpha_{-n_1}^{i_1}\tilde{\alpha}_{-n_2}^{i_2}\alpha_{-n_3}^{i_3}e^{2\pi i (-n_1\sigma_1+n_2\sigma_2-n_3\sigma_3)}\\
&\left.+\alpha_{-n_1}^{i_1}\alpha_{-n_2}^{i_2}\tilde{\alpha}_{-n_3}^{i_3}e^{2\pi i (-n_1\sigma_1-n_2\sigma_2+n_3\sigma_3)}+\alpha_{-n_1}^{i_1}\alpha_{-n_2}^{i_2}\alpha_{-n_3}^{i_3}e^{2\pi i (-n_1\sigma_1-n_2\sigma_2-n_3\sigma_3)}\right\}\\
\end{split}
\end{equation*}
\begin{equation*}
\begin{split}
A^{i_1i_2i_3}_{3,1}(\sigma_1,\sigma_2,\sigma_3)=\sum_{n,m>0}\frac{8m}{g}&\left\{ \delta^{i_2i_3}\cos 2\pi m\sigma_{23}(\tilde{\alpha}_{-n}^{i_1}e^{2\pi i n \sigma_1}+\alpha_{-n}^{i_1}e^{-2\pi i n\sigma_1}) \right.\\
&+\delta^{i_1i_3}\cos 2\pi m\sigma_{13}(\tilde{\alpha}_{-n}^{i_2}e^{2\pi i n \sigma_2}+\alpha_{-n}^{i_2}e^{-2\pi i n\sigma_2})\\ 
&\left.+\delta^{i_1i_2}\cos 2\pi m\sigma_{12}(\tilde{\alpha}_{-n}^{i_3}e^{2\pi i n \sigma_3}+\alpha_{-n}^{i_3}e^{-2\pi i n\sigma_3}) \right\}
\end{split}
\end{equation*}

\begin{equation*}
\begin{split}
&A^{0000}_{4,4}(\sigma_1,\sigma_2,\sigma_3,\sigma_4)=\sum_{n_1,n_2,n_3,n_4>0}\frac{16}{n_1n_2n_3n_4}\\
&\left\{\tilde{\alpha}_{-n_1}^{0}\tilde{\alpha}_{-n_2}^{0}\tilde{\alpha}_{-n_3}^{0}\tilde{\alpha}_{-n_4}^{0}e^{2\pi i(+n_1\sigma_1+n_2\sigma_2+n_3\sigma_3+n_4\sigma_4)}\right.\\
\\
&+\alpha_{-n_1}^{0}\tilde{\alpha}_{-n_2}^{0}\tilde{\alpha}_{-n_3}^{0}\tilde{\alpha}_{-n_4}^{0}e^{2\pi i(-n_1\sigma_1+n_2\sigma_2+n_3\sigma_3+n_4\sigma_4)}\\
&+\tilde{\alpha}_{-n_1}^{0}\alpha_{-n_2}^{0}\tilde{\alpha}_{-n_3}^{0}\tilde{\alpha}_{-n_4}^{0}e^{2\pi i(+n_1\sigma_1-n_2\sigma_2+n_3\sigma_3+n_4\sigma_4)}\\
&+\tilde{\alpha}_{-n_1}^{0}\tilde{\alpha}_{-n_2}^{0}\alpha_{-n_3}^{0}\tilde{\alpha}_{-n_4}^{0}e^{2\pi i(+n_1\sigma_1+n_2\sigma_2-n_3\sigma_3+n_4\sigma_4)} \\
&+\tilde{\alpha}_{-n_1}^{0}\tilde{\alpha}_{-n_2}^{0}\tilde{\alpha}_{-n_3}^{0}\alpha_{-n_4}^{0}e^{2\pi i(+n_1\sigma_1+n_2\sigma_2+n_3\sigma_3-n_4\sigma_4)}\\
\\
&+\alpha_{-n_1}^{0}\alpha_{-n_2}^{0}\tilde{\alpha}_{-n_3}^{0}\tilde{\alpha}_{-n_4}^{0}e^{2\pi i(-n_1\sigma_1-n_2\sigma_2+n_3\sigma_3+n_4\sigma_4)}\\
&+\alpha_{-n_1}^{0}\tilde{\alpha}_{-n_2}^{0}\alpha_{-n_3}^{0}\tilde{\alpha}_{-n_4}^{0}e^{2\pi i(-n_1\sigma_1+n_2\sigma_2-n_3\sigma_3+n_4\sigma_4)}\\
&+\alpha_{-n_1}^{0}\tilde{\alpha}_{-n_2}^{0}\tilde{\alpha}_{-n_3}^{0}\alpha_{-n_4}^{0}e^{2\pi i(-n_1\sigma_1+n_2\sigma_2+n_3\sigma_3-n_4\sigma_4)} \\
&+\tilde{\alpha}_{-n_1}^{0}\alpha_{-n_2}^{0}\alpha_{-n_3}^{0}\tilde{\alpha}_{-n_4}^{0}e^{2\pi i(+n_1\sigma_1-n_2\sigma_2-n_3\sigma_3+n_4\sigma_4)}\\
&+\tilde{\alpha}_{-n_1}^{0}\alpha_{-n_2}^{0}\tilde{\alpha}_{-n_3}^{0}\alpha_{-n_4}^{0}e^{2\pi i(+n_1\sigma_1-n_2\sigma_2+n_3\sigma_3-n_4\sigma_4)} \\
&+\tilde{\alpha}_{-n_1}^{0}\tilde{\alpha}_{-n_2}^{0}\alpha_{-n_3}^{0}\alpha_{-n_4}^{0}e^{2\pi i(+n_1\sigma_1+n_2\sigma_2-n_3\sigma_3-n_4\sigma_4)}\\
\\
&+\alpha_{-n_1}^{0}\alpha_{-n_2}^{0}\alpha_{-n_3}^{0}\tilde{\alpha}_{-n_4}^{0}e^{2\pi i(-n_1\sigma_1-n_2\sigma_2-n_3\sigma_3+n_4\sigma_4)} \\
&+\alpha_{-n_1}^{0}\alpha_{-n_2}^{0}\tilde{\alpha}_{-n_3}^{0}\alpha_{-n_4}^{0}e^{2\pi i(-n_1\sigma_1-n_2\sigma_2+n_3\sigma_3-n_4\sigma_4)}\\
&+\alpha_{-n_1}^{0}\tilde{\alpha}_{-n_2}^{0}\alpha_{-n_3}^{0}\alpha_{-n_4}^{0}e^{2\pi i(-n_1\sigma_1+n_2\sigma_2-n_3\sigma_3-n_4\sigma_4)} \\
&+\tilde{\alpha}_{-n_1}^{0}\alpha_{-n_2}^{0}\alpha_{-n_3}^{0}\alpha_{-n_4}^{0}e^{2\pi i(+n_1\sigma_1-n_2\sigma_2-n_3\sigma_3-n_4\sigma_4)}\\
\\
&\left.+\alpha_{-n_1}^{0}\alpha_{-n_2}^{0}\alpha_{-n_3}^{0}\alpha_{-n_4}^{0}e^{2\pi i(-n_1\sigma_1-n_2\sigma_2-n_3\sigma_3-n_4\sigma_4)}\right\}
\end{split}
\end{equation*}

\begin{equation*}
\begin{split}
A^{0000}_{4,2}(\sigma_1,\sigma_2,\sigma_3,\sigma_4)=&\sum_{n,m,l>0}\frac{16}{gnml}\\
&\Bigl\{ \cos2\pi l\sigma_{34}\left(\tilde{\alpha}^{0}_{-n}\tilde{\alpha}^{0}_{-m}e^{2\pi i (+n\sigma_1+m\sigma_2)}+\tilde{\alpha}^{0}_{-n}\alpha^{0}_{-m}e^{2\pi i (+n\sigma_1-m\sigma_2)}\right.\\
&\phantom{ \cos2\pi l\sigma_{34}\ }\left.+\alpha^{0}_{-n}\tilde{\alpha}^{0}_{-m}e^{2\pi i (-n\sigma_1+m\sigma_2)}+\alpha^{0}_{-n}\alpha^{0}_{-m}e^{2\pi i (-n\sigma_1-m\sigma_2)}\right)\\
&\!\!\!+ \cos2\pi l\sigma_{24}\left(\tilde{\alpha}^{0}_{-n}\tilde{\alpha}^{0}_{-m}e^{2\pi i (+n\sigma_1+m\sigma_3)}+\tilde{\alpha}^{0}_{-n}\alpha^{0}_{-m}e^{2\pi i (+n\sigma_1-m\sigma_3)}\right.\\
&\phantom{ \cos2\pi l\sigma_{24}\ }\left.+\alpha^{0}_{-n}\tilde{\alpha}^{0}_{-m}e^{2\pi i (-n\sigma_1+m\sigma_3)}+\alpha^{0}_{-n}\alpha^{0}_{-m}e^{2\pi i (-n\sigma_1-m\sigma_2)}\right)\\
&\!\!\!+ \cos2\pi l\sigma_{23}\left(\tilde{\alpha}^{0}_{-n}\tilde{\alpha}^{0}_{-m}e^{2\pi i (+n\sigma_1+m\sigma_4)}+\tilde{\alpha}^{0}_{-n}\alpha^{0}_{-m}e^{2\pi i (+n\sigma_1-m\sigma_4)}\right.\\
&\phantom{ \cos2\pi l\sigma_{23}\ }\left.+\alpha^{0}_{-n}\tilde{\alpha}^{0}_{-m}e^{2\pi i (-n\sigma_1+m\sigma_4)}+\alpha^{0}_{-n}\alpha^{0}_{-m}e^{2\pi i (-n\sigma_1-m\sigma_4)}\right)\\
&\!\!\!+ \cos2\pi l\sigma_{14}\left(\tilde{\alpha}^{0}_{-n}\tilde{\alpha}^{0}_{-m}e^{2\pi i (+n\sigma_2+m\sigma_3)}+\tilde{\alpha}^{0}_{-n}\alpha^{0}_{-m}e^{2\pi i (+n\sigma_2-m\sigma_3)}\right.\\
&\phantom{ \cos2\pi l\sigma_{14}\ }\left.+\alpha^{0}_{-n}\tilde{\alpha}^{0}_{-m}e^{2\pi i (-n\sigma_2+m\sigma_3)}+\alpha^{0}_{-n}\alpha^{0}_{-m}e^{2\pi i (-n\sigma_2-m\sigma_3)}\right)\\
&\!\!\!+ \cos2\pi l\sigma_{24}\left(\tilde{\alpha}^{0}_{-n}\tilde{\alpha}^{0}_{-m}e^{2\pi i (+n\sigma_1+m\sigma_3)}+\tilde{\alpha}^{0}_{-n}\alpha^{0}_{-m}e^{2\pi i (+n\sigma_1-m\sigma_3)}\right.\\
&\phantom{ \cos2\pi l\sigma_{24}\ }\left.+\alpha^{0}_{-n}\tilde{\alpha}^{0}_{-m}e^{2\pi i (-n\sigma_1+m\sigma_3)}+\alpha^{0}_{-n}\alpha^{0}_{-m}e^{2\pi i (-n\sigma_1-m\sigma_3)}\right)\\
&\!\!\!+ \cos2\pi l\sigma_{12}\left(\tilde{\alpha}^{0}_{-n}\tilde{\alpha}^{0}_{-m}e^{2\pi i (+n\sigma_3+m\sigma_4)}+\tilde{\alpha}^{0}_{-n}\alpha^{0}_{-m}e^{2\pi i (+n\sigma_3-m\sigma_4)}\right.\\
&\phantom{ \cos2\pi l\sigma_{12}\ }\left.+\alpha^{0}_{-n}\tilde{\alpha}^{0}_{-m}e^{2\pi i (-n\sigma_3+m\sigma_4)}+\alpha^{0}_{-n}\alpha^{0}_{-m}e^{2\pi i (-n\sigma_3-m\sigma_4)}\right)\Bigr\}\\
\end{split}
\end{equation*}

\begin{equation*}
\begin{split}
A^{0000}_{4,0}(\sigma_1,\sigma_2,\sigma_3,\sigma_4)&= \sum_{n,m>0}\frac{16}{g^2nm}\Bigl\{  \cos2\pi n \sigma_{12} \cos2\pi m \sigma_{34}\\
&+  \cos2\pi n \sigma_{13} \cos2\pi m \sigma_{24}+  \cos2\pi n \sigma_{14} \cos2\pi m \sigma_{23}\Bigr\}
\end{split}
\end{equation*}

\begin{equation*}
\begin{split}
&A^{i_1i_2i_3i_4}_{4,4}(\sigma_1,\sigma_2,\sigma_3,\sigma_4)=\sum_{n_1,n_2,n_3,n_4>0}16\\
&\left\{\tilde{\alpha}_{-n_1}^{i_1}\tilde{\alpha}_{-n_2}^{i_2}\tilde{\alpha}_{-n_3}^{i_3}\tilde{\alpha}_{-n_4}^{i_4}e^{2\pi i(+n_1\sigma_1+n_2\sigma_2+n_3\sigma_3+n_4\sigma_4)}\right.\\
\\
&+\alpha_{-n_1}^{i_1}\tilde{\alpha}_{-n_2}^{i_2}\tilde{\alpha}_{-n_3}^{i_3}\tilde{\alpha}_{-n_4}^{i_4}e^{2\pi i(-n_1\sigma_1+n_2\sigma_2+n_3\sigma_3+n_4\sigma_4)}\\
&+\tilde{\alpha}_{-n_1}^{i_1}\alpha_{-n_2}^{i_2}\tilde{\alpha}_{-n_3}^{i_3}\tilde{\alpha}_{-n_4}^{i_4}e^{2\pi i(+n_1\sigma_1-n_2\sigma_2+n_3\sigma_3+n_4\sigma_4)}\\
&+\tilde{\alpha}_{-n_1}^{i_1}\tilde{\alpha}_{-n_2}^{i_2}\alpha_{-n_3}^{i_3}\tilde{\alpha}_{-n_4}^{i_4}e^{2\pi i(+n_1\sigma_1+n_2\sigma_2-n_3\sigma_3+n_4\sigma_4)} \\
&+\tilde{\alpha}_{-n_1}^{i_1}\tilde{\alpha}_{-n_2}^{i_2}\tilde{\alpha}_{-n_3}^{i_3}\alpha_{-n_4}^{i_4}e^{2\pi i(+n_1\sigma_1+n_2\sigma_2+n_3\sigma_3-n_4\sigma_4)}\\
\\
&+\alpha_{-n_1}^{i_1}\alpha_{-n_2}^{i_2}\tilde{\alpha}_{-n_3}^{i_3}\tilde{\alpha}_{-n_4}^{i_4}e^{2\pi i(-n_1\sigma_1-n_2\sigma_2+n_3\sigma_3+n_4\sigma_4)}\\
&+\alpha_{-n_1}^{i_1}\tilde{\alpha}_{-n_2}^{i_2}\alpha_{-n_3}^{i_3}\tilde{\alpha}_{-n_4}^{i_4}e^{2\pi i(-n_1\sigma_1+n_2\sigma_2-n_3\sigma_3+n_4\sigma_4)}\\
&+\alpha_{-n_1}^{i_1}\tilde{\alpha}_{-n_2}^{i_2}\tilde{\alpha}_{-n_3}^{i_3}\alpha_{-n_4}^{i_4}e^{2\pi i(-n_1\sigma_1+n_2\sigma_2+n_3\sigma_3-n_4\sigma_4)} \\
&+\tilde{\alpha}_{-n_1}^{i_1}\alpha_{-n_2}^{i_2}\alpha_{-n_3}^{i_3}\tilde{\alpha}_{-n_4}^{i_4}e^{2\pi i(+n_1\sigma_1-n_2\sigma_2-n_3\sigma_3+n_4\sigma_4)}\\
&+\tilde{\alpha}_{-n_1}^{i_1}\alpha_{-n_2}^{i_2}\tilde{\alpha}_{-n_3}^{i_3}\alpha_{-n_4}^{i_4}e^{2\pi i(+n_1\sigma_1-n_2\sigma_2+n_3\sigma_3-n_4\sigma_4)} \\
&+\tilde{\alpha}_{-n_1}^{i_1}\tilde{\alpha}_{-n_2}^{i_2}\alpha_{-n_3}^{i_3}\alpha_{-n_4}^{i_4}e^{2\pi i(+n_1\sigma_1+n_2\sigma_2-n_3\sigma_3-n_4\sigma_4)}\\
\\
&+\alpha_{-n_1}^{i_1}\alpha_{-n_2}^{i_2}\alpha_{-n_3}^{i_3}\tilde{\alpha}_{-n_4}^{i_4}e^{2\pi i(-n_1\sigma_1-n_2\sigma_2-n_3\sigma_3+n_4\sigma_4)} \\
&+\alpha_{-n_1}^{i_1}\alpha_{-n_2}^{i_2}\tilde{\alpha}_{-n_3}^{i_3}\alpha_{-n_4}^{i_4}e^{2\pi i(-n_1\sigma_1-n_2\sigma_2+n_3\sigma_3-n_4\sigma_4)}\\
&+\alpha_{-n_1}^{i_1}\tilde{\alpha}_{-n_2}^{i_2}\alpha_{-n_3}^{i_3}\alpha_{-n_4}^{i_4}e^{2\pi i(-n_1\sigma_1+n_2\sigma_2-n_3\sigma_3-n_4\sigma_4)} \\
&+\tilde{\alpha}_{-n_1}^{i_1}\alpha_{-n_2}^{i_2}\alpha_{-n_3}^{i_3}\alpha_{-n_4}^{i_4}e^{2\pi i(+n_1\sigma_1-n_2\sigma_2-n_3\sigma_3-n_4\sigma_4)}\\
\\
&\left.+\alpha_{-n_1}^{i_1}\alpha_{-n_2}^{i_2}\alpha_{-n_3}^{i_3}\alpha_{-n_4}^{i_4}e^{2\pi i(-n_1\sigma_1-n_2\sigma_2-n_3\sigma_3-n_4\sigma_4)}\right\}
\end{split}
\end{equation*}

\begin{equation*}
\begin{split}
A^{i_1i_2i_3i_4}_{4,2}&(\sigma_1,\sigma_2,\sigma_3,\sigma_4)=\sum_{n,m,l>0}\frac{16l}{g}\\
&\Bigl\{\delta^{i_3i_4}\cos2\pi l\sigma_{34}\left(\tilde{\alpha}^{i_1}_{-n}\tilde{\alpha}^{i_2}_{-m}e^{2\pi i (+n\sigma_1+m\sigma_2)}+\tilde{\alpha}^{i_1}_{-n}\alpha^{i_2}_{-m}e^{2\pi i (+n\sigma_1-m\sigma_2)}\right.\\
&\phantom{\delta^{i_3i_4}\cos2\pi l\sigma_{34}\ }\left.+\alpha^{i_1}_{-n}\tilde{\alpha}^{i_2}_{-m}e^{2\pi i (-n\sigma_1+m\sigma_2)}+\alpha^{i_1}_{-n}\alpha^{i_2}_{-m}e^{2\pi i (-n\sigma_1-m\sigma_2)}\right)\\
&\!\!\!+\delta^{i_2i_4}\cos2\pi l\sigma_{24}\left(\tilde{\alpha}^{i_1}_{-n}\tilde{\alpha}^{i_3}_{-m}e^{2\pi i (+n\sigma_1+m\sigma_3)}+\tilde{\alpha}^{i_1}_{-n}\alpha^{i_3}_{-m}e^{2\pi i (+n\sigma_1-m\sigma_3)}\right.\\
&\phantom{\delta^{i_2i_4}\cos2\pi l\sigma_{24}\ }\left.+\alpha^{i_1}_{-n}\tilde{\alpha}^{i_3}_{-m}e^{2\pi i (-n\sigma_1+m\sigma_3)}+\alpha^{i_1}_{-n}\alpha^{i_3}_{-m}e^{2\pi i (-n\sigma_1-m\sigma_2)}\right)\\
&\!\!\!+\delta^{i_2i_3}\cos2\pi l\sigma_{23}\left(\tilde{\alpha}^{i_1}_{-n}\tilde{\alpha}^{i_4}_{-m}e^{2\pi i (+n\sigma_1+m\sigma_4)}+\tilde{\alpha}^{i_1}_{-n}\alpha^{i_4}_{-m}e^{2\pi i (+n\sigma_1-m\sigma_4)}\right.\\
&\phantom{\delta^{i_2i_3}\cos2\pi l\sigma_{23}\ }\left.+\alpha^{i_1}_{-n}\tilde{\alpha}^{i_4}_{-m}e^{2\pi i (-n\sigma_1+m\sigma_4)}+\alpha^{i_1}_{-n}\alpha^{i_4}_{-m}e^{2\pi i (-n\sigma_1-m\sigma_4)}\right)\\
&\!\!\!+\delta^{i_1i_4}\cos2\pi l\sigma_{14}\left(\tilde{\alpha}^{i_2}_{-n}\tilde{\alpha}^{i_3}_{-m}e^{2\pi i (+n\sigma_2+m\sigma_3)}+\tilde{\alpha}^{i_2}_{-n}\alpha^{i_3}_{-m}e^{2\pi i (+n\sigma_2-m\sigma_3)}\right.\\
&\phantom{\delta^{i_1i_4}\cos2\pi l\sigma_{14}\ }\left.+\alpha^{i_2}_{-n}\tilde{\alpha}^{i_3}_{-m}e^{2\pi i (-n\sigma_2+m\sigma_3)}+\alpha^{i_2}_{-n}\alpha^{i_3}_{-m}e^{2\pi i (-n\sigma_2-m\sigma_3)}\right)\\
&\!\!\!+\delta^{i_2i_4}\cos2\pi l\sigma_{24}\left(\tilde{\alpha}^{i_1}_{-n}\tilde{\alpha}^{i_3}_{-m}e^{2\pi i (+n\sigma_1+m\sigma_3)}+\tilde{\alpha}^{i_1}_{-n}\alpha^{i_3}_{-m}e^{2\pi i (+n\sigma_1-m\sigma_3)}\right.\\
&\phantom{\delta^{i_2i_4}\cos2\pi l\sigma_{24}\ }\left.+\alpha^{i_1}_{-n}\tilde{\alpha}^{i_3}_{-m}e^{2\pi i (-n\sigma_1+m\sigma_3)}+\alpha^{i_1}_{-n}\alpha^{i_3}_{-m}e^{2\pi i (-n\sigma_1-m\sigma_3)}\right)\\
&\!\!\!+\delta^{i_1i_2}\cos2\pi l\sigma_{12}\left(\tilde{\alpha}^{i_3}_{-n}\tilde{\alpha}^{i_4}_{-m}e^{2\pi i (+n\sigma_3+m\sigma_4)}+\tilde{\alpha}^{i_3}_{-n}\alpha^{i_4}_{-m}e^{2\pi i (+n\sigma_3-m\sigma_4)}\right.\\
&\phantom{\delta^{i_1i_2}\cos2\pi l\sigma_{12}\ }\left.+\alpha^{i_3}_{-n}\tilde{\alpha}^{i_4}_{-m}e^{2\pi i (-n\sigma_3+m\sigma_4)}+\alpha^{i_3}_{-n}\alpha^{i_4}_{-m}e^{2\pi i (-n\sigma_3-m\sigma_4)}\right)\Bigr\}\\
\end{split}
\end{equation*}

\begin{equation*}
\begin{split}
A^{i_1i_2i_3i_4}_{4,0}&(\sigma_1,\sigma_2,\sigma_3,\sigma_4)= \sum_{n,m>0}\frac{16nm}{g^2}\Bigl\{\delta^{i_1i_2}\delta^{i_3i_4}\cos2\pi n \sigma_{12} \cos2\pi m \sigma_{34}\\
&+\delta^{i_1i_3}\delta^{i_2i_4}\cos2\pi n \sigma_{13} \cos2\pi m \sigma_{24}+\delta^{i_1i_4}\delta^{i_2i_3}\cos2\pi n \sigma_{14} \cos2\pi m \sigma_{23}\Bigr\}
\end{split}
\end{equation*}
\subsection*{Worldsheet fermion}
\begin{equation*}
\begin{split}
B^{0i}_2(\sigma,\sigma) & = 4 \sum_{r_1,r_2>0}\left(\tilde{\psi}_{-r_1}^{0}\tilde{\psi}_{-r_2}^{i}e^{2\pi i(r_1+r_2)\sigma}+i\psi_{-r_1}^{0}\tilde{\psi}_{-r_2}^{i}e^{2\pi i(-r_1+r_2)\sigma}\right.\\
&\phantom{=\sum_{r_1,r_2>0}4}
\left. -\ i\tilde{\psi}_{-r_1}^{0}\psi_{-r_2}^{i}e^{2\pi i(r_1-r_2)\sigma}+\psi_{-r_1}^{0}\psi_{-r_2}^{i}e^{2\pi i(-r_1-r_2)\sigma}\right)
\end{split}
\end{equation*}
\begin{equation*}
\begin{split}
B^{i_1i_2}_2(\sigma,\sigma) & = (-4) \sum_{r_1,r_2>0}\left(\tilde{\psi}_{-r_1}^{i_1}\tilde{\psi}_{-r_2}^{i_2}e^{2\pi i(r_1+r_2)\sigma}+i\psi_{-r_1}^{i_1}\tilde{\psi}_{-r_2}^{i_2}e^{2\pi i(-r_1+r_2)\sigma}\right.\\
&\phantom{=\sum_{r_1,r_2>0}(-4)}
\left. +\ i\tilde{\psi}_{-r_1}^{i_1}\psi_{-r_2}^{i_2}e^{2\pi i(r_1-r_2)\sigma}-\psi_{-r_1}^{i_1}\psi_{-r_2}^{i_2}e^{2\pi i(-r_1-r_2)\sigma}\right)
\end{split}
\end{equation*}
\begin{equation*}
\begin{split}
&B^{0i0j}_{4,4}(\sigma_1\,\sigma_1,\sigma_2,\sigma_2)=\sum_{r_1,r_2,r_3,r_4>0}16\Bigl\{\\
&\phantom{+\ i} \tilde{\psi}_{-r_1}^0\tilde{\psi}_{-r_2}^i\tilde{\psi}_{-r_3}^0\tilde{\psi}_{-r_4}^j e^{2\pi i \left\{(+r_1+r_2)\sigma_1+(+r_3+r_4)\sigma_2\right\}}\\
&+i\tilde{\psi}_{-r_1}^0\tilde{\psi}_{-r_2}^i\tilde{\psi}_{-r_3}^0\psi_{-r_4}^j e^{2\pi i \left\{(+r_1+r_2)\sigma_1+(+r_3-r_4)\sigma_2\right\}}\\
&-i\tilde{\psi}_{-r_1}^0\tilde{\psi}_{-r_2}^i\psi_{-r_3}^0\tilde{\psi}_{-r_4}^j e^{2\pi i \left\{(+r_1+r_2)\sigma_1+(-r_3+r_4)\sigma_2\right\}}\\
&+i\tilde{\psi}_{-r_1}^0\psi_{-r_2}^i\tilde{\psi}_{-r_3}^0\tilde{\psi}_{-r_4}^j e^{2\pi i \left\{(+r_1-r_2)\sigma_1+(+r_3+r_4)\sigma_2\right\}}\\
&-i\psi_{-r_1}^0\tilde{\psi}_{-r_2}^i\tilde{\psi}_{-r_3}^0\tilde{\psi}_{-r_4}^j e^{2\pi i \left\{(-r_1+r_2)\sigma_1+(+r_3+r_4)\sigma_2\right\}}\\
&+\phantom{i}\tilde{\psi}_{-r_1}^0\tilde{\psi}_{-r_2}^i\psi_{-r_3}^0\psi_{-r_4}^j e^{2\pi i \left\{(+r_1+r_2)\sigma_1+(-r_3-r_4)\sigma_2\right\}}\\
&-\phantom{i}\tilde{\psi}_{-r_1}^0\psi_{-r_2}^i\tilde{\psi}_{-r_3}^0\psi_{-r_4}^j e^{2\pi i \left\{(+r_1-r_2)\sigma_1+(+r_3-r_4)\sigma_2\right\}}\\
&+\phantom{i}\psi_{-r_1}^0\tilde{\psi}_{-r_2}^i\tilde{\psi}_{-r_3}^0\psi_{-r_4}^j e^{2\pi i \left\{(-r_1+r_2)\sigma_1+(+r_3-r_4)\sigma_2\right\}}\\
&+\phantom{i}\tilde{\psi}_{-r_1}^0\psi_{-r_2}^i\psi_{-r_3}^0\tilde{\psi}_{-r_4}^j e^{2\pi i \left\{(+r_1-r_2)\sigma_1+(-r_3+r_4)\sigma_2\right\}}\\
&-\phantom{i}\psi_{-r_1}^0\tilde{\psi}_{-r_2}^i\psi_{-r_3}^0\tilde{\psi}_{-r_4}^j e^{2\pi i \left\{(-r_1+r_2)\sigma_1+(-r_3+r_4)\sigma_2\right\}}\\
&+\phantom{i}\psi_{-r_1}^0\psi_{-r_2}^i\tilde{\psi}_{-r_3}^0\tilde{\psi}_{-r_4}^j e^{2\pi i \left\{(-r_1-r_2)\sigma_1+(+r_3+r_4)\sigma_2\right\}}\\
&+i\tilde{\psi}_{-r_1}^0\psi_{-r_2}^i\psi_{-r_3}^0\psi_{-r_4}^j e^{2\pi i \left\{(+r_1-r_2)\sigma_1+(-r_3-r_4)\sigma_2\right\}}\\
&-i\psi_{-r_1}^0\tilde{\psi}_{-r_2}^i\psi_{-r_3}^0\psi_{-r_4}^j e^{2\pi i \left\{(-r_1+r_2)\sigma_1+(-r_3-r_4)\sigma_2\right\}}\\
&+i\psi_{-r_1}^0\psi_{-r_2}^i\tilde{\psi}_{-r_3}^0\psi_{-r_4}^j e^{2\pi i \left\{(-r_1-r_2)\sigma_1+(+r_3-r_4)\sigma_2\right\}}\\
&-i\psi_{-r_1}^0\psi_{-r_2}^i\psi_{-r_3}^0\tilde{\psi}_{-r_4}^j e^{2\pi i \left\{(-r_1-r_2)\sigma_1+(-r_3+r_4)\sigma_2\right\}}\\
&+\phantom{i}\psi_{-r_1}^0\psi_{-r_2}^i\psi_{-r_3}^0\psi_{-r_4}^j e^{2\pi i \left\{(-r_1-r_2)\sigma_1+(-r_3-r_4)\sigma_2\right\}}\Bigr\}
\end{split}
\end{equation*}
\begin{equation*}
\begin{split}
&B^{0i0j}_{4,2}(\sigma_1\,\sigma_1,\sigma_2,\sigma_2)=\sum_{r,s,t>0}\frac{16i}{g}\sin2\pi t\sigma_{12}\Bigl\{\\
&\delta^{ij}\left(\tilde{\psi}_{-r}^0\tilde{\psi}_{-s}^0e^{2\pi i(r\sigma_1+s\sigma_2)}
-i\tilde{\psi}_{-r}^0\psi_{-s}^0e^{2\pi i(r\sigma_1-s\sigma_2)}\right.\\
&\hspace{10em}\ \left.-i\psi_{-r}^0\tilde{\psi}_{-s}^0e^{2\pi i(-r\sigma_1+s\sigma_2)}
-\psi_{-r}^0\psi_{-s}^0e^{2\pi i(-r\sigma_1-s\sigma_2)}\right)\\
&-\left(\tilde{\psi}_{-r}^i\tilde{\psi}_{-s}^je^{2\pi i(r\sigma_1+s\sigma_2)}
+i\tilde{\psi}_{-r}^i\psi_{-s}^je^{2\pi i(r\sigma_1-s\sigma_2)}\right.\\
&\hspace{10em}\ \left.+i\psi_{-r}^i\tilde{\psi}_{-s}^je^{2\pi i(-r\sigma_1+s\sigma_2)}
-\psi_{-r}^i\psi_{-s}^je^{2\pi i(-r\sigma_1-s\sigma_2)}\right)\Bigr\}
\end{split}
\end{equation*}
\begin{equation*}
\begin{split}
 B^{0i0j}_{4,0}(\sigma_1,\sigma_1,\sigma_2,\sigma_2)
= \sum_{r,s>0}\frac{16}{g^2}\delta^{ij}\sin 2\pi r \sigma_{12} \sin 2\pi s \sigma_{12}
\end{split}
\end{equation*}
\begin{equation*}
\begin{split}
& B^{0ijk}_{4,4}(\sigma_1,\sigma_1,\sigma_2,\sigma_2)=\sum_{r_1,r_2,r_3,r_4>0}16\Bigl\{\\
&\phantom{+i}\tilde{\psi}_{-r_1}^{0}\tilde{\psi}_{-r_2}^{i}\tilde{\psi}_{-r_3}^{j}\tilde{\psi}_{-r_4}^{k}e^{2\pi i \left\{(+r_1+r_2)\sigma_1+(+r_3+r_4)\sigma_2)\right\}}\\
&+i\tilde{\psi}_{-r_1}^{0}\tilde{\psi}_{-r_2}^{i}\tilde{\psi}_{-r_3}^{j}\psi_{-r_4}^{k}e^{2\pi i \left\{(+r_1+r_2)\sigma_1+(+r_3-r_4)\sigma_2)\right\}}\\
&+i\tilde{\psi}_{-r_1}^{0}\tilde{\psi}_{-r_2}^{i}\psi_{-r_3}^{j}\tilde{\psi}_{-r_4}^{k}e^{2\pi i \left\{(+r_1+r_2)\sigma_1+(-r_3+r_4)\sigma_2)\right\}}\\
&+i\tilde{\psi}_{-r_1}^{0}\psi_{-r_2}^{i}\tilde{\psi}_{-r_3}^{j}\tilde{\psi}_{-r_4}^{k}e^{2\pi i \left\{(+r_1-r_2)\sigma_1+(+r_3+r_4)\sigma_2)\right\}}\\
&-i\psi_{-r_1}^{0}\tilde{\psi}_{-r_2}^{i}\tilde{\psi}_{-r_3}^{j}\tilde{\psi}_{-r_4}^{k}e^{2\pi i \left\{(-r_1+r_2)\sigma_1+(+r_3+r_4)\sigma_2)\right\}}\\
&-\phantom{i}\tilde{\psi}_{-r_1}^{0}\tilde{\psi}_{-r_2}^{i}\psi_{-r_3}^{j}\psi_{-r_4}^{k}e^{2\pi i \left\{(+r_1+r_2)\sigma_1+(-r_3-r_4)\sigma_2)\right\}}\\
&-\phantom{i}\tilde{\psi}_{-r_1}^{0}\psi_{-r_2}^{i}\tilde{\psi}_{-r_3}^{j}\psi_{-r_4}^{k}e^{2\pi i \left\{(+r_1-r_2)\sigma_1+(+r_3-r_4)\sigma_2)\right\}}\\
&+\phantom{i}\psi_{-r_1}^{0}\tilde{\psi}_{-r_2}^{i}\tilde{\psi}_{-r_3}^{j}\psi_{-r_4}^{k}e^{2\pi i \left\{(-r_1+r_2)\sigma_1+(+r_3-r_4)\sigma_2)\right\}}\\
&-\phantom{i}\tilde{\psi}_{-r_1}^{0}\psi_{-r_2}^{i}\psi_{-r_3}^{j}\tilde{\psi}_{-r_4}^{k}e^{2\pi i \left\{(+r_1-r_2)\sigma_1+(-r_3+r_4)\sigma_2)\right\}}\\
&+\phantom{i}\psi_{-r_1}^{0}\tilde{\psi}_{-r_2}^{i}\psi_{-r_3}^{j}\tilde{\psi}_{-r_4}^{k}e^{2\pi i \left\{(-r_1+r_2)\sigma_1+(-r_3+r_4)\sigma_2)\right\}}\\
&+\phantom{i}\psi_{-r_1}^{0}\psi_{-r_2}^{i}\tilde{\psi}_{-r_3}^{j}\tilde{\psi}_{-r_4}^{k}e^{2\pi i \left\{(-r_1-r_2)\sigma_1+(+r_3+r_4)\sigma_2)\right\}}\\
&-i\tilde{\psi}_{-r_1}^{0}\psi_{-r_2}^{i}\psi_{-r_3}^{j}\psi_{-r_4}^{k}e^{2\pi i \left\{(+r_1-r_2)\sigma_1+(-r_3-r_4)\sigma_2)\right\}}\\
&+i\psi_{-r_1}^{0}\tilde{\psi}_{-r_2}^{i}\psi_{-r_3}^{j}\psi_{-r_4}^{k}e^{2\pi i \left\{(-r_1+r_2)\sigma_1+(-r_3-r_4)\sigma_2)\right\}}\\
&+i\psi_{-r_1}^{0}\psi_{-r_2}^{i}\tilde{\psi}_{-r_3}^{j}\psi_{-r_4}^{k}e^{2\pi i \left\{(-r_1-r_2)\sigma_1+(+r_3-r_4)\sigma_2)\right\}}\\
&+i\psi_{-r_1}^{0}\psi_{-r_2}^{i}\psi_{-r_3}^{j}\tilde{\psi}_{-r_4}^{k}e^{2\pi i \left\{(-r_1-r_2)\sigma_1+(-r_3+r_4)\sigma_2)\right\}}\\
&+\phantom{i}\psi_{-r_1}^{0}\psi_{-r_2}^{i}\psi_{-r_3}^{j}\psi_{-r_4}^{k}e^{2\pi i \left\{(-r_1-r_2)\sigma_1+(-r_3-r_4)\sigma_2)\right\}}\Bigr\}
\end{split}
\end{equation*}
\begin{equation*}
\begin{split}
&B^{0ijk}_{4,2}(\sigma_1,\sigma_1,\sigma_2,\sigma_2)=\sum_{r,s,t>0}\frac{16i}{g}\sin 2\pi t \sigma_{12}\Bigl\{\\
&\phantom{+\ }\delta^{ij}\left(\tilde{\psi}_{-r}^{0}\tilde{\psi}_{-s}^{k}e^{2\pi i(r\sigma_1+s\sigma_2)}+i\tilde{\psi}_{-r}^{0}\psi_{-s}^{k}e^{2\pi i(r\sigma_1-s\sigma_2)}\right.\\
&\hspace{11em}\ \ \left.-i\psi_{-r}^{0}\tilde{\psi}_{-s}^{k}e^{2\pi i(-r\sigma_1+s\sigma_2)}+\psi_{r}^{0}\psi_{r}^{k}e^{2\pi i(-r\sigma_1-s\sigma_2)}\right)\\
&-\delta^{ik}\left(\tilde{\psi}_{-r}^{0}\tilde{\psi}_{-s}^{j}e^{2\pi i(r\sigma_1+s\sigma_2)}+i\tilde{\psi}_{-r}^{0}\psi_{-s}^{j}e^{2\pi i(r\sigma_1-s\sigma_2)}\right.\\
&\hspace{11em}\ \ \left.-i\psi_{-r}^{0}\tilde{\psi}_{-s}^{j}e^{2\pi i(-r\sigma_1+s\sigma_2)}+\psi_{r}^{0}\psi_{r}^{j}e^{2\pi i(-r\sigma_1-s\sigma_2)}\right)\Bigr\}
\end{split}
\end{equation*}
\begin{equation*}
B^{0ijk}_{4,0}(\sigma_1,\sigma_1,\sigma_2,\sigma_2)=0 
\end{equation*}
\begin{equation*}
\begin{split}
& B^{i_1i_2i_3i_4}_{4,4}(\sigma_1,\sigma_1,\sigma_2,\sigma_2)=\sum_{r_1,r_2,r_3,r_4>0}16\Bigl\{\\
&\phantom{+i}\tilde{\psi}_{-r_1}^{i_1}\tilde{\psi}_{-r_2}^{i_2}\tilde{\psi}_{-r_3}^{i_3}\tilde{\psi}_{-r_4}^{i_4}e^{2\pi i \left\{(+r_1+r_2)\sigma_1+(+r_3+r_4)\sigma_2)\right\}}\\
&+i\tilde{\psi}_{-r_1}^{i_1}\tilde{\psi}_{-r_2}^{i_2}\tilde{\psi}_{-r_3}^{i_3}\psi_{-r_4}^{i_4}e^{2\pi i \left\{(+r_1+r_2)\sigma_1+(+r_3-r_4)\sigma_2)\right\}}\\
&+i\tilde{\psi}_{-r_1}^{i_1}\tilde{\psi}_{-r_2}^{i_2}\psi_{-r_3}^{i_3}\tilde{\psi}_{-r_4}^{i_4}e^{2\pi i \left\{(+r_1+r_2)\sigma_1+(-r_3+r_4)\sigma_2)\right\}}\\
&+i\tilde{\psi}_{-r_1}^{i_1}\psi_{-r_2}^{i_2}\tilde{\psi}_{-r_3}^{i_3}\tilde{\psi}_{-r_4}^{i_4}e^{2\pi i \left\{(+r_1-r_2)\sigma_1+(+r_3+r_4)\sigma_2)\right\}}\\
&+i\psi_{-r_1}^{i_1}\tilde{\psi}_{-r_2}^{i_2}\tilde{\psi}_{-r_3}^{i_3}\tilde{\psi}_{-r_4}^{i_4}e^{2\pi i \left\{(-r_1+r_2)\sigma_1+(+r_3+r_4)\sigma_2)\right\}}\\
&-\phantom{i}\tilde{\psi}_{-r_1}^{i_1}\tilde{\psi}_{-r_2}^{i_2}\psi_{-r_3}^{i_3}\psi_{-r_4}^{i_4}e^{2\pi i \left\{(+r_1+r_2)\sigma_1+(-r_3-r_4)\sigma_2)\right\}}\\
&-\phantom{i}\tilde{\psi}_{-r_1}^{i_1}\psi_{-r_2}^{i_2}\tilde{\psi}_{-r_3}^{i_3}\psi_{-r_4}^{i_4}e^{2\pi i \left\{(+r_1-r_2)\sigma_1+(+r_3-r_4)\sigma_2)\right\}}\\
&-\phantom{i}\psi_{-r_1}^{i_1}\tilde{\psi}_{-r_2}^{i_2}\tilde{\psi}_{-r_3}^{i_3}\psi_{-r_4}^{i_4}e^{2\pi i \left\{(-r_1+r_2)\sigma_1+(+r_3-r_4)\sigma_2)\right\}}\\
&-\phantom{i}\tilde{\psi}_{-r_1}^{i_1}\psi_{-r_2}^{i_2}\psi_{-r_3}^{i_3}\tilde{\psi}_{-r_4}^{i_4}e^{2\pi i \left\{(+r_1-r_2)\sigma_1+(-r_3+r_4)\sigma_2)\right\}}\\
&-\phantom{i}\psi_{-r_1}^{i_1}\tilde{\psi}_{-r_2}^{i_2}\psi_{-r_3}^{i_3}\tilde{\psi}_{-r_4}^{i_4}e^{2\pi i \left\{(-r_1+r_2)\sigma_1+(-r_3+r_4)\sigma_2)\right\}}\\
&-\phantom{i}\psi_{-r_1}^{i_1}\psi_{-r_2}^{i_2}\tilde{\psi}_{-r_3}^{i_3}\tilde{\psi}_{-r_4}^{i_4}e^{2\pi i \left\{(-r_1-r_2)\sigma_1+(+r_3+r_4)\sigma_2)\right\}}\\
&-i\tilde{\psi}_{-r_1}^{i_1}\psi_{-r_2}^{i_2}\psi_{-r_3}^{i_3}\psi_{-r_4}^{i_4}e^{2\pi i \left\{(+r_1-r_2)\sigma_1+(-r_3-r_4)\sigma_2)\right\}}\\
&-i\psi_{-r_1}^{i_1}\tilde{\psi}_{-r_2}^{i_2}\psi_{-r_3}^{i_3}\psi_{-r_4}^{i_4}e^{2\pi i \left\{(-r_1+r_2)\sigma_1+(-r_3-r_4)\sigma_2)\right\}}\\
&-i\psi_{-r_1}^{i_1}\psi_{-r_2}^{i_2}\tilde{\psi}_{-r_3}^{i_3}\psi_{-r_4}^{i_4}e^{2\pi i \left\{(-r_1-r_2)\sigma_1+(+r_3-r_4)\sigma_2)\right\}}\\
&-i\psi_{-r_1}^{i_1}\psi_{-r_2}^{i_2}\psi_{-r_3}^{i_3}\tilde{\psi}_{-r_4}^{i_4}e^{2\pi i \left\{(-r_1-r_2)\sigma_1+(-r_3+r_4)\sigma_2)\right\}}\\
&+\phantom{i}\psi_{-r_1}^{i_1}\psi_{-r_2}^{i_2}\psi_{-r_3}^{i_3}\psi_{-r_4}^{i_4}e^{2\pi i \left\{(-r_1-r_2)\sigma_1+(-r_3-r_4)\sigma_2)\right\}}\Bigr\}
\end{split}
\end{equation*}
\begin{equation*}
\begin{split}
&B^{i_1i_2i_3i_4}_{4,2}(\sigma_1,\sigma_1,\sigma_2,\sigma_2)=\sum_{r,s,t>0}\frac{16i}{g}\sin 2\pi t \sigma_{12}\Bigl\{\\
&\phantom{+\ }\delta^{i_1i_3}\left(\tilde{\psi}_{-r}^{i_2}\tilde{\psi}_{-s}^{i_4}e^{2\pi i(r\sigma_1+s\sigma_2)}+i\tilde{\psi}_{-r}^{i_2}\psi_{-s}^{i_4}e^{2\pi i(r\sigma_1-s\sigma_2)}\right.\\
&\hspace{12em}\ \left.+i\psi_{-r}^{i_2}\tilde{\psi}_{-s}^{i_4}e^{2\pi i(-r\sigma_1+s\sigma_2)}-\psi_{r}^{i_2}\psi_{r}^{i_4}e^{2\pi i(-r\sigma_1-s\sigma_2)}\right)\\
&+\delta^{i_2i_4}\left(\tilde{\psi}_{-r}^{i_1}\tilde{\psi}_{-s}^{i_3}e^{2\pi i(r\sigma_1+s\sigma_2)}+i\tilde{\psi}_{-r}^{i_1}\psi_{-s}^{i_3}e^{2\pi i(r\sigma_1-s\sigma_2)}\right.\\
&\hspace{12em}\ \left.+i\psi_{-r}^{i_1}\tilde{\psi}_{-s}^{i_3}e^{2\pi i(-r\sigma_1+s\sigma_2)}-\psi_{r}^{i_1}\psi_{r}^{i_3}e^{2\pi i(-r\sigma_1-s\sigma_2)}\right)\\
&-\delta^{i_1i_4}\left(\tilde{\psi}_{-r}^{i_2}\tilde{\psi}_{-s}^{i_3}e^{2\pi i(r\sigma_1+s\sigma_2)}+i\tilde{\psi}_{-r}^{i_2}\psi_{-s}^{i_3}e^{2\pi i(r\sigma_1-s\sigma_2)}\right.\\
&\hspace{12em}\ \left.+i\psi_{-r}^{i_2}\tilde{\psi}_{-s}^{i_3}e^{2\pi i(-r\sigma_1+s\sigma_2)}-\psi_{r}^{i_2}\psi_{r}^{i_3}e^{2\pi i(-r\sigma_1-s\sigma_2)}\right)\\
&-\delta^{i_2i_3}\left(\tilde{\psi}_{-r}^{i_1}\tilde{\psi}_{-s}^{i_4}e^{2\pi i(r\sigma_1+s\sigma_2)}+i\tilde{\psi}_{-r}^{i_1}\psi_{-s}^{i_4}e^{2\pi i(r\sigma_1-s\sigma_2)}\right.\\
&\hspace{12em}\ \left.+i\psi_{-r}^{i_1}\tilde{\psi}_{-s}^{i_4}e^{2\pi i(-r\sigma_1+s\sigma_2)}-\psi_{r}^{i_1}\psi_{r}^{i_4}e^{2\pi i(-r\sigma_1-s\sigma_2)}\right)\Bigr\}
\end{split}
\end{equation*}
\begin{equation*}
B^{i_1i_2i_3i_4}_{4,0}(\sigma_1,\sigma_1,\sigma_2,\sigma_2)=\sum_{r,s>0}\frac{16}{g^2}(\delta^{i_1i_4}\delta^{i_2i_3}-\delta^{i_1i_3}\delta^{i_2i_4})\sin2\pi r \sigma_{12}\sin 2\pi s\sigma_{12}
\end{equation*}

\newpage
\section{Details of proof of BRST invariance}\label{boundary:appendix:b}
Detailed calculations omitted in the proof of BRST invariance of the boundary state are shown in this appendix. First we show detailed computations of \eqref{boundary:eq:remma}.
\begin{equation*}
\begin{split}
&\left[\int_0^{2\pi}\!\!d\sigma\ \hat{c}\hat{\Pi}_\mu\partial\hat{X}^\mu(\sigma), \int_0^{\sigma_{n-1}}\!\!d\sigma_n\ i\hat{\Pi}_i\mathbf{\Phi}^i(\hat{X}^0) (\sigma_n)f(\sigma_n)\right]\\
%=&-\left[\int_0^{2\pi}\!\!d\sigma\ \hat{c}\hat{\Pi}_0\partial\hat{X}^0(\sigma), \int_0^{\sigma_{n-1}}\!\!d\sigma_n\ i\hat{\Pi}_i\mathbf{\Phi}^i (\sigma_n)f(\sigma_n)\right]\\
%&+\left[\int_0^{2\pi}\!\!d\sigma\ \hat{c}\hat{\Pi}_i\partial\hat{X}^i(\sigma), \int_0^{\sigma_{n-1}}\!\!d\sigma_n\ i\hat{\Pi}_j\mathbf{\Phi}^j (\sigma_n)f(\sigma_n)\right]\\
%&+\int_0^{\sigma_{n-1}}\!\!d\sigma_n i\hat{\Pi}_i\mathbf{\Phi}^i(\hat{X}^0(\sigma_n))\left[\int_0^{2\pi}\!\!d\sigma\ \hat{c}\hat{\Pi}_\mu\partial\hat{X}^\mu(\sigma),f(\sigma_n)\right]\\
=&-\int_0^{2\pi}\!\!d\sigma\int_0^{\sigma_{n-1}}\!\!d\sigma_n\ \hat{c}(\sigma)\hat{\Pi}_i(\sigma_n)[\hat{\Pi}_0(\sigma),\mathbf{\Phi(\hat{X}^0(\sigma_n))}]\partial_\sigma \hat{X}^0 f(\sigma_n)\\
&+\int_0^{2\pi}\!\!d\sigma\int_0^{\sigma_{n-1}}\!\!d\sigma_n\ \hat{c}(\sigma)\hat{\Pi}(\sigma)[\partial_\sigma \hat{X}^i(\sigma), i\hat{\Pi}_j(\sigma_n)]\mathbf{\Phi}(\hat{X}^0(\sigma_n))f(\sigma_n)\\
&+\int_0^{\sigma_{n-1}}\!\!d\sigma_n i\hat{\Pi}_i\mathbf{\Phi}^i(\hat{X}^0(\sigma_n))\left[\int_0^{2\pi}\!\!d\sigma\ \hat{c}\hat{\Pi}_\mu\partial\hat{X}^\mu(\sigma),f(\sigma_n)\right]\\
\end{split}
\end{equation*}
By using the commutation relation $[\hat{X}^\mu(\sigma),\hat{\Pi}\nu(\sigma')]=i\delta(\sigma-\sigma')\eta^{\mu\nu}$ this can be rewritten as
\begin{equation*}
\begin{split}
&-i(-i)\eta^{00}\int_0^{2\pi}\!\!d\sigma\int_0^{\sigma_{n-1}}\!\!d\sigma_n\ \hat{c}(\sigma)\hat{\Pi}_i(\sigma_n)\delta(\sigma-\sigma_n)\dot{\mathbf{\Phi}}(\hat{X}^0(\sigma_n))\partial_\sigma \hat{X}^0(\sigma)f(\sigma_n)\\
&+i^2\int_0^{2\pi}\!\!d\sigma\int_0^{\sigma_{n-1}}\!\!d\sigma_n\ \hat{c}(\sigma)\hat{\Pi}_i(\sigma)\partial_\sigma \delta(\sigma-\sigma_n)\mathbf{\Phi(\hat{X}^0(\sigma_n))}f(\sigma_n)\\
&+\int_0^{\sigma_{n-1}}\!\!d\sigma_n i\hat{\Pi}_i\mathbf{\Phi}^i(\hat{X}^0(\sigma_n))\left[\int_0^{2\pi}\!\!d\sigma\ \hat{c}\hat{\Pi}_\mu\partial\hat{X}^\mu(\sigma),f(\sigma_n)\right]\\
\end{split}
\end{equation*}
Integrating over $\sigma$ in the first and the second line
\begin{equation*}
\begin{split}
&\int_0^{\sigma_{n-1}}\!\!d\sigma_n\ \hat{c}\hat{\Pi}_i\mathbf{\Phi}^i(\hat{X}^0)\partial\hat{X}^0(\sigma_n) f(\sigma_n)\\
&+\int_0^{\sigma_{n-1}}\!\!d\sigma_n\ \partial\left(\hat{c}\hat{\Pi}\right)\mathbf{\Phi}(\hat{X}^0)(\sigma_n)f(\sigma_n)\\
&-\int_0^{\sigma_{n-1}}\!\!d\sigma_n\ \hat{c}(2\pi)\hat{\Pi}_i(2\pi)\delta(2\pi-\sigma_n)\mathbf{\Phi}(\hat{X}^0(\sigma_n))f(\sigma_n)\\ 
&+\int_0^{\sigma_{n-1}}\!\!d\sigma_n\ \hat{c}(0)\hat{\Pi}_i(0)\delta(\sigma_n)\mathbf{\Phi}(\hat{X}^0(\sigma_n))f(\sigma_n)\\
&+\int_0^{\sigma_{n-1}}\!\!d\sigma_n\ i\hat{\Pi}_i\mathbf{\Phi}^i(\hat{X}^0(\sigma_n))\left[\int_0^{2\pi}\!\!d\sigma\ \hat{c}\hat{\Pi}_\mu\partial\hat{X}^\mu(\sigma),f(\sigma_n)\right]\\
\end{split}
\end{equation*}
We can collect the first and second line by using $\partial\left(\hat{c}\hat{\Pi}\mathbf{\Phi}\right)(\sigma_n)$. Taking account that the region of integration $\int_0^{\sigma_{n-1}}\!\!d\sigma_n$ in the fourth line includes a point $\sigma_n=0$, however, not necessarily $\sigma_n=2\pi$, we have
\begin{equation*}
\begin{split}
&\int_0^{\sigma_{n-1}}\!\!d\sigma_n\ \partial\left(\hat{c}\hat{\Pi}\mathbf{\Phi}\right)(\sigma_n)f(\sigma_n)\\
&-\int_0^{\sigma_{n-1}}\!\!d\sigma_n\ \hat{c}(2\pi)\hat{\Pi}_i(2\pi)\delta(\sigma_n-2\pi)\mathbf{\Phi}(\hat{X}^0(\sigma_n))f(\sigma_n)\\
&+\ \hat{c}\hat{\Pi}\mathbf{\Phi}(\hat{X}^0)(0)f(0)\\
&+\int_0^{\sigma_{n-1}}\!\!d\sigma_n i\hat{\Pi}_i\mathbf{\Phi}^i(\hat{X}^0(\sigma_n))\left[\int_0^{2\pi}\!\!d\sigma\ \hat{c}\hat{\Pi}_\mu\partial\hat{X}^\mu(\sigma),f(\sigma_n)\right]\\
\end{split}
\end{equation*}
Integrate by part in the first line to have
\begin{equation}
\begin{split}
=&\ \hat{c}\hat{\Pi}\mathbf{\Phi}(\sigma_{n-1})f(\sigma_{n-1})-\ \hat{c}\hat{\Pi}\mathbf{\Phi}(0)f(0)\\
&-\int_0^{\sigma_{n-1}}\!\!d\sigma_n\ \hat{c}\hat{\Pi}\mathbf{\Phi}(\sigma_n)\partial_{\sigma_n}f(\sigma_n)\\
&-\int_0^{\sigma_{n-1}}\!\!d\sigma_n\ \hat{c}(2\pi)\hat{\Pi}_i(2\pi)\delta(\sigma_n-2\pi)\mathbf{\Phi}(\hat{X}^0(\sigma_n))f(\sigma_n)\\
&+\ \hat{c}\hat{\Pi}\mathbf{\Phi}(\hat{X}^0)(0)f(0)\\
&+\int_0^{\sigma_{n-1}}\!\!d\sigma_n\ i\hat{\Pi}_i\mathbf{\Phi}^i(\hat{X}^0(\sigma_n))\left[\int_0^{2\pi}\!\!d\sigma\ \hat{c}\hat{\Pi}_\mu\partial\hat{X}^\mu(\sigma),f(\sigma_n)\right]\\ 
\end{split}
\end{equation}
The second term in the first line and the fourth line cancel out each other. Finally we get
\begin{equation*}
\begin{split}
&\left[\int_0^{2\pi}\!\!d\sigma\ \hat{c}\hat{\Pi}_\mu\partial\hat{X}^\mu(\sigma), \int_0^{\sigma_{n-1}}\!\!d\sigma_n\ i\hat{\Pi}_i\mathbf{\Phi}^i(\hat{X}^0) (\sigma_n)f(\sigma_n)\right]\\
=&\ \hat{c}\hat{\Pi}\mathbf{\Psi} (\sigma_{n-1})f(\sigma_{n-1})\\
&-\int_0^{\sigma_{n-1}}\!\!d\sigma_n\ \hat{c}\hat{\Pi}\mathbf{\Psi} (\sigma_n)\partial_{\sigma_{n}} f(\sigma_n)\\
&-\int_0^{\sigma_{n-1}}\!\!d\sigma_n\ \hat{c}(2\pi)\hat{\Pi}_i(2\pi)\delta(\sigma_n-2\pi)\mathbf{\Psi}(\hat{X}^0(\sigma_n))f(\sigma_n)\\
&+\int_0^{\sigma_{n-1}}\!\!d\sigma_n\ i\hat{\Pi}\mathbf{\Psi} (\sigma_n)\left[\int_0^{2\pi}\!\!d\sigma\ \hat{c}\hat{\Pi}_\mu\partial\hat{X}^\mu,f(\sigma_n)\right].
\end{split}
\end{equation*}

Next we confirm that \eqref{boundary:eq:theorem} is satisfied for $n=1$ and $n=2$ to complete the inductive method. 
\subsection*{$n=1$}
From \eqref{boundary:eq:remma} we have
\begin{equation*}
\begin{split}
&\left[\int_0^{2\pi}\!\!d\sigma\ \hat{c}\hat{\Pi}_\mu\partial_\sigma\hat{X}^\mu, \int_0^{2\pi}\!\!d\sigma_1\ i \hat{\Pi}_i\mathbf{\Phi}^i(\sigma_1)f(\sigma_1)\right]\\
=&\ \hat{c}\hat{\Pi}\mathbf{\Phi} (2\pi)\partial f(2\pi)\\
&-\int_0^{2\pi}\!\!d\sigma_1\ \hat{c}\hat{\Pi}\mathbf{\Phi} (\sigma_1)\partial f(\sigma_1)\\
&-\ \hat{c}\hat{\Pi}\mathbf{\Phi}(\hat{X})(2\pi)f(2\pi)\\
&+\int_0^{2\pi}\!\!d\sigma_1\ i\hat{\Pi}\mathbf{\Phi} (\sigma_1)\left[\int_0^{2\pi}\!\!d\sigma\ \hat{c}\hat{\Pi}_\mu\partial\hat{X}^\mu,f(\sigma_1)\right]\\
=&-\int_0^{2\pi}\!\!d\sigma_1\ \hat{c}\hat{\Pi}\mathbf{\Phi} (\sigma_1)\partial f(\sigma_1)\\
&+\int_0^{2\pi}\!\!d\sigma_1\ i\hat{\Pi}\mathbf{\Phi} (\sigma_1)\left[\int_0^{2\pi}\!\!d\sigma\ \hat{c}\hat{\Pi}_\mu\partial\hat{X}^\mu,f(\sigma_1)\right].
\end{split}
\end{equation*}

\subsection*{$n=2$}
We set $f(\sigma_1)=\int_0^{\sigma_1}\!\!d\sigma_1 i\hat{\Pi}\mathbf{\Phi}(\sigma_2)g(\sigma_2)$ in \eqref{boundary:eq:remma} to have
\begin{equation*}
\begin{split}
&\left[\int_0^{2\pi}\!\!d\sigma\ \hat{c}\hat{\Pi}_\mu\partial_\sigma\hat{X}^\mu, \int_0^{2\pi}\!\!d\sigma_1\ i \hat{\Pi}_i\mathbf{\Phi}^i(\sigma_1)\int_0^{\sigma_1}\!\!d\sigma_2\ i \hat{\Pi}_i\mathbf{\Phi}^i(\sigma_2)g(\sigma_2)\right]\\
=& -\int_0^{2\pi}\!\!d\sigma_{1}\ \hat{c}\hat{\Pi}\mathbf{\Phi}(\sigma_{1})\  i\hat{\Pi}\mathbf{\Phi}(\sigma_{1})\ g(\sigma_{n})\\
&+\int_0^{2\pi}\!\!d\sigma_{1}\ i\hat{\Pi}\mathbf{\Phi}(\sigma_{1})\ \hat{c}\hat{\Pi}\mathbf{\Phi}(\sigma_{1})\ g(\sigma_{n})\\
&-\int_0^{2\pi}\!\!d\sigma_1\ i\hat{\Pi}\mathbf{\Phi}(\sigma_1)\int_0^{\sigma_1}\!\!d\sigma_{2}\ \hat{c}\hat{\Pi}\mathbf{\Phi}(\sigma_{2})\partial_{\sigma_{2}} g(\sigma_{2}) \\
&+\int_0^{2\pi}\!\!d\sigma_1\ i\hat{\Pi}\mathbf{\Phi}(\sigma_1)\int_0^{\sigma_{1}}\!\!d\sigma_{2}\ \hat{c}(2\pi)\hat{\Pi}_i(2\pi)\delta(\sigma_{2}-2\pi)\mathbf{\Phi}(\hat{X}^0(\sigma_{2}))\partial_{\sigma_{2}} g(\sigma_{2})\\
&+\int_0^{2\pi}\!\!d\sigma_1\ i\hat{\Pi}\mathbf{\Phi}(\sigma_1)\int_0^{\sigma_{1}}\!\!d\sigma_{2}\ i\hat{\Pi}\mathbf{\Phi}(\sigma_{2})\partial \left[\int_0^{2\pi}\!\!d\sigma\ \hat{c}\hat{\Pi}\partial\hat{X}(\sigma), g(\sigma_{n+1})\right] \\
=&-\int_0^{2\pi}\!\!d\sigma_1\ i\hat{\Pi}\mathbf{\Phi}(\sigma_1)\int_0^{\sigma_1}\!\!d\sigma_{2}\ \hat{c}\hat{\Pi}\mathbf{\Phi}(\sigma_{2})\partial_{\sigma_{2}} g(\sigma_{2}) \\
&+\int_0^{2\pi}\!\!d\sigma_1\ i\hat{\Pi}\mathbf{\Phi}(\sigma_1)\int_0^{\sigma_{1}}\!\!d\sigma_{2}\ i\hat{\Pi}\mathbf{\Phi}(\sigma_{2})\left[\int_0^{2\pi}\!\!d\sigma\ \hat{c}\hat{\Pi}\partial\hat{X}(\sigma), g(\sigma_{n+1})\right].
\end{split}
\end{equation*}
Here we have used
\begin{equation*}
\begin{split}
&\int_0^{2\pi}\!\!d\sigma_1\ i\hat{\Pi}\mathbf{\Phi}(\sigma_1)\int_0^{\sigma_{1}}\!\!d\sigma_{2}\ \hat{c}(2\pi)\hat{\Pi}_i(2\pi)\delta(\sigma_{2}-2\pi)\mathbf{\Phi}(\hat{X}^0(\sigma_{2}))\partial_{\sigma_{2}} g(\sigma_{2})\\
=&\int_0^{2\pi}\!\!d\sigma_1\ i\hat{\Pi}\mathbf{\Phi}(\sigma_1) \theta(\sigma_1-2\pi) \hat{c}\hat{\Pi}\mathbf{\Phi}(2\pi)\partial g(2\pi)=0.
\end{split}
\end{equation*}

\section{Closed string coupling of multiple D0-branes}\label{paper:section:emtensor}
The results of calculations of couplings of multiple D0-branes to massless closed strings derived from disk amplitudes \cite{0103124}, Matrix theory potential \cite{9711078,9712185,9904095}, and non-Abelian DBI action \cite{9910053} are coincides:\begin{equation}
\begin{split}
    I_\phi(t,k) & = T_0\,\mathrm{Str}\left[\left(1-\frac{1}{2}\dot{\mathbf{X}}^i\dot{\mathbf{X}}^i-\frac{1}{4(2\pi\alpha')^2}[\mathbf{X}^i,\mathbf{X}^j][\mathbf{X}^i,\mathbf{X}^j]\right)^{ik\mathbf{X}}\right] \\
    I_h^{00}(t,k) & = T_0\,\mathrm{Str}\left[\left(1+\frac{1}{2}\dot{\mathbf{X}}^i\dot{\mathbf{X}}^i-\frac{1}{4(2\pi\alpha')^2}[\mathbf{X}^i,\mathbf{X}^j][\mathbf{X}^i,\mathbf{X}^j]\right)e^{ik\mathbf{X}}\right] \\
    I_h^{0i}(t,k) & = T_0\,\mathrm{Str}\left[\left(\dot{\mathbf{X}}^i\right)e^{ik\mathbf{X}} \right] \\
    I_h^{ij}(t,k) & = T_0\,\mathrm{Str}\left[\left(\dot{\mathbf{X}}^i\dot{\mathbf{X}}^i-\frac{1}{4}[\mathbf{X}^i,\mathbf{X}^j][\mathbf{X}^i,\mathbf{X}^j]\right)e^{ik\mathbf{X}}\right] \\
    I_b^{0i}(t,k) &= T_0\,\mathrm{Str}\left[\left(\frac{i}{2\pi\alpha'}[\mathbf{X}^i,\mathbf{X}^j]\dot{\mathbf{X}}^j\right)e^{ik\mathbf{X}}\right] \\
    I_b^{ij}(t,k) &= T_0\,\mathrm{Str}\left[\left(\frac{-i}{2\pi\alpha'}[\mathbf{X}^i,\mathbf{X}^j]\right)e^{ik\mathbf{X}}\right] \label{paper:eq:emtensor-super}
\end{split}
\end{equation}
up to order ${\alpha'}^2$. In the case of bosonic string, the results obtained from the disk amplitudes \cite{0103124} are
\begin{equation}
\begin{split}
   T^{00}(t,k^i) &= -\frac{2{\alpha'}^2}{g^2}\int_0^1\!\!d\sigma\ \mathrm{tr}\left[e^{i(1-\sigma)k\mathbf{X}}ik\dot{\mathbf{X}}e^{i\sigma k\mathbf{X}}ik\dot{\mathbf{X}}\right]\cos 2\pi \sigma - \frac{2{\alpha'}^2}{g^2}\mathrm{tr}\left[ik\ddot{\mathbf{X}}e^{ik\mathbf{X}}\right]\\
   T^{0i}(t,k^i) &= \frac{\alpha'}{g}\int_0^1\!\!d\sigma\ \mathrm{tr}\left[e^{i(1-\sigma)k\mathbf{X}}ik\dot{\mathbf{X}}e^{i\sigma k\mathbf{X}}\mathbf{X}^j\right]e^{-2\pi i \sigma}+\frac{2g}{\alpha'}\mathrm{tr}\left[\mathbf{X}^ie^{ik\mathbf{X}}\right]\\
   T^{ij}(t,k^i)& = -  2\int_0^1\!\!d\sigma\ \mathrm{tr}\left[e^{i(1-\sigma)k\mathbf{X}}\mathbf{X}^ie^{i\sigma k\mathbf{X}}\mathbf{X}^j\right]e^{-2\pi i \sigma}.
\end{split}\label{paper:eq:emtensor-bosonic}
\end{equation}

\section{Non-Abelian extension of boundary state}\label{paper:part:review}
In this appendix we review a boundary state of a single D-brane and consider their extension to multiple D-branes in order to see the reason why we think the formulas \eqref{paper:eq:pdim-nonabelian-boundarystate}  for non-Abelian boundary state presented in \cite{NPB308221} is correct. A boundary state of a D-brane is defined by a state which reproduces the correct disk amplitude with one closed string. An effect of open string fields on a D-brane is accounted for by introducing the Wilson loop factor. The boundary state including the effect of open string background fields describes closed string emission via open strings on the D-brane. In the basis of these studies, we consider non-Abelian extension of boundary states.   

In section \ref{paper:section:review-2} we review a boundary state of a single D-brane. In section \ref{paper:section:review-3} effects of gauge and scalar fields on a single D-brane is incorporated into the boundary state by using the Wilson loop factor. In section \ref{paper:section:review-4} the non-Abelian extension is considered. 

\subsection{Boundary state of single D-brane}\label{paper:section:review-2}
A D$p$-brane is a hypersurface on which open strings have their endpoints. An open string with the endpoints at $\sigma=0$ on D$p$-brane satisfies the Neumann boundary conditions 
\begin{subequations}
\begin{equation}
    \partial_\sigma X^a(\tau,\sigma=0)=0 \qquad a=0,1,\cdots,p  \label{review:eq:openboundary1}
\end{equation}
along the longitudinal directions to the brane, and Dirichlet boundary conditions
\begin{equation}
    X^i(\tau,\sigma=0)=\xi^i \qquad i=p+1,\cdots,D-1.\label{review:eq:openboundary2}
\end{equation}
\end{subequations}
along the transverse directions to the brane. Here $X^\mu (\mu=0,\cdots,D-1)$ is a string coordinate, $(\tau,\sigma)$ is a worldsheet coordinate, $\xi^i$ is the position of D-brane along the Dirichlet directions, and $D$ is the dimension of the spacetime, namely $D=26$ in case of bosonic string  and $D=10$ in case of superstring. A cylindrical worldsheet of which boundary attaches to a D-brane can be seen as the one-loop diagram of open strings. By exchanging the worldsheet coordinate $\tau$ and $\sigma$, such a worldsheet can be considered as the tree diagram of a closed string created by the D-brane as illustrated in figure \ref{review:fig:openclosed}. Therefore we can say D-branes act as a source of closed strings. These two descriptions are equivalent due to the conformal invariance of string theory. 

\psfrag{tau}{$\tau$}\psfrag{sigma}{$\sigma$}\psfrag{a}{$a$}\psfrag{i}{$i$}
\EPSFIGURE[ht]{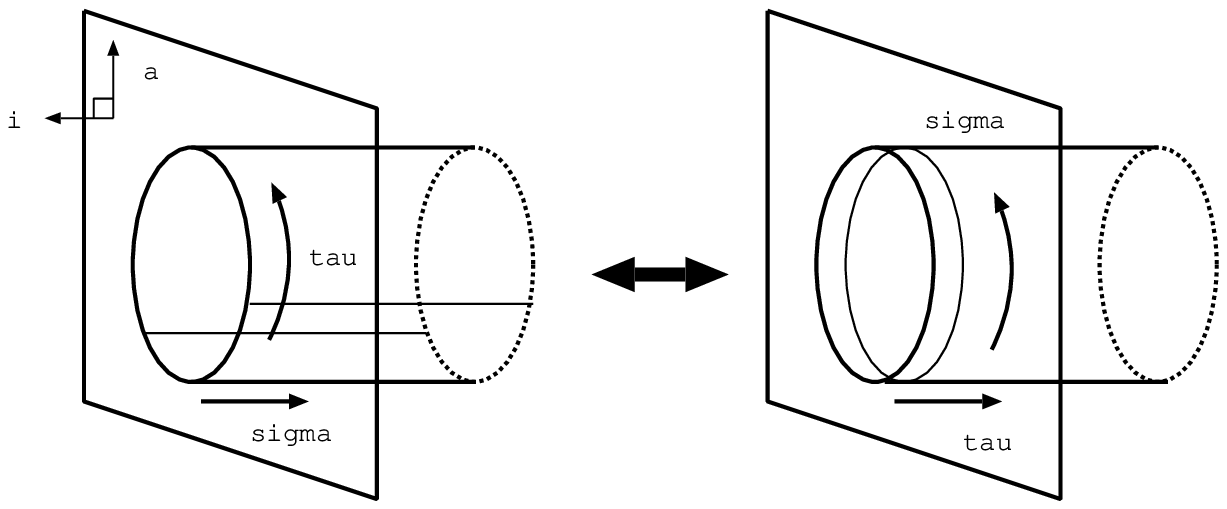,width=0.6\textwidth}{Open channel and closed channel\label{review:fig:openclosed}}

Consider a conformal transformation
\begin{equation*}
   \zeta=\sigma+i\tau  \to -i\zeta =\tau-i\sigma.
\end{equation*}
After a conversion $\sigma\to -\sigma$, the transformation becomes exchange of the worldsheet coordinates:
\begin{equation}
     \sigma \leftrightarrow \tau. \label{review:eq:modular-transformation}
\end{equation} 
This transformation leads us from the open string channel to the closed string channel. In terms of closed strings, the boundary conditions \eqref{review:eq:openboundary1} and \eqref{review:eq:openboundary2} can be expressed by
\begin{equation}
\begin{split}
     \partial_\tau \hat{X}^a (\sigma,\tau=0)|Dp\rangle &=0 \\
     \hat{X}^i (\sigma,\tau=0)|Dp\rangle &=\xi^i
\end{split}\label{review:eq:closedboundary1}
\end{equation}
where $|Dp\rangle$, a boundary state, is a closed string state into which the effect of boundary is incorporated \cite{NPB29383}. We can say that $|Dp\rangle$ is the eigenstate of $\Pi^a(\sigma)$ and $\hat{X}^i(\sigma)$ defined by
\begin{equation*}
\begin{split}
\hat{\Pi}^a(\sigma)&=\frac{1}{2\pi\alpha'}\partial_\tau\hat{X}(\sigma,\tau=0)\\
\hat{X}^i(\sigma)&=\hat{X}^i(\sigma,\tau=0)\\
\end{split}
\end{equation*}
In other words
\begin{equation*}
|Dp\rangle=|\Pi^a(\sigma)=0,X^i(\sigma)=\xi^i\rangle.
\end{equation*}
where $|\Pi^a(\sigma),X^i(\sigma)\rangle$ is defined by the eigenstates of $\hat{\Pi}^a(\sigma),\hat{X}^i(\sigma)$
\begin{equation}
\begin{split}
     \hat{\Pi}^a(\sigma)|\Pi^a(\sigma)\rangle &=  \Pi(\sigma)^a|\Pi^a(\sigma)\rangle \\
     \hat{X}^i(\sigma)|X^i(\sigma)\rangle &= X^i(\sigma)|X^i(\sigma)\rangle.
\end{split}\label{review:eq:string-eigenstate}
\end{equation}
Therefore we can express the boundary state by using a functional integral
\begin{equation}
      |Dp\rangle=\int\!\!\mathcal{D}X^a(\sigma)|X^a(\sigma),X^i(\sigma)=\xi^i\rangle. \label{review:eq:superposition}
\end{equation}
This represents that $|Dp\rangle$ is superposition of closed strings with various shape of the loop on the D-brane. Recall that he string embedding function $X^i(\sigma)$ represents spacetime coordinate of a closed string at the boundary $\tau=0$.

The worldsheet bosonic field $\hat{X}^\mu(\sigma,\tau)$ can be expanded in oscillators as
\begin{equation*} 
     \hat{X}^\mu(\sigma,\tau)=\hat{x}^\mu +\alpha'\hat{p}^\mu\tau+i\sqrt{\frac{\alpha'}{2}}\sum_{n\ne 0} \left(\frac{\alpha_n^\mu}{n}e^{-in(\tau-\sigma)}+\frac{\tilde{\alpha}_n^\mu}{n}e^{-in(\tau+\sigma)}\right).%\label{review:eq:modeexpansion1}
\end{equation*}
Substituting this expansion, \eqref{review:eq:closedboundary1} becomes
\begin{align}
\begin{split}
   (\alpha_n^a+\tilde{\alpha}_{-n}^a)|Dp\rangle &=0\\
   (\alpha_n^i-\tilde{\alpha}_{-n}^i)|Dp\rangle &=0
\end{split}\label{review:eq:closedboundary2}\\
\begin{split}
   \hat{p}^a|Dp\rangle&=0\\
   \hat{x}^i|Dp\rangle &=\xi^i . 
\end{split}\label{review:eq:closedboundary3}
\end{align}
The solution to these conditions can be found by utilizing coherent states for the harmonic oscillator. The mode operators satisfy the following commutation relations:
\begin{equation*}
\begin{split}
    [\alpha_{n}^\mu,\alpha_{m}^\nu]&=[\tilde{\alpha}_{n}^\mu,\tilde{\alpha}_{m}^\nu]=m\delta_{m+n,0}\eta^{\mu\nu}\\
    [\hat{x}^\mu,\hat{p}^\mu]&=i\eta^{\mu\nu}\\
    \mathrm{(others)}&=0
\end{split}
\end{equation*}
We define
\begin{equation*}
\begin{split}
   a_{(\mu,n,+)}&=\sqrt{n}\alpha^\nu_{n}\eta_{\mu\nu} \quad  a_{(\mu,n,+)}^\dagger=\sqrt{n}\alpha^\mu_{-n}\\
   a_{(\mu,n,-)}&=\sqrt{n}\tilde{\alpha}^\mu_{n}\phantom{\eta_{\mu\nu}} \quad  a_{(\mu,n,-)}^\dagger=\sqrt{n}\tilde{\alpha}^\mu_{-n} \qquad n>0\\
\end{split}
\end{equation*}
The operators defined in this way satisfy 
\begin{equation*}
   \left[a_{(\mu,n,\eta)},a^\dagger_{(\mu',n',\eta')}\right]=\delta_{\mu\mu'}\delta_{nn'}\delta_{\eta\eta'}, \quad a_{(\mu,n,\eta)}|0\rangle=0.
\end{equation*}
where $\eta,\eta'=\pm$. Therefore we have a series of creation and annihilation operators of the harmonic oscillator labeled by $({\mu,m,\eta})$. By using these operators we can rewrite the conditions \eqref{review:eq:closedboundary2} as
\begin{equation}
\begin{split}
   a_{(n,a,+)}|Dp\rangle &=-a_{(n,a,-)}^\dagger|Dp\rangle\\
   a_{(n,i,+)}|Dp\rangle &=\phantom{-}a_{(n,i,-)}^\dagger|Dp\rangle\\
   a_{(n,a,-)}|Dp\rangle &=-a_{(n,a,+)}^\dagger|Dp\rangle\\
   a_{(n,i,-)}|Dp\rangle &=\phantom{-}a_{(n,i,+)}^\dagger|Dp\rangle \qquad n>0.
\end{split}\label{review:eq:closedboundary3.5}
\end{equation}
The coherent state of harmonic oscillator is defined by
\begin{equation*}
   |z\rangle=e^{za^\dagger}|0\rangle.
\end{equation*}
We introduce a  operation which shift $a$ by $z$:
\begin{equation*}
   a(z)=e^{-za^\dagger}ae^{za^\dagger}=a+z.
\end{equation*}
By using this equation, we can see the coherent state is an eigenstate of the annihilation operator $a$. In fact
\begin{equation*}
    a|z\rangle = e^{za^\dagger}e^{-za^\dagger}ae^{za^\dagger}|0\rangle=e^{za^\dagger}(a+z)|0\rangle = z | z\rangle.
\end{equation*}
Hence the solution to \eqref{review:eq:closedboundary1}, or equivalently \eqref{review:eq:closedboundary3} and \eqref{review:eq:closedboundary3.5}, is
\begin{equation*}
\begin{split}
    |Dp\rangle & = f_p(\hat{x},\hat{p})\exp\left\{-\sum_{a=0}^{p}\sum_{n=1}^\infty a^\dagger_{(a,n,+)}a_{(a,n,-)}^\dagger+\sum_{i=p+1}^{D-1}\sum_{n=1}^\infty a^\dagger_{(i,n,+)}a_{(i,n,-)}^\dagger\right\}|0\rangle\\
              & = f_p(\hat{x},\hat{p})\exp\left\{-\sum_{n>0}\frac{1}{n}\alpha^a_{-n}\eta_{ab}\tilde{\alpha}_{-n}^b+\sum_{n>0}\frac{1}{n}\alpha^i_{-n}\delta_{ij}\tilde{\alpha}_{-n}^j\right\}.
\end{split}
\end{equation*}
The boundary condition \eqref{review:eq:closedboundary3} determines the function $N(\hat{x})$ so that
\begin{equation*}
    f_p(\hat{x},\hat{p})=N_p\delta(\hat{x}^i)    
\end{equation*}
where $N_p$ is a normalization constant which will be determined later (see equation \eqref{review:eq:normalization}). After all $|Dp\rangle$ is given by
\begin{equation}
\begin{split}
   |Dp\rangle &= N_p \exp\left\{-\sum_{n>0}\frac{1}{n}\alpha^\mu_{-n}S_{\mu\nu}\tilde{\alpha}^\nu_{-n}\right\}\delta(\hat{x}^i)|0\rangle\\
            S_{\mu\nu} &= (\eta^{ab},-\delta^{ij})            
\end{split}\label{review:eq:boundarystate1}
\end{equation}
We extract an operator which creates the boundary state out of the vacuum:
\begin{equation*}
\begin{split}
   V&=N_p\exp\left\{-\sum_{n>0}\frac{1}{n}\alpha^\mu_{-n}S_{\mu\nu}\tilde{\alpha}^\nu_{-n}\right\}\delta(\hat{x}^i)
%  \\ |Dp\rangle &=V |0\rangle.
\end{split}
\end{equation*}
This is the same operator as that deduced by factorization of open string loop amplitudes \cite{NPB50222,NPB57490,NPB94221}. The generalization to the Dirichlet boundary conditions is studied in \cite{NPB431131}. 

\DOUBLEFIGURE[ht]{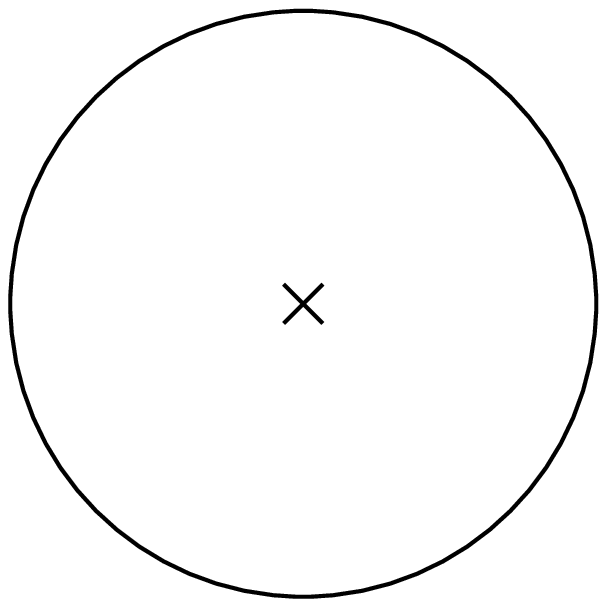,width=0.3\textwidth}{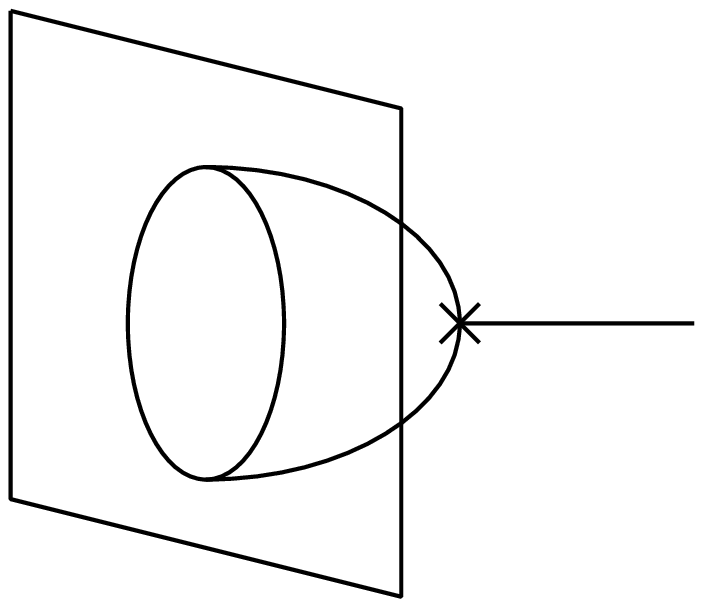,width=0.4\textwidth}{Disk with a closed string vertex\label{review:fig:disk}}{Emission (absorption) of a closed string from (into) a D-brane\label{review:fig:disk2}}

The boundary state gives a disk amplitude with a closed string vertex (see figure \ref{review:fig:disk}), or equivalently tree amplitude of a closed string which appears out of (annihilates into) the D-brane (see figure \ref{review:fig:disk2}). In what follows, we will find the boundary state $|B\rangle $which reproduces the disk amplitude with a single closed string vertex $V_\Psi$ for an arbitrary state $\Psi$ \cite{NPB288525}:
\begin{equation}
   \langle V_\Psi \rangle_{\mathrm{disk}} = \langle \Psi | B\rangle. \label{review:eq:disk}
\end{equation}
First we consider a harmonic oscillator with frequency $\omega$ and unite mass. The action of this oscillator is
\begin{equation*}
   S=\int_0^T\!\!dt\ \left(\frac{1}{2}\dot{\phi}(t)^2+\frac{1}{2}\omega^2\phi(t)^2\right). 
\end{equation*}
The normalized eigenfunction with an occupation number $n$ in position space is given by
\begin{equation*}
   u_n(x)=\left(\frac{\omega}{\pi}\right)^{1/4}2^{-n/2}(n!)^{1/2}H_n(\omega^{1/2}x)e^{-\omega x^2/2} \quad n=0,1,\cdots
\end{equation*} 
where $H_n(x)$ is the Hermite polynomial. %Note that $u_n$ satisfies a boundary condition
%\begin{equation*}
%   \lim_{x\to\pm\infty}u_n(x) =0.         
%\end{equation*}
We look for an operator $V(a^\dagger,S_0)$ which satisfy
\begin{align} 
       \int\!\!dx\ u_n(x)e^{-S_0[x]} &= \frac{1}{\sqrt{n!}}\langle 0 | a^n V (a^\dagger,S_0) |0\rangle. \label{review:eq:condition1}
%\\       S[x]&= \int_0^T\!\!dt\ \left(\frac{1}{2}\dot{x}^2+\frac{1}{2}\omega^2x^2\right) +S_0[x(0)].\nonumber
\end{align}
Here $S_0[\phi(0)=q]$ is a boundary action at $t=0$ which serve to impose the Neumann boundary condition
\begin{equation*}
     \dot{\phi}(0)=0. \label{review:eq:neumann1}
\end{equation*}
We operate $\sum_{n=0}^\infty\frac{1}{\sqrt{n!}}z^n$ on both sides of \eqref{review:eq:condition1}. Using the generating function of Hermite polynomial, the left hand side gives
\begin{equation*}
    \sum_{n=0}^\infty \frac{1}{\sqrt{n!}}u_n(q)z^n = \left(\frac{\omega}{\pi}\right)^{1/4}\exp\left\{-\frac{1}{2}\omega q^2+(2\omega)^{1/2}qz-\frac{1}{2}z^2\right\}.
\end{equation*}
The right hand side becomes
\begin{equation*}
\begin{split}
    \sum_{n=0}^\infty\frac{1}{n!}\langle 0 | a^n V(a^\dagger,S_0)|0\rangle &= 
    \langle 0 | a^{za} V(a^\dagger,S_0)|0\rangle \\&=  V(z,S_0)
\end{split}
\end{equation*}
This can be seen by utilizing a relation
\begin{equation*}
   e^{za}a^\dagger e^{-za}=a^\dagger+z.
\end{equation*}
Therefore the operation of $\sum_{n=0}^\infty\frac{1}{\sqrt{n!}}z^n$ on  \eqref{review:eq:condition1} gives 
\begin{equation*}
 V(z,S_0)=\left(\frac{\omega}{\pi}\right)^{1/4}\int\!\!dq\ \exp\left\{-S[q]-\frac{1}{2}\omega q^2+(2\omega)^{1/2}qz-\frac{1}{2}z^2\right\}.    
\end{equation*}
After all we find the operator $V$ is given by
\begin{equation}
 V(a^\dagger,S_0)=\left(\frac{\omega}{\pi}\right)^{1/4}\int\!\!dq\ \exp\left\{-S[q]-\frac{1}{4}\omega q^2+(2\omega)^{1/2}a^\dagger q-\frac{1}{2}(a^\dagger)^2\right\}. \label{review:eq:vertex1}
\end{equation}
We change the variable so that $q=(2\omega)^{-1/2}x$, then \eqref{review:eq:vertex1} becomes
 \begin{equation*}
 V(a^\dagger,S_0)=(4\pi\omega)^{-1/4}\int\!\!dx\ \exp\left\{-S[x]-\frac{1}{4}x^2+a^\dagger x-\frac{1}{2}(a^\dagger)^2\right\}.
\end{equation*}
It can be shown that the eigenstate of $a^\dagger+a = \phi(0)$ with the eigenvalue $x$ is given by
\begin{equation*}
    |x\rangle = (2\pi)^{-1/4}\exp\left\{-\frac{1}{4}x^2+xa^\dagger-\frac{1}{2}(a^\dagger)^2\right\}|0\rangle.
\end{equation*}
In fact the following relation holds:
\begin{equation*}
   (a+a^\dagger)|x\rangle = x|x\rangle.
\end{equation*}
Using $|x\rangle$ defined in this way, \eqref{review:eq:vertex1} operating on the vacuum can be rewritten as
\begin{equation*}
     |B\rangle = V(a^\dagger,S_0)|0\rangle = (2\omega)^{-1/4}\int\!\!dx\ e^{-S_0[x]}|x\rangle.
\end{equation*}

By utilizing this operator we rewrite a functional integral
\begin{equation*}
  I=\int\!\!\mathcal{D}\phi(t) \exp\left\{-\frac{1}{2}\int_0^T\!\!dt \left(\dot{\phi}(t)^2+\omega^2\phi(t)^2 \right)-S_0[\phi(0)]-S_T[\phi(T)] \right\}.
\end{equation*}
With insertion of 
\begin{equation*}
     \int\!\!dx\ \delta(\phi(0)-x) \int\!\!dy\ \delta(\phi(T)-y)
\end{equation*}
we can rewrite $I$ by using standard methods for the path integral:
\begin{equation}
\begin{split}
 I &=\sum_{n=0}^\infty e^{-(n+\frac{1}{2})\omega t}\int\!\!dx\ e^{-S_0[x]}u_n(x)\int\!\!dy\ e^{-S_T[y]}u_n(y)\\
   &=\langle 0 | V(a,S_T) e^{-\omega t (a^\dagger a+\frac{1}{2})}V(a^\dagger,S_0) |0\rangle. 
\end{split}\label{review:eq:cylinder1}
\end{equation}
This indicates the factorization of two boundaries.
\\ \\
We will extend these formulas to the bosonic string theory. We consider a scalar field $\hat{X}^\mu(\tau,\sigma)$ on a worldsheet with the coordinate region
\begin{equation*}
    0 \le \sigma \le 2\pi, \quad 0 \le t \le T = -\log \epsilon.
\end{equation*}
This region is considered with Euclidean signature by taking $t=i\tau$. On the complex plane of $z=e^{-(t-i\sigma)}$, this region corresponds 
\begin{equation*}
    \epsilon \le  |z| \le 1. 
\end{equation*}
\psfrag{1}{1}\psfrag{sigma}{$\sigma$}\psfrag{tau}{$t$}\psfrag{z}{$z$}\psfrag{T}{$T$}
\EPSFIGURE[ht]{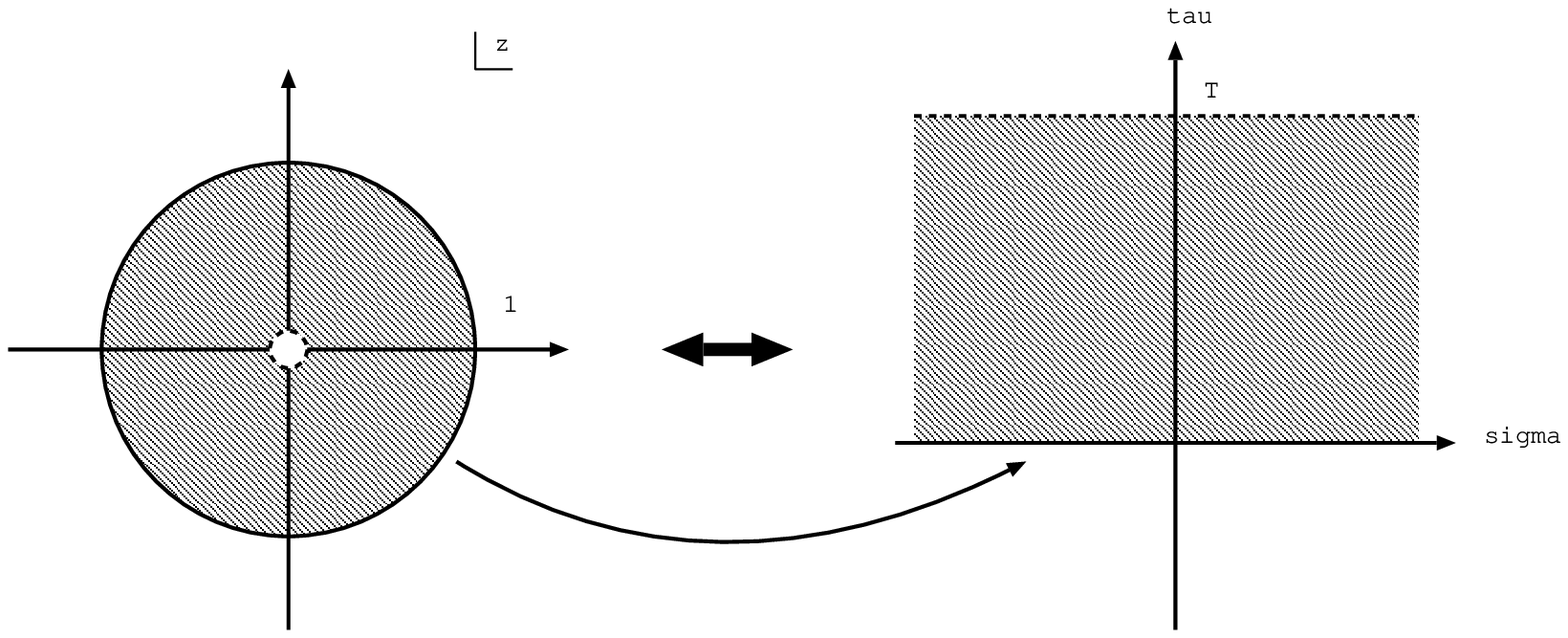,width=0.6\textwidth}{Map between disk and rectangle\label{review:fig:map}}
The worldsheet field $\hat{X}^\mu(\tau,\sigma)$ is expanded as
\begin{equation}
\begin{split}
   \hat{X}^\mu(\tau,\sigma) 
&=  \hat{q}^\mu +\alpha'\hat{p}^\mu\tau+i\sqrt{\frac{\alpha'}{2}}\sum_{n\ne0}\left(\frac{\alpha_{n}^\mu}{n}e^{-in(\tau-\sigma)}+\frac{\tilde{\alpha}_{n}^\mu}{n}e^{-in(\tau+\sigma)}\right) \\
&= \hat{\xi}^\mu(\tau) + \sum_{n>0}\sqrt{\alpha'}\left(\hat{\psi}^\mu_{n}(\tau)e^{-in\sigma}+\hat{\tilde{\psi}}^\mu_{n}(\tau)e^{in\sigma}\right).
\end{split}\label{review:eq:modeexpansion1}
\end{equation}
Here $\psi^\mu_n(\tau), \tilde{\psi}^\mu_n(\tau)$ and $\xi^\mu(\tau)$ relate to $\hat{x}^\mu,\hat{p}^\mu,\alpha_n^\mu$ and $\tilde{\alpha}_n^\mu$ through
\begin{equation*}
\begin{split}
    \hat{\xi}^\mu(\tau) &= \hat{q}^\mu+\alpha'\hat{p}^\mu\tau\\
    \hat{\psi}^\mu_n(\tau) &= \frac{i}{n\sqrt{2}}\left(\tilde{\alpha}_{n}^\mu e^{-in\tau}-\alpha_{-n}^\mu e^{in\tau}\right) \\
    \hat{\tilde{\psi}}^\mu_n(\tau) &= \frac{i}{n\sqrt{2}}\left(\alpha_{n}^\mu e^{-in\tau}-\tilde{\alpha}_{-n}^\mu e^{in\tau}\right).
\end{split}
\end{equation*}
It is convenient to introduce $a_n^\mu,a_{-n}^\mu, \tilde{a}^\mu_{n}$ and $\tilde{a}^\mu_{-n}$ such that 
\begin{equation}
\begin{split}
\alpha_{n\phantom{-}}^\mu = -i\sqrt{n}a_{n\phantom{-}}^\mu \quad &\leftrightarrow \quad  a_{n\phantom{-}}^\mu=\phantom{-}\frac{i}{\sqrt{n}}\alpha^\mu_{n\phantom{-}} \\
\alpha_{-n}^\mu = \phantom{-}i\sqrt{n}a_{-n}^\mu \quad &\leftrightarrow \quad  a_{-n}^\mu=-\frac{i}{\sqrt{n}}\alpha^\mu_{-n} \\
\tilde{\alpha}_{n\phantom{-}}^\mu = -i\sqrt{n}\tilde{a}_{n\phantom{-}}^\mu \quad &\leftrightarrow\quad   \tilde{a}_{n\phantom{-}}^\mu=\phantom{-}\frac{i}{\sqrt{n}}\tilde{\alpha}^\mu_{n\phantom{-}} \\
\tilde{\alpha}_{-n}^\mu =\phantom{-} i\sqrt{n}\tilde{a}_{-n}^\mu \quad &\leftrightarrow \quad  \tilde{a}_{-n}^\mu=-\frac{i}{\sqrt{n}}\tilde{\alpha}^\mu_{-n} \qquad n>0. 
\end{split}\label{review:eq:normalized-creation-operator}  
\end{equation}
These creation and annihilation operators introduced in this way satisfy
\begin{equation*}
\begin{split}
   [a^\mu_n,a^\nu_m]=\eta^{\mu\nu}\delta_{n+m,0}, \quad (a^\mu_n)^\dagger = a^\mu_{-n}\\
   [\tilde{a}^\mu_n,\tilde{a}^\nu_m]=\eta^{\mu\nu}\delta_{n+m,0}, \quad (\tilde{a}^\mu_n)^\dagger = \tilde{a}^\mu_{-n}
\end{split}
\end{equation*}
By using these operators, $\hat{\psi}^\mu_n(\tau)$ and $\hat{\tilde{\psi}}^\mu_n$ can be rewritten as
\begin{equation}
\begin{split}
%    \xi^\mu(\tau) &= \hat{q}^\mu-i\alpha'\hat{p}^\mu\tau\\
    \hat{\psi}^\mu_n(\tau) &= \frac{1}{\sqrt{2n}}\left(\tilde{a}_{n}^\mu e^{-in\tau}+a_{-n}^\mu e^{in\tau}\right) \\
    \hat{\tilde{\psi}}^\mu_n(\tau) &= \frac{1}{\sqrt{2n}}\left(a_{n}^\mu e^{-in\tau}+\tilde{a}_{-n}^\mu e^{in\tau}\right).
\end{split}\label{review:eq:eigenstate1}
\end{equation}
Substituting \eqref{review:eq:modeexpansion1}, the action becomes
\begin{equation*}
\begin{split}
   S&=\frac{1}{4\pi\alpha'}\int_0^{2\pi}\!\!d\sigma\int_0^T\!\!dt\ \left(\partial_t\hat{X}^\mu\partial_t\hat{X}^\mu+\partial_\sigma\hat{X}^\mu\partial_\sigma\hat{X}^\mu \right) \\
    &=\int_0^T\!\!dt\ \left\{\frac{1}{2\alpha'}\partial_t \hat{\xi}^\mu\partial_t \hat{\xi}^\mu+\sum_{n=1}^\infty\sum_{\mu=0}^{D-1}\left(\frac{d\hat{\tilde{\psi}}_n^\mu}{dt} \frac{d\hat{\psi}_n^\mu}{dt} +n^2\hat{\tilde{\psi}}_n^\mu\hat{\psi}_n^\mu\right)\right\} .
\end{split}
\end{equation*}
where $D$ is the dimension of spacetime, and $t=i\tau$. The first term gives $\hat{p}^2$, hence does not affect the operator $V$. We can see that there are $2D$ real oscillators for each frequency $\omega=n=1,2,\cdots$. Decompose $\psi^\mu_n(\tau)$ and $\tilde{\psi}^\mu_n(\tau)$ into real and imaginary parts:
\begin{equation*}
\begin{split}
       \hat{\psi}_n^\mu(\tau)&=\sqrt{\frac{1}{2}}\left(\hat{\phi}_n^\mu(\tau)+i\hat{\chi}_n^\mu(\tau)\right)\\
       \hat{\tilde{\psi}}_n^\mu(\tau)&=\sqrt{\frac{1}{2}}\left(\hat{\phi}_n^\mu(\tau)-i\hat{\chi}_n^\mu(\tau)\right)
\end{split}
\end{equation*}
where $\hat{\phi}^\mu_n(\tau)$ and $\hat{\chi}^\mu_n(^tau)$ are real fields expressed as
\begin{equation*}
\begin{split}
\hat{\phi}_n^\mu(\tau) &=\frac{1}{\sqrt{2n}}\left(\frac{\tilde{a}_n^\mu+a_{n}^\mu}{\sqrt{2}}e^{-in\tau}+\frac{a_{-n}^\mu+\tilde{a}_{-n}^\mu}{\sqrt{2}}e^{in\tau}\right) \\
\hat{\chi}_n^\mu(\tau) &=\frac{1}{\sqrt{2n}}\left(\frac{\tilde{a}_n^\mu-a_{n}^\mu}{\sqrt{2}i}e^{-in\tau}+\frac{a_{-n}^\mu-\tilde{a}_{-n}^\mu}{\sqrt{2}i}e^{in\tau}\right).
\end{split}
\end{equation*}
Now we calculate the corresponding terms of \eqref{review:eq:vertex1} in the string theory:
\begin{align*}
\begin{split}
    -\frac{1}{2}\omega q^2 
&\to -\frac{1}{2}\sum_{n=1}^\infty\sum_{\mu=0}^{D-1}n\left[ (\phi_n^\mu)^2 +  (\chi_n^\mu)^2\right]\\
&=-\sum_{n=1}^\infty\sum_{\mu=0}^{D-1}n\psi_n^\mu\tilde{\psi}_n^\mu
\end{split}
\intertext{}
\begin{split}
    -(2\omega)^{1/2}q a^\dagger 
&\to -\sum_{n=1}^\infty\sum_{\mu=0}^{D-1}(2n)^{1/2}\left[ \phi_n^\mu\frac{a_{-n}^\mu+\tilde{a}_{-n}^\mu}{\sqrt{2}}+\chi_n^\mu\frac{a_{-n}^\mu-\tilde{a}_{-n}^\mu}{\sqrt{2}i}\right] \\
&=-\sum_{n=1}^\infty\sum_{\mu=0}^{D-1}(2n)^{1/2}\left(a_{-n}^\mu\psi_n^\mu+\tilde{a}_{-n}^\mu\tilde{\psi}_n^\mu\right)
\end{split}
\intertext{}
\begin{split}
-\frac{1}{2}(a^\dagger)^2 &\to -\frac{1}{2}\sum_{n=1}^\infty\sum_{\mu=0}^{D-1}\left[
\left(\frac{a_{-n}^\mu+\tilde{a}_{-n}^\mu}{\sqrt{2}}\right)^2+\left(\frac{a_{-n}^\mu-\tilde{a}_{-n}^\mu}{\sqrt{2}i}\right)^2\right]\\
&=-\sum_{n=1}^\infty\sum_{\mu=0}^{D-1}a_{-n}^\mu \tilde{a}_{-n}^\mu
\end{split}
\intertext{}
\begin{split}
  \left(\frac{\omega}{\pi}\right)^{1/4} &\to \prod_{n=1}^\infty \left(\frac{n}{\pi}\right)^{D/2}
\end{split}
\end{align*}
We change the variables so that $\psi_n^\mu = (2n)^{-1/2}x_n^\mu, \tilde{\psi}_n^\mu = (2n)^{-1/2}\tilde{x}_n^\mu$. After this variable change, the operator \eqref{review:eq:vertex1} becomes
\begin{multline}
  V(a_{-n}^\mu,\tilde{a}_{-n}^\mu,\hat{q}^\mu,S_0)=
\left[\prod_{n=1}^\infty(4\pi n)^{-D/2}\int\!\!dx_n^\mu\int\!\!d\tilde{x}_n^\mu\right]\\
\exp\Bigg\{-S_0\left[x_n^\mu,\tilde{x}_n^\mu,\hat{q}^\mu\right]\\
+\sum_{n=1}^\infty\sum_{\mu=0}^{D-1}\left(-x_n^\mu\tilde{x}_n^\mu+a_{-n}^\mu x_{n}^\mu+\tilde{a}_{-n}^\mu \tilde{x}_n^\mu-a_{-n}^\mu\tilde{a}_{-n}^\mu
\right)\Bigg\}
\label{review:eq:vertex2}
\end{multline}
We can rewrite this formula in a simple form by introducing notations
\begin{equation*}
\begin{split}
    (y|x)&=\sum_{n=1}^\infty\sum_{\mu=0}^{D-1}y_n^\mu x_n^\mu\\
    |x,\tilde{x}\rangle&=\exp\left\{-\frac{1}{2}(x|\tilde{x})+(a^\dagger|x)+(\tilde{a}^\dagger|\tilde{x})-(a^\dagger|\tilde{a}^\dagger)\right\}|0\rangle\\
    \int\!\!\mathcal{D}x\mathcal{D}\tilde{x}&= \prod_{n=1}^\infty\int\!\!dx_n^\mu \int\!\!\tilde{x}_n^\mu.
\end{split}
\end{equation*} 
Using these notations, the boundary state created by operating \eqref{review:eq:vertex2} on the vacuum is expressed as
\begin{equation}
    |B[S]\rangle=\int\!\!\mathcal{D}x\mathcal{D}\tilde{x}\exp\left\{-S\left[x,\tilde{x},\hat{q}\right]\right\}|x,\tilde{x}\rangle. \label{review:eq:vertex3}
\end{equation}
Here we have neglected the overall factor. It can be shown \cite{NPB308221} that $|x,\tilde{x}\rangle$ is the eigenstate of $\hat{x}_n^\mu=(2n)^{-1/2}\hat{\psi}_n^\mu(0)$ and $\hat{\tilde{x}}_n^\mu=(2n)^{-1/2}\hat{\tilde{\psi}}_n^\mu(0)$ and $\hat{p}^\mu$:
\begin{equation*}
\begin{split}
    \hat{x}_m^\mu |x,\tilde{x}\rangle &= (\tilde{a}_n^\mu+a_{-n}^\mu)|x,\tilde{x}\rangle = x_n^\mu|x,\tilde{x}\rangle\\
  \hat{\tilde{x}}_m^\mu |x,\tilde{x}\rangle &= (a_n^\mu+\tilde{a}_{-n}^\mu)|x,\tilde{x}\rangle = \tilde{x}_n^\mu|x,\tilde{x}\rangle\\
    \hat{p}^\mu|x,\tilde{x}\rangle &= 0
\end{split}
\end{equation*}
In order to show the first equal in each line, we have used \eqref{review:eq:eigenstate1}. By utilizing an relation
\begin{equation*}
   |0\rangle=\int\!\!\mathcal{D}\xi |\xi\rangle, \quad \hat{q}^\mu|\xi\rangle = \xi^\mu | \xi\rangle.
\end{equation*}
we can express the eigenstate $|x,\tilde{x}\rangle$ by
\begin{equation*}
    |x,\tilde{x}\rangle = \int\!\!\mathcal{D}\xi |x,\tilde{x},\xi\rangle.
\end{equation*}
Hence the boundary state \eqref{review:eq:vertex3} becomes
\begin{equation}
\begin{split}
    |B[S]\rangle&=\int\!\!\mathcal{D}x\mathcal{D}\tilde{x}\mathcal{D}\xi\exp\left\{-S\left[x,\tilde{x},\xi\right]\right\}|x,\tilde{x},\xi\rangle\\
    &=\exp\left\{-S\left[\hat{x},\hat{\tilde{x}},\hat{q}\right]\right\}\int\!\!\mathcal{D}x\mathcal{D}\tilde{x}\mathcal{D}\xi|x,\tilde{x},\xi\rangle\\
&=\exp\left\{-S\left[\hat{x},\hat{\tilde{x}},\hat{q}\right]\right\}|B[S=0]\rangle.
\end{split}\label{review:eq:vertex4}
\end{equation}
In terms of $\hat{x}_n^\mu$ and $\hat{\tilde{x}}_n^\mu$, the worldsheet field \eqref{review:eq:modeexpansion1} at the boundary becomes
\begin{equation}
\begin{split}
   \hat{X}^\mu(\sigma)=\hat{X}^\mu(\tau=0,\sigma) 
&= \hat{q}^\mu + \sum_{n>0}\sqrt{\frac{\alpha'}{2n}}\left(\hat{x}^\mu_{n}e^{-in\sigma}+\hat{\tilde{x}}^\mu_{n}e^{in\sigma}\right).
\end{split}\label{review:eq:modeexpansion2}
\end{equation}
Therefore we can see that $|x,\tilde{x},\xi\rangle$ is the eigenstate of $\hat{X}^\mu(\sigma)$:
\begin{equation*}
\begin{split}
   \hat{X}^\mu(\sigma)|x,\tilde{x},\xi\rangle = &X^\mu(\sigma)|x,\tilde{x},\xi\rangle\\
   X^\mu(\sigma) =& \xi^\mu + \sum_{n>0}\sqrt{\alpha'}\left(x^\mu_{n}e^{-in\sigma}+\tilde{x}^\mu_{n}e^{in\sigma}\right).
\end{split}
\end{equation*}
This means that $|x,\tilde{x},\xi$ is identical to $|X^\mu(\sigma)\rangle$ which defined by the eigenstate of $\hat{X}^\mu(\sigma)$ in \eqref{review:eq:string-eigenstate}.
Considering this fact, we introduce the following notations:
\begin{equation*}
\begin{split}
 |X^\mu(\sigma)\rangle   &=  |x,\tilde{x},\xi\rangle \\
 \int\!\!\mathcal{D}X^\mu(\sigma)&=\int\!\!\mathcal{D}x\mathcal{D}\tilde{x}\mathcal{D}\xi\\
\end{split}
\end{equation*}
%where $|X^\mu(\sigma)\rangle$ is the eigenstate of $\hat{X}^\mu(\sigma)$:
%\begin{equation*}
%\begin{split}
%    \hat{X}^\mu(\sigma) |X^\mu(\sigma)\rangle &= X^\mu(\sigma)|X^\mu(\sigma)\rangle.\\
%\end{split}
%\end{equation*}
%In addition we define the conjugate momentum and its eigenstate
%\begin{equation*}
%\begin{split}
%   \hat{\Pi}^\mu(\sigma)&=\frac{1}{2\pi\alpha'}\partial_\tau \hat{X}^\mu(\tau=0,\sigma)\\
%   \hat{\Pi}^\mu |\Pi^\mu(\sigma)\rangle &= \Pi^\mu(\sigma)|\Pi(\sigma)^\mu\rangle. 
%\end{split}
%\end{equation*}
These satisfy a relation
\begin{equation*}
  |\Pi^\mu(\sigma)=0\rangle =   \int\!\!\mathcal{D}X^\mu(\sigma)|X^\mu(\sigma)\rangle.  
\end{equation*}
By using $\hat{X}^\mu(\sigma)$ and $\hat{\Pi}^\mu(\sigma)$, the boundary state \eqref{review:eq:vertex4} becomes
\begin{equation}
\begin{split}
  |B[S]\rangle&=\int\!\!\mathcal{D}X^\mu(\sigma)e^{-S\left[X^\mu(\sigma)\right]}|X^\mu(\sigma)\rangle\\
&=e^{-S\left[\hat{X}^\mu(\sigma)\right]}|Dp\rangle.
\end{split}\label{review:eq:vertex5}
\end{equation}
where 
\begin{equation*}
  |Dp\rangle = |B[S=0]\rangle=\int\!\!\mathcal{D}X^\mu(\sigma)|X^\mu(\sigma)\rangle=|\Pi^\mu(\sigma)=0\rangle.
\end{equation*}
Note that the boundary condition 
\begin{equation*}
    \hat{\Pi}^\mu(\sigma)|Dp\rangle = \frac{1}{2\pi\alpha'}\partial_\tau \hat{X}^\mu(\tau=0,\sigma)|Dp\rangle =0
\end{equation*}
represents the Neumann boundary conditions \eqref{review:eq:closedboundary1}.

In the case of a single D-brane without open string background fields, the boundary action is \[S_0=0.\] In this case, \eqref{review:eq:vertex2} can be calculated by using the Gaussian integral. We evaluate $\prod_n n^{-D/2}$ in the overall factor by zeta-function regularization:
\begin{equation*}
\begin{split}
   \log \left(\prod_{n=1}^\infty xn^{-D/2}\right)&=\lim_{s\to 0}\frac{d}{ds}\left(\sum_{n=1}^\infty(xn^{-D/2})^s\right)\\
       &= \zeta(0)\log x +\frac{1}{2}D\zeta'(0)\\
       &= -\frac{1}{2}\log x -\frac{1}{4}D\log 2\pi \\
\therefore \quad \prod_{n=1}^\infty n^{-D/2} &= (2\pi)^{-D/4}. 
\end{split}  
\end{equation*}
In this way the overall factor can be regularize. Here we do not determine the overall normalization, which will be given later in \eqref{review:eq:normalization}. We do the Gaussian integral, and then we have
\begin{equation*}
  V=N_p\exp\left\{\sum_{n>0}a_{-n}^\mu\tilde{a}_{-n}^\mu\right\}.
\end{equation*}
By substituting \eqref{review:eq:normalized-creation-operator} into this, we find that the boundary state is given by \cite{NPB288525}
\begin{equation}
\begin{split}
  |B\rangle &= V|0\rangle \\
&= \exp\left\{-\sum_{n>0}\frac{1}{n}\alpha_{-n}^\mu\tilde{\alpha}_{-n}^\mu\right\}|0\rangle.
\end{split}
\end{equation}
This result is identical to \eqref{review:eq:boundarystate1} for the Neumann directions. The extension to the Dirichlet directions was considered in \cite{NPB431131,9604091}. 

By using the operator $V$, we can express the functional integral $I$, which represents the cylinder amplitude as illustrated in figure \ref{review:fig:cylinder-btw}. In \eqref{review:eq:cylinder1}, the corresponding term to $ \omega a^\dagger a +\frac{1}{2}$ in the string theory is
\begin{equation*}
\begin{split}
   \omega a^\dagger a +\frac{1}{2} &\to \sum_{n=1}^\infty\sum_{\mu=0}^{D-1}n \left(a_{-n}^\mu a_n^\mu+\tilde{a}_{-n}^\mu \tilde{a}_n^\mu \right)-2a \\
&= \sum_{n=1}^\infty\sum_{\mu=0}^{D-1}\left(\alpha_{-n}^\mu\alpha_{n}^\mu+\tilde{\alpha}_{-n}^\mu\tilde{\alpha}_{n}^\mu\right) -2
%\\&= L_0+\tilde{L}_0-2.
\end{split}
\end{equation*}
$a=1$ comes from the vacuum energy. $L_0$ and $\tilde{L}_0$ are the Virasoro generators given by
\begin{equation*}
\begin{split}
   L_0 &= \frac{\alpha'}{4}\hat{p}^2+\sum_{n=1}^\infty\alpha_{-n}^\mu\alpha_n^\mu\\
   \tilde{L}_0 &= \frac{\alpha'}{4}\hat{p}^2+\sum_{n=1}^\infty\tilde{\alpha}_{-n}^\mu\tilde{\alpha}_n^\mu
\end{split}  
\end{equation*}
Note that $\hat{p}|B\rangle =0$ for Neumann directions. Thus the functional integral \eqref{review:eq:cylinder1} becomes
\begin{equation*}
 I  = \langle B | e^{-(L_0+\tilde{L}_0-2)T} |B\rangle.
\end{equation*}
%where we parameterize propagations of closed string by $l$. 
This result factorizes into two boundaries with appropriate weight $(M_\Psi^2)^{-1}$:
\begin{equation*}
 I \sim \sum_{\Psi}\langle B| \Psi \rangle (M_\Psi^2)^{-1}\langle \Psi | B\rangle. 
\end{equation*}
where $\Psi$ represents a closed string state \cite{NPB288525}.

\EPSFIGURE[ht]{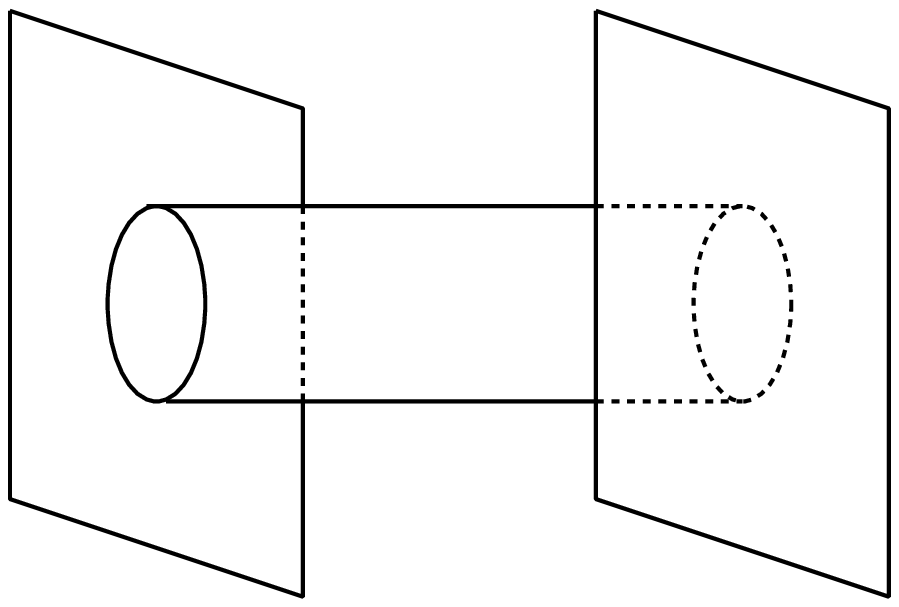,width=0.6\textwidth}{Cylinder stretched between two D-branes\label{review:fig:cylinder-btw}}
~ \\ \\
The normalization constant $N_p$ is determined by comparing the cylinder amplitude of closed channel with that of open channel. The amplitude of a cylinder stretched between two parallel D$p$-branes is
\begin{equation}
    A_\mathrm{closed}=\langle Dp |\Delta  |Dp\rangle \label{review:eq:cylinder2}
\end{equation}
where $|Dp\rangle$ is given by \eqref{review:eq:boundarystate1}:
\begin{equation*}
   |Dp\rangle=N_p\exp\left\{-\frac{1}{n}\sum_{n>0}\alpha_{-n}^\mu S_{\mu\nu}\tilde{\alpha}_{-n}^\nu\right\}|\delta(\hat{x}^i)0\rangle.
\end{equation*}
We place the second D-bane at $\hat{x}^i=\xi^i$. Hence $\langle Dp|$ is given by
\begin{equation*}
   \langle Dp|=N_p\langle 0| \delta(\hat{x}^i-\xi^i)\exp\left\{-\frac{1}{n}\sum_{n>0}\alpha_{n}^\mu S_{\mu\nu}\tilde{\alpha}_{n}^\nu\right\}.
\end{equation*}
$\Delta$ is the closed string propagator which is given by
\begin{equation*}
    \Delta =\frac{\alpha'}{2}\int_0^1\!\!d\rho\ \rho^{L_0+\tilde{L}_0-3}.
\end{equation*}
Take care that we should integrate over the modulus $l=-\log \rho$ of the cylinder. A physical state is annihilated by $L_0-\tilde{L}_0$. Therefore we can modify the propagator so that it propagates only physical states:
\begin{equation*} 
   \Delta = \frac{\alpha'}{2}\int_0^1\!\!d\rho\int_0^{2\pi}\!\!\frac{d\phi}{2\pi}\ \rho^{L_0+\tilde{L}_0-3}e^{i\phi(L_0-\tilde{L}_0)}.
\end{equation*}
After the change of variable $z=\rho e^{i\phi}$, the closed string propagator is
\begin{equation*}
   \Delta = \frac{\alpha'}{4\pi}\int_{|z|\le 1}\!\!\frac{dzd\bar{z}}{|z|^2}z^{L_0-1}\bar{z}^{\tilde{L}_0-1}.
\end{equation*}
We define $d_\bot=D-p-1$. The zero mode part in \eqref{review:eq:cylinder2} becomes 
\begin{equation*}
\begin{split}
&   (N_p)^2\frac{\alpha'}{4\pi}\int_{|z|\le 1}\frac{d^2z}{|z|^4}\langle p^\mu=0| \delta(\hat{x}^i-\xi^i)|z|^{\frac{\alpha'}{2}\hat{p}^2}\delta(\hat{x}^i)|p^\mu=0\rangle\\
&=(N_p)^2\frac{\alpha'}{4\pi}\int_{|z|\le 1}\frac{d^2z}{|z|^4}\int\!\!\frac{dk^i}{(2\pi)^{d_\bot}}\int\!\!\frac{d{k'}^i}{(2\pi)^{d_\bot}}
\langle p^\mu=0|(e^{ik'_i(\hat{x}^i-\xi^i)}|z|^{\frac{\alpha'}{2}\hat{p}^2}e^{ik_i\hat{x}^i})|p^\mu=0\rangle \\
&=(N_p)^2\frac{\alpha'}{4\pi}\int_{|z|\le 1}\frac{d^2z}{|z|^4}\int\!\!\frac{dk^i}{(2\pi)^{d_\bot}}\int\!\!\frac{d{k'}^i}{(2\pi)^{d_\bot}}\\
&\hspace{10em}\langle p^a=0,p^i=-{k'}^i|(e^{-ik'_i\xi^i}|z|^{\frac{\alpha'}{2}\hat{p}^2})|p^a=0,p^i=k^i\rangle \\
&=V_{p+1}(N_p)^2\frac{\alpha'}{4\pi}\int_{|z|\le 1}\frac{d^2z}{|z|^4}\int\!\!\frac{dk^i}{(2\pi)^{p+1}}e^{ik_i\xi^i}|z|^{\frac{\alpha'}{2}k_i^2}.
\end{split}
\end{equation*}
where we have used
\begin{equation*}
  \langle k^i | {k'}^i \rangle = (2\pi)^{p+1}\delta^{p+1}(k^i-{k'}^i), \quad V_{p+1}(2\pi)^{p+1}\delta^{p+1}(0).
\end{equation*}
Performing the Gaussian integral, the zero mode part gives
\begin{equation}
   (N_p)^2\frac{\alpha'}{4\pi}V_{p+1}e^{-\frac{\xi^2}{2\pi\alpha't}}(2\pi^2t\alpha')^{-d_\bot/2}, \quad |z|=e^{-\pi t}.\label{review:eq:cylinder-zeromode}
\end{equation}

The oscillator part in \eqref{review:eq:cylinder2} gives
\begin{equation}
\begin{split}
&   \int_{|z|\le 1}\frac{d^2z}{|z|^4}\langle 0| e^{-\sum_{n>0}\frac{1}{n}\alpha_nS\tilde{\alpha}_n}z^{N}\bar{z}^{\tilde{N}}e^{-\sum_{n>0}\frac{1}{n}\alpha_{-n}S\tilde{\alpha}_{-n}}|0\rangle.
\end{split}\label{review:eq:cylinder-nonzeromode2}
\end{equation}
Here we have introduced
\begin{equation*}
   N=\sum_{n>0}\alpha_{-n}\alpha_n \quad \tilde{N}=\sum_{n>0}\tilde{\alpha}_{-n}\tilde{\alpha}_n.
\end{equation*}
These operators satisfy the following relations:
\begin{equation*}
\begin{split}
  z^Ne^{\alpha_{-n}}z^{-N}&=e^{\alpha_{-n}z^n} \quad   \bar{z}^Ne^{\alpha_{-n}}\bar{z}^{-N}=e^{\alpha_{-n}\bar{z}^n}\\
z^{\tilde{N}}e^{\tilde{\alpha}_{-n}}z^{-\tilde{N}}&=e^{\tilde{\alpha}_{-n}z^n} \quad   \bar{z}^{\tilde{N}}e^{\tilde{\alpha}_{-n}}\bar{z}^{-\tilde{N}}=e^{\tilde{\alpha}_{-n}\bar{z}^n} \quad n\ne 0.
\end{split}
\end{equation*}
By using these relations, \eqref{review:eq:cylinder-nonzeromode2} becomes
\begin{equation*}
\begin{split}
&   \int_{|z|\le 1}\frac{d^2z}{|z|^4}\langle 0| e^{-\sum_{n>0}\frac{1}{n}\alpha_nS\tilde{\alpha}_n}e^{-\sum_{n>0}\frac{1}{n}\alpha_{-n}S\tilde{\alpha}_{-n}|z|^{2n}}|0\rangle.
\end{split}
\end{equation*}
Contracting oscillators in this equation, we have
\begin{equation}
    \prod_{n=1}^\infty \left(\frac{1}{1-|z|^{2n}}\right)^{D-2} \label{review:eq:cylinder-nonzeromode1}
\end{equation}
The extra power $(-2)$ comes from contributions of the ghost fields \cite{9912161}. However, we omit detailed calculations of the contractions. 

We change integral variables so that
\begin{equation*}
    |z|=e^{-\pi t} \quad d^2z=-\pi e^{-2\pi t}dt d\varphi.
\end{equation*}
Multiplying the zero mode part \eqref{review:eq:cylinder-zeromode} and the oscillator part \eqref{review:eq:cylinder-nonzeromode1}, the cylinder amplitude $A_\mathrm{closed}$ in \eqref{review:eq:cylinder2} gives
\begin{equation*}
\begin{split}
   A_\mathrm{closed}&=(N_p)^2V_{p+1}\frac{\alpha'\pi}{2}(2\pi^2\alpha')^{-d_\bot/2}\int_0^\infty\!\!dt\ t^{-d_\bot/2}e^{-\frac{\xi_i^2}{2\pi\alpha't}}\left(f_1(e^{-\pi t})^{-24}\right)\\
   f_1(q)&=q^{\frac{1}{12}}\prod_{n=1}^\infty(1-q^{2n}).
\end{split}
\end{equation*}
In order to compare this amplitude with that obtained in open channel, we perform change of modulus $t=\frac{1}{\tau}$. After this change, $A_\mathrm{closed}$ becomes
\begin{equation}
      A_\mathrm{closed}=(N_p)^2V_{p+1}\frac{\alpha'\pi}{2}(2\pi^2\alpha')^{-d_\bot/2}\int_0^\infty\!\!\frac{d\tau}{\tau}\ \tau^{12-\frac{p+1}{2}}e^{-\frac{\xi_i^2\tau}{2\pi\alpha'}}\left(f_1(e^{-\frac{\pi}{\tau}})^{-24}\right).\label{review:eq:cylinder-closedchannel}
\end{equation}
On the other hand, the same amplitude is obtained from the viewpoint of open string one-loop calculations:
\begin{equation}
     A_\mathrm{open}=V_{p+1}(8\pi\alpha')^{-\frac{p+1}{2}}\int_0^\infty\!\!\frac{d\tau}{\tau}\ \tau^{12-\frac{p+1}{2}}e^{-\frac{\xi_i^2\tau}{2\pi\alpha'}}\left(f_1(e^{-\frac{\pi}{\tau}})^{-24}\right).\label{review:eq:cylinder-openchannel}
\end{equation}
These two results \eqref{review:eq:cylinder-closedchannel} and \eqref{review:eq:cylinder-openchannel} are identical with the choice of $N_p$ such that
\begin{equation}
    N_p=\frac{T_p}{2}, \quad T_p=\frac{\sqrt{\pi}}{2^{\frac{D-10}{2}}}(2\pi\sqrt{\alpha'})^{\frac{D}{2}-2-p} \label{review:eq:normalization}.
\end{equation}
Therefore we can see that the boundary state $|Dp\rangle$ reproduces the cylinder amplitude $A$ correctly. Here $T_p$ represents the tension of D$p$-brane. 
\\ \\
To summarize this subsection, we have seen that the boundary state $Dp\rangle$ satisfies the boundary condition
\begin{equation}
\begin{split}
 \partial_\tau \hat{X}^a (\sigma,\tau=0)|Dp\rangle &=0 \qquad a=0,1,\cdots,p\\
     \hat{X}^i (\sigma,\tau=0)|Dp\rangle &=\xi^i \qquad i=p+1,\dots,D-1
\end{split}\tag{\ref{review:eq:closedboundary1}}
\end{equation}
and reproduces the correct disk and cylinder amplitudes:
\begin{align}
  \langle V_\Psi(k) \rangle_\mathrm{disk} &=\langle \Psi(k) | Dp\rangle \tag{\ref{review:eq:disk}}\\ 
  A_\mathrm{closed}& = \langle Dp | \Delta | Dp \rangle \tag{\ref{review:eq:cylinder2}}. 
\end{align}
The boundary state can be considered as superposition of emitted closed strings of various shape
\begin{equation}
      |Dp\rangle=\int\!\!\mathcal{D}X^a(\sigma)|X^a(\sigma),X^i(\sigma)=\xi^i\rangle. \tag{\ref{review:eq:superposition}}
\end{equation}
The BRST invariant formulation and the ghost sector of the boundary state in superstring theory was studies in \cite{NPB29383,NPB308221}. We note that the boundary state $|Dp\rangle$ is BRST invariant considering the contribution of matter and ghost sectors. This represents the boundary state is physical state of closed strings.

\subsection{Boundary state of single D-brane with Abelian field}\label{paper:section:review-3}
In the previous subsection, we have ignored open strings attached to D-branes. In this subsection we will consider a boundary state which includes excitation of massless opens strings. In presence of  open string excitation on a D-brane, the system is altered as a source of closed strings from the D-brane with no excitation. In other words closed strings emitted by the D-brane is affected by open string background fields. Therefore we consider to construct a boundary state in which the contribution of open string background fields is incorporated. A boundary state with an Abelian gauge field was studied by adding a Wilson loop factor in \cite{NPB288525,NPB308221}. In case of constant field strength, the boundary state can be calculated explicitly \cite{NPB288525,NPB308221,9906214}. A boundary state with the general gauge field was studied in \cite{9909027,9909095}. An another approach to construct a boundary state with the general open string background fields was adopted in \cite{0211232,0312260}. Less in known about non-Abelian extension, which is one of the main theme of this paper.  

An open string couples to a gauge field on a D-brane through the point-like charge at its endpoints. In presence of a background Abelian gauge field $A^a(\hat{X}^i(\tau))$, we should add the boundary action at $\sigma=0$
\begin{equation*}
   S_\mathrm{boundary}=-\frac{i}{2\pi\alpha'}\int\!\!dt\ A_a(\hat{X}(t)) \partial_\tau \hat{X}^a(t).
\end{equation*}
Note that the argument of the gauge field $A^\mu$ is the spacetime coordinates $\hat{X}^a(t)$ of open string along the Neumann directions. This boundary action leads to the boundary condition
\begin{equation*}
\begin{split}
   (\partial_\sigma \hat{X}_a-F_{ab}(\hat{X})\partial_\tau\hat{X}^b)\big|_{\sigma=0}&=0 \qquad a=0,1,\cdots,p\\
   \hat{X}^i\big|_\mathrm{\sigma=0}&=\xi^i \qquad i=p+1,\dots,D-1
\end{split}
\end{equation*}
where $F_{ab}=\partial_a A_b-\partial_b A_a$ is the field strength, and $\tau=it$. We translate these conditions in closed channel, as \eqref{review:eq:modular-transformation} in the previous subsection. In this way we have the boundary conditions for a D$p$-brane with background gauge field:
\begin{equation}
\begin{split}
      (\partial_\tau \hat{X}_a-F_{ab}(\hat{X})\partial_\sigma\hat{X}^b)\big|_{\tau=0}|Dp[A]\rangle&=0 \\
   \hat{X}^i\big|_\mathrm{\tau=0}|Dp[A]\rangle&=\xi^i.
\end{split}\label{review:eq:closedboundary4}
\end{equation}
Here $|Dp[A]\rangle$ is a boundary state of D$p$-brane with a gauge field $A^a(\hat{X}(\sigma))$. The condition along the Neumann directions is modified, and becomes nonlinear in $\hat{X}$. Therefore it is difficult to solve this condition in general. In case of constant field strength, namely
\begin{equation}
  F_{ab}(\hat{X})=F_{ab} \quad A_a(\hat{X})=-\frac{1}{2}F_{ab}\hat{X}^b  \label{review:eq:constant-f}, 
\end{equation}
these boundary conditions can be easily solved \cite{9912161}. Substituting the mode expansion \eqref{review:eq:modeexpansion1}, \eqref{review:eq:closedboundary4} becomes
\begin{align*}
\begin{split}
   ((1+F)_{ab}\alpha_n^b+(1-F)_{ab}\tilde{\alpha}_{-n}^b)|Dp[A]\rangle &=0\\
   (\alpha_n^i-\tilde{\alpha}_{-n}^i)|Dp[A]\rangle &=0
\end{split}
\\
\begin{split}
   \hat{p}^a|Dp[A]\rangle&=0\\
   \hat{x}^i|Dp[A]\rangle &=\xi^i . 
\end{split}
\end{align*}
These are satisfied by the following boundary state:
\begin{equation}
\begin{split}
 |Dp[A]\rangle&=\frac{T_p}{2}\sqrt{-\det(\eta+F)}\exp\left\{-\sum_{n>0}\frac{1}{n}\alpha^\mu_{-n}M_{\mu\nu}\tilde{\alpha}^\nu_{-n}\right\}\delta(\hat{x}^i-\xi^i)|0\rangle\\
M_{\mu\nu}&=\left(\left(\frac{1-F}{1+F}\right)_{ab},-\delta_{ij}\right).
\end{split}\label{review:eq:boundary-constant-f}
\end{equation}
We can see that the boundary state of D$p$-brane with a gauge field $A^a$ reduces to that without a gauge field when $A^a=0$. Explicitly we can show
\begin{equation*}
|Dp[A=0]\rangle=|Dp\rangle.
\end{equation*}
The normalization constant can be determined by comparing couplings to closed massless closed strings with the DBI action \cite{9912161}. We will show the extra factor $\sqrt{-\det(\eta+F)}$ appears later in \eqref{review:eq:boundary-constant-f2}.
\\ \\
In the previous subsection, we have seen that a boundary which includes a boundary action $S$ is given by
\begin{equation}
\begin{split}
  |B[S]\rangle&=\int\!\!\mathcal{D}X^\mu(\sigma)e^{iS\left[X^\mu(\sigma)\right]}|X^\mu(\sigma)\rangle\\
&=e^{iS\left[\hat{X}^\mu(\sigma)\right]}|Dp\rangle.
\end{split}\label{review:eq:vertex6}
\end{equation}
which is obtained from \eqref{review:eq:vertex5} with the Wick rotation $i\sigma\to \sigma$.
This formula represents that the contribution of the boundary action $S$ is accounted for by including a factor $e^{iS[X^\mu(\sigma)]}$ in the functional integral \eqref{review:eq:superposition}. In other words, $e^{iS[X^\mu(\sigma)]}$ gives a weight function in the functional integral for $X^\mu(\sigma)$. In addition, we can say that the boundary state which incorporates the boundary action $S$ is obtained by operating $e^{iS[\hat{X}^\mu(\sigma)]}$ on the boundary state $|Dp\rangle$ of D$p$-brane. 

We suppose that contribution of open string fields on a D-brane to the boundary state is accounted for by choosing the action $S[\hat{X}(\sigma)]$ so that
\begin{equation*}
    S[\hat{X}^\mu(\sigma)]= S_\mathrm{boundary}[\hat{X}^\mu(\sigma)].
\end{equation*}
Here $S_\mathrm{boundary}$ is the boundary term in the open string action as functional, while the argument is replaced by the worldsheet field of closed strings at the boundary $\hat{X}(\sigma).$
%a boundary action $S_\mathrm{bounary}[\hat{X}^\mu]$ in string theories as the action $S[\hat{X}^\mu]$ in \eqref{review:eq:vertex6}. 
%We note that in \eqref{review:eq:vertex} $\hat{X}^\mu$ represents a worldsheet of closed string, but not open string. 
This is natural from the viewpoint of the modular transformation between open and closed channels.
In what follows, we concentrate on the gauge field. $S$ in the boundary state \eqref{review:eq:vertex6} is
\begin{equation}
     S[\hat{X}^\mu(\sigma)]=\int_0^{2\pi}\!\!d\sigma\ A_a(\hat{X}(\sigma)) \partial_\sigma \hat{X}^a(\sigma).\label{review:eq:gaugeboundaryaction}
\end{equation}
Here we have rescaled the gauge so that $\frac{1}{2\pi\alpha'}A_a \to A_a$.
The exponential of this action gives the Wilson loop operator
\begin{equation}
      e^{iS[\hat{X}^\mu(\sigma)]}=\exp\left\{i\int_0^{2\pi}\!\!d\sigma\ A_a(\hat{X}(\sigma)) \partial_\sigma \hat{X}^a(\sigma) \right\}.
\label{review:eq:gaugeboundaryaction-exponent}
\end{equation}
Note that $\hat{X}^\mu(\tau)$ is periodic with respect to $\sigma$, namely $\hat{X}^\mu(\sigma+2\pi)=\hat{X}^\mu(\sigma)$.
We define a vertex operator of the gauge field $A$ by
\begin{equation*}
\begin{split}
   U_A(k)&=A_a(k)\partial_\sigma \hat{X}^a(\sigma) e^{ik\hat{X}(\sigma)}\\
   U_A(\hat{X}(\sigma)) &= A_a(\hat{X}(\sigma))\partial_\sigma \hat{X}^a(\sigma) =
\int\!\!\frac{d^{p+1}k}{(2\pi)^{p+1}} A_a(k)\partial_\sigma \hat{X}^a e^{ik\hat{X}(\sigma)}
\end{split}.
\end{equation*}
By using this operator, \eqref{review:eq:gaugeboundaryaction-exponent} can be expressed by
\begin{equation*}
    e^{iS[\hat{X}^\mu(\sigma)]}=\sum_{n=1}^\infty \left(i\int\!\!d\sigma\ U_A(\hat{X}(\sigma))\right)^n.
\end{equation*}
Take the product of $n$ vertex operators $U_A[\hat{X}(\sigma_i)] (i=1,\cdots,n)$, integrate over the position $\sigma_i$ of each vertex on the D-brane, sum up over the the number of vertices $n$, then we get this factor $e^{-S[\hat{X}^\mu(\sigma)]}$. It is worthwhile to note that the boundary state defined in this way, namely 
\begin{equation}
   |Dp[A]\rangle = \exp\left\{i\int\!\!d\sigma A_a(\hat{X}(\sigma))\partial \sigma \hat{X}^a(\sigma)\right\}|Dp\rangle, \label{review:eq:boundary-general-gauge}
\end{equation}
satisfies the required boundary conditions \cite{9909027}: 
\begin{equation}
\begin{split}
      (\partial_\tau \hat{X}_a-F_{ab}(\hat{X})\partial_\sigma\hat{X}^b)\big|_{\tau=0}|Dp[A]\rangle&=0 \\
   \hat{X}^i\big|_\mathrm{\tau=0}|Dp[A]\rangle&=\xi^i.
\end{split}\tag{\ref{review:eq:closedboundary4}}
\end{equation}
This follows from a property \cite{9909027}
\begin{equation*}
    e^{i\int\!\!d\sigma A\partial\sigma \hat{X}(\sigma)} \partial_\tau \hat{X}_a(\sigma) e^{-i\int\!\!d\sigma A\partial\sigma \hat{X}(\sigma)}= \partial_\tau \hat{X}_a(\sigma)  -F_{ab}(\hat{X}(\sigma))\partial_\sigma \hat{X}^b(\sigma).
\end{equation*}
and the boundary condition \eqref{review:eq:closedboundary1} for $|Dp\rangle$.

We consider what amplitude a boundary state containing the boundary action reproduces. The disk amplitude which should be reproduced by the boundary states is
\begin{equation}
   \langle V_\Psi(k) e^{-S_\mathrm{boundary}} \rangle_\mathrm{disk}. \label{review:eq:disk-amplitude1}
\end{equation}
$V_\Psi(k)$ is a closed string vertex of state $\Psi$ with momentum $k$, and $S_\mathrm{boundary}[A]$ is the boundary action in the string action
\begin{equation*}
    S_\mathrm{boundary}[A] =   \int\!\!d\tau\ A_a(\hat{X}(\tau)) \partial_\tau \hat{X}^a(\tau).
\end{equation*}  
We denote the vertex operator of the gauge field $A$ by
\begin{equation*}
\begin{split}
   V_A(k)&=A_a(k)\partial_\tau \hat{X}^a(\tau) e^{ik\hat{X}(\tau)}\\
   V_A(\hat{X}(\tau)) &= A_a(\hat{X}(\tau))\partial_\tau \hat{X}^a(\tau) =
\int\!\!\frac{d^{p+1}k}{(2\pi)^{p+1}} A_a(k)\partial_\tau \hat{X}^a e^{ik\hat{X}(\tau)}
\end{split}.
\end{equation*}
By using this vertex, the exponential of the boundary action action in \eqref{review:eq:disk-amplitude1} becomes 
\begin{equation*}
    e^{iS_\mathrm{boundary}[A]}=\sum_{n=1}^\infty \left(i\int\!\!d\sigma\ V_A(\hat{X}(\sigma))\right)^n.
\end{equation*}
Take the product of $n$ vertex operators $V_A[\hat{X}(\tau_i)] (i=1,\cdots,n)$, integrate over the position $\tau_i$ of each vertex at the boundary, sum up over the the number of vertices $n$, then we get this factor $e^{iS_\mathrm{boundary}[A]}$. In figure \ref{review:fig:disk3-abelian} we illustrate the disk amplitude \eqref{review:eq:disk-amplitude1} with a closed string vertex of $h_{\mu\nu},b_{\mu\nu},\phi$ and two open string vertices of $A^a,X^i$. 

\psfrag{h_munu,b_munu,phi}{$h_{\mu\nu},b_{\mu\nu},\phi$}
\psfrag{A^a,X^i}{$A^a,X^i$}
\EPSFIGURE[htb]{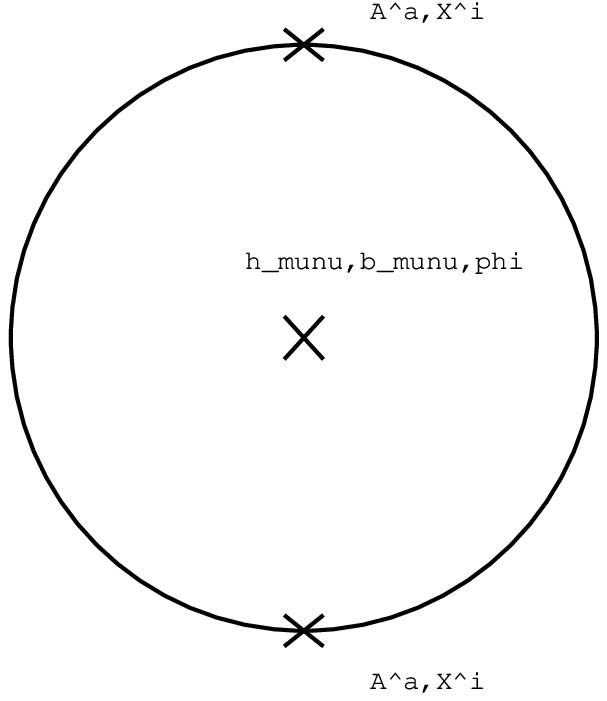,width=0.7\textwidth}{Disk with one closed and two open string vertices\label{review:fig:disk3-abelian}}

We illustrate what insertion of $U_A$ and $V_A$ indicates. Figure \ref{review:fig:disk3-abelian}  shows the disk amplitude with one closed string vertex on the disk and two open string vertices at the boundary. Because a boundary of worldsheet should attach to a D-brane, this amplitude corresponds to figure \ref{review:fig:disk2-abelian}. Open string vertices at the boundary means strips, and a closed string vertex represents a cylinder. Therefore figure \ref{review:fig:disk-abelian} is equivalent to figure \ref{review:fig:disk-abelian} from the viewpoint of worldsheet. Figure \ref{review:fig:disk-abelian} illustrates a process in which one incoming open string splits into outgoing one closed string and one open string, or two incoming open strings create one closed string. $e^{i\int\!\!d\tau U_A}$ and $e^{i\int\!\!d\sigma V_A}$ indicates that we should sum up the number of open string vertices, and integrating the position of those vertices at the boundary. Such a Wilson loop factor obtained in this way represents the effect of open string background fields on closed string emission. The formula \eqref{review:eq:vertex6} indicates that 
 open string fields can be incorporated into the boundary state by including the Wilson loop factor.

\DOUBLEFIGURE[htb]{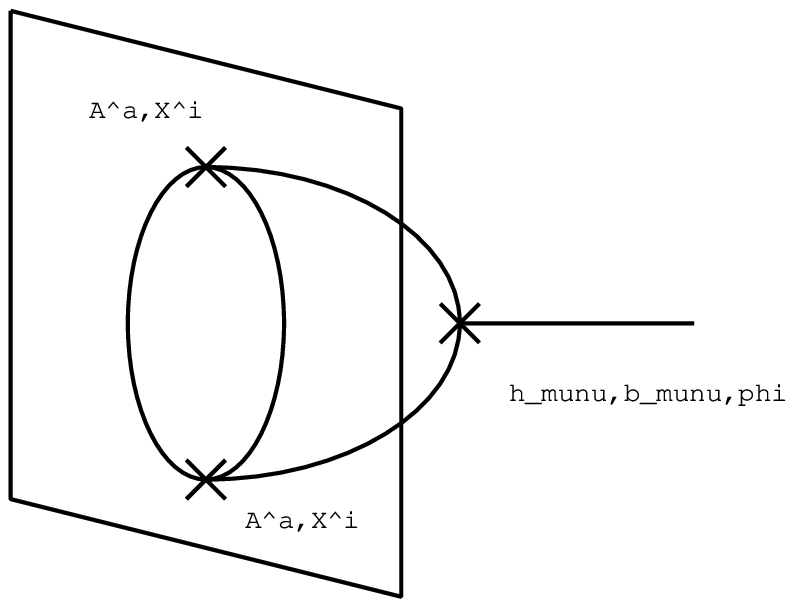,width=0.5\textwidth}{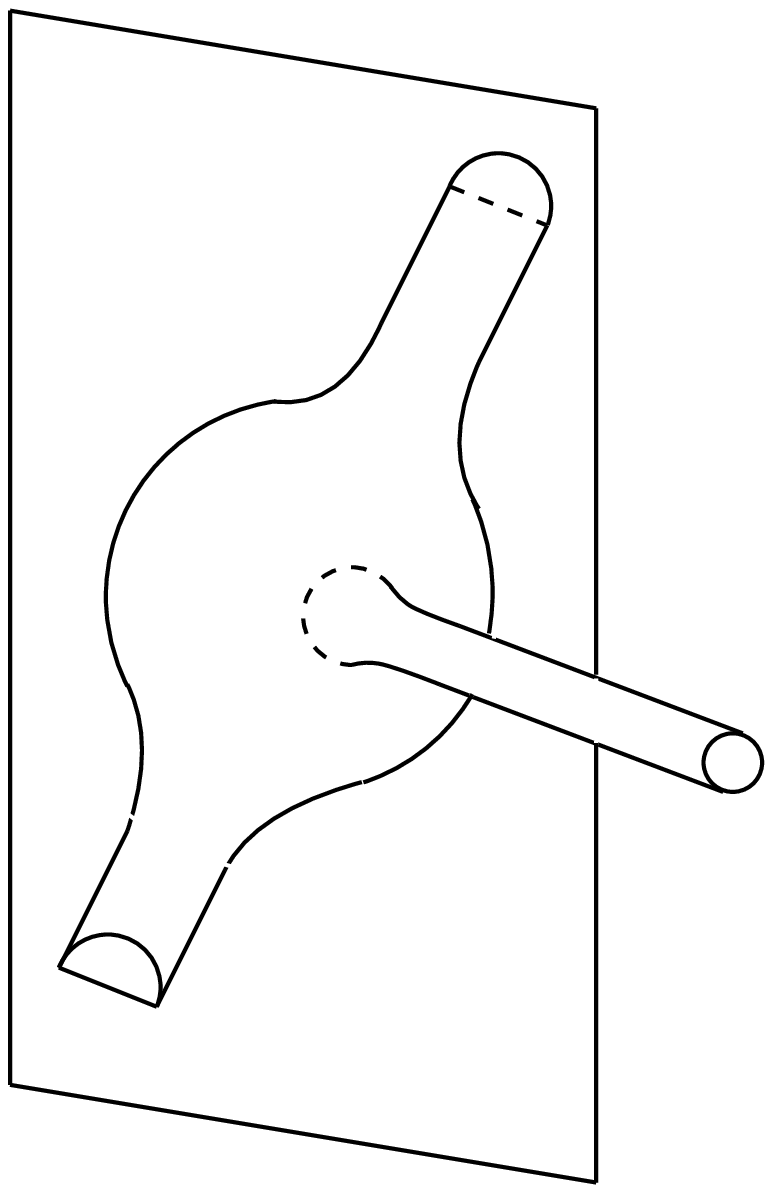,width=0.3\textwidth}{Worldsheet with one closed and two open string vertices\label{review:fig:disk2-abelian}}{Closed string emission via two open strings\label{review:fig:disk-abelian}}

In the case of constant field strength \eqref{review:eq:constant-f}, we can calculate the boundary state explicitly \cite{NPB288525,NPB308221}. The boundary action is
\begin{equation*}
\begin{split}
 S[\hat{X}^\mu(\sigma)]& =\frac{i}{4\pi\alpha'}\int_0^{2\pi}\!\!d\sigma\ F_{\mu\nu}\hat{X}^\nu(\sigma)\partial_\sigma \hat{X}^\mu(\sigma)\\
  &= \frac{1}{2}\sum_{n=1}^\infty \sum_{\mu,\nu=0}^{D-1} \hat{x}_n^\mu F_{\mu\nu}\hat{\tilde{x}}_n^\nu\\
  &= \frac{1}{2}(\hat{x} | F | \tilde{\hat{x}})
\end{split}
\end{equation*}
Here we have substituted \eqref{review:eq:modeexpansion2}, and defined $(y| f |x)=\sum_{n,m,\mu,\nu}y_n^\mu f_{\mu\nu}\delta_{mn} x_n^\nu$ for $f_{\mu\nu}$. From \eqref{review:eq:vertex2}, the boundary state becomes
\begin{equation*}
\begin{split}
&   |B[A_\mu=F_{\mu\nu}X^\mu]\rangle = \\
&\int\!\!\mathcal{D}x\mathcal{D}\tilde{x}\exp\left\{-\frac{1}{2}(x|F|\tilde{x})-\frac{1}{2}(x|\tilde{x})+(\tilde{a}^\dagger|\tilde{x})+(a^\dagger|x)-(a^\dagger|\tilde{a}^\dagger)\right\}|0\rangle.
\end{split}
\end{equation*}
The functional integral is Gaussian, and can be performed to give
\begin{equation*}
\begin{split}
    |B[A_\mu=F_{\mu\nu}X^\mu]\rangle &= \frac{T_p}{2}\prod_{n=1}^\infty[\det(1+F)]^{-1}\exp\left\{-\sum_{n>0}\frac{1}{n}\alpha{-n}^\mu M_{\mu\nu}\tilde{\alpha}_{-n}^\nu\right\}|0\rangle\\
   M_{\mu\nu}&=\left(\frac{1-F}{1+F}\right)_{\mu\nu}.
\end{split}
\end{equation*}
We use the zeta function regularization to evaluate the determinant:
\begin{equation*}
    \sum_{n=1}^\infty \ 1 = \zeta(0) = -\frac{1}{2}. 
\end{equation*}
Thus we get
\begin{equation}
    |B[A_\mu=F_{\mu\nu}X^\mu]\rangle = \frac{T_p}{2}\sqrt{\det(1+F)}\exp\left\{-\sum_{n>0}\frac{1}{n}\alpha_{-n}^\mu M_{\mu\nu}\tilde{\alpha}_{-n}^\nu\right\}|0\rangle.\label{review:eq:boundary-constant-f2}
\end{equation}
This is identical to \eqref{review:eq:boundary-constant-f} obtained by requiring the boundary conditions. The overall normalization is determined so that the boundary state reduces $Dp\rangle$ when $F=0$, or vanishing gauge field. Therefore we can say the formula \eqref{review:eq:vertex6} reproduces the boundary state \eqref{review:eq:boundary-constant-f} which determined by the boundary conditions in case of constant field strength. We note that the boundary state \eqref{review:eq:boundary-constant-f}, or equivalently \eqref{review:eq:boundary-constant-f}, of D$p$-brane with constant magnetic field $F_{ab}$ is transformed by a T-dual transformation along $X^a$ or $X^b$ to that of a D$(p-1)$ brane rotated in $(a,b)$ plane \cite{9912275}. Therefore the presence of constant field strength does not break the BRST invariance of the boundary state $|Dp\rangle$ of D$p$-brane.
\\ \\

The boundary state with non-constant field strength was studied in \cite{9909027,9909095,0211232,0312260}. It is difficult to perform  the functional integral in \eqref{review:eq:vertex6} in general. However, we can calculate the formula \eqref{review:eq:vertex6} in terms of oscillators in $\alpha'$ expansion for the general configuration of open string fields. In \cite{9909027,9909095} the following properties were shown in such a way. The boundary state \eqref{review:eq:boundary-general-gauge} satisfies the nonlinear boundary condition
\begin{equation}
 \left(\partial_\tau \hat{X}_a(\sigma)-F_{ab}(\hat{X})\hat{X}^b(\sigma)\right)|Dp[A]\rangle=0 \tag{\ref{review:eq:closedboundary4}}.
\end{equation}
where $F_{ab}(\hat{X}(\sigma))=\partial_a A_b(\hat{X}(\sigma))-\partial_b A_a(\hat{X}(\sigma))$. Formally $|Dp[A]\rangle$ is BRST invariant for an arbitrary configuration of the gauge field $A_a$ on the D-brane. However, it contains divergences which cannot be regularized by the zeta function regularization. Such singularities vanish when the gauge field satisfies the equation of motions. Therefore the boundary state $|Dp[A]\rangle$ is well-defined when the gauge field $A_a$ in on-shell. The couplings to massless closed string, namely
\begin{equation*}
   \langle h_{\mu\nu} | Dp[A]\rangle ,\quad \langle b_{\mu\nu} | Dp[A]\rangle , \quad \langle \phi | Dp[A]\rangle,
\end{equation*}
gives a part of DBI action linear in $h_{\mu\nu},b_{\mu\nu},\phi$ at the leading order in $\alpha'$. We follow this study in order to develop a non-Abelian extension of boundary states in section \ref{paper:section:3}.

\cite{0211232,0312260} takes an another approach. The boundary state is defined with prescription to remove singularities. Thus the constructed boundary state includes no singularity. However, the boundary state is BRST invariant only when the equation of motion for the open string field is satisfied. This approach is essentially same as that shown above except the difference of prescription for divergences.

Incorporation of scalar fields $\phi^i$ on a D-brane was studied in \cite{9909095}. The boundary state \eqref{review:eq:boundary-general-gauge} is modified to 
\begin{equation}
\begin{split}
   |Dp[A,\phi]\rangle &= \exp\left\{
i\int\!\!d\sigma \left(A_a(\hat{X}(\sigma))\partial_\sigma \hat{X}^a(\sigma)+\phi_i(\hat{X}(\sigma))\partial_\tau \hat{X}^i(\sigma)\right)\right\}|Dp\rangle\\
&= \exp\left\{
i\int\!\!d\sigma \left(A_a(\hat{X}(\sigma))\partial_\sigma \hat{X}^a(\sigma)+ \Xi_i(\hat{X}(\sigma))\hat{\Pi}^i(\sigma)\right)\right\}|Dp\rangle.
\end{split}\label{review:eq:boundary-general-scalar}
\end{equation}
Here we have defined
\begin{equation*}
    \Xi^i = 2\pi\alpha' \phi^i, \quad \hat{\Pi}^i=\frac{1}{2\pi\alpha'}\partial_\tau \hat{X}^i(\sigma).  
\end{equation*}
This boundary state satisfies the following boundary conditions identical to those imposed by the boundary action: 
\begin{equation}
\begin{split}
      (\partial_\tau \hat{X}_a+\partial_\tau \hat{X}_i\partial_a\Xi^i(X) -F_{ab}(\hat{X})\partial_\sigma\hat{X}^b)\big|_{\tau=0}|Dp[A]\rangle&=0 \\
      (\hat{X}^i-\Xi^i(\hat{X}))\big|_\mathrm{\tau=0}|Dp[A]\rangle&=0.
\end{split}
\end{equation}
We note that in case of a constant scalar field $\Xi^i(\hat{X}(\sigma))=\xi^i$, the zero mode part of the boundary state becomes
\begin{equation}
  e^{i\int\!\!d\sigma\xi_i\hat{\Pi}^i(\sigma)}\delta(\hat{x}^i)|0\rangle = e^{i\xi_i\hat{p}^i}|x^i=0\rangle = |x^i=\xi^i\rangle. \label{review:eq:position-shift}
\end{equation}
This indicates that the scalar field $\phi=2\pi\alpha'\xi$ shifts the position of the D$p$-brane. Therefore we can say that a scalar field on a D-brane represents deformation of shape in the transverse directions. Inclusion of the scalar field can be understood through the T-duality \cite{9909095}. 
\\ \\

We summarize this subsection, in particular what \eqref{review:eq:disk-amplitude1} and \eqref{review:eq:boundary-general-scalar} indicate. We define a boundary state $|Dp[A,\phi]\rangle$ of a D$p$-brane with gauge and scalar fields $A^a,\phi^i$ so that
\begin{equation}
      \langle \Psi(k)| Dp[A,\phi]\rangle = \langle V_\Psi(k) e^{iS_\mathrm{boundary}[A,\phi]} \rangle_\mathrm{disk}\label{review:eq:disk-amplitude2}
\end{equation}
where the boundary action $S_\mathrm{boundary}$ is
\begin{equation}
   S_\mathrm{boundary} = \int\!\!d\tau\ \left(A_a(\hat{X})\partial_\tau \hat{X}^a(\tau)+\phi_i(\hat{X}) \partial_\sigma \hat{X}^i(\tau)\right).\label{review:eq:boundaryaction-open}
\end{equation}
It is supported from the considerations above that $|Dp[A,\phi]\rangle$ is given by
\begin{equation}
\begin{split}
  |B[S]\rangle&=\int\!\!\mathcal{D}X^\mu(\sigma)e^{iS\left[X^\mu(\sigma)\right]}|X^\mu(\sigma)\rangle\\
&=e^{iS\left[\hat{X}^\mu(\sigma)\right]}|Dp\rangle
\end{split}\label{review:eq:boundarystate}
\end{equation}
where the boundary action $S$ is
\begin{equation}
   S[\hat{X}^\mu(\sigma)] = \int\!\!d\sigma\ \left(A_a(\hat{X})\partial_\sigma \hat{X}^a(\sigma)+\phi_i(\hat{X}) \partial_\tau \hat{X}^i(\sigma)\right).\label{review:eq:boundaryaction-closed}
\end{equation}
The open and closed string descriptions relate through the modular transformation \eqref{review:eq:modular-transformation}, namely $\sigma \leftrightarrow \tau$.

%The disk amplitude was calculated for emission of a massless closed string from single D-brane in \cite{}

\subsection{Boundary state of multiple D-branes with non-Abelian field}\label{paper:section:review-4}
In this subsection, we consider how to extend the boundary state of a single D-brane to that of multiple D-branes. In case of multiple D-branes, open strings stretched between different D-branes cause difficulty. A worldsheet which represents closed string emission via such open string have many separated boundaries on different D-branes in general. These worldsheet does not allow a simple relation between open and closed string channels. Therefore it becomes difficult to reveal how effects of open strings are encoded into a boundary state. However, the result in the previous subsection that the Wilson loop factor incorporates the contribution of open strings into a boundary state suggest a naive extension by using the non-Abelian Wilson loop factor. In this subsection we consider what this way of extension implicates. In section \ref{paper:section:3}, we will show that a formula for the boundary state of multiple D-branes obtained in this way gives the couplings to massless closed strings reviewed in section \ref{paper:section:2}.

In $N$ D-brane system, an open string fields has a Chan-Paton factor representing D-branes on which the open string end. In other words, open string fields are $N\times N$ matrices. Therefore open string fields on multiple D-branes are non-Abelian. In the low energy limit, $(p+1)$-dimensional  $U(N)$ supersymmetric Yang-Mills theory is realized on $N$ D$p$-branes. Recall that a scalar field on a D-brane means displacement of position in the transverse directions. An open string vertex operator also has a Chang-Patron factor, hence is matrix-valued. A diagonal component means an open string with both endpoints on the same D-brane. A off-diagonal component represents an open string stretched between two different D-branes. The presence of off-diagonal components, or open strings stretched between different D-branes, causes several difficulties. One of the difficulties is concerning noncommutivity and fuzziness. In the case of multiple D-branes, matrix-valued scalar fields $\mathbf{X}^i$ are noncommutative in presence of the off-diagonal component. This indicates that we cannot determine definite positions of each D-brane. In other words, the shape of D-branes become fuzzy.  Therefore it is difficult to determine boundary conditions that a boundary state of multiple D-branes with the general configuration of open string fields satisfies.
\EPSFIGURE{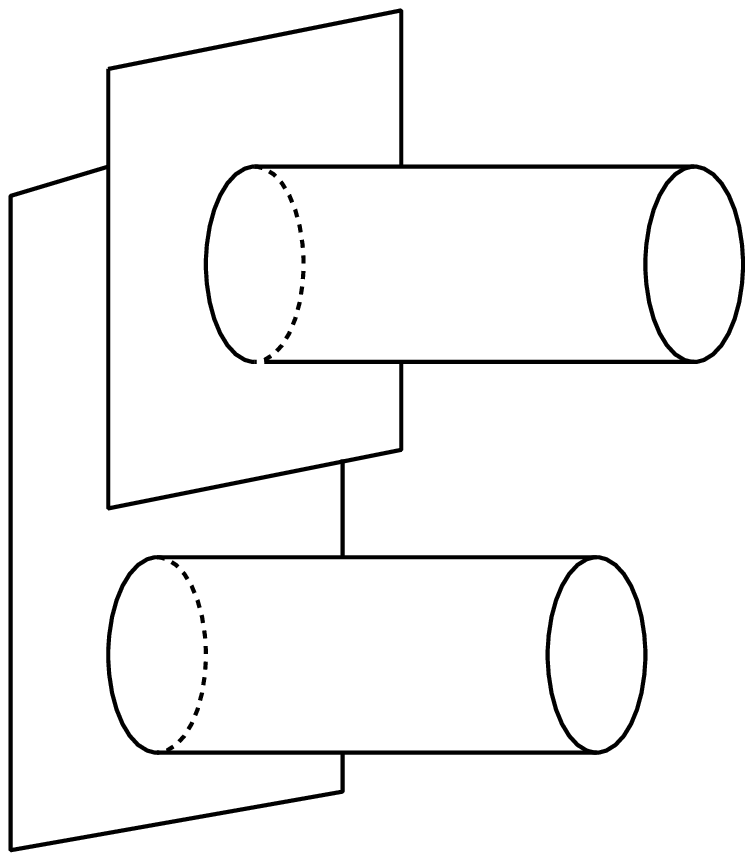,width=0.7\textwidth}{Superposition of closed string emitted from each D-brane\label{review:fig:review-disk-diagonal}}

The other difficulty caused by open strings stretched between different D-branes concerns shape of worldsheets. We consider a worldsheet which indicates a process in which an closed string is created via open strings. In absence of open strings stretched between different D-branes, closed string emission from multiple D-branes is superposition of that from each D-brane as illustrated in figure \ref{review:fig:review-disk-diagonal}. Therefore a boundary state $|B\rangle_\mathrm{diaonal}$ of $N$ D-branes is the sum of each D-brane:
\begin{equation*}
  |B\rangle_\mathrm{diagonal} = \sum_{m=1}^N |Dp\rangle_m
\end{equation*} 
where $|Dp\rangle_m$ is the boundary state of $i$-th D-brane. Consider $N$ D-branes located at $x^i=\xi^i_m (m=1,\cdots, N)$ without excitation of open strings. In this case, the sum boundary states of each D-brane is given by
\begin{equation*}
\begin{split}
  |B\rangle_\mathrm{diagonal}&=\sum_{m=1}^N |Dp[\xi^i_m]\rangle\\
                          |Dp[\xi^i_m]\rangle&= \frac{T_p}{2}\left\{-\frac{1}{n}\sum_{n>0}\alpha_{-m}^\mu S_{\mu\nu}\tilde{\alpha}^\nu_{-n}\right\}\delta(\hat{x}^i-\xi^i_m)|0\rangle,\quad S_{\mu\nu}=(\eta_{ab},-\delta_{ij}). 
\end{split}
\end{equation*}
By using the operator $e^{\hat{p}_i\xi^i_m}$ 
which moves the position of D-brane to $x^i=\xi_m^i$ operating on the boundary state $|Dp\rangle$ of a D$p$-brane at $x^i=0$ as shown in \eqref{review:eq:position-shift}, this boundary state can be rewritten as
\begin{equation}
\begin{split}
  |B\rangle_\mathrm{diagonal}&=\sum_{m=1}^Ne^{i\hat{p}_i\xi^i_m}|Dp\rangle = \sum_{m=1}^Ne^{i\int\!\!d\sigma\hat{\Pi}_i(\sigma)\xi^i}|Dp\rangle = \mathrm{tr}e^{i\int\!\!d\sigma\hat{\Pi}_i(\sigma)\mathbf{X}^i}|Dp\rangle \\
\mathbf{X}^i&=\mathrm{diag}(\xi_1^i,\cdots,\xi^i_N).
\end{split}\label{review:eq:nonabelian-diagonal}
\end{equation}

\DOUBLEFIGURE{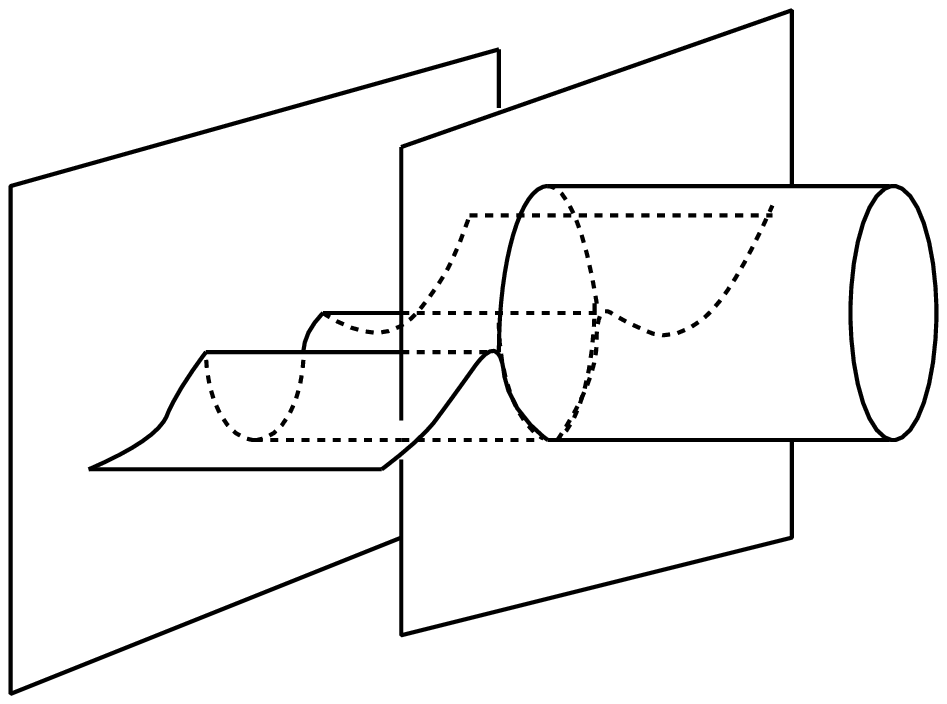,width=0.45\textwidth}{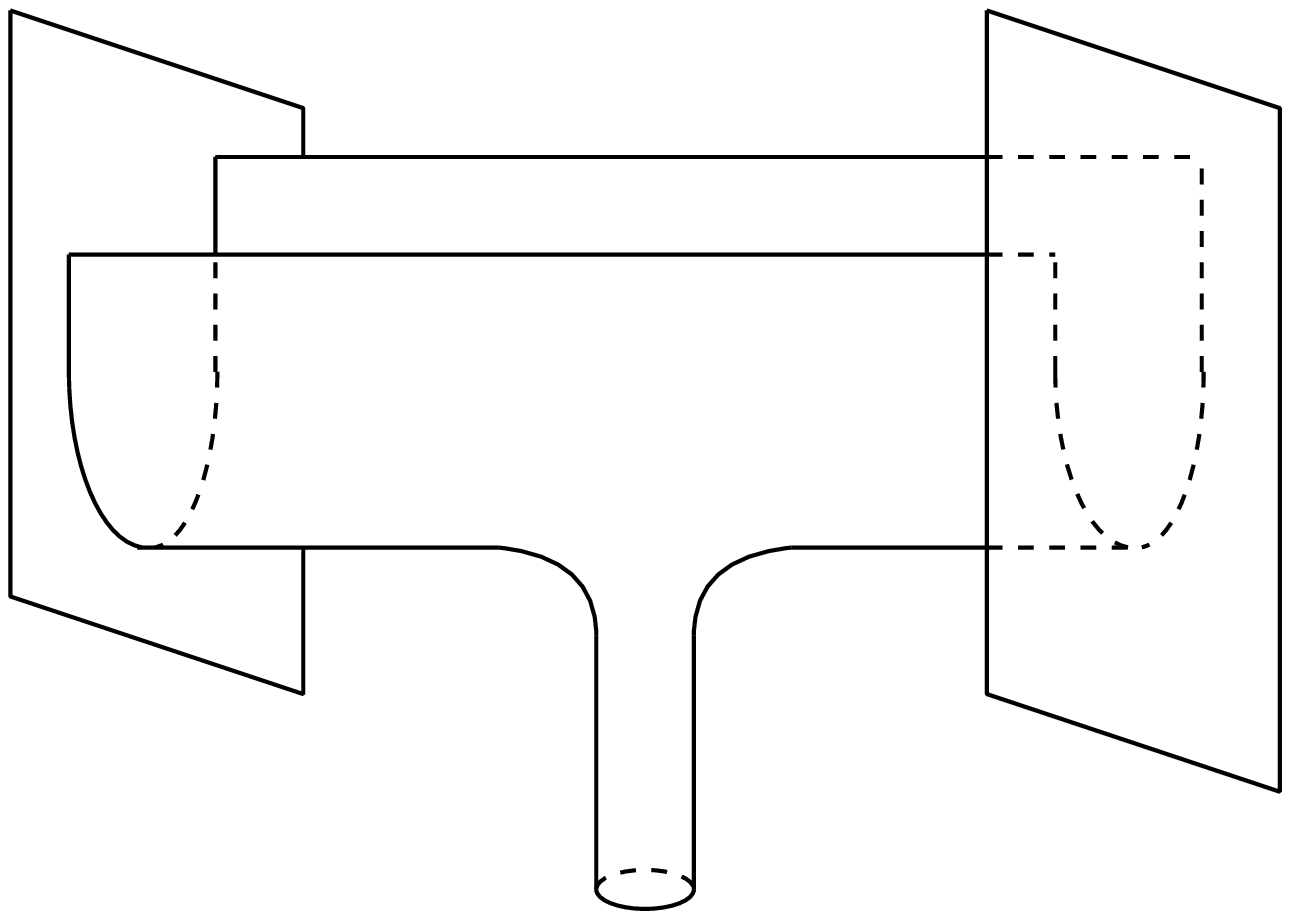,width=0.45\textwidth}{Emission of one closed string via two open strings\label{review:fig:review-disk5-nonabelian}}{Emission of one closed string via two open strings\label{review:fig:review-disk-nonabelian}} 
 
In presence of open strings stretched between different D-branes, the situation becomes complex. We consider two coincident D-branes. The scalar field  $\mathbf{X}^i$ in this system is $2\times 2$ matrix. The vertex operator of $\mathbf{X}^i$ is also $2\times 2$ matrix. For example $(\mathbf{X}^i)_{12}$ corresponds to an open string with one endpoint on the first D-brane and the other endpoint on the second D-brane. A worldsheet which represents closed string emission via open strings stretched between two D-branes is illustrated in figure \ref{review:fig:review-disk5-nonabelian} and \ref{review:fig:review-disk-nonabelian}. 
Although in these two figures D-branes are illustrated as they are distant each other for simplicity of figures, we consider they are coincident. 
The boundaries of these worldsheets do not draw a circle on a single D-brane. Therefore it is not clear whether there is a simple relation between open and closed channel as the modular transformation \eqref{review:eq:modular-transformation} for a cylindrical worldsheet in case of a single D-brane. This indicates that the studies in subsection \ref{paper:section:review-2} and \cite{NPB288525,NPB29383,NPB308221} are not applicable strictly in case of multiple D-branes with open string fields.

A cylinder is replaced by a closed string vertex, and an open string between D-branes is replaced by an off-diagonal component of an open string vertex. Therefore the worldsheets in figure \ref{review:fig:review-disk5-nonabelian} and figure \ref{review:fig:review-disk-nonabelian} correspond to that in figure \ref{review:fig:review-disk2-nonabelian}. Furthermore this worldsheet corresponds to figure \ref{review:fig:review-disk3-nonabelian} on the disk. In the disk amplitude shown in figure \ref{review:fig:review-disk3-nonabelian}, we need to take trace on Chan-Paton factors, because the boundary of worldsheet cannot jump the D-brane on which they exists except at the open string vertices. In addition, we need to determine the ordering of open string fields because they are noncommutative. By requiring the invariance under the gauge transformation for the open string fields, we can see it is natural to adopt the path ordering prescription. From these considerations, we define the boundary state of multiple D-branes $|B[\mathrm{A},\mathbf{X}^i]\rangle$by a closed string state which satisfies
\begin{equation}
   \langle \Psi | B[\mathrm{A}^a,\mathbf{X}^i]\rangle = \langle V_\Psi \mathrm{trP}e^{iS_\mathrm{boundary}[\mathbf{A}^a(\hat{X}(\tau)),\mathbf{X}^i(\hat{X}(\tau))]} \rangle_\mathrm{disk} \label{review:eq:disk-nonabelian}
\end{equation}
for an arbitrary closed string state $|\Psi\rangle$, where $\mathrm{P}$ is the path ordering with respect to $\tau$, and $\mathrm{tr}$ is trace on Chan-Paton factors. The $S_\mathrm{boundary}$ is given by
\begin{equation}
   S_\mathrm{boundary}[\mathbf{A}^a(\hat{X}(\tau)),\mathbf{X}^i(\hat{X}(\tau))] = \int\!\!d\tau\ \left(\mathbf{A}_a(\hat{X})\partial_\tau \hat{X}^a(\tau)+\mathbf{X}_i(\hat{X}(\tau)) \hat{\Pi}^i(\tau)\right).\label{review:eq:boundaryaction-open-nonabelian}
\end{equation}
which are non-Abelian extension of \eqref{review:eq:boundaryaction-open}. Note that $S_\mathrm{boundary}$ is matrix-valued, and becomes a boundary action after we take trace with a prescription for ordering. This amplitude was calculated in \cite{0103124} for multiple D0-branes. The results coincide with that obtained from Matrix theory potential \cite{9711078,9712185,9904095} and non-Abelian DBI action \cite{9910053} as shown in section \ref{paper:section:2}.  

\psfrag{(A^a)_mn,(X^i)_mn}{$(\mathbf{A}^a)_{mn},(\mathbf{X}^i)_{mn}$}
\psfrag{(A^a)_nm,(X^i)_nm}{$(\mathbf{A}^a)_{nm},(\mathbf{X}^i)_{nm}$}
\DOUBLEFIGURE{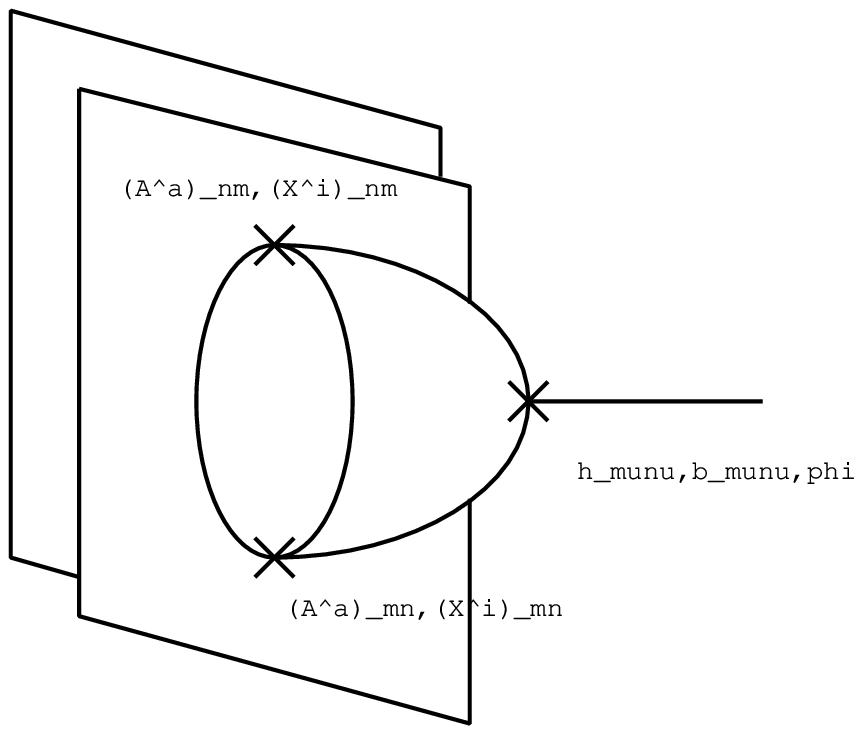,width=0.6\textwidth}{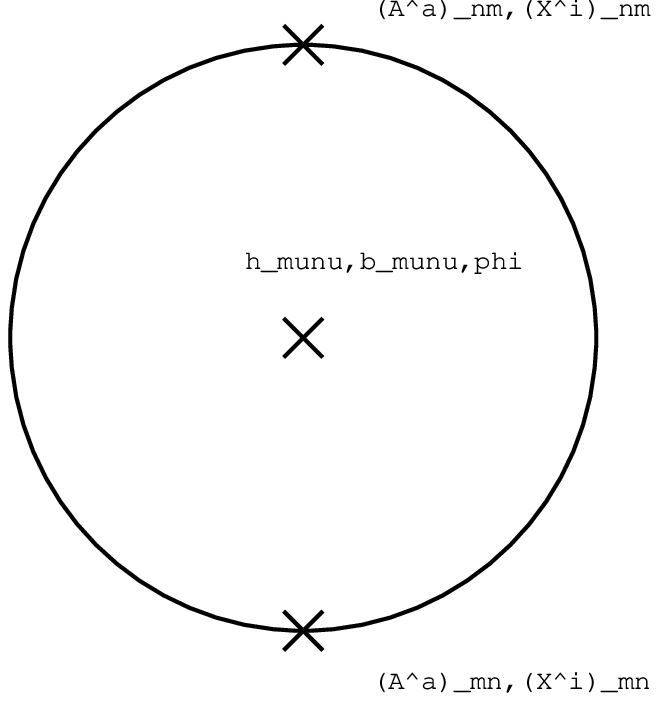,width=0.4\textwidth}{Worldsheet with one closed and two open string vertices on D-brane\label{review:fig:review-disk2-nonabelian}}{Disk with one closed and two open string vertices\label{review:fig:review-disk3-nonabelian}}

Recall that in case of a single D-brane, the effect of open string fields on the boundary state is accounted for by including the Wilson loop factor in subsection \ref{paper:section:review-3}. This implies the boundary state of multiple D-branes with open string fields is obtained simply simply by the non-Abelian extension of the Wilson loop factor as discussed in \cite{NPB308221}. The form of the boundary state $|B\rangle_\mathrm{diagonal}$ in \eqref{review:eq:nonabelian-diagonal} also suggests that the same form is valid even in presence of off-diagonal components of matrix-valued open string fields with a prescription for ordering. In order to impose the gauge invariance on the boundary state, we choose the path ordering prescription. Considering them all, we suppose the non-Abelian boundary state which satisfies the condition \eqref{review:eq:disk-nonabelian} is given by
\begin{equation}
     |B[\mathbf{A}^a,\mathbf{X}^i]\rangle = \mathrm{trP}\exp\left\{iS[\mathbf{A}^a(\hat{X}(\sigma)),\mathbf{X}^i(\hat{X}(\sigma))]\right\}|Dp\rangle \label{review:eq:nonabelian-boundarystate}
\end{equation}
where the path ordering $P$ with respect to $\sigma$ is required by invariance under gauge transformation of open string fields. $S$ is given by
\begin{equation}
   S[\mathbf{A}^a(\hat{X}(\sigma)),\mathbf{X}^i(\hat{X}(\sigma))] = \int\!\!d\sigma\ \left(\mathbf{A}_a(\hat{X})\partial_\sigma \hat{X}^a(\sigma)+\mathbf{X}_i(\hat{X}(\sigma)) \hat{\Pi}^i(\sigma)\right).\label{review:eq:boundaryaction-closed-nonabelian}
\end{equation}
In section \ref{paper:section:3}, we show that the boundary state \eqref{review:eq:nonabelian-boundarystate} satisfied the condition \eqref{review:eq:disk-nonabelian} at least for couplings to massless closed strings  $\Psi=h_{\mu\nu},b_{\mu\nu},\phi$ up to order ${\alpha'}^2$.


\begin{thebibliography}{36}
%\cite{NPB288525}
\bibitem{NPB288525}
C.~G.~.~Callan, C.~Lovelace, C.~R.~Nappi and S.~A.~Yost,
``String Loop Corrections To Beta Functions,''
Nucl.\ Phys.\ B {\bf 288}, 525 (1987).
%%CITATION = NUPHA,B288,525;%%

%\cite{NPB29383}
\bibitem{NPB29383}
C.~G.~.~Callan, C.~Lovelace, C.~R.~Nappi and S.~A.~Yost,
``ADDING HOLES AND CROSSCAPS TO THE SUPERSTRING,''
Nucl.\ Phys.\ B {\bf 293}, 83 (1987).
%%CITATION = NUPHA,B293,83;%%

%\cite{NPB308221}
\bibitem{NPB308221}
C.~G.~.~Callan, C.~Lovelace, C.~R.~Nappi and S.~A.~Yost,
``Loop Corrections To Superstring Equations Of Motion,''
Nucl.\ Phys.\ B {\bf 308}, 221 (1988).
%%CITATION = NUPHA,B308,221;%%

%\cite{9909027}
\bibitem{9909027}
K.~Hashimoto,
``Corrections to D-brane action and generalized boundary state,''
Phys.\ Rev.\ D {\bf 61}, 106002 (2000)
arXiv:hep-th/9909027].
%%CITATION = HEP-TH 9909027;%%

%\cite{9909095}
\bibitem{9909095}
K.~Hashimoto,
``Generalized supersymmetric boundary state,''
JHEP {\bf 0004}, 023 (2000)
[arXiv:hep-th/9909095].
%%CITATION = HEP-TH 9909095;%%

%\cite{0103124}
\bibitem{0103124}
Y.~Okawa and H.~Ooguri,
``Energy-momentum tensors in matrix theory and in noncommutative gauge
theories,''
arXiv:hep-th/0103124.
%%CITATION = HEP-TH 0103124;%%

%\cite{9711078}
\bibitem{9711078}
D.~Kabat and W.~I.~Taylor,
``Spherical membranes in matrix theory,''
Adv.\ Theor.\ Math.\ Phys.\  {\bf 2}, 181 (1998)
[arXiv:hep-th/9711078].
%%CITATION = HEP-TH 9711078;%%

%\cite{9712185}
\bibitem{9712185}
D.~Kabat and W.~I.~Taylor,
``Linearized supergravity from matrix theory,''
Phys.\ Lett.\ B {\bf 426}, 297 (1998)
[arXiv:hep-th/9712185].
%%CITATION = HEP-TH 9712185;%%

%\cite{9904095}
\bibitem{9904095}
W.~I.~Taylor and M.~Van Raamsdonk,
``Multiple D0-branes in weakly curved backgrounds,''
Nucl.\ Phys.\ B {\bf 558}, 63 (1999)
[arXiv:hep-th/9904095].
%%CITATION = HEP-TH 9904095;%%

%\cite{9910053}
\bibitem{9910053}
R.~C.~Myers,
``Dielectric-branes,''
JHEP {\bf 9912}, 022 (1999)
[arXiv:hep-th/9910053].
%%CITATION = HEP-TH 9910053;%%

%\cite{9701190}
\bibitem{9701190}
M.~Billo, P.~Di Vecchia and D.~Cangemi,
``Boundary states for moving D-branes,''
Phys.\ Lett.\ B {\bf 400}, 63 (1997)
[arXiv:hep-th/9701190].
%%CITATION = HEP-TH 9701190;%%

%\cite{9906214}
\bibitem{9906214}
P.~Di Vecchia, M.~Frau, A.~Lerda and A.~Liccardo,
``(F,Dp) bound states from the boundary state,''
Nucl.\ Phys.\ B {\bf 565}, 397 (2000)
[arXiv:hep-th/9906214].
%%CITATION = HEP-TH 9906214;%%

%\cite{9701125}
\bibitem{9701125}
A.~A.~Tseytlin,
``On non-abelian generalisation of the Born-Infeld action in string theory,''
Nucl.\ Phys.\ B {\bf 501}, 41 (1997)
[arXiv:hep-th/9701125].
%%CITATION = HEP-TH 9701125;%%

%\cite{9703217}
\bibitem{9703217}
A.~Hashimoto and W.~I.~Taylor,
``Fluctuation spectra of tilted and intersecting D-branes from the
Born-Infeld action,''
Nucl.\ Phys.\ B {\bf 503}, 193 (1997)
[arXiv:hep-th/9703217].
%%CITATION = HEP-TH 9703217;%%

%\cite{0002180}
\bibitem{0002180}
F.~Denef, A.~Sevrin and J.~Troost,
``Non-Abelian Born-Infeld versus string theory,''
Nucl.\ Phys.\ B {\bf 581}, 135 (2000)
[arXiv:hep-th/0002180].
%%CITATION = HEP-TH 0002180;%%

%\cite{NPB431131}
\bibitem{NPB431131}
M.~B.~Green and P.~Wai,
``The Insertion of boundaries in world sheets,''
Nucl.\ Phys.\ B {\bf 431}, 131 (1994).
%%CITATION = NUPHA,B431,131;%%

%\cite{9912161}
\bibitem{9912161}
P.~Di Vecchia and A.~Liccardo,
``D branes in string theory. I,''
NATO Adv.\ Study Inst.\ Ser.\ C.\ Math.\ Phys.\ Sci.\  {\bf 556}, 1 (2000)
[arXiv:hep-th/9912161].
%%CITATION = HEP-TH 9912161;%%

%\cite{9912275}
\bibitem{9912275}
P.~Di Vecchia and A.~Liccardo,
``D-branes in string theory. II,''
arXiv:hep-th/9912275.
%%CITATION = HEP-TH 9912275;%%

%\cite{9707068}
\bibitem{9707068}
P.~Di Vecchia, M.~Frau, I.~Pesando, S.~Sciuto, A.~Lerda and R.~Russo,
``Classical p-branes from boundary state,''
Nucl.\ Phys.\  B {\bf 507}, 259 (1997)
[arXiv:hep-th/9707068].
%%CITATION = NUPHA,B507,259;%%

%\cite{9608024}
\bibitem{9608024}
M.~R.~Douglas, D.~Kabat, P.~Pouliot and S.~H.~Shenker,
``D-branes and short distances in string theory,''
Nucl.\ Phys.\ B {\bf 485}, 85 (1997)
[arXiv:hep-th/9608024].
%%CITATION = HEP-TH 9608024;%%

%\cite{9903165}
\bibitem{9903165}
J.~Polchinski,
``M-theory and the light cone,''
Prog.\ Theor.\ Phys.\ Suppl.\  {\bf 134}, 158 (1999)
[arXiv:hep-th/9903165].
%%CITATION = HEP-TH 9903165;%%

%\cite{9804163}
\bibitem{9804163}
N.~Ishibashi,
``p-branes from (p-2)-branes in the bosonic string theory,''
Nucl.\ Phys.\ B {\bf 539}, 107 (1999)
[arXiv:hep-th/9804163].
%%CITATION = HEP-TH 9804163;%%

%\cite{0501086}
\bibitem{0501086}
I.~Ellwood,
``Relating branes and matrices,''
JHEP {\bf 0508}, 078 (2005)
[arXiv:hep-th/0501086].
%%CITATION = HEP-TH 0501086;%%

%\cite{9704125}
\bibitem{9704125}
K.~Hashimoto and H.~Hata,
``D-brane and gauge invariance in closed string field theory,''
Phys.\ Rev.\ D {\bf 56}, 5179 (1997)
[arXiv:hep-th/9704125].
%%CITATION = HEP-TH 9704125;%%

%\cite{0312260}
\bibitem{0312260}
T.~Maeda, T.~Nakatsu and T.~Oonishi,
``Non-linear field equation from boundary state formalism,''
arXiv:hep-th/0312260.
%%CITATION = HEP-TH 0312260;%%

%\cite{9908142}
\bibitem{9908142}
N.~Seiberg and E.~Witten,
``String theory and noncommutative geometry,''
JHEP {\bf 9909}, 032 (1999)
[arXiv:hep-th/9908142].
%%CITATION = HEP-TH 9908142;%%

%\cite{0505184}
\bibitem{0505184}
S.~Terashima,
``Noncommutativity and tachyon condensation,''
JHEP {\bf 0510}, 043 (2005)
[arXiv:hep-th/0505184].
%%CITATION = HEP-TH 0505184;%%

%\cite{0309074}
\bibitem{0309074}
T.~Asakawa, S.~Kobayashi and S.~Matsuura,
``Closed string field theory with dynamical D-brane,''
JHEP {\bf 0310}, 023 (2003)
[arXiv:hep-th/0309074].
%%CITATION = HEP-TH 0309074;%%

%\cite{9705111}
\bibitem{9705111}
Y.~Kazama,
``Scattering of quantized Dirichlet particles,''
Nucl.\ Phys.\ B {\bf 504}, 285 (1997)
[arXiv:hep-th/9705111].
%%CITATION = HEP-TH 9705111;%%

%\cite{0603152}
\bibitem{0603152}
Y.~Baba, N.~Ishibashi and K.~Murakami,
``D-branes and closed string field theory,''
JHEP {\bf 0605}, 029 (2006)
[arXiv:hep-th/0603152].
%%CITATION = HEP-TH 0603152;%%

%\cite{0306189}
\bibitem{0306189}
I.~Kishimoto, Y.~Matsuo and E.~Watanabe,
``Boundary states as exact solutions of (vacuum) closed string field theory,''
Phys.\ Rev.\ D {\bf 68}, 126006 (2003)
[arXiv:hep-th/0306189].
%%CITATION = HEP-TH 0306189;%%

%\cite{NPB50222}
\bibitem{NPB50222}
E.~Cremmer and J.~Scherk,
``Factorization Of The Pomeron Sector And Currents In The Dual Resonance Model,''
Nucl.\ Phys.\ B {\bf 50}, 222 (1972).
%%CITATION = NUPHA,B50,222;%%

%\cite{NPB57490}
\bibitem{NPB57490}
L.~Clavelli and J.~A.~Shapiro,
``Pomeron Factorization In General Dual Models,''
Nucl.\ Phys.\ B {\bf 57}, 490 (1973).
%%CITATION = NUPHA,B57,490;%%

%\cite{NPB94221}
\bibitem{NPB94221}
M.~Ademollo, A.~D'Adda, R.~D'Auria, F.~Gliozzi, E.~Napolitano, S.~Sciuto and P.~Di Vecchia,
``Soft Dilations And Scale Renormalization In Dual Theories,''
Nucl.\ Phys.\ B {\bf 94}, 221 (1975).
%%CITATION = NUPHA,B94,221;%%

%\cite{9604091}
\bibitem{9604091}
M.~B.~Green and M.~Gutperle,
``Light-cone supersymmetry and D-branes,''
Nucl.\ Phys.\ B {\bf 476}, 484 (1996)
[arXiv:hep-th/9604091].
%%CITATION = HEP-TH 9604091;%%

%\cite{0211232}
\bibitem{0211232}
K.~Murakami and T.~Nakatsu,
``Open Wilson lines as states of closed string,''
Prog.\ Theor.\ Phys.\  {\bf 110}, 285 (2003)
[arXiv:hep-th/0211232].
%%CITATION = HEP-TH 0211232;%%


\end{thebibliography}
\end{document}